\documentclass[12pt]{article}

\usepackage{setspace, graphicx, caption, subcaption, float, relsize, rotating, color, verbatim, multirow}
\usepackage[round, comma, longnamesfirst]{natbib}
\usepackage{graphicx}    % needed for including graphics e.g. EPS, PS
\usepackage[colorlinks=true, urlcolor=blue,linkcolor=blue, citecolor=magenta]{hyperref}

 \renewcommand{\thefootnote}{\fnsymbol{footnote}}

\usepackage{empheq}

\usepackage{enumerate} 

\usepackage{dsfont}
\usepackage{mathtools}
\usepackage{bm}

\usepackage{stmaryrd}
\usepackage{amssymb} %to use \triangleq
\usepackage{amsmath}
\usepackage{amsfonts}
\usepackage{amsthm} %to use proof
\usepackage{mathrsfs}
\usepackage{bigints} %big integral
\usepackage{enumerate} %to use enumerate
\usepackage{authblk} %to use affil
\usepackage{color} %to use color
\usepackage{sectsty}

\usepackage{amsfonts}
\usepackage{graphicx}
\usepackage{times}
\usepackage{bm}
\usepackage{natbib}

\usepackage[plain,noend]{algorithm2e}

\newtheorem{thm}{Theorem}[section]

\newtheorem{prop}[thm]{Proposition}

\newtheorem{theorem}[thm]{Theorem}
\newtheorem{assumption}[thm]{Assumption}

\newcommand{\nc}{\newcommand}

\nc{\dps}{\displaystyle}
\nc{\tr}{\text{tr}}

\title{\Large  A Unified Framework for Efficient Estimation of General
	Treatment Models \thanks{
We thank Xiaohong Chen, Shigeyuki Hamori, and Toru Kitagawa for helpful comments. The third
author is grateful for financial supports from Japan Center for Economic
Research.
}} 
\author[$^\star$]{Chunrong Ai \thanks{E-mail:  chunrong.ai@warrington.ufl.edu}}
\author[$^\star$]{Oliver Linton \thanks{E-mail:  obl20@cam.ac.uk}}
\author[$^\dagger$]{Kaiji Motegi \thanks{E-mail: motegi@econ.kobe-u.ac.jp }}
\author[$^\ddagger$]{Zheng Zhang \thanks{E-mail: zhengzhang@ruc.edu.cn}}
\affil[$\star$]{Department of Economics, University of Florida}
\affil[$\star$]{Faculty of Economics, University of Cambridge}
\affil[$\dagger$]{Graduate School of Economics, Kobe University}
\affil[$\ddagger$]{Institute of Statistics \& Big Data, Renmin University of China}

\numberwithin{equation}{section}
 
\allowdisplaybreaks 

\begin{document}
 
\setcounter{section}{0}

\maketitle

\baselineskip0.6cm

\begin{abstract}
This paper presents a weighted optimization framework that unifies the
binary, multi-valued, continuous, as well as mixture of discrete and
continuous treatment, under the unconfounded treatment assignment. With a
general loss function, the framework includes the average, quantile and
asymmetric least squares causal effect of treatment as special cases. For
this general framework, we first derive the semiparametric efficiency bound
for the causal effect of treatment, extending the existing bound results to
a wider class of models. We then propose a generalized optimization
estimation for the causal effect with weights estimated by solving an
expanding set of equations. Under some sufficient conditions, we establish
consistency and asymptotic normality of the proposed estimator of the causal
effect and show that the estimator attains our semiparametric efficiency
bound, thereby extending the existing literature on efficient estimation of
causal effect to a wider class of applications. Finally, we discuss
etimation of some causal effect functionals such as the treatment effect
curve and the average outcome. To evaluate the finite sample performance of
the proposed procedure, we conduct a small scale simulation study and find
that the proposed estimation has practical value. To illustrate the
applicability of the procedure, we revisit the literature on campaign
advertise and campaign contributions. Unlike the existing procedures which
produce mixed results, we find no evidence of campaign advertise on campaign
contribution.
\end{abstract}

\renewcommand\thefootnote{\arabic{footnote}}

\section{Introduction}
Modeling and estimating the causal effect of treatment have received
considerable attention from both econometrics and statistics literature (see 
\cite{Hirano03}, \cite{imbens2004nonparametric}, \cite{heckman2005structural}%
, \cite{angrist2008mostly}, \cite{imbens2009recent}, \cite{fan2010sharp}, 
\cite{chernozhukov2013inference}, \cite{sloczynski2017general}, and \cite%
{rothe2017robust} for examples). Most existing studies focus on the binary
treatment where an individual either receives the treatment or does not,
ignoring the treatment intensity. In many applications, however, the
treatment intensity is a part of the treatment, and its causal effect is
also of great interest to decision makers. For example, in evaluating how
financial incentives affect health care providers, the causal effect may
depend on not only the introduction of incentive but also the level of
incentive. For example, in evaluating the effect of corporate bond purchase
schemes on market quality, the causal effect may depend not just on whether
the bond is selected into the scheme but how much of it is purchased, \cite{linton2018}. Similarly, in studying how taxes affect addictive substance
usages, the causal effect may depend on the imposition of tax as well as the
tax rate.  In recognition of the importance of the treatment intensity, the
binary treatment literature has been extended to the multi-valued treatment
(e.g., \cite{imbens2000role} and \cite{cattaneo2010efficient}) and
continuous treatment (e.g., {\cite{hirano2004propensity}, \cite{imai2004causal}, \cite{florens2008identification}}, \cite{Fong_Hazlett_Imai_2017} and \cite{yiu2018covariate}).

The parameter of primary interest in this literature is the average causal
effect of treatment, defined as the difference in response to two levels of
treatment by the same individual, averaged over a set of individuals. The
identification and estimation difficulty is that each individual only
receives one level of treatment. To overcome the difficulty, researchers
impose the\textit{\ unconfounded treatment assignment} condition, which
allows them to find statistical matches for each observed individual from
all other treatment levels. Under this condition, the average causal effect
is estimated in the binary treatment by the difference of the weighted
average responses with the propensity scores as weights (e.g., \cite%
{rosenbaum1983central}). The efficiency bound of the average causal effect
in this model is derived by \cite{robins1994estimation} and \cite%
{hahn1998role}, and efficient estimation is proposed by \cite%
{robins1994estimation}, \cite{hahn1998role}, \cite{Hirano03}, \cite%
{bang2005doubly}, \cite{qin2007empirical}, \cite{cao2009improving}, \cite%
{tan2010bounded}, \cite{graham2012inverse}, \cite{vansteelandt2010model},
and \cite{chan2016globally}. Of particular interest in this literature is
the study by \cite{Hirano03} which shows that the weighted average
difference estimator attains the semiparametric efficiency bound if the
weights are estimated by the empirical likelihood estimation. In the
multi-valued treatment, \cite{imbens2000role} generalizes the propensity
score, and \cite{cattaneo2010efficient} derives the efficiency bound and
proposes an estimator that attains the efficiency bound. In the continuous
treatment, \cite{hirano2004propensity} and \cite{imai2004causal}
parameterize the generalized propensity score function and propose a
consistent estimation of the average causal effect. Their estimators are not
efficient and could be biased if the generalized propensity score function
is misspecified. \cite{florens2008identification} use a control function
approach to identify the average causal effect in the continuous treatment
and propose a consistent estimation. It is unclear if their estimation is
efficient.

In addition to the average causal effect\ of treatment (ATE), it is also
important to investigate the distributional impact of treatment. For
instance, a decision maker may be interested in the causal effect of a
treatment on the outcome dispersion or on the lower tail of the outcome
distribution. \cite{Doksum1974Empirical} and \cite{lehmann1974} introduce
the quantile causal effect of treatment (QTE). \cite{Firpo2007Efficient}
computes the efficiency bound and proposes an efficient estimation of QTE
for the binary treatment. For additional studies on QTE, we refer to \cite%
{Abadie1998Instrumental}, \cite{Chernozhukov2010An}, \cite%
{Markus2013Unconditional}, \cite{angrist2008mostly} and \cite%
{Donald2014Estimation}.

To the best of our knowledge, we are unaware of existence of any studies
that compute the efficiency bound and propose efficient estimation of the
causal effect in the continuous or mixture of discrete and continuous
treatment under general loss function that permits ATE and QTE. The few
studies that come close are \cite{Fong_Hazlett_Imai_2017} and \cite%
{yiu2018covariate}. \cite{Fong_Hazlett_Imai_2017} propose an estimation of
the average causal effect of continuous treatment but do not establish
consistency of their estimation. In fact, their simulation results indicate
their estimation could be seriously biased. \cite{yiu2018covariate} study
the average causal effect of both discrete and continuous treatment by
parameterizing propensity scores. Their estimation could be biased if their
parameterization is incorrect.

The main objective of this paper is to present a weighted optimization
framework that unifies the binary, multi-valued, continuous as well as
mixture of discrete and continuous treatment and allows for general loss
function, where the weights are called the \emph{stabilized weights} by \cite%
{robins2000marginal} and are defined as the ratio of the marginal
probability distribution of the treatment status over the conditional
probability distribution of the treatment status given covariates. For this
general framework, we first apply the approach of \cite%
{Bickel_Klaassen_Ritov_Wellner_1993} to compute the efficiency bound of the
causal effect of treatment, extending the semiparametric efficiency bound
results of \cite{hahn1998role}, \cite{cattaneo2010efficient}, and \cite%
{Firpo2007Efficient} from the binary treatment to a variety of treatments
and to the general loss function. Our bound reveals that the weighted
optimization with known {\color{black} stabilized} weights does not produce
efficient estimation since it fails to account for the information
restricting the {\color{black}stabilized} weights. Similar observation is
also noted by \cite{Hirano03} in the binary treatment. Here we show that
their observation holds true for much wider class of treatment models. We
exploit the information that the {\color{black} stabilized} weights satisfy
some (finite but expanding number of) equations by estimating the {%
	\color{black}stabilized} weights from those equations and then estimate the
causal effect by the generalized optimization with the true {\ \color{black}%
	stabilized} weights replaced by the estimated weights. Under some sufficient
conditions, we show that our proposed estimator is consistent and
asymptotically normally distributed and, more importantly, it attains our
semiparametric efficiency bound, thereby extending the efficient estimation
work of \cite{Hirano03}, \cite{cattaneo2010efficient} and \cite%
{Firpo2007Efficient} to a much wider class of treatment models. 

The paper is organized as follows. Section \ref{sec:framework} sets up the
basic framework, Section \ref{sec:efficiency_theorem} computes the
semiparametric efficiency bound of the causal effect of treatment, Section %
\ref{sec:point estimation} presents a generalized optimization estimator,
Section \ref{sec:asymptotics} establishes large sample properties of the
proposed estimator, Section \ref{sec:variance} presents a consistent
covariance matrix, {\color{black} Section \ref{sec:extension} discusses some
	extensions}, Section \ref{sec:simulation} reports on a simulation study,   Section \ref{sec:application} presents an application,
followed by some concluding remarks in Section \ref{sec:conclusion}. All
technical proofs and extra simulation results are relegated to the
supplemental material.

\section{Basic framework and notation}\label{sec:framework}

Let $T$ denote the observed treatment status variable with support $\mathcal{%
	T}\subset \mathbb{R}$, where $\mathcal{T}$ \ is either a discrete or a
continuous or a mixture of discrete and continuous subset, and has a
marginal probability distribution function $F_{T}(t)$. Let $Y^{\ast }(t)$
denote the potential response when treatment $T=t$ is assigned. Let $L(\cdot
)$ denote a known convex loss function whose derivative, denoted by $
L^{\prime }(\cdot )$, exists almost everywhere. For the leading part of the
paper, we shall maintain that there exists a parametric causal effect
function $g(t;\beta )$ with the unknown value ${\boldsymbol{\beta }}_{0}\in 
\mathbb{R}^{p}$ (with $p\in \mathbb{N}$) uniquely solving: 
\begin{equation}
{\boldsymbol{\beta }}_{0}=\arg \min_{\boldsymbol{\beta} }\int_{\mathcal{T}}\mathbb{E}%
\left[ L\left( Y^{\ast }(t)-g(t;\boldsymbol{\beta})\right) \right] dF_{T}(t).
\label{ident}
\end{equation}
The parameterization of the causal effect is restrictive. Some extensions to
the unspecified causal effect function shall be discussed later in the paper.

The generality of model \eqref{ident} permits many important models and much
more. For example, it includes the average causal effect of binary treatment
studied in \cite{hahn1998role} and \cite{Hirano03} (i.e., $\mathcal{T=}%
\{0,1\}$, $L(v)=v^{2}$ and $g(t;\boldsymbol{\ \beta }_{0})=\beta _{0}+\beta
_{1}t$), the quantile causal effect of binary treatment studied in \cite%
{Firpo2007Efficient} (i.e., $\mathcal{T=}\{0,1\}$, $L(v)=v(\tau -I(v\leq 0))$  is an almost everywhere differentiable function
with $\tau \in (0,1)$ and $g(t;\boldsymbol{\beta }_{0})=t\beta
_{1}+(1-t)\beta _{0}$), the average causal effect of multi-valued treatment
studied in \cite{cattaneo2010efficient} (i.e., $\mathcal{T=}\{0,1,\ldots
,J\} $ for some $J\in \mathbb{N}$, $L(v)=v^{2}$ and $g(t;\boldsymbol{\beta }%
_{0})=\sum_{j=0}^{J}\beta _{j}I(t=j)$), and the average causal effect of
continuous treatment studied in \cite{hirano2004propensity} (i.e., $%
L(v)=v^{2}$ and $\mathbb{E}[Y^{\ast }(t)]=g(t;\boldsymbol{\beta }_{0})$). It
also includes the quantile causal effect of multi-valued (i.e., $L(v)=v(\tau
-I(v\leq 0))$ with $\tau \in (0,1)$ and $g(t;\boldsymbol{\beta }%
_{0})=\sum_{j=0}^{J}\beta _{j}I(t=j)$) and continuous treatment (i.e., $%
L(v)=v(\tau -I(v\leq 0))$ and $\inf \left\{ q:\mathbb{P}(Y^{\ast }(t)\geq
q)\leq \tau \right\} =g(t;\boldsymbol{\beta }_{0})$). The latter has never
been studied in the existing literature. Moreover, with $L(v)=v^{2}\left%
\vert \tau -I(v\leq 0)\right\vert $, it covers asymmetric least squares
estimation of the causal effect of (binary, multi-valued, continuous,
mixture of discrete and continuous) treatment. The asymmetric least squares
regression received attention from some noted econometricians (see \cite%
{newey1987asymmetric}) but zero attention in the causal effect literature.

The problem with \eqref{ident} is that the potential outcome $Y^{\ast }(t)$
is not observed for all $t$. Let $Y:=Y^{\ast }(T)$ denote the observed
response. One may attempt to solve: 
\begin{equation*}
\min_{\boldsymbol{\beta } }\mathbb{E}[L(Y-g(T;\boldsymbol{\beta }))].
\end{equation*}
However, if there exists a selection into treatment, the true value ${%
	\boldsymbol{\beta }}_{0}$ does not solve the above minimization problem.
Indeed, in this case, the observed response and treatment assignment data
alone cannot identify ${\boldsymbol{\beta }}_{0}$. To address this
identification issue, most studies in the literature impose a selection on
observable condition (e.g., \cite{Hirano03}, \cite{imai2004causal} and \cite%
{Fong_Hazlett_Imai_2017}). Specifically, let $\boldsymbol{X}$ denote a
vector of covariates. The following condition shall be maintained throughout
the paper.

\begin{assumption}[\emph{Unconfounded Treatment Assignment}]
	\label{as:TYindep} For all $t\in \mathcal{T}$, given $\boldsymbol{X}$ , $T$
	is independent of $Y^{\ast }(t)$, i.e., $Y^{\ast }(t)\perp T|\boldsymbol{X,}$
	for all $t\in \mathcal{T}$.
\end{assumption}

Let $F_{T|X}$ denote the conditional probability distribution of $T$ given
the observed covariates $\boldsymbol{X}$ and let $dF_{T|X}$\ denote the
probability measure. In the literature, $dF_{T|X}$ is called the \emph{\
	generalized propensity score} \citep{hirano2004propensity,
	imai2004causal}. Suppose that $dF_{T|X}(T|\boldsymbol{X})$ is positive
everywhere and denote 
\begin{equation*}
\pi _{0}(T,\boldsymbol{X}):=\frac{dF_{T}(T)}{dF_{T|X}(T|\boldsymbol{\ X})}.
\end{equation*}%
The function $\pi _{0}(T,\boldsymbol{X})$ is called the \emph{{\color{black}
		stabilized} weight} in \cite{robins2000marginal}. Under Assumption \ref%
{as:TYindep}, we obtain 
\begin{equation}
\mathbb{E}[\pi _{0}(T,\boldsymbol{X})L(Y-g(T;\boldsymbol{\beta }))]=\int \mathbb{E}\left[
L(Y^{\ast }(t)-g(t;\boldsymbol{\beta } ))\right] dF_{T}(t)  \label{eq:CIA}
\end{equation}%
(see Appendix \ref{sec:proof_identification}), and hence the true value $%
\boldsymbol{\beta }_{0}$ solves the weighted optimization problem: 
\begin{equation}
\boldsymbol{\beta }_{0}=\arg \min_{\boldsymbol{\beta } }\mathbb{E}[\pi _{0}(T,\boldsymbol{X%
})L(Y-g(T;\boldsymbol{\beta }))].  \label{newmodel}
\end{equation}%
This result is very insightful. It tells us that the selection bias in the 
\textit{unconfounded treatment assignment} can be corrected through
covariate-balancing. More importantly, it says that the true value $%
\boldsymbol{\beta }_{0}$ can be identified from the observed data. The
weighted optimization (\ref{newmodel}) provides a unified framework for
estimating the causal effect of a variety of treatments, including binary,
multi-level, continuous and mixture of discrete and continuous treatment,
and under general loss function. The goal of this paper is to compute the
semiparametric efficiency bound and present an efficient estimation of ${%
	\boldsymbol{\beta }}_{0}$ under this general framework.

\section{Efficiency bound}\label{sec:efficiency_theorem}
We begin by applying the approach of \cite%
{Bickel_Klaassen_Ritov_Wellner_1993} to compute the semiparametric
efficiency bound of the parameter ${\boldsymbol{\beta }}_{0}$ defined by (%
\ref{ident}) under Assumption \ref{as:TYindep}. This gives the least
possible variance achievable by a regular estimator in the semiparametric
model. The result is presented in the following theorem.

\begin{theorem}
	\label{efficiency theorem} Suppose that $g(T,\boldsymbol{\beta })$ is twice
	differentiable with respect to $\boldsymbol{\beta }$ in the parameter space $\Theta\subset \mathbb{R}^p$, with $m(T;\boldsymbol{ \beta }_{0}):=\nabla _{\beta }g(T;\boldsymbol{\beta}_{0})$, and $
	\mathbb{E}\left[ L^{\prime }(Y-g(T;\boldsymbol{\beta }))|Y,\boldsymbol{X}\right] $ is
	differentiable with respect to $\boldsymbol{\beta }\in \Theta$. Denote 
	\begin{align*}
	& \varepsilon (T,\boldsymbol{X};\boldsymbol{\beta }_{0}):=\mathbb{E}%
	[L^{\prime }(Y-g(T;\boldsymbol{\beta }_{0}))|T,\boldsymbol{X}]\ , \\[2mm]
	& H_{0}:=-\nabla _{\beta }\mathbb{E}\left[ \pi _{0}(T,\boldsymbol{X}%
	)L^{\prime }(Y-g(T;\boldsymbol{\beta }))m(T;\boldsymbol{\beta })\right] \Big|%
	_{\boldsymbol{\beta }=\boldsymbol{\beta }_{0}}\ , \\[2mm]
	& \psi (Y,T,\boldsymbol{X};\boldsymbol{\beta }_{0}):=\pi _{0}(T,\boldsymbol{ X})m(T;\boldsymbol{\beta }_{0})L^{\prime }(Y-g(T;\boldsymbol{\beta }%
	_{0}))-\pi _{0}(T,\boldsymbol{X})m(T;\boldsymbol{\beta }_{0})\varepsilon (T,%
	\boldsymbol{X};\boldsymbol{\beta }_{0}) \\
	& \qquad +\mathbb{E}\left[ \varepsilon (T,\boldsymbol{X};\boldsymbol{\beta }%
	_{0})\pi _{0}(T,\boldsymbol{X})m(T;\boldsymbol{\beta }_{0})|\boldsymbol{X}%
	\right] +\mathbb{E}\left[ \varepsilon (T,\boldsymbol{X};\boldsymbol{\beta }%
	_{0})\pi _{0}(T,\boldsymbol{X})m(T;\boldsymbol{\beta }_{0})|T\right] .
	\end{align*}%
	Suppose that ${H_{0}}$ is nonsingular and $\mathbb{E}\left[ \psi (Y,T,%
	\boldsymbol{X};\boldsymbol{\beta }_{0})\psi (Y,T,\boldsymbol{X};\boldsymbol{%
		\beta }_{0})^{\top }\right] $ exists and is finite. Under Assumption \ref%
	{as:TYindep} and model (\ref{ident}), the efficient influence function of $%
	\boldsymbol{\beta }_{0}$ is given by 
	\begin{equation*}
	S_{eff}(Y,T,\boldsymbol{X};\boldsymbol{\beta }_{0})=H_{0}^{-1}\psi (Y,T,%
	\boldsymbol{X};\boldsymbol{\beta }_{0}).
	\end{equation*}%
	Consequently, the efficient variance bound of $\boldsymbol{\beta }_{0}$ is 
	\begin{equation*}
	V_{eff}=\mathbb{E}\left[ S_{eff}(Y,T,\boldsymbol{X};\boldsymbol{\beta }%
	_{0})S_{eff}(Y,T,\boldsymbol{X};\boldsymbol{\beta }_{0})^{\top }\right] .
	\end{equation*}
\end{theorem}

The proof of Theorem \ref{efficiency theorem} is given in Section 3.1 of
supplemental material. It is worth noting that our bound $V_{eff}$ is equal
to the bound of the average causal effect derived by \cite{hahn1998role} for
the binary treatment (see Section 3.2 of supplemental material), equal to
the bound of the average causal effect derived by \cite%
{cattaneo2010efficient} for the multi-valued treatment (see Section 3.3 of
supplemental material), and equal to the bound of the quantile causal effect
derived by \cite{Firpo2007Efficient} for the binary treatment (see Section
3.4 of supplemental material). Moreover, our bound applies to a much wider
class of models, including quantile causal effect of multi-valued,
continuous and mixture of discrete and continuous treatment as well as the
asymmetric least squares estimation of the causal effect of all kinds of
treatments.

It is also worth noting that, if the {\color{black}stabilized weights are
	known and $g(t;\boldsymbol{\beta }_{0})$ is correctly specified, one can
	estimate }$\boldsymbol{\beta }_{0}${\ by solving the sample analogue of the
	weighted optimization }(\ref{newmodel}). The asymptotic variance of the
estimator is 
\begin{equation*}
V_{ineff}=\mathbb{E}\left[ S_{ineff}(Y,T,\boldsymbol{X};\boldsymbol{\beta }%
_{0})S_{ineff}(Y,T,\boldsymbol{X};\boldsymbol{\beta }_{0})^{\top }\right] ,
\end{equation*}%
with
	\begin{equation*}
	S_{ineff}(Y,T,\boldsymbol{X};\boldsymbol{\beta }_{0})=H_{0}^{-1}\cdot \pi
	_{0}(T,\boldsymbol{X})m(T;\boldsymbol{\beta }_{0})L^{\prime }\left\{ Y-g(T;
	\boldsymbol{\beta }_{0})\right\} .
	\end{equation*}
 It is easy to show that $V_{ineff}>V_{eff}$ (see Proposition \ref{prop:var_pi_known} of Appendix \ref{sec:pi_known}), implying that the
weighted optimization estimator is not efficient. This follows because the
weighted optimization does not account for the restriction on the {\ %
	\color{black}stabilized} weight $\pi _{0}(t,\boldsymbol{x})$: 
\begin{equation}
\mathbb{E}\left[ \pi _{0}(T,\boldsymbol{X})u(T)v(\boldsymbol{X})\right] =%
\mathbb{E}[u(T)]\cdot \mathbb{E}[v(\boldsymbol{X})]\   \label{moment1}
\end{equation}%
holds for any suitable functions $u(t)$ and $v(\boldsymbol{x})$.
Incorporating restriction (\ref{moment1}) into the estimation of the causal
effect can improve efficiency. Similar observation is also noted by \cite%
{Hirano03} in the binary treatment. Exactly how to incorporate restriction (%
\ref{moment1}) into the estimation is the subject of the next section.

\section{Efficient estimation} \label{sec:point estimation}
One way to incorporate (\ref{moment1}) into the
estimation is to estimate the {\color{black}stabilized} weights from (\ref%
{moment1}) and then implement (\ref{newmodel}) with the estimated weights.
But before so doing, we must verify that (\ref{moment1}) uniquely identifies 
$\pi _{0}(T,\boldsymbol{X})$. After some manipulations, it is
straightforward to show that 
\begin{equation*}
\mathbb{E}\left[ \pi (T,\boldsymbol{X})u(T)v(\boldsymbol{X})\right] =\mathbb{%
	\ \ E}[u(T)]\cdot \mathbb{E}[v(\mathbf{X})]
\end{equation*}%
holds for any suitable functions $u(t)$ and $v(\boldsymbol{x})$ if and only
if ${\small \pi (T,\mathbf{X})=\pi }_{0}{\small (T,\mathbf{X})}$. Therefore,
condition (\ref{moment1}) identifies the {\color{black}stabilized} weights.
The challenge now is that (\ref{moment1}) implies infinite number of
equations. With a finite sample of observations, it is impossible to solve
infinite number of equations. To overcome this difficulty, we approximate
the (infinite dimensional) function space with the (finite dimensional)
sieve space. Specifically, let $u_{K_{1}}(T)=(u_{K_{1},1}(T),\ldots $ $%
,u_{K_{1},K_{1}}(T))^{\top }$ and $v_{K_{2}}(\boldsymbol{X})=\left(
v_{K_{2},1}(\boldsymbol{X}),\ldots ,v_{K_{2},K_{2}}(\boldsymbol{X})\right)
^{\top }$ denote the known basis functions with dimensions $K_{1}\in \mathbb{%
	\ N}$ and $K_{2}\in \mathbb{N}$ respectively. The functions $u_{K_{1}}(t)$
and $v_{K_{2}}(\boldsymbol{x})$ are called the \emph{\ approximation sieves}
that can approximate any suitable functions $u(t)$ and $v(\boldsymbol{\ x})$
arbitrarily well (see \cite{chen2007large} and Appendix \ref{sec:sieve} for
more discussion on sieve approximation). Since the sieve approximating space
is also a subspace of the functional space, $\pi _{0}(T,\boldsymbol{X})$
satisfies 
\begin{equation}
\mathbb{E}\left[ \pi _{0}(T,\boldsymbol{X})u_{K_{1}}(T)v_{K_{2}}(\boldsymbol{X})^{\top }\right] =\mathbb{E}[u_{K_{1}}(T)]\cdot \mathbb{E}[v_{K_{2}}(\boldsymbol{X})]^{\top }.  \label{sievemoment}
\end{equation}%
Unfortunately, it is not the only solution. Indeed, for any monotonic
increasing and globally concave function $\rho (v)$, with 
\begin{equation}
\Lambda _{K_{1}\times K_{2}}^{\ast }=\arg \max_{\Lambda \in \mathbb{R}%
	^{K_{1}\times K_{2}}}\mathbb{E}\left[ \rho (u_{K_{1}}(T)^{\top }\Lambda
v_{K_{2}}(\boldsymbol{X}))\right] -\mathbb{E}[u_{K_{1}}(T)]^{\top }\Lambda 
\mathbb{E}[v_{K_{2}}(\boldsymbol{X})],  \label{def:Lambda*}
\end{equation}%
$\pi _{K}^{\ast }(T,\mathbf{X})=\rho ^{\prime }\left( u_{K_{1}}(T)^{\top
}\Lambda _{K_{1}\times K_{2}}^{\ast }v_{K_{2}}(\boldsymbol{X})\right) $ also
solves (\ref{sievemoment}), {\color{black}where $\rho ^{\prime }(v)$ denotes
	the first derivative.} Let $\pi _{K}(T,\mathbf{X})=\rho ^{\prime }\left(
u_{K_{1}}(T)^{\top }\Lambda _{K_{1}\times K_{2}}v_{K_{2}}(\boldsymbol{X}%
)\right) $ denote the {\color{black}best} approximation of $\pi _{0}(T,%
\boldsymbol{X})$ {\color{black}under the }$\sup ${\ norm $L_{\infty }$ and
	suppose that }$\Vert \pi _{K}(T,\mathbf{X})-\pi _{0}(T,\mathbf{X})\Vert
_{\infty }=O(K^{-\alpha })$ {for some $\alpha >0$}.{\ Then, 
	\begin{align*}
	& \Vert \pi _{K}^{\ast }(T,\mathbf{X})-\pi _{0}(T,\mathbf{X})\Vert _{\infty }
	\\
	=& O\left( \max \left\{ \Vert \pi _{K}^{\ast }(T,\mathbf{X})-\pi _{K}(T,%
	\mathbf{X})\Vert _{\infty },\Vert \pi _{K}(T,\mathbf{X})-\pi _{0}(T,\mathbf{X%
	})\Vert _{\infty }\right\} \right) \\
	=& \max \left\{ O\left( \zeta (K)K^{-\alpha }\right) ,O(K^{-\alpha
	})\right\} =O\left( \zeta (K)K^{-\alpha }\right) ,
	\end{align*}%
	where $\zeta (K)=\zeta _{1}(K_{1})\zeta _{2}(K_{2})$, $\zeta
	_{1}(K_{1})=\sup_{t\in \mathcal{T}}\Vert u_{K_{1}}(t)\Vert $, $\zeta
	_{2}(K_{2})=\sup_{\boldsymbol{x}\in \mathcal{X}}\Vert v_{K_{2}}(\boldsymbol{x%
	})\Vert $ (see Lemma 4.1 in supplemental material}). \newline

Let $\{T_{i},\boldsymbol{X}_{i},Y_{i}\}_{i=1}^{N}$ denote an independently
and identically distributed sample of observations drawn from the joint
distribution of $(T,\mathbf{X},Y)$. {\color{black}We propose to estimate the
	stabilized weights $\pi _{i}=\pi _{0}(T_{i},\mathbf{X}_{i})$ by solving the
	entropy maximization problem: 
	\begin{equation}
	\left\{ 
	\begin{array}{cc}
	& \max \left\{ -\sum_{i=1}^{N}\pi _{i}\log \pi _{i}\right\} \\[2mm] 
	& \text{subject to}\ \frac{1}{N}\sum_{i=1}^{N}\pi
	_{i}u_{K_{1}}(T_{i})v_{K_{2}}(\boldsymbol{X}_{i})^{\top }=\left( \frac{1}{N}%
	\sum_{i=1}^{N}u_{K_{1}}(T_{i})\right) \left( \frac{1}{N}%
	\sum_{j=1}^{N}v_{K_{2}}(\boldsymbol{X}_{j})^{\top }\right) .%
	\end{array}%
	\right.  \label{E:cm1}
	\end{equation}%
	The primal problem \eqref{E:cm1} is difficult to compute. We instead
	consider its dual problem that can be solved by numerically efficient and
	stable algorithms. Specifically, let $\rho (v):=-e^{-v-1}$ for any $v\in 
	\mathbb{R}$, \cite{tseng1991relaxation} showed that the dual solution is
	given by: 
	\begin{equation*}
	\hat{\pi}_{K}(T_{i},\boldsymbol{X}_{i}):=\rho ^{\prime }\left(
	u_{K_{1}}(T_{i})^{\top }\hat{\Lambda}_{K_{1}\times K_{2}}v_{K_{2}}(%
	\boldsymbol{X}_{i})\right) ,
	\end{equation*}%
	where $\hat{\Lambda}_{K_{1}\times K_{2}}$ is the maximizer of the strictly
	concave function $\hat{G}_{K_{1}\times K_{2}}$ defined by{\footnotesize \ 
		\begin{equation}
		\widehat{\Lambda }_{K_{1}\times K_{2}}=\arg \max_{\Lambda }\hat{G}%
		_{K_{1}\times K_{2}}(\Lambda ):=\frac{1}{N}\sum_{i=1}^{N}\rho \left(
		u_{K_{1}}(T_{i})^{\top }\Lambda v_{K_{2}}(\boldsymbol{X}_{i})\right) -\left( 
		\frac{1}{N}\sum_{i=1}^{N}u_{K_{1}}(T_{i})\right) ^{\top }\Lambda \left( 
		\frac{1}{N}\sum_{j=1}^{N}v_{K_{2}}(\boldsymbol{X}_{j})\right) .
		\label{def:G^hat}
		\end{equation}%
	} The duality between \eqref{E:cm1} and \eqref{def:G^hat} is shown in
	Appendix \ref{sec:dual_appendix}. }By Corollary 4.3 in Section 4 of
supplemental material, we have 
\begin{equation*}
\left\Vert \hat{\pi}_{K}(T,\mathbf{X})-\pi _{K}^{\ast }(T,\mathbf{X}%
)\right\Vert _{\infty }=O_{p}\left( \zeta (K)\sqrt{\frac{K}{N}}\right) .
\end{equation*}%
Having estimated the weights, we now estimate $\beta _{0}$ by applying the generalized optimization: 
\begin{equation}
\hat{\boldsymbol{\beta }}=\arg \min_{\boldsymbol{\beta} }\sum_{i=1}^{N}\hat{\pi}
_{K}(T_{i},\boldsymbol{X}_{i})L\left( Y_{i}-g(T_{i};\boldsymbol{\beta} )\right) .
\label{def:estimator}
\end{equation}
\textbf{Remarks}:

\begin{enumerate}
	\item Alternatively, one can estimate the {\color{black}stabilized} weights
	by estimating the generalized propensity score function as well as the
	marginal distribution of the treatment variable nonparametrically (e.g.,
	kernel estimation). But these alternatively estimated weights do not satisfy
	empirical moment in \eqref{E:cm1} and may not result in efficient estimation
	of the causal effect.
	
	\item Notice that $\rho (v)=-e^{-v-1}$ satisfies the invariance property
	(i.e., $-\rho ^{\prime \prime }(v)=\rho ^{\prime }(v)$). It turns out that
	this invariance property is critical for establishing consistency of the
	generalized optimization estimator. Any other choice of $\rho (\cdot )$ that
	does not have the invariance property may result in biased causal effect
	estimate.
\end{enumerate}

\section{Large sample properties}

\label{sec:asymptotics} To establish the large sample properties of the
generalized optimization estimator, we first show that the estimated weight
function $\hat{\pi}_{K}(t,\boldsymbol{x})$ is consistent and compute its
convergence rates under both {$L_{\infty }$} norm and the $L_{2}$ norm. The
following conditions shall be imposed.

\begin{assumption}
	\label{as:suppX} The support $\mathcal{X}$ of $\boldsymbol{X}$ is a compact
	subset of $\mathbb{R}^{r}$. The support $\mathcal{T}$ of the treatment
	variable $T$ is a compact subset of $\mathbb{R}$.
\end{assumption}

\begin{assumption}
	\label{as:pi0} There exist two positive constants $\eta_1$ and $\eta_2$ such
	that 
	\begin{equation*}
	0 < \eta_1 \leq \pi_0(t,\boldsymbol{x}) \leq \eta_2 <\infty \ ,\ \forall (t, 
	\boldsymbol{x}) \in \mathcal{T}\times\mathcal{X}\ .
	\end{equation*}
\end{assumption}

\begin{assumption}
	\label{as:smooth_pi} There exist $\Lambda _{K_{1}\times K_{2}}\in \mathbb{R}
	^{K_{1}\times K_{2}}$ and a positive constant $\alpha >0$ such that 
	\begin{equation*}
	\sup_{(t,\boldsymbol{x})\in \mathcal{T}\times \mathcal{X}}\left\vert (\rho
	^{\prime -1}\left( \pi _{0}(t,\boldsymbol{x})\right) -u_{K_{1}}(t)^{\top
	}\Lambda _{K_{1}\times K_{2}}v_{K_{2}}(\boldsymbol{x})\right\vert
	=O(K^{-\alpha }).
	\end{equation*}
\end{assumption}

\begin{assumption}
	\label{as:u&v} For every $K_1$ and $K_2$, the smallest eigenvalues of $%
	\mathbb{E}\left[u_{K_1}(T)u_{K_1}(T)^\top\right]$ and $\mathbb{E}\left[%
	v_{K_2}(\boldsymbol{X})v_{K_2}(\boldsymbol{X})^\top\right]$ are bounded away
	from zero uniformly in $K_1$ and $K_2$.
\end{assumption}

\begin{assumption}
	\label{as:K&N_consistency}There are two sequences of constants $\zeta
	_{1}(K_{1})$ and $\zeta _{2}(K_{2})$ satisfying\newline
	$\sup_{t\in \mathcal{T}}\Vert u_{K_{1}}(t)\Vert \leq \zeta _{1}(K_{1})$ and $%
	\sup_{\boldsymbol{x}\in \mathcal{X}}\Vert v_{K_{2}}(\boldsymbol{x})\Vert
	\leq \zeta _{2}(K_{2})$, $K=K_{1}(N)K_{2}(N)$ and $\zeta (K)=\zeta
	_{1}(K_{1})\zeta _{2}(K_{2})$, such that $\zeta (K)K^{-\alpha }\rightarrow 0$
	and $\zeta (K)\sqrt{K/N}\rightarrow 0$ as $N\rightarrow \infty $.
\end{assumption}

Assumption \ref{as:suppX} restricts both the covariates and treatment level
to be bounded. This condition is restrictive but convenient for computing
the convergence rate under $L_{\infty}$ norm. It is commonly imposed in the
nonparametric regression literature. This condition can be relaxed, however,
if we restrict the tail distribution of $(\boldsymbol{ X},T)$. Assumption \ref{as:pi0}
restricts the weight function to be bounded and bounded away from zero. Given Assumption \ref{as:suppX}, this condition is equivalent to $dF_{T|X}(T|%
\boldsymbol{X})$ being bounded away from zero, meaning that each type of
individuals (denoted by $\boldsymbol{ X}$) always have a sufficient portion participating
in each level of treatment. This restriction is important for our analysis
since each individual participates only in one level of treatment and this
condition allows us to construct her statistical counterparts from all other
treatments. Although Assumption \ref{as:pi0} is useful in causal analysis and  establishing the convergence rates,  it is not  essential  and could be relaxed by allowing $\eta_1$ (resp. $\eta_2$) depend on $N$ and go to zero (resp. infinity) slowly, as $N\to \infty$. Notice that $u_{K_{1}}(t)^{\top }\Lambda v_{K_{2}}(\boldsymbol{x}%
)$ is a linear sieve approximation to any suitable function of $(\boldsymbol{ X},T)$.
Assumption \ref{as:smooth_pi} requires the sieve approximation error of $%
\rho ^{\prime -1}\left( \pi _{0}(t,\boldsymbol{x})\right) $ to shrink at a
polynomial rate. This condition is satisfied for a variety of sieve basis
functions. For example, if both $\boldsymbol{ X}$ and $T$ are discrete, then the
approximation error is zero for sufficient large $K$ and in this case
Assumption \ref{as:smooth_pi} is satisfied with $\alpha =+\infty $. If some
components of $(\boldsymbol{ X},T)$ are continuous, the polynomial rate depends positively
on the smoothness of $\rho ^{\prime -1}\left( \pi _{0}(t,\boldsymbol{x}
)\right) $ in continuous components and negatively on the number of the
continuous components. We will show that the convergence rate of the
estimated weight function (and consequently the rate of the generalized
optimization estimator) is bounded by this polynomial rate. Assumption \ref
	{as:u&v} essentially requires the sieve basis functions to be orthogonal,   see Appendix \ref{sec:sieve} for a thorough discussion.
	Similar condition is common in the sieve regression literature (see \cite{andrews1991asymptotic} and \cite{Newey97}). If the approximation error is
	nonzero, Assumption \ref{as:K&N_consistency} requires it to shrink to zero
	at an appropriate rate as sample size increases. 

Under these conditions, we are able to establish the following theorem:

\begin{theorem}
	\label{rate_pi}Suppose that Assumptions \ref{as:suppX}-\ref%
	{as:K&N_consistency} hold. Then, we obtain the following:%
	\begin{equation*}
	\sup_{(t,\boldsymbol{x})\in \mathcal{T}\times \mathcal{X}}|\hat{\pi}_{K}(t,%
	\boldsymbol{x})-\pi _{0}(t,\boldsymbol{x})|=O_{p}\left( K^{-\alpha }\zeta
	(K)+\zeta (K)\sqrt{\frac{K}{N}}\right) \ ,
	\end{equation*}
	\begin{equation*}
	\int_{\mathcal{T}\times \mathcal{X}}|\hat{\pi}_{K}(t,\boldsymbol{x})-\pi
	_{0}(t,\boldsymbol{x})|^{2}dF_{T,X}(t,\boldsymbol{x})=O_{p}\left(
	K^{-2\alpha }+\frac{K}{N}\right) ,
	\end{equation*}
	\begin{equation*}
	\frac{1}{N}\sum_{i=1}^{N}|\hat{\pi}_{K}(T_{i},\boldsymbol{X}_{i})-\pi
	_{0}(T_{i},\boldsymbol{X}_{i})|^{2}=O_{p}\left( K^{-2\alpha }+\frac{K}{N}%
	\right) .
	\end{equation*}
\end{theorem}

The proof of Theorem \ref{rate_pi} immediately follows from applying Lemma
4.1 and Corollary 4.3 in Section 4 of supplemental material. \newline

The following additional conditions are needed to establish the consistency
of the proposed estimator $\hat{\boldsymbol{\beta }}$.

\begin{assumption}
	\label{as:Theta} The parameter space $\Theta\subset\mathbb{R} ^{p}$ is a
	compact set and the true parameter $\boldsymbol{\beta}_0$ is in the interior
	of $\Theta$ , where $p\in\mathbb{N}$.
\end{assumption}

\begin{assumption}
	\label{as:solution} There exists a unique solution $\boldsymbol{\beta }_{0}$
	for the optimization problem 
	\begin{equation*}
	\min_{\boldsymbol{\beta }\in \Theta }\int_{\mathcal{T}}\mathbb{E}\left[
	L(Y^{\ast }(t)-g(t;\boldsymbol{\beta }))\right] dF_{T}(t)\ .
	\end{equation*}
\end{assumption}

\begin{assumption}
	\label{as:EY2}  $\mathbb{E}\left[\sup_{\boldsymbol{\beta} \in
		\Theta }|L\left( Y-g(T;\boldsymbol{\beta} )\right)|^{2}\right]<\infty $.
\end{assumption}

Assumption \ref{as:Theta} restricts the parameter to a compact set. This
condition is commonly imposed in the nonlinear regression but can be relaxed
if $g(t;\boldsymbol{\beta})$ is linear in $\boldsymbol{\beta} $. Assumption \ref{as:solution} is an
identification condition. This condition cannot be relaxed. Assumption \ref%
{as:EY2} is an envelope condition that is sufficient for the applicability
of the uniform law of large number:  
\begin{eqnarray*}
	\frac{1}{N}\sum_{i=1}^{N}\pi _{0}(T_{i},\boldsymbol{X}_{i})L\left(
	Y_{i}-g(T_{i};\boldsymbol{\beta} )\right)  =\mathbb{E}[\pi _{0}(T,\boldsymbol{X}%
	)L\left( Y-g(T;\boldsymbol{\beta})\right) ]+o_{p}(1)\text{ uniformly over }\boldsymbol{\beta} \in
	\Theta ; \\
	\frac{1}{N}\sum_{i=1}^{N}L\left( Y_{i}-g(T_{i};\boldsymbol{\beta})\right) ^{2}=
	\mathbb{E}[L\left( Y-g(T;\boldsymbol{\beta})\right) ^{2}]+o_{p}(1)\text{ uniformly over }\boldsymbol{\beta} \in \Theta .
\end{eqnarray*}
Under these and other conditions, we establish the consistency of the
generalized optimization estimator.

\begin{theorem}
	\label{thm:consistency} Suppose that Assumptions \ref{as:TYindep}, \ref%
	{as:suppX}-\ref{as:K&N_consistency}, and \ref{as:Theta}-\ref{as:EY2} hold.
	Then, $\Vert \hat{\boldsymbol{\beta }}-\boldsymbol{\beta }_{0}\Vert %
	\xrightarrow{p}0$.
\end{theorem}

To establish the asymptotic distribution of the proposed estimator, we need
some smoothness condition on the regression function and some
under-smoothing condition on the sieve approximation (i.e., larger $K$ than
needed for consistency). We also have to address the possibility of a
nonsmooth loss function. These conditions are presented below.

\begin{assumption}
	\label{as:m_smooth} \
	\begin{enumerate}
		\item $g(t;\boldsymbol{\beta })$ is twice continuously differentiable in $%
		\boldsymbol{\beta }\in \Theta $;
		
		\item $L(Y-g(T;\boldsymbol{\beta }))$ is differentiable in $\boldsymbol{\
			\beta }$ with probability one, i.e., for any directional vector $\boldsymbol{%
			\ \eta }\in \mathbb{R}^{p}$, there exists an integrable random variable $%
		L^{\prime }(Y-g(T;\boldsymbol{\beta }))$ such that{\footnotesize \ 
			\begin{equation*}
			\mathbb{P}\left( \lim_{\epsilon \rightarrow 0}\frac{L(Y-g(T;\boldsymbol{\
					\beta }+\epsilon \boldsymbol{\eta }))-L(Y-g(T;\boldsymbol{\beta }))}{%
				\epsilon }=L^{\prime }(Y-g(T;\boldsymbol{\beta }))\cdot \left\langle m(T;%
			\boldsymbol{\beta }),\boldsymbol{\eta }\right\rangle _{\mathbb{R}%
				^{p}}\right) =1,
			\end{equation*}%
		} where $\left\langle \cdot ,\cdot \right\rangle _{\mathbb{R}^{p}}$ is the
		inner product in Euclidean space $\mathbb{R}^{p}$;
		
		\item $\mathbb{E}\left[ L^{\prime }(Y-g(T;\boldsymbol{\beta }_{0}))^{2} %
		\right] <\infty $.
	\end{enumerate}
\end{assumption}

\begin{assumption}
	\label{as:first_order} Suppose that 
	\begin{equation*}
	\frac{1}{N}\sum_{i=1}^{N}\hat{\pi}_{K}(T_{i},\boldsymbol{X}_{i})L^{\prime
	}\left( Y_{i}-g(T_{i};\hat{\boldsymbol{\beta }})\right) m(T_{i};\hat{%
		\boldsymbol{\beta }})=o_{P}(N^{-1/2})
	\end{equation*}%
	holds with probability approaching one.
\end{assumption}

\begin{assumption}
	\label{as:H} $\mathbb{E}\left[ \pi _{0}(T,\boldsymbol{X})L^{\prime }(Y-g(T;%
	\boldsymbol{\beta }))m(T;\boldsymbol{\beta })\right] $ is differentiable
	with respect to $\boldsymbol{\beta}$ and $H_{0}:=-\nabla _{\beta }\mathbb{E}\left[ \pi
	_{0}(T,\boldsymbol{X})L^{\prime }(Y-g(T;\boldsymbol{\beta }))m(T;\boldsymbol{%
		\beta })\right] \Big|_{\boldsymbol{\beta }=\boldsymbol{\beta }_{0}}$ is
	nonsingular.
\end{assumption}

\begin{assumption}
	\label{as:smooth_varepsilon} $\varepsilon (t,\boldsymbol{x};{\boldsymbol{
			\beta }}_{0}):=\mathbb{E}[L^{\prime }(Y-g(T;\boldsymbol{\beta }_{0}))|T=t, 
	\boldsymbol{X}=\boldsymbol{x}]$ is continuously differentiable in $(t, 
	\boldsymbol{x})$.
\end{assumption}

\begin{assumption}
	\label{as:entropy} \ 
	
	\begin{enumerate}
	\item $\mathbb{E}\left[ \sup_{\boldsymbol{\beta} \in \Theta }|L^{\prime }(Y-g(T;\boldsymbol{\beta}
	))^{2+\delta }\right] <\infty $ for some $\delta >0$;
	
	\item The function class $\{L^{\prime }(y-g(t;\boldsymbol{\beta} )):\boldsymbol{\beta} \in \Theta \}$
	satisfies: 
	\begin{equation*}
	\mathbb{E}\left[ \sup_{\boldsymbol{\beta} _{1}:\Vert \boldsymbol{\beta}_{1}-\boldsymbol{\beta} \Vert <\delta
	}\left\vert L^{\prime }(Y-g(T;\boldsymbol{\beta} _{1}))-L^{\prime }(Y-g(T;\boldsymbol{\beta}
	))\right\vert ^{2}\right] ^{1/2}\leq a\cdot \delta ^{b}
	\end{equation*}%
	for any $\forall \boldsymbol{\beta}\in \Theta $ and any small $\delta >0$ and for some
	finite positive constants $a$ and $b$.
\end{enumerate}
\end{assumption}

\begin{assumption}
	\label{as:K&N_c} $\zeta (K)\sqrt{K^{4}/N}\rightarrow 0$ and $\sqrt{N}%
	K^{-\alpha}\to 0$ as $N\to \infty$.
\end{assumption}

Assumption \ref{as:m_smooth} imposes sufficient regularity conditions on
both regression function and loss function. These conditions permit
nonsmooth loss functions and are satisfied by the example loss functions
mentioned in previous sections. Assumption \ref{as:first_order} is
essentially the first order condition, similar to the one imposed in $Z$
-estimation. Again, this first order condition is satisfied by popular
nonsmooth loss functions. Assumptions \ref{as:H} ensures that the efficient
variance to be finite. Assumption \ref{as:entropy} is a stochastic
equicontinuity condition which is needed for establishing weak convergence,
see \cite{andrews1994empirical}. Again, it is satisfied by widely used
nonsmooth loss functions. Assumption \ref{as:K&N_c} imposes further
restriction on the smoothing parameter ($K$) so that the sieve approximation
is under-smoothed. This condition is stronger than Assumption \ref%
{as:K&N_consistency} but it is commonly imposed in the semiparametric
regression literature.

\begin{theorem}
	\label{main_theorem} Suppose that Assumptions \ref{as:TYindep}, \ref%
	{as:suppX}-\ref{as:u&v}, \ref{as:Theta}-\ref{as:EY2}, and \ref{as:m_smooth}-%
	\ref{as:K&N_c} hold. Then, 
	\begin{equation*}
	\sqrt{N}\left( \hat{\boldsymbol{\beta }}-\boldsymbol{\beta }_{0}\right) %
	\xrightarrow{d}\mathcal{N}(0,V_{eff}),
	\end{equation*}%
	where $V_{eff}=\mathbb{E}\left[ S_{eff}(T,\boldsymbol{X},Y;\boldsymbol{\beta 
	}_{0})S_{eff}(T,\boldsymbol{X},Y;\boldsymbol{\beta }_{0})^{\top }\right] $.
	Therefore, $\hat{\boldsymbol{\beta }}$ attains the semi-parametric
	efficiency bound of Theorem \ref{efficiency theorem}.
\end{theorem}

The proof of Theorem \ref{main_theorem} is given in Section 5 of
supplemental material.

\section{Variance estimation}\label{sec:variance}
In order to conduct statistical inference, a consistent
covariance matrix is needed. Theorem \ref{efficiency theorem} suggests that
such consistent covariance can be obtained by replacing $H_{0}$ and $\psi
(Y,T,\boldsymbol{X};\boldsymbol{\beta }_{0})$ with some consistent
estimates. Since the nonsmooth loss function may fail the exchangeability
between the expectation and derivative operator, some care in the estimation
of $H_{0}$ is warranted. Using the tower property of conditional
expectation, we rewrite $H_{0}$ as: 
\begin{align*}
H_{0}=& -\nabla _{\beta }\mathbb{E}\left[ \pi _{0}(T,\boldsymbol{X})\mathbb{E%
}\left[ L^{\prime }(Y-g(T;\boldsymbol{\beta }))|T,\boldsymbol{X}\right] m(T;%
\boldsymbol{\beta })\right] \Big|_{\boldsymbol{\beta }=\boldsymbol{\beta }%
	_{0}} \\
=& -\mathbb{E}\left[ \pi _{0}(T,\boldsymbol{X})\nabla _{\beta }\mathbb{E}%
\left[ L^{\prime }(Y-g(T;\boldsymbol{\beta }))|T,\boldsymbol{X}\right] \Big|%
_{\boldsymbol{\beta }=\boldsymbol{\beta }_{0}}m(T;\boldsymbol{\beta }%
_{0})^{\top }\right] \\
& -\mathbb{E}\left[ \pi _{0}(T,\boldsymbol{X})\mathbb{E}\left[ L^{\prime
}(Y-g(T;\boldsymbol{\beta }_{0}))|T,\boldsymbol{X}\right] \nabla _{\beta
}m(T;\boldsymbol{\beta }_{0})\right] .
\end{align*}%
Applying integration by part (see Appendix \ref{appendix:proof_integration}), we obtain 
\begin{align}\label{appendix:integration_by_parts}
& \nabla _{\beta }\mathbb{E}\left[ L^{\prime }(Y-g(T;\boldsymbol{\beta }
))|T=t,\boldsymbol{X}=\boldsymbol{x}\right] \Big|_{\boldsymbol{\beta }=%
	\boldsymbol{\beta }_{0}} \notag \\
=& \mathbb{E}\left[ L^{\prime }(Y-g(T;\boldsymbol{\beta }_{0}))\frac{\frac{%
		\partial }{\partial y}f_{Y,T,\boldsymbol{X}}(Y,T,\boldsymbol{X})}{f_{Y,T,%
		\boldsymbol{X}}(Y,T,\boldsymbol{X})}\bigg|T=t,\boldsymbol{X}=\boldsymbol{x}%
\right] m(t;\boldsymbol{\beta }_{0})
\end{align}%
and consequently 
\begin{equation*}
H_{0}=-\mathbb{E}\left[ \pi _{0}(T,\boldsymbol{X})L^{\prime }(Y-g(T;%
\boldsymbol{\beta }_{0}))\left\{ \frac{\frac{\partial }{\partial y}f_{Y,T,%
		\boldsymbol{X}}(Y,T,\boldsymbol{X})}{f_{Y,T,\boldsymbol{X}}(Y,T,\boldsymbol{X%
	})}m(T;\boldsymbol{\beta }_{0})m(T;\boldsymbol{\beta })^{\top }+\nabla
_{\beta }m(T;\boldsymbol{\beta }_{0})^{\top }\right\} \right] .
\end{equation*}%
Since the density $f_{Y,T,\boldsymbol{X}}(y,t,\boldsymbol{x})$ can be
estimated via the usual kernel estimator: 
\begin{equation*}
\hat{f}_{Y,T,\boldsymbol{X}}(y,t,\boldsymbol{x})=\frac{1}{N}\sum_{i=1}^{N}%
\frac{1}{h_{Y}\cdot h_{T}\cdot \prod_{j=1}^{r}h_{j}}k\left( \frac{Y_{i}-y}{%
	h_{Y}}\right) k\left( \frac{T_{i}-t}{h_{T}}\right) \prod_{j=1}^{r}k\left( 
\frac{X_{ij}-x_{j}}{h_{j}}\right)
\end{equation*}%
where $k(\cdot )$ is the kernel function, and $\{h_{Y},h_{T},h_{j},j=1,%
\ldots ,r\}$ are the bandwidths. Then $H_{0}$ can be estimated by 
\begin{equation*}
\hat{H}=-\frac{1}{N}\sum_{i=1}^{N}\hat{\pi}_{K}(T_{i},\boldsymbol{X}%
_{i})L^{\prime }(Y_{i}-g(T_{i};\hat{\boldsymbol{\beta }}))\left\{ \frac{%
	\frac{\partial }{\partial y}\hat{f}_{Y,T,\boldsymbol{X}}(Y_{i},T_{i},%
	\boldsymbol{X}_{i})}{\hat{f}_{Y,T,\boldsymbol{X}}(Y_{i},T_{i},\boldsymbol{X}%
	_{i})}m(T_{i};\hat{\boldsymbol{\beta }})m(T_{i};\hat{\boldsymbol{\beta }}%
)^{\top }+\nabla _{\beta }m(T_{i};\hat{\boldsymbol{\beta }})\right\} .
\end{equation*}%
Also, $\psi (Y,T,\boldsymbol{X};\boldsymbol{\beta }_{0})$ can be directly
estimated by the kernel method: 
\begin{align*}
\hat{\psi}(y,t,\boldsymbol{x};\hat{\boldsymbol{\beta }})=& \hat{\pi}_{K}(t,%
\boldsymbol{x})L^{\prime }(y-g(t;\hat{\boldsymbol{\beta }}))m(t;\hat{%
	\boldsymbol{\beta }}) \\
& -\hat{\pi}_{K}(t,\boldsymbol{x})\hat{\mathbb{E}}\left[ L^{\prime }(Y-g(T;%
\hat{\boldsymbol{\beta }}))|T=t,\boldsymbol{X}=\boldsymbol{x}\right] m(t;%
\hat{\boldsymbol{\beta }}) \\
& +\widehat{\mathbb{E}}\left[ \hat{\pi}_{K}(T_{i},\boldsymbol{X}%
_{i})L^{\prime }(Y-g(T;\hat{\boldsymbol{\beta }}))|T=t\right] m(t;\hat{%
	\boldsymbol{\beta }}) \\
& +\widehat{\mathbb{E}}\left[ \hat{\pi}_{K}(T_{i},\boldsymbol{X}%
_{i})L^{\prime }(Y-g(T;\hat{\boldsymbol{\beta }}))|\boldsymbol{X}=%
\boldsymbol{x}\right] m(t;\hat{\boldsymbol{\beta }}),
\end{align*}%
where 
\begin{align*}
& \widehat{\mathbb{E}}\left[ L^{\prime }(Y-g(T;\hat{\boldsymbol{\beta }}%
))|T=t,\boldsymbol{X}=\boldsymbol{x}\right] :=\frac{\sum_{i=1}^{N}L^{\prime
	}(Y_{i}-g(T_{i};\hat{\boldsymbol{\beta }}))k\left( \frac{T_{i}-t}{h_{T}}%
	\right) \prod_{j=1}^{r}k\left( \frac{X_{ij}-x_{j}}{h_{j}}\right) }{%
	\sum_{i=1}^{N}k\left( \frac{T_{i}-t}{h_{T}}\right) \prod_{j=1}^{r}k\left( 
	\frac{X_{ij}-x_{j}}{h_{j}}\right) }, \\
& \widehat{\mathbb{E}}\left[ \hat{\pi}_{K}(T_{i},\boldsymbol{X}%
_{i})L^{\prime }(Y-g(T;\hat{\boldsymbol{\beta }}))|T=t\right] :=\frac{%
	\sum_{i=1}^{N}\hat{\pi}_{K}(T_{i},\boldsymbol{X}_{i})L^{\prime
	}(Y_{i}-g(T_{i};\hat{\boldsymbol{\beta }}))k\left( \frac{T_{i}-t}{h_{T}}%
	\right) }{\sum_{i=1}^{N}k\left( \frac{T_{i}-t}{h_{T}}\right) }, \\
& \widehat{\mathbb{E}}\left[ \hat{\pi}_{K}(T_{i},\boldsymbol{X}%
_{i})L^{\prime }(Y-g(T;\hat{\boldsymbol{\beta }}))|\boldsymbol{X}=%
\boldsymbol{x}\right] :=\frac{\sum_{i=1}^{N}\hat{\pi}_{K}(T_{i},\boldsymbol{X%
	}_{i})L^{\prime }(Y_{i}-g(T_{i};\hat{\boldsymbol{\beta }}))\prod_{j=1}^{r}k%
	\left( \frac{X_{ij}-x_{j}}{h_{j}}\right) }{\sum_{i=1}^{N}\prod_{j=1}^{r}k%
	\left( \frac{X_{ij}-x_{j}}{h_{j}}\right) .}.
\end{align*}%
The asymptotic covariance is estimated by 
\begin{equation*}
\hat{V}=\hat{H}^{-1}\left\{ \frac{1}{N}\sum_{i=1}^{N}\hat{\psi}(Y_{i},T_{i},%
\boldsymbol{X}_{i};\hat{\boldsymbol{\beta }})\hat{\psi}(Y_{i},T_{i},%
\boldsymbol{X}_{i};\hat{\boldsymbol{\beta }})^{\top }\right\} (\hat{H}^{\top
})^{-1}.
\end{equation*}%
Under the standard conditions imposed in the kernel regression literature,  
see Section  6 of supplemental material or Theorem 6 of \cite{masry1996multivariate},
it follows that the kernel estimators are uniformly strong consistent (in
almost surely sense). Also from Theorems \ref{rate_pi} and \ref%
{thm:consistency}, we have $\sup_{(t,\boldsymbol{x})\in \mathcal{T}\times 
	\mathcal{X}}|\hat{\pi}_{K}(t,\boldsymbol{x})-\pi _{0}(t,\boldsymbol{x}%
)|=o_{p}(1)$ and $\Vert \hat{\boldsymbol{\beta }}-\boldsymbol{\beta }%
_{0}\Vert \rightarrow 0$. With these results, we obtain the consistency of $%
\hat{V}$.

\begin{theorem}
	\label{var_consistency} Suppose that Assumptions \ref{as:TYindep}, \ref%
	{as:suppX}-\ref{as:K&N_consistency}, \ref{as:Theta}-\ref{as:EY2} hold, and
 the conditions in Section  6 of supplemental material hold.
	Then, $\hat{V}$ converges to $V_{eff}$ in probability.
\end{theorem}

\textbf{Remark}: If the exchangeability of expectation and derivative
operator is valid, we can estimate $H_{0}$ and $\psi (Y,T,\boldsymbol{X};%
\boldsymbol{\beta }_{0})$ by direct differentiation and construct a much
simpler variance estimator based on the first order condition %
\eqref{def:G^hat} and \eqref{def:estimator}, see Appendix \ref%
{sec:variance_L''}.

\section{Some extensions}\label{sec:extension}
The condition that the causal effect is parameterized
may be restrictive for some applications. To relax this condition, we now
consider the nonparametric specification $\theta (t)=\mathbb{E}[Y^{\ast
}(t)] $. We shall study estimation of $\theta (t)$ as well as the average
effect $\psi =\int_{\mathcal{T}}\mathbb{E}[\theta (t)]dF_{T}(t)$ under the
loss function $L(v)=v^{2}$. Extension to the general loss function requires
considerable derivation and shall be dealt with in a separate paper.

\subsection{Estimation of effect curve}

We begin with estimation of $\theta (t)$. Note that, for all $t\in \mathcal{T%
}$ and under Assumption \ref{as:TYindep}, we can rewrite $\theta (t)$ as 
\begin{equation*}
\theta (t)=\mathbb{E}\left[ \pi _{0}(T,\boldsymbol{X})Y|T=t\right] .
\end{equation*}%
With $\pi _{0}(T,\boldsymbol{X})$ replaced by $\hat{\pi}_{K}(T,\boldsymbol{X}%
)$, we estimate $\theta (t)$ by regressing $\hat{\pi}_{K}(T,\boldsymbol{X})Y$
on $u_{K_{1}}(T)$: 
\begin{equation*}
\hat{\theta}_{K}(t)=\left[ \sum_{i=1}^{N}\hat{\pi}_{K}(T_{i},\boldsymbol{X}%
_{i})Y_{i}u_{K_{1}}(T_{i})^{\top }\right] \left[
\sum_{i=1}^{N}u_{K_{1}}(T_{i})u_{K_{1}}(T_{i})^{\top }\right]
^{-1}u_{K_{1}}(t).
\end{equation*}%
To aid presentation of the asymptotic properties of $\hat{\theta}_{K}(t)$,
we denote 
\begin{align*}
& \varepsilon _{i}={\pi }_{0}(T_{i},\boldsymbol{X}_{i})Y_{i}-\mathbb{E}\left[
{\pi }_{0}(T_{i},\boldsymbol{X}_{i})Y_{i}|T_i,\boldsymbol{X}_{i}\right]+%
\mathbb{E}\left[ {\pi }_{0}(T_{i},\boldsymbol{X}_{i})Y_{i}|\boldsymbol{X}_{i}%
\right]-\mathbb{E}\left[ {\pi }_{0}(T_{i},\boldsymbol{X}_{i})Y_{i}\right], \\
&\sigma ^{2}(T_{i})=Var(\varepsilon _{i}|T_{i}), \ \Sigma _{K_{1}\times
	K_{1}}=\mathbb{E}[u_{K_{1}}(T)u_{K_{1}}^{\top }(T)\sigma ^{2}(T)]\ ,\ \Phi
_{K_{1}\times K_{1}}=\mathbb{E}[u_{K_{1}}(T)u_{K_{1}}^{\top }(T)],
\end{align*}%
and 
\begin{equation*}
V_{K}(t)=u_{K_{1}}^{\top }(t)\Phi _{K_{1}\times K_{1}}^{-1}\Sigma
_{K_{1}\times K_{1}}\Phi _{K_{1}\times K_{1}}^{-1}u_{K_{1}}(t).
\end{equation*}

\begin{theorem}
	\label{thm:curve} Suppose that for some $\tilde{\alpha}>0$, there exists
	some $\gamma^*\in \mathbb{R}^{K_{1}}$ such that $\sup_{t\in \mathcal{T}}\left\vert \theta (t)-(\gamma^*)^{\top }u_{K_{1}}(t)\right\vert
	=O(K_{1}^{-\tilde{\alpha}})$ and $\sigma ^{2}(t)$ is bounded, and that
	Assumptions \ref{as:TYindep}, \ref{as:suppX}-\ref{as:smooth_varepsilon}
	hold. Then:
	
	\begin{enumerate}
		\item (Consistency) 
		\begin{equation*}
		\int_{\mathcal{T}}|\hat{\theta}_{K}(t)-\theta (t)|^{2}dF_{T}(t)=O_{p}\left( 
		\frac{\zeta (K)^{2}K}{N}+\zeta (K)^{2}K^{-2\alpha }+K_{1}^{-2\tilde{\alpha}%
		}\right)
		\end{equation*}
		and 
		\begin{equation*}
		\sup_{t\in \mathcal{T}}|\hat{\theta}_{K}(t)-\theta (t)|=O_{p}\left[ \zeta
		_{1}(K_{1})\left( \zeta (K)\sqrt{K/N}+\zeta (K)K^{-2\alpha }+K_{1}^{-\tilde{%
				\alpha}}\right) \right] .
		\end{equation*}
		
		\item (Asymptotic Normality) suppose $\sqrt{N}K^{-\tilde{\alpha}}\to 0$ for any fixed $t\in \mathcal{T}$, 
		\begin{equation*}
		\sqrt{N}{V}_{K}(t)^{-1/2}\left[ \hat{\theta}_{K}(t)-\theta (t) \right] %
		\xrightarrow{d}N(0,1).
		\end{equation*}
	\end{enumerate}
\end{theorem}
See Section 7.1 in supplemental material for a detailed proof.

\subsection{Efficient estimation of average effect}

To estimate the average effect $\psi $, we notice that 
\begin{equation*}
\psi =\mathbb{E}[\pi _{0}(T,\boldsymbol{X})Y].
\end{equation*}%
Hence, we estimate $\psi $ by 
\begin{equation*}
\hat{\psi}_{K}=\frac{1}{N}\sum_{i=1}^{N}\hat{\pi}_{K}(T_{i},\boldsymbol{X}%
_{i})Y_{i}.
\end{equation*}%
The following theorem establishes the asymptotic distribution of $\hat{\psi}%
_{K}$ and shows that $\hat{\psi}_{K}$ attains the efficiency bound of $\psi $
derived by \cite{kennedy2017non}. \newline

\begin{theorem}
	\label{thm:average_outcome} Suppose that Assumptions \ref{as:TYindep}, \ref%
	{as:suppX}-\ref{as:K&N_c} hold. Then:
	
	\begin{enumerate}
		\item (Consistency) $\hat{\psi}_K\xrightarrow{p}\psi$;
		
		\item (Asymptotic Efficiency) $\sqrt{N}(\hat{\psi}_{K}-\psi )\xrightarrow{d}
		N(0,V_{eff}^{\psi })$, where $V_{eff}^{\psi }=\mathbb{E}\left[ \left(
		S_{eff}^{\psi }\right) ^{2}\right] $ and 
		\begin{align*}
		S_{eff}^{\psi }(T,\boldsymbol{X},Y)=\pi _{0}(T,\boldsymbol{X})& \left\{ Y- 
		\mathbb{E}[Y|\boldsymbol{X},T]\right\} +\left\{ \mathbb{E}[\pi _{0}(T, 
		\boldsymbol{X})Y|\boldsymbol{X}]-\mathbb{E}[\pi _{0}(T,\boldsymbol{X}
		)Y]\right\} \\
		& +\left\{ \mathbb{E}[\pi _{0}(T,\boldsymbol{X})Y|T]-\mathbb{E}[\pi _{0}(T, 
		\boldsymbol{X})Y]\right\} .
		\end{align*}
	\end{enumerate}
\end{theorem}
See Section 7.2 in supplemental material for a proof.
\section{ Monte Carlo simulations \label{sec:simulation}}

The large sample properties established in previous sections do not indicate
how the generalized optimization estimator behave in finite samples. To
evaluate its finite sample performance, we conduct a small scale simulation
study on a continuous treatment. We consider two data generating processes
(DGPs):
\begin{description}
	\item[DGP-1] {\small $T_{i}=X_{1i}+X_{2i}+0.2X_{3i}+0.2X_{4i}+\xi _{i}$ and $%
		Y_{i}=1+X_{1i}+0.1X_{2i}+0.1X_{3i}+0.1X_{4i}+T_{i}+\epsilon _{i}$, }$%
	i=1,2,\ldots ,N${\small . }
	
	\item[DGP-2] {\small $T_{i}=(X_{1i}+0.5)^{2}+0.4X_{2i}+0.4X_{3i}+0.4X_{4i}+%
		\xi _{i}$ and $Y_{i}=0.75X_{1i}^{2}+0.2(X_{2i}+0.5)^{2}+T_{i}+\epsilon _{i}$
		, }$i=1,2,\ldots ,N${\small . }
\end{description}

The error terms are drawn from $\xi _{i}\overset{i.i.d.}{\sim }N(0,9)$ and $%
\epsilon _{i}\overset{i.i.d.}{\sim }N(0,25)$ independently. The covariates
are drawn from $\boldsymbol{X}_{i}=[X_{1i},X_{2i},X_{3i},X_{4i}]^{\top }%
\overset{i.i.d.}{\sim }N(0,\boldsymbol{\Sigma })$, where the diagonal
elements of $\boldsymbol{\Sigma }$ are all 1 and the off-diagonal elements
are all $\rho \in \{0.0,0.2,0.4\}$. These designs are similar to those of 
\cite{Fong_Hazlett_Imai_2017}. For instance, under DGP-1, $T_{i}$ is linear
in $\boldsymbol{X}_{i}$, and $Y_{i}$ is linear in $\boldsymbol{X}_{i}$ and $%
T_{i}$; under DGP-2, $T_{i}$ is nonlinear in $\boldsymbol{X}_{i}$, and $%
Y_{i} $ is nonlinear in $\boldsymbol{X}_{i}$ and linear in $T_{i}$.\footnote{%
	In the supplemental material \cite{Ai_Motegi_Zhang_cts_treat_supp_mat}, we
	consider two extra DGPs following \cite{Fong_Hazlett_Imai_2017}. One has a
	nonlinear component in the DGP of $T_{i}$ only, and the other has a
	nonlinear component in the DGP of $Y_{i}$ only. Those cases serve as
	intermediate scenarios between DGP-1 and DGP-2 considered in the present
	paper. Not surprisingly, the simulation results of those extra cases also
	turn out to be intermediate results between DGP-1 and DGP-2.} Since $Y_{i}$
is linear in $T_{i}$ under both DGPs, we obtain $\mathbb{E}[Y(t)]=\beta
_{1}+\beta _{2}t$ with the true value $(\beta _{10},\beta _{20})=(1,1).$ 

To estimate the stabilized weights, we use the following approximating basis
functions 
\begin{equation*}
u_{K_{1}}(t_{i})=[1,t_{i},t_{i}^{2}]^{\top },\ (\text{i.e.}\ K_{1}=3)\text{
	and}
\end{equation*}%
\begin{equation*}
u_{K_{2}}(\boldsymbol{X}%
_{i})=[1,X_{1i},X_{2i},X_{3i},X_{4i},X_{1i}^{2},X_{2i}^{2},X_{3i}^{2},X_{4i}^{2}]^{\top },\ (%
\text{i.e.}\ K_{2}=9),
\end{equation*}%
or 
\begin{equation*}
u_{K_{2}}(\boldsymbol{X}_{i})=\text{first and second order polynomials of }%
\boldsymbol{X}_{i}\text{, }(i.e.,K_{2}=15).
\end{equation*}%
\newline
We generate a random sample of size $N\in \{100,500,1000\}$ from each DGP.
For each sample, we compute the generalized estimator as well as the
estimator suggested in Fong, Hazlett, and Imai 
\citeyearpar{Fong_Hazlett_Imai_2017} under the quadratic loss function.
Fong, Hazlett, and Imai \citeyearpar{Fong_Hazlett_Imai_2017} use a linear
model specification. Hence, their specification is correct under DGP-1 and
incorrect under DGP-2. After computing both estimators, another sample is
generated and both estimators are computed again. This exercise is repeated
for $J=1000$ times.

The bias, standard deviation, root mean squared error, and coverage
probability based on the 95\% confidence band of both estimators from $%
J=1000 $ simulations are calculated and reported in Tables \ref%
{table:sim_results_beta1_DGP1}-\ref{table:sim_results_beta2_DGP1} for DGP-1
and in Tables \ref{table:sim_results_beta1_DGP4}-\ref%
{table:sim_results_beta2_DGP4} for DGP-2. 

Glancing at these tables, we find that, under DGP-1, both the generalized
optimization estimator (labeled as SW) and Fong, Hazlett, and Imai's %
\citeyearpar{Fong_Hazlett_Imai_2017} estimator (labeled as CBGPS) are
unbiased. CBGPS has smaller bias and standard deviation than the generalized
optimization estimator across all sample sizes. This is not surprising since
CBGPS has a correct parametric specification. The coverage probability,
however, is comparable between the two estimators. See, for example, Table %
\ref{table:sim_results_beta2_DGP1} ($\beta _{2}$) with $\rho =0.2$ and $%
N=500 $, where the coverage probability is 0.926 for the SW estimator with $%
K_{2}=9 $ and 0.890 for the CBGPS estimator. These results are not sensitive
to the correlation coefficient $\rho \in \{0.0,0.2,0.4\}$ and inclusion of
interaction terms of $\boldsymbol{X}_{i}$. 

Under DGP-2, SW dominates CBGPS. For example, in Table \ref%
{table:sim_results_beta1_DGP4} ($\beta _{1}$) with $\rho =0.4$ and $N=1000$,
SW has a bias 0.016 but CBGPS has a larger bias -0.182. Although SW has
larger standard deviation (0.605) than CBGPS (0.194), SW's coverage
probability is 0.923, which is better than CBGPS' coverage probability of
0.800. Table \ref{table:sim_results_beta2_DGP4} ($\beta _{2}$) with $\rho
=0.4$ and $N=1000$ displays a similar pattern. 

We notice that, under DGP-2, CBGPS always has a negative bias in $\beta _{1}$
and a positive bias in $\beta _{2}$, and the biases do not shrink as sample
size $N$ increases. SW, on the other hand, is approximately unbiased for all
sample sizes. Moreover, SW's coverage probability is closer to the true
coverage probability in larger samples (i.e., $N\in \{500,1000\})$. \newline

To summarize, the generalized optimization estimator performs well in finite
samples, and the performance is still good even when the model is nonlinear.
In contrast, the existing alternative estimator of Fong, Hazlett, and Imai %
\citeyearpar{Fong_Hazlett_Imai_2017} is sensitive to model misspecification.

\section{Empirical application \label{sec:application}}

To illustrate the applicability of the generalized optimization procedure,
we revisit U.S. presidential campaign data analyzed by \cite%
{Urban_Niebler_2014} and \cite{Fong_Hazlett_Imai_2018}. The motivation of
the original study, \cite{Urban_Niebler_2014}, is well summarized in 
\citet[Sec.
2]{Fong_Hazlett_Imai_2018}:

\begin{quote}
	\cite{Urban_Niebler_2014} explored the potential causal link between
	advertising and campaign contributions. Presidential campaigns ordinarily
	focus their advertising efforts on competitive states, but if political
	advertising drives more donations, then it may be worthwhile for candidates
	to also advertise in noncompetitive states. The authors exploit the fact
	that media markets sometimes cross state boundaries. This means that
	candidates may inadvertently advertise in noncompetitive states when they
	purchase advertisements for media markets that mainly serve competitive
	states. By restricting their analysis to noncompetitive states, the authors
	attempt to isolate the effect of advertising from that of other campaigning,
	which do not incur these media market spillovers.
\end{quote}

The treatment of interest, the number of political advertisements aired in
each zip code, can be regarded as a continuous variable since it takes a
range of values from 0 to 22379 across $N=16265$ zip codes. \cite%
{Urban_Niebler_2014} restricted themselves to a binary treatment framework,
and they dichotomized the treatment variable by examining whether a zip code
received more than 1000 advertisements or not. Their empirical results
suggest that advertising in non-competitive states had a significant impact
on the level of campaign contributions.  

Dichotomizing a continuous treatment variable requires an ad-hoc choice of a
cut-off value, and it makes an empirical result hard to interpret. \cite%
{Fong_Hazlett_Imai_2018} analyzed the continuous version of the treatment
variable, taking advantage of their proposed CBGPS method. Their empirical
results suggest, contrary to \cite{Urban_Niebler_2014}, that advertising in
non-competitive states did \textit{not} have a significant impact on the
level of campaign contributions \citep[cf. ][Table 2]{Fong_Hazlett_Imai_2018}.

As shown in Section \ref{sec:simulation}, our generalized optimization
estimator has a better performance than Fong, Hazlett, and Imai's %
\citeyearpar{Fong_Hazlett_Imai_2018} CBGPS estimator. Our estimator exhibits
a solid performance even if a DGP of treatment $T_{i}$ or outcome $Y_{i}$ is
nonlinear in covariate $\boldsymbol{X}_{i}$. It is thus of interest to apply
our approach to the continuous version of the treatment variable in order to
see how results change.

\subsection{Fong, Hazlett, and Imai's \citeyearpar{Fong_Hazlett_Imai_2018}
	CBGPS approach}

We begin with Fong, Hazlett, and Imai's \citeyearpar{Fong_Hazlett_Imai_2018}
CBGPS estimator as a benchmark. It requires a choice of pre-treatment
covariates $\boldsymbol{X}_{i}$ in a generalized propensity score model.
There are eight covariates 
\begin{equation}
\boldsymbol{X}_{1}=%
\begin{bmatrix}
\log (\text{Population}) \\ 
\%\text{Over 65} \\ 
\log (\text{Income}+1) \\ 
\%\text{Hispanic} \\ 
\%\text{Black} \\ 
\text{Population Density} \\ 
\%\text{College Graduates} \\ 
\text{Can Commute}%
\end{bmatrix}%
.  \label{eq:pretreat_covariates_first}
\end{equation}%
Subscript $i$ is omitted for brevity, but (\ref{eq:pretreat_covariates_first}%
) is defined for each zip code $i\in \{1,\dots ,N\}$. The definition of each
covariate is almost self-explanatory 
\citep[see][Sec. 5 for more
details]{Fong_Hazlett_Imai_2018}. Following 
\citet[Table
1]{Fong_Hazlett_Imai_2018}, we add squared terms to construct a $15\times 1$
vector of pre-treatment covariates: 
\begin{equation}
\boldsymbol{X}=%
\begin{bmatrix}
\boldsymbol{X}_{1} \\ 
\{\log (\text{Population})\}^{2} \\ 
\{\%\text{Over 65}\}^{2} \\ 
\{\log (\text{Income}+1)\}^{2} \\ 
\{\%\text{Hispanic}\}^{2} \\ 
\{\%\text{Black}\}^{2} \\ 
\{\text{Population Density}\}^{2} \\ 
\{\%\text{College Graduates}\}^{2}%
\end{bmatrix}%
.  \label{eq:pretreat_covariates}
\end{equation}%
The square of \textquotedblleft Can Commute\textquotedblright\ is not added
since it is a binary indicator of whether it is possible to commute to zip
code $i$ from a competitive state so that $\text{Can Commute}=\{\text{Can
	Commute}\}^{2}$. 

\begin{figure}[h!]	
	\caption{Empirical Distributions of Political Advertisements \label{fig:hist_treat}}
	\begin{center}
		\begin{subfigure}[t]{48mm}
			\centering
			\includegraphics[height = 46mm, width = 48mm]{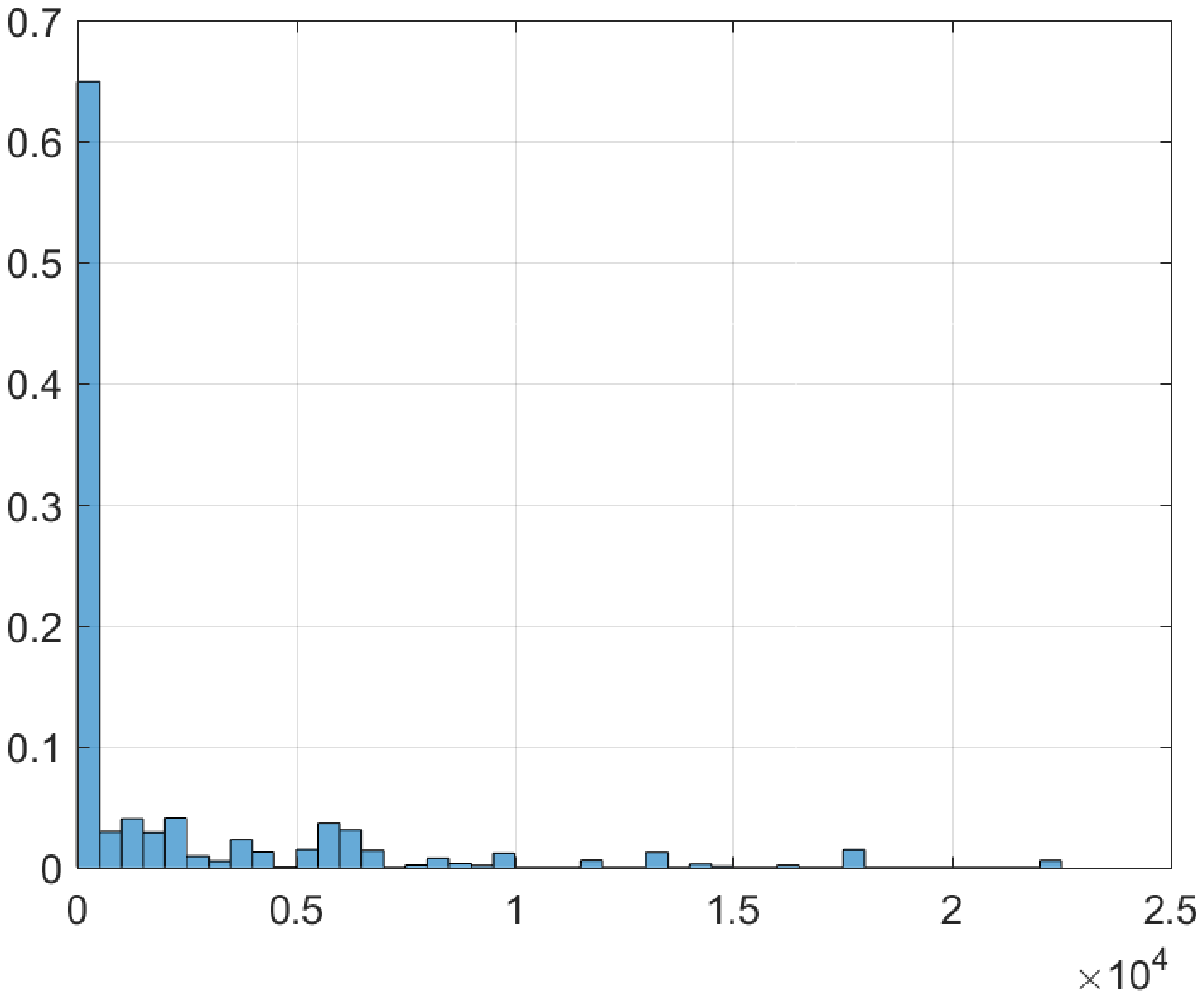}
			\caption{1. Original $T$}	
		\end{subfigure}
		\ \ 
		\begin{subfigure}[t]{48mm}
			\centering
			\includegraphics[height = 46mm, width = 48mm]{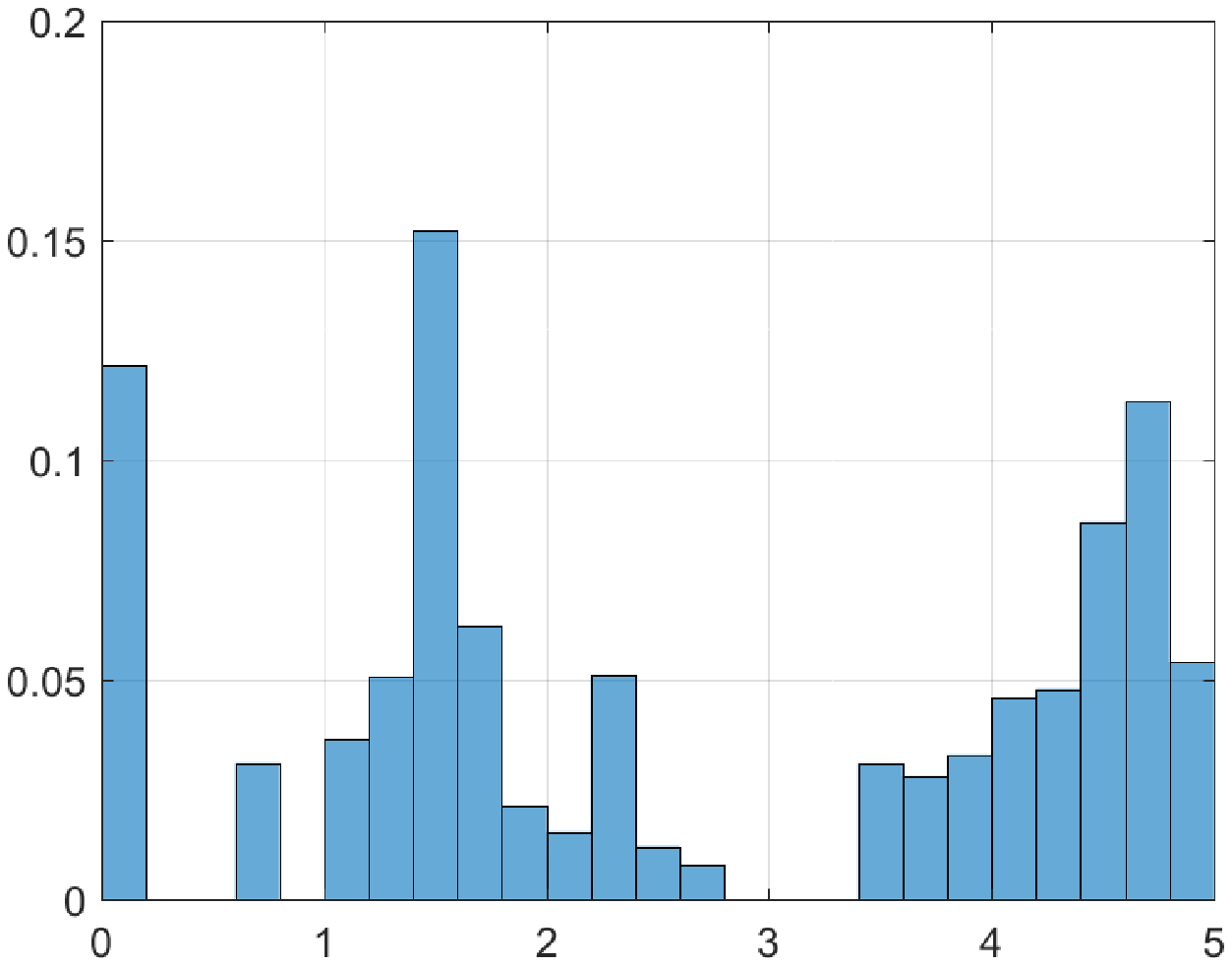}
			\caption{2. Box-Cox}
		\end{subfigure}
		\ \ 
		\begin{subfigure}[t]{48mm}
			\centering
			\includegraphics[height = 46mm, width = 48mm]{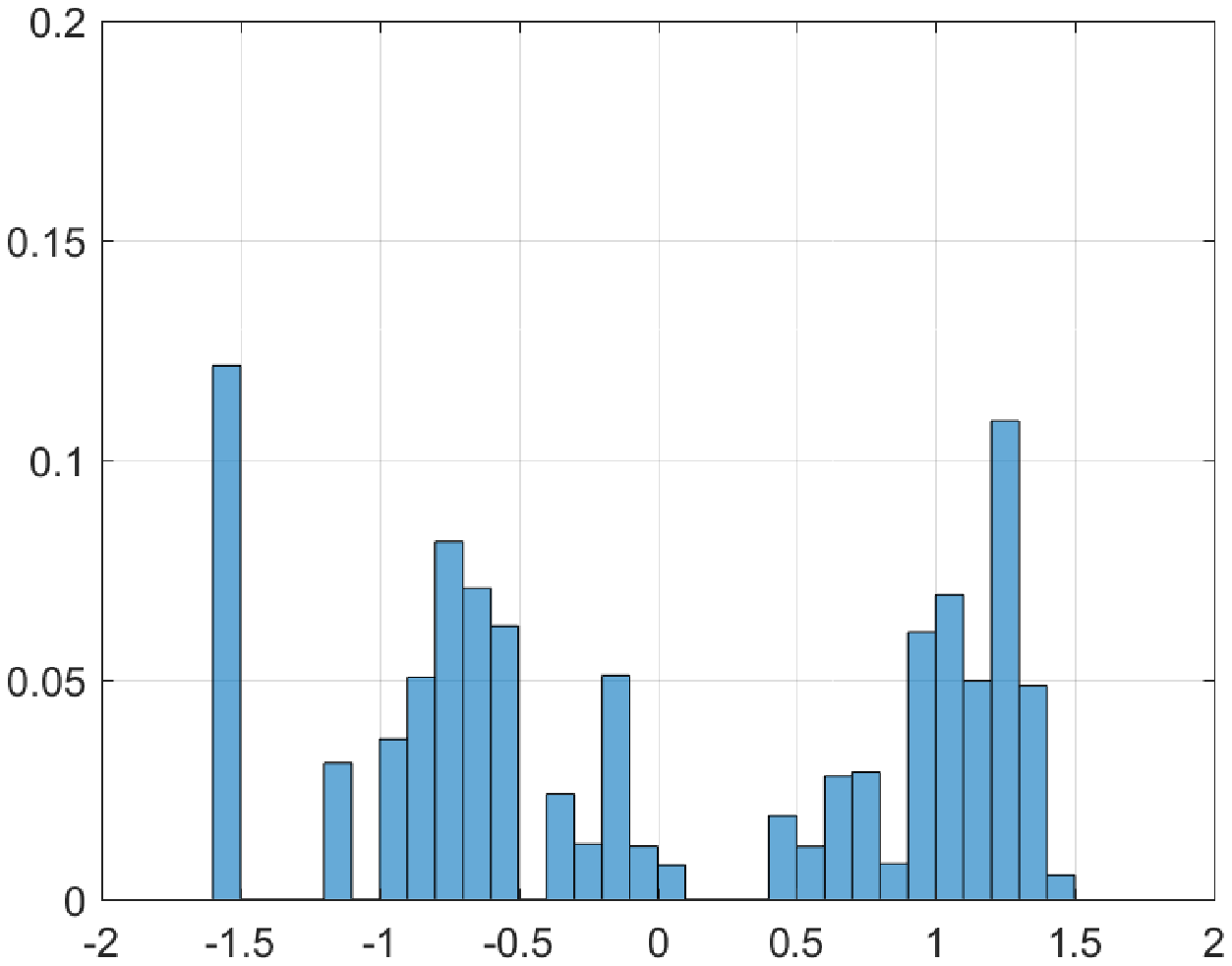}
			\caption{3. Box-Cox \& Std.}
		\end{subfigure} \\ 
		
	\end{center}
	{\fontsize{10pt}{20pt} \selectfont
		In this figure we draw histograms of the treatment variable studied in \cite{Fong_Hazlett_Imai_2018} (i.e. the number of political advertisements aired in each zip code).
		Panel 1 plots the original treatment $T$; 
		Panel 2 plots $T'$, namely the treatment after running the Box-Cox transformation with $\lambda = -0.16$;
		Panel 3 plots $T^{*}$, namely the standardized version of $T'$.
		For each histogram, the sum of the heights of bars is normalized to 1 so that the vertical axis is associated with empirical probability. 
	}
\end{figure}

Let $T_{i}$ be the treatment of interest (i.e. the number of political
advertisements aired in each zip code). The CBGPS approach assumes that the
standardized treatment variable 
\begin{equation}
T_{i}^{\ast }=s_{T}^{-1/2}(T_{i}-\bar{T})  \label{eq:std_treat}
\end{equation}%
follows the standard normal distribution, where $\bar{T}=(1/N)%
\sum_{i=1}^{N}T_{i}$ and $s_{T}=(1/(N-1))\sum_{i=1}^{N}(T_{i}-\bar{T})^{2}$.
Given the data of political advertisements, the normality assumption is far
from satisfied (see Panel 1 of Figure \ref{fig:hist_treat}). \cite%
{Fong_Hazlett_Imai_2018} therefore run a Box-Cox transformation $%
T_{i}^{\prime }=\{(T_{i}+1)^{\lambda }-1\}/\lambda $ with $\lambda =-0.16$
and then standardize $T_{i}^{\prime }$ according to (\ref{eq:std_treat}).
They choose $\lambda =-0.16$ since it yields the greatest correlation
between the sample quantiles of the standardized treatment and the
corresponding theoretical quantiles of the standard normal distribution. As %
\citet[p.15]{Fong_Hazlett_Imai_2018} admit, the Gaussian approximation is
very poor even after running the Box-Cox transformation (see Panels 2-3 of
Figure \ref{fig:hist_treat}). This result suggests that the normality of a
standardized treatment is often a too strong assumption to make in practice.

For an outcome model, we consider four cases for covariates $\boldsymbol{Z}%
_{i}$:
\begin{description}
	\item \setlength{\leftskip}{2.0cm}
	
	\item[Case \#1.] $\boldsymbol{Z}_{i} = [T_{i}, T_{i}^{2}, 1]^{\top}$.
	
	\item[Case \#2.] $\boldsymbol{Z}_{i} = [T_{i}, T_{i}^{2}, \boldsymbol{SD}
	_{i}^{\top}]^{\top}$.
	
	\item[Case \#3.] $\boldsymbol{Z}_{i} = [T_{i}, T_{i}^{2}, 1, \boldsymbol{X}
	_{1i}^{\top}]^{\top}$.
	
	\item[Case \#4.] $\boldsymbol{Z}_{i} = [T_{i}, T_{i}^{2}, \boldsymbol{SD}
	_{i}^{\top}, \boldsymbol{X}_{1i}^{\top}]^{\top}$.
\end{description}
Note that $\boldsymbol{SD}_{i} = [SD_{1i}, SD_{2i}, \dots, SD_{24i}]^{\top}$%
, where $SD_{ji}$ is a binary indicator that equals 1 if zip code $i$
belongs to state $j$ and equals 0 otherwise. Any zip code contained in the
dataset belongs to one and only one of 24 states (e.g. Alabama, Arkansas,
\dots, Wyoming). 

For each of Cases \#1--\#4, we compute the CBGPS estimator and its
asymptotic 95\% confidence bands 
\citep[see][Sec. 3.2 for
procedures]{Fong_Hazlett_Imai_2018}. Our main interest lies in the
parameters of $(T_{i}, T_{i}^{2})$ and their statistical significance. See
Table \ref{table:empirical_result_CBGPS} for results. It is evident that the
empirical results depend critically on a specification of $\boldsymbol{Z}%
_{i} $. In Case \#2, $T_{i}$ has a significantly positive impact on $Y_{i}$
and $T_{i}^{2}$ has a significantly negative impact on $Y_{i}$. In the other
three cases, both $T_{i}$ and $T_{i}^{2}$ have \textit{insignificant}
impacts on $Y_{i}$.

\subsection{Generalized Optimization approach}

One practical advantage of our proposed approach over the CBGPS approach is
that we do not require the normality assumption for the treatment variable $%
T $. As indicated in Figure \ref{fig:hist_treat}, the normality assumption
is too strong for the number of political advertisements aired in each zip
code whether or not the Box-Cox transformation is implemented. The
stabilized weighting approach allows us to work with the original treatment
variable (Panel 1 of Figure \ref{fig:hist_treat}).  

We assume that the link function is quadratic with $p = 3$: 
\begin{equation*}
g(T, \boldsymbol{\beta}) = \beta_{0} + \beta_{1} T + \beta_{2} T^{2}.
\end{equation*}
Our covariates $\boldsymbol{X}$ are chosen to be identical to Eq. (\ref%
{eq:pretreat_covariates}). Given that the dimension of $\boldsymbol{X}$ is
as large as 15, we use simple specifications for the polynomials: 
\begin{equation*}
u_{K_{1}} (T) = [1, T, T^{2}]^{\top}, \quad v_{K_{2}} (\boldsymbol{X}) = [1, 
\boldsymbol{X}^{\top}]^{\top}
\end{equation*}
so that $K_{1} = 3$ and $K_{2} = 16$. 

We are interested in $\beta _{1}$ and $\beta _{2}$, the loadings of $T$ and $%
T^{2}$, respectively. Their point estimates and asymptotic 95\% confidence
bands are:%
\begin{equation*}
\hat{\beta}_{1}=-0.004,\quad CB(\beta _{1})=[-0.122,0.115]
\end{equation*}
\begin{equation*}
\hat{\beta}_{2}=1.5\times 10^{-7},\quad CB(\beta _{2})=[-1.2\times
10^{-5},1.3\times 10^{-5}].
\end{equation*}%
Neither $\hat{\beta}_{1}$ nor $\hat{\beta}_{2}$ is different from 0 at the
5\% level. Hence there do not exist statistically significant impacts of the
political advertisements on the level of campaign contributions $Y$.

\section{Conclusions \label{sec:conclusion}}

In this paper we present a weighted optimization framework that unifies the
binary, multi-valued, continuous and a mixture of discrete and continuous
treatment, under the condition of unconfounded treatment assignment. Under
this general framework, we first apply the result of \cite%
{Bickel_Klaassen_Ritov_Wellner_1993} to compute the semiparametric
efficiency bound for the causal effect of treatment under a general loss
function. We then propose a generalized optimization estimation with the
weights estimated by solving an expanding set of equations. These equations
impose restriction on the weights and extract valuable information about the
causal effect. Under some sufficient conditions, we establish consistency
and asymptotic normality of the generalized optimization estimator and show
that it attains the semiparametric efficiency bound. Since none of the
existing studies has investigated efficient estimation of the continuous
treatment model, our efficiency bound result and efficient estimation extend
the existing literature on the binary and multi-valued treatment to the
continuous treatment model.

There are several extensions worth pursuing in future projects. First,
estimation of the nonparametric causal effect function under general loss
function is not dealt with in this paper. But this is an important extension
since it removes the burden of parameterizing the causal effect. Second,
panel data are common in empirical literature. Our approach is readily
applicable to those data, though efficiency issue is more difficult. All
these extensions shall be taken up in future studies.

\singlespacing

\bibliographystyle{econometrica}
\bibliography{Semiparametric}
%%%%%%%%%%%%%%%%%%%%%%%%%%%%%%%%%%%%%%%%%%%%%%%%%%%%%%%%%%%%%%%%%%%%%%%

\appendix

\section{Appendix \label{sec:appendix}}
\subsection{Proof of \eqref{eq:CIA} \label{sec:proof_identification}}
Using the law of iterated expectation and Assumption \ref{as:TYindep}, we can deduce that
\begin{align*}
&\mathbb{E}\left[\pi_0 (T,\boldsymbol{X})L\left(Y-g(T;\boldsymbol{\beta})\right)\right]\\
=&\mathbb{E}\left[ \mathbb{E}[\pi (T,\boldsymbol{X})L(Y^{\ast
}(T)-g(T;\boldsymbol{\beta} ))|T,\boldsymbol{X}]\right] \\
=&\int \pi_0 (t,\boldsymbol{x})\cdot \mathbb{E}[L(Y^{\ast }(T)-g(T;\boldsymbol{\beta}
))|T=t,\boldsymbol{X}=\boldsymbol{x}]\ dF_{T|X}(t|\boldsymbol{x})dF_{X}(\boldsymbol{x}) \\
=&\int  \mathbb{E}\left[L (Y^{\ast }(t)-g(t;\boldsymbol{\beta} ))|T=t,\boldsymbol{X}=\boldsymbol{x}
\right] dF_{T}(t)dF_{X}(\boldsymbol{x})\\
=&\int  \mathbb{E}\left[ L(Y^{\ast }(t)-g(t;\boldsymbol{\beta} ))|\boldsymbol{X}=\boldsymbol{x}
\right] dF_{T}(t)dF_{X}(\boldsymbol{x})\quad (\text{using Assumption}\ \ref{as:TYindep})\\
=&\int {\small \mathbb{E}}\left[ L
(Y^{\ast }(t)-g(t;\boldsymbol{\beta}))\right] dF_{T}(t).
\end{align*}

\subsection{Asymptotic result when $\pi_0(T,\bold{X})$ is known \label{sec:pi_known}}
Suppose the stabilized weights $\pi_0(T,\bold{X})$ is known, the weighted optimization estimator of $\boldsymbol{\beta}_0$, denoted by $\boldsymbol{\beta}^*$, is
\begin{equation*}
\boldsymbol{\beta}^*=\min_{\beta }\sum_{i=1}^N \pi_0 (T_{i},\boldsymbol{X}_{i})L(Y
_{i} -g(T_{i};\beta )).
\end{equation*}
We also assume the asymptotic first order condition 
\begin{align}\label{eq:aymptotic_first_order_pi0}
\frac{1}{N}\sum_{i=1}^N\pi_0 (T_{i},\boldsymbol{X}_{i})L'(Y
_{i} -g(T_{i};{\boldsymbol{\beta}}^* ))m(T_i;{\boldsymbol{\beta}}^*)=o_{P}(N^{-1/2})\ 
\end{align}
holds with probability approaching to one.
\begin{prop}\label{prop:var_pi_known}
	Suppose Assumptions  \ref{as:solution},  \ref{as:EY2}, \ref{as:m_smooth} and \ref{as:entropy} hold, and \eqref{eq:aymptotic_first_order_pi0} holds, then we have
	\begin{enumerate}
		\item $\boldsymbol{\beta}^*\xrightarrow{p} \boldsymbol{\beta}_0$;
		\item $\sqrt{N}(\boldsymbol{\beta}^*-\boldsymbol{\beta}_0)\xrightarrow{d}\mathcal{N}(0,V_{ineff})$, where 
		$$V_{ineff}:= H_0^{-1}\cdot \mathbb{E}\left[ \pi_0 (T,\boldsymbol{X})^2L'(Y
		-g(T;\boldsymbol{\beta }_0 ))^2m(T;\boldsymbol{\beta }_0)m(T;\boldsymbol{\beta }_0)^{\top}\right]\cdot H_0^{-1} ; $$
		\item furthermore, if  $\mathbb{E}\left[L'(Y(t)-g(t;\boldsymbol{\beta}_0))\right]=0$ holds for all $t\in\mathcal{T}$, then  $V_{ineff}\geq  V_{eff}$ in the sense of that $c^{\top}\cdot  V_{ineff} \cdot c\geq  c^{\top} \cdot V_{eff} \cdot c$ for any vector $c\in\mathbb{R}^{p}$.
	\end{enumerate}
\end{prop}

\begin{proof}
	By Assumption \ref{as:EY2} and the uniform law of large number, we obtain
	\begin{equation*}
	\frac{1}{N}\sum_{i=1}^{N}\pi_0 (T_{i},\boldsymbol{X}_{i})L\left\{
	Y_{i}-g(T_{i};\boldsymbol{\beta})\right\} \rightarrow \mathbb{E}\left[\pi_0 (T,\boldsymbol{X}%
	)L\left\{ Y-g(T;\boldsymbol{\beta})\right\}\right] \text{in probability uniformly
		over} \ \boldsymbol{\beta},
	\end{equation*}
	then in light of Assumption \ref{as:solution} we can obtain the consistency result $\|\boldsymbol{\beta}^*- \boldsymbol{\beta}_0\|\xrightarrow{p}0$.
	
	The first order condition \eqref{eq:aymptotic_first_order_pi0}
	holds with probability approaching to one.	
	Note that $L'(\cdot)$ may not be a differentiable function, e.g. $L'(v)=\tau-I(v<0)$ in quantile regression, we cannot simply apply Mean Value Theorem on \eqref{eq:aymptotic_first_order_pi0} to obtain the expression for $\sqrt{N}(\boldsymbol{\beta}^*-\boldsymbol{\beta}_0)$. To solve this problem, we resort to the empirical process theory in \cite{andrews1994empirical}. Define 
	\begin{align*}
	f(\boldsymbol{\beta}):=\mathbb{E}\left[\pi_0 (T,\boldsymbol{X})L'(Y
	-g(T;\boldsymbol{\beta} ))m(T;\boldsymbol{\beta})\right],
	\end{align*} 	
	which is a differentiable function in $\boldsymbol{\beta}$ and by Assumption \ref{as:solution} $f(\boldsymbol{\beta}_0)=0$. Using Mean Value Theorem, we can obtain
	\begin{align*}
	0=\sqrt{N}f(\boldsymbol{\beta}_0)=\sqrt{N}f(\boldsymbol{\beta}^*)-\nabla_{\beta}f(\bar{\boldsymbol{\beta}})\cdot \sqrt{N}(\boldsymbol{\beta}^*-\boldsymbol{\beta}_0)\ ,
	\end{align*}
	where $\bar{\boldsymbol{\beta}}$ lies on the line joining ${\boldsymbol{\beta}}^*$ and $\boldsymbol{\beta}_0$. Because $\nabla_{\beta}f(\boldsymbol{\beta})$ is continuous in $\boldsymbol{\beta}$ at $\boldsymbol{\beta}_0$, and $\|{\boldsymbol{\beta}}^*-{\boldsymbol{\beta}}_0\|\xrightarrow{p}0$, then we have
	\begin{align*}
	\sqrt{N}(\boldsymbol{\beta}^*-\boldsymbol{\beta}_0)=&\left[\nabla_{\beta}f(\boldsymbol{\beta}_0)\right]^{-1}\cdot \sqrt{N}f(\boldsymbol{\beta}^*).
	\end{align*}
	Define the empirical process{
		\begin{align*}
		\nu_N(\boldsymbol{\beta})=\frac{1}{\sqrt{N}}\sum_{i=1}^N \left\{\pi_0 (T_{i},\boldsymbol{X}_{i})L'(Y
		_{i} -g(T_{i};\boldsymbol{\beta} ))m(T_i;\boldsymbol{\beta})-\mathbb{E}\left[\pi_0 (T,\boldsymbol{X})L'(Y
		-g(T;\boldsymbol{\beta} ))m(T;\boldsymbol{\beta})\right]\right\}\ .
		\end{align*}}
	By  \eqref{eq:aymptotic_first_order_pi0} and the definition of $\nu_N(\boldsymbol{\beta})$,   we  have{
		\begin{align*}
		\sqrt{N}(\boldsymbol{\beta}^*-\boldsymbol{\beta}_0)
		=&\nabla_{\beta}f(\boldsymbol{\beta}_0)^{-1}\cdot \Bigg\{ \sqrt{N}f(\boldsymbol{\beta}^*)-\frac{1}{\sqrt{N}}\sum_{i=1}^N\pi_0 (T_{i},\boldsymbol{X}_{i})L'(Y
		_{i} -g(T_{i};{\boldsymbol{\beta}}^* ))m(T_i;{\boldsymbol{\beta}}^*)\\
		&\qquad \qquad\qquad  +\frac{1}{\sqrt{N}}\sum_{i=1}^N\pi_0 (T_{i},\boldsymbol{X}_{i})L'(Y
		_{i} -g(T_{i};{\boldsymbol{\beta}}^* ))m(T_i;{\boldsymbol{\beta}}^*)\Bigg\}\\
		=&-\nabla_{\beta}f(\boldsymbol{\beta}_0)^{-1}\cdot \nu_N(\boldsymbol{\beta}^*)+o_p(1)\\
		=&H_0^{-1}\cdot \bigg\{\left(\nu_N(\boldsymbol{\beta}^*) -\nu_N(\boldsymbol{\beta}_0)\right) +\nu_N(\boldsymbol{\beta}_0)\bigg\} +o_p(1) \ .
		\end{align*}}
	{\color{black} By Assumptions \ref{as:m_smooth}, \ref{as:entropy}, Theorems 4 and 5 of \cite{andrews1994empirical}, we have that $\nu_N(\cdot)$ is stochastically equicontinuous, which implies $\nu_N(\boldsymbol{\beta}^*) -\nu_N(\boldsymbol{\beta}_0)\xrightarrow{p}0$}. Therefore,
	\begin{align*}
	\sqrt{N}(\boldsymbol{\beta}^*-\boldsymbol{\beta}_0)=H_0^{-1}\frac{1}{\sqrt{N}}\sum_{i=1}^N \pi_0 (T_{i},\boldsymbol{X}_{i})L'(Y
	_{i} -g(T_{i};\boldsymbol{\beta}_0 ))m(T_i;\boldsymbol{\beta}_0)+o_p(1)\ ,
	\end{align*}
	then we can conclude that the  asymptotic variance of $\sqrt{N}(\boldsymbol{\beta}^*-\boldsymbol{\beta}_0)$ is $V_{_{ineff}}$.
	
	We next show $V_{ineff}\geq V_{eff}$. From Theorem \ref{efficiency theorem}, we have{\small
		\begin{align*}
		&V_{eff}
		=H_0^{-1}\cdot \bigg\{\mathbb{E}\left[ \pi_0 (T,\boldsymbol{X})^2L'(Y
		-g(T;\boldsymbol{\beta}_0 ))^2m(T;\boldsymbol{\beta}_0)m(T;\boldsymbol{\beta}_0)^{\top}\right]\\
		&\qquad +\mathbb{E}\left[\mathbb{E}[\pi_0 (T,\boldsymbol{X})L'(Y
		-g(T;\boldsymbol{\beta}_0 ))m(T;\boldsymbol{\beta}_0)|T,\bold{X}]\cdot \mathbb{E}[\pi_0 (T,\boldsymbol{X})L'(Y
		-g(T;\boldsymbol{\beta}_0 ))m(T;\boldsymbol{\beta}_0)|T,\bold{X}]^{\top}\right]\\
		&\qquad +\mathbb{E}\left[\mathbb{E}[\pi_0 (T,\boldsymbol{X})L'(Y
		-g(T;\boldsymbol{\beta}_0 ))m(T;\boldsymbol{\beta}_0)|\bold{X}]\cdot \mathbb{E}[\pi_0 (T,\boldsymbol{X})L'(Y
		-g(T;\boldsymbol{\beta}_0 ))m(T;\boldsymbol{\beta}_0)|\bold{X}]^{\top}\right]\\
		&\qquad +\mathbb{E}\left[\mathbb{E}[\pi_0 (T,\boldsymbol{X})L'(Y
		-g(T;\boldsymbol{\beta}_0 ))m(T;\boldsymbol{\beta}_0)|T]\cdot \mathbb{E}[\pi_0 (T,\boldsymbol{X})L'(Y
		-g(T;\boldsymbol{\beta}_0 ))m(T;\boldsymbol{\beta}_0)|T]^{\top}\right]\\
		&\qquad-2\cdot \mathbb{E}\left[\mathbb{E}[\pi_0 (T,\boldsymbol{X})L'(Y
		-g(T;\boldsymbol{\beta}_0 ))m(T;\boldsymbol{\beta}_0)|T,\bold{X}]\cdot \pi_0 (T,\boldsymbol{X})L'(Y
		-g(T;\boldsymbol{\beta}_0 ))m(T;\boldsymbol{\beta}_0)^{\top}\right] \\
		&\qquad-2\cdot \mathbb{E}\left[\mathbb{E}[\pi_0 (T,\boldsymbol{X})L'(Y
		-g(T;\boldsymbol{\beta}_0 ))m(T;\boldsymbol{\beta}_0)|T,\bold{X}]\cdot \mathbb{E}[\pi_0 (T,\boldsymbol{X})L'(Y
		-g(T;\boldsymbol{\beta}_0 ))m(T;\boldsymbol{\beta}_0)|\bold{X}]^{\top} \right]\\
		&\qquad-2\cdot \mathbb{E}\left[\mathbb{E}[\pi_0 (T,\boldsymbol{X})L'(Y
		-g(T;\boldsymbol{\beta}_0 ))m(T;\boldsymbol{\beta}_0)|T,\bold{X}]\cdot \mathbb{E}[\pi_0 (T,\boldsymbol{X})L'(Y
		-g(T;\boldsymbol{\beta}_0 ))m(T;\boldsymbol{\beta}_0)|T]^{\top} \right]\\
		&\qquad+ 2\cdot \mathbb{E}\left[\pi_0 (T,\boldsymbol{X})L'(Y
		-g(T;\boldsymbol{\beta}_0 ))m(T;\boldsymbol{\beta}_0)\cdot \mathbb{E}[\pi_0 (T,\boldsymbol{X})L'(Y
		-g(T;\boldsymbol{\beta}_0 ))m(T;\boldsymbol{\beta}_0)|\bold{X}]^{\top}\right] \\
		&\qquad+ 2\cdot \mathbb{E}\left[\pi_0 (T,\boldsymbol{X})L'(Y
		-g(T;\boldsymbol{\beta}_0 ))m(T;\boldsymbol{\beta}_0)\cdot \mathbb{E}[\pi_0 (T,\boldsymbol{X})L'(Y
		-g(T;\boldsymbol{\beta}_0 ))m(T;\boldsymbol{\beta}_0)|T]^{\top}\right] \\
		&\qquad+2\cdot \mathbb{E}\left[\mathbb{E}[\pi_0 (T,\boldsymbol{X})L'(Y
		-g(T;\boldsymbol{\beta}_0 ))m(T;\boldsymbol{\beta}_0)|T]\cdot \mathbb{E}[\pi_0 (T,\boldsymbol{X})L'(Y
		-g(T;\boldsymbol{\beta}_0 ))m(T;\boldsymbol{\beta}_0)|\bold{X}]^{\top} \right]\bigg\} H_0^{-1}\\
		=& H_0^{-1} \bigg\{\mathbb{E}\left[ \pi_0 (T,\boldsymbol{X})^2L'(Y
		-g(T;\boldsymbol{\beta}_0 ))^2m(T;\boldsymbol{\beta}_0)m(T;\boldsymbol{\beta}_0)^{\top}\right]\\
		&\qquad -\mathbb{E}\left[\mathbb{E}[\pi_0 (T,\boldsymbol{X})L'(Y
		-g(T;\boldsymbol{\beta}_0 ))m(T;\boldsymbol{\beta}_0)|T,\bold{X}]\cdot \mathbb{E}[\pi_0 (T,\boldsymbol{X})L'(Y
		-g(T;\boldsymbol{\beta}_0 ))m(T;\boldsymbol{\beta}_0)|T,\bold{X}]^{\top}\right]\\
		&\qquad + \mathbb{E}\left[\mathbb{E}[\pi_0 (T,\boldsymbol{X})L'(Y
		-g(T;\boldsymbol{\beta}_0 ))m(T;\boldsymbol{\beta}_0)|\bold{X}]\cdot \mathbb{E}[\pi_0 (T,\boldsymbol{X})L'(Y
		-g(T;\boldsymbol{\beta}_0 ))m(T;\boldsymbol{\beta}_0)|\bold{X}]^{\top}\right]\bigg\} H_0^{-1},
		\end{align*}}
	where the last equality holds by noting  $$\mathbb{E}[\pi_0 (T,\boldsymbol{X})L'(Y
	-g(T;\boldsymbol{\beta}_0 ))m(T;\boldsymbol{\beta}_0)|T=t]=\mathbb{E}\left[L'(Y^*(t)-g(t;\boldsymbol{\beta}_0 ))\right]\cdot m(t;\boldsymbol{\beta}_0)=0\ ,$$ 
	since the model is correctly specified, i.e. $\mathbb{E}\left[L'(Y^*(t)-g(t;\boldsymbol{\beta}_0 ))\right]=0$ for $t\in\mathcal{T}$. Therefore, {\footnotesize
		\begin{align*}
		&V_{ineff}-V_{eff} \\
		=&H_0^{-1} 
		\bigg\{\mathbb{E}\left[\mathbb{E}[\pi_0 (T,\boldsymbol{X})L'(Y
		-g(T;\boldsymbol{\beta}_0 ))m(T;\boldsymbol{\beta}_0)|T,\bold{X}]\cdot \mathbb{E}[\pi_0 (T,\boldsymbol{X})L'(Y
		-g(T;\boldsymbol{\beta}_0 ))m(T;\boldsymbol{\beta}_0)|T,\bold{X}]^{\top}\right] \\
		& - \mathbb{E}\left[\mathbb{E}[\pi_0 (T,\boldsymbol{X})L'(Y
		-g(T;\boldsymbol{\beta}_0 ))m(T;\boldsymbol{\beta}_0)|\bold{X}]\cdot \mathbb{E}[\pi_0 (T,\boldsymbol{X})L'(Y
		-g(T;\boldsymbol{\beta}_0 ))m(T;\boldsymbol{\beta}_0)|\bold{X}]^{\top}\right]\bigg\}H_0^{-1}  \geq 0\ ,
		\end{align*}}
	where the last inequality holds by using Jensen's inequality:{\footnotesize
		\begin{align*}
		&\mathbb{E}\left[\mathbb{E}[\pi_0 (T,\boldsymbol{X})(Y
		-g(T;\boldsymbol{\beta}_0 ))m(T;\boldsymbol{\beta}_0)|\bold{X}]\cdot \mathbb{E}[\pi_0 (T,\boldsymbol{X})(Y
		-g(T;\boldsymbol{\beta}_0 ))m(T;\boldsymbol{\beta}_0)|\bold{X}]^{\top}\right]\\
		=&\mathbb{E}\left[\mathbb{E}\left[\mathbb{E}[\pi_0 (T,\boldsymbol{X})(Y
		-g(T;\boldsymbol{\beta}_0 ))m(T;\boldsymbol{\beta}_0)|T,\bold{X}]|\boldsymbol{X}\right]\cdot \mathbb{E}\left[\mathbb{E}[\pi_0 (T,\boldsymbol{X})(Y
		-g(T;\boldsymbol{\beta}_0 ))m(T;\boldsymbol{\beta}_0)|T,\bold{X}]|\boldsymbol{X}\right]^{\top}\right]\\
		< &\mathbb{E}\left[\mathbb{E}[\pi_0 (T,\boldsymbol{X})(Y
		-g(T;\boldsymbol{\beta}_0 ))m(T;\boldsymbol{\beta}_0)|T,\bold{X}]\cdot \mathbb{E}[\pi_0 (T,\boldsymbol{X})(Y
		-g(T;\boldsymbol{\beta}_0 ))m(T;\boldsymbol{\beta}_0)|T,\bold{X}]^{\top}\right].
		\end{align*}}
\end{proof}

\subsection{Discussion on sieve basis}\label{sec:sieve}

To construct our  estimator, we need to specify  $u_{K_1}(T)$ and $v_{K_2}(\boldsymbol{X})$. Although the approximation theory is derived for general sequences of approximating functions, the most common class of functions are power series, B-splines, and wavelets, see also \cite{Newey97} and \cite{chen2007large}. For example, we use the power series for demonstration.
Suppose the dimension of covariate $\boldsymbol{X}$ is $r\in\mathbb{N}$, namely $\boldsymbol{X}=(X_1,...,X_r)^{\top}$.  
Let $\lambda :=(\lambda _{1} ,\ldots  \lambda _{r})^{\top }$ be an $r$-dimensional vector of nonnegative integers (multi-indices), 
with norm $\vert \lambda \vert  :=\sum _{j =1}^{r}\lambda _{j}$. 
Let $(\lambda  (k))_{k =1}^{\infty }$ be a sequence that includes all distinct multi-indices and satisfies $\vert \lambda  (k)\vert  \leq \vert \lambda  (k +1)\vert $, 
and let $\boldsymbol{X}^{\lambda } :=\prod _{j =1}^{r}X_{j}^{\lambda _{j}}$. 
For a sequence $\lambda  (k)$ we consider the series function ${v}_{K_2}(\boldsymbol{X})=({v}_{K_2,1}(\boldsymbol{X}),...,v_{K_2,K_2}(\boldsymbol{X}))^{\top}$,  
where  ${v}_{K,k} (\boldsymbol{X}) =\boldsymbol{X}^{\lambda  (k)}$, $k\in\{1,...,K_2\}$. 
Similarly, $u_{K_1}(t)=(1,t,...,t^{K_1-1})^{\top}$.  With $u_{K_1}(t)$ and $v_{K_2}(\boldsymbol{x})$, we can approximate a suitable function $f:\mathbb{R}^{r+1}\to \mathbb{R}$
by $u_{K_1}(t)^{\top}\Lambda_{K_1\times K_2}v_{K_2}(\boldsymbol{x})$ for some deterministic matrix $\Lambda_{K_1\times K_2}\in\mathbb{R}^{K_1\times K_2}$. 

Notice that for non-degenerate matrices $B_{K_1\times K_1}$ and $C_{K_2\times K_2}$, we have  $${u}_{K_1}(t)^{\top}{\Lambda}_{K_1\times K_2}v_{K_2}(\boldsymbol{x})=u_{K_1}(t)^{\top}B^{\top}_{K_1\times K_1}(B_{K_1\times K_1}^{\top})^{-1}\Lambda_{K_1\times K_2}C_{K_2\times K_2}^{-1}C_{K_2\times K_2}v_{K_2}(\boldsymbol{x}).$$ Hence we can use  $\tilde{u}_{K_1}(t)=B_{K_1\times K_1}u_{K_1}(t)$ and $\tilde{v}_{K_2}(\boldsymbol{x})=C_{K_2\times K_2}v_{K_2}(\boldsymbol{x})$  as the new basis for approximation. In particular, we can choose $B_{K_1\times K_1}:=\mathbb{E}[u_{K_1}(T)u_{K_1}(T)^\top]^{-1/2}$ and $C_{K_2\times K_2}:=\mathbb{E}[v_{K_2}(\boldsymbol{ X})v_{K_2}(\boldsymbol{ X})^\top]^{-1/2}$, which are non-degenerate matrices by Assumption \ref{as:u&v}, such that the new basises $\tilde{u}_{K_1}(T)$ and $\tilde{v}_{K_2}(\boldsymbol{ X})$ are orthonormalized, namely, $\mathbb{E}[\tilde{u}_{K_1}(T)\tilde{u}_{K_1}^{\top}(T)^\top]=I_{K_1\times K_1}$, and $\mathbb{E}[\tilde{v}_{K_2}(\boldsymbol{X})\tilde{v}_{K_2}^{\top}(\boldsymbol{X})]=I_{K_2\times K_2}$. Therefore, without loss of generality, we always assume the original sieve basises ${u}_{K_1}(T)$ and ${v}_{K_2}(\boldsymbol{ X})$ are orthonormalized, 
\begin{align}\label{eq:orthonormal_basis}
\mathbb{E}\left[u_{K_1}(T)u_{K_1}^{\top}(T)\right]=I_{K_1\times K_1}, \ \mathbb{E}\left[v_{K_2}(\boldsymbol{X})v_{K_2}^{\top}(\boldsymbol{X})\right]=I_{K_2\times K_2}. 
\end{align}
In pratical application, we can construct othonormal basis by using empirical Gram-Schmidt on the original basis.

We use Frobenius norm $\|A\|:= \sqrt{\tr(AA^{\top})}$ for an arbitrary matrix $A$.  Define 
\begin{align*}
\zeta_1(K_1) := \sup_{t\in\mathcal{T}}\|u_{K_1}(t)\|, \
\zeta_2(K_2):= \sup_{\boldsymbol{x}\in\mathcal{X}}\|v_{K_2}(\boldsymbol{x})\|.
\end{align*}
This bound depends on the array of approximating functions that is used. \cite{Newey94, Newey97} showed that $ \zeta_1(K_1) \leq CK_1$, $\zeta_2(K_2)  \leq CK_2$ for power series, and $ \zeta_1(K_1) \leq C\sqrt{K_1}$, $\zeta_2(K_2)  \leq C\sqrt{K_2}$ for B-splines.

\subsection{Duality of primal problem \eqref{E:cm1} \label{sec:dual_appendix}}
Let
$$ \boldsymbol{\pi}:= \left(\pi_1,\ldots,\pi_N\right)^\top, \  f(\boldsymbol{\pi}):= \sum_{i=1}^N\pi_i\log\pi_i\ , $$
the primal problem \eqref{E:cm1} can be written as:
\begin{align}\label{E:cm2}
\left\{\begin{array}{cc}
&\min f(\boldsymbol{\pi}) \\[2mm]
&\text{subject to} \ N^{-1}\sum_{i=1}^N\pi_i{u}_{K_1,k}(T_i){v}_{K_2,k'}(\bold{X}_i) =\bar{u}_{K_1,k}\cdot \bar{v}_{K_2,k'}\ ,\\[2mm] 
&k\in\{1,...,K_1\}\ , k'\in\{1,...,K_2\}\ , 
\end{array} \right.
\end{align}
where
\begin{align*}
\bar{u}_{K_1,k}:=\frac{1}{N}\sum_{j=1}^N{u}_{K_1,k}(T_j) \ , \  \bar{v}_{K_2,k'}:=\frac{1}{N}\sum_{j=1}^N{v}_{K_2,k'}(\bold{X}_j)\ ,
\end{align*}
and $u_{K_1,k}(T)$ (resp. $v_{K_2,k'}$) is the $k^{th}$ (resp. ${k'}^{th}$) component of $u_{K_1}(T)$ (resp. $v_{K_2}(\bold{X})$). Let $m_{K_1K_2}(T,\bold{X})$ be a $K=K_1\cdot K_2$ dimensional vector function whose elements are $\{u_{K_1,k}(T)v_{K_2,k'}(\bold{X});k=1,...,K_1, k'=1,...,K_2\}$,  and we define
$$E_{{K_1K_2}\times N}:= \left(m_{K_1K_2}(T_1,\bold{X}_1),\ldots,m_{K_1K_2}(T_N,\bold{X}_N)\right).$$ Let  $b_{K_1K_2}$ be a $K=K_1\cdot K_2$ dimensional vector function whose elements are $\{\bar{u}_{K_1,k}\bar{v}_{K_2,k'};k=1,...,K_1, k'=1,...,K_2\}$. The optimization problem  (\ref{E:cm2}) becomes
\begin{align}\label{E:cm3}
\left\{\begin{array}{cc}
&\min_{\boldsymbol{\pi}}  f(\boldsymbol{\pi}) \\[2mm]
&\text{subject to} ~~~~  E_{{K_1K_2}\times N} \cdot\boldsymbol{\pi} =N\cdot b_{K1K_2}
\end{array}\right..
\end{align}
By \cite{tseng1991relaxation}, the conjugate convex function of $f(\cdot)$ to be
\begin{align*}
f^*(\boldsymbol{z})=&\sup_{\boldsymbol{\pi}}\sum_{i=1}^N\left\{z_i\pi_i-\pi_i\log\pi_i \right\}=\sum_{i=1}^N\left\{z_i\pi^*_i-\pi^*_i\log\pi^*_i\right\}\ ,
\end{align*}
where $\pi_j^*$ satisfies the first order condition:
\begin{align*}
z_j=\log\pi_j^* +1 \Rightarrow \pi_j^*=e^{z_j-1}\ ;
\end{align*}
then we have
\begin{align*}
f^*(\boldsymbol{z})=&\sum_{i=1}^N\left\{z_ie^{z_i-1}-e^{z_i-1}(z_i-1)\right\}=\sum_{i=1}^Ne^{z_i-1}=\sum_{i=1}^N-\rho(-z_i) \ ,
\end{align*}
where $\rho\left(z_j\right):= -e^{-z_j-1}$. By \cite{tseng1991relaxation}, the dual problem of (\ref{E:cm3}) is
\begin{align}
&\max_{\lambda\in \mathbb{R}^{K}} \left\{\lambda^{\top}\left(N\cdot b_{K1K_2}\right)- f^*\left(\lambda^\top E_{K_1 K_2,N}\right)\right\} \notag\\
=&\max_{\Lambda\in \mathbb{R}^{K_1\times K_2}} \sum_{j=1}^N \left\{ \bar{u}_{K_1}^{\top}\Lambda\bar{v}_{K_2}+\rho\left(-u_{K_1}(T_j)^{\top}\Lambda v_{K_2}(\bold{X}_j)\right)\right\}\notag\\
=&\max_{\Lambda\in \mathbb{R}^{K_1\times K_2}} \sum_{j=1}^N\left\{ \rho \left(u_{K_1}(T_j)^{\top}\Lambda v_K(\bold{X}_j)\right)- \bar{u}_{K_1}^{\top}\Lambda\bar{v}_{K_2}\right\}\notag\\
=&\max_{\Lambda\in \mathbb{R}^{K_1\times K_2}} \hat{G}_{K_1\times K_2}(\Lambda),\label{def:Ghat}
\end{align}
where $E_{K_1K_2,N}^{(j)}$ is the $j^{th}$ column of $E_{K_1K_2\times N}$ and
\begin{align*}
G_{K_1\times K_2}(\Lambda)=\frac{1}{N}\sum_{j=1}^N\rho (u_{K_1}(T_j)^{\top}\Lambda v_K(\bold{X}_j))-\bar{u}_{K_1}^{\top}\Lambda\bar{v}_{K_2}.
\end{align*}
Therefore, the dual solution of \eqref{E:cm1} is given by
\begin{align}\label{def:pihat}
\hat{\pi}_K(T_i,\bold{X}_i)=\rho'\left({u}_{K_1}(T_i)^{\top}\hat{\Lambda}_{K_1\times K_2}v_{K_2}(\bold{X}_i)\right)\ ,
\end{align}
where $\hat{\Lambda}_{K_1\times K_2}$ is the maximizer of the strictly concave objective function $\hat{G}_{K_1\times K_2}$.

\subsection{Proof of \eqref{appendix:integration_by_parts}}\label{appendix:proof_integration}
\begin{align*}
&\nabla _{\beta }\mathbb{E}\left[ L^{\prime }(Y-g(T;\boldsymbol{\beta }
))|T=t,\boldsymbol{X}=\boldsymbol{x}\right] \Big|_{\boldsymbol{\beta }=%
	\boldsymbol{\beta }_{0}} \notag \\
=&\nabla _{\beta }\left[\int_{\mathbb{R}}L'(y-g(t;\boldsymbol{ \beta }))f_{Y|T,X}(y|t,\boldsymbol{x})dy\right]\Big|_{\boldsymbol{\beta }=%
	\boldsymbol{\beta }_{0}} \\
=& \nabla _{\beta }\left[\int_{\mathbb{R}}L'(z)f_{Y|T,X}(z+g(t;\boldsymbol{ \beta })|t,\boldsymbol{x})dz\right]\Big|_{\boldsymbol{\beta }=%
	\boldsymbol{\beta }_{0}}\  (\text{use}\ z=y-g(t;\boldsymbol{ \beta })) \\
=& \int_{\mathbb{R}}L'(z)\cdot \frac{\partial}{\partial y}f_{Y|T,X}(z+g(t;\boldsymbol{ \beta }_0)|t,\boldsymbol{x})dz\cdot m(t;\boldsymbol{ \beta }_0)\\
=& \int_{\mathbb{R}}L'(y-g(t;\boldsymbol{ \beta }))\cdot \frac{\partial}{\partial y}f_{Y|T,X}(y|t,\boldsymbol{x})dy\cdot m(t;\boldsymbol{ \beta }_0) \\
=& \int_{\mathbb{R}}L'(y-g(t;\boldsymbol{ \beta }))\cdot \frac{\frac{\partial}{\partial y}f_{Y|T,X}(y|t,\boldsymbol{x})}{f_{Y|T,X}(y|t,\boldsymbol{x})}f_{Y|T,X}(y|t,\boldsymbol{x})dy\cdot m(t;\boldsymbol{ \beta }_0) \\
=& \int_{\mathbb{R}}L'(y-g(t;\boldsymbol{ \beta }))\cdot \frac{\frac{\partial}{\partial y}f_{Y,T,X}(y,t,\boldsymbol{x})}{f_{Y,T,X}(y,t,\boldsymbol{x})}f_{Y|T,X}(y|t,\boldsymbol{x})dy\cdot m(t;\boldsymbol{ \beta }_0) \\
=& \mathbb{E}\left[ L^{\prime }(Y-g(T;\boldsymbol{\beta }_{0}))\frac{\frac{%
		\partial }{\partial y}f_{Y,T,\boldsymbol{X}}(Y,T,\boldsymbol{X})}{f_{Y,T,%
		\boldsymbol{X}}(Y,T,\boldsymbol{X})}\bigg|T=t,\boldsymbol{X}=\boldsymbol{x}%
\right] m(t;\boldsymbol{\beta }_{0}).
\end{align*}

\subsection{A variance estimator when $L''$ exists}\label{sec:variance_L''}
We present another simple variance estimator if the loss function $L(\cdot)$ is twice differentiable.
By definition, $\hat{\Lambda}_{K_1\times K_2}$ and $\hat{\boldsymbol{\beta}}$ satisfy the following equation:{
	\begin{align}\label{eq:first_order}
	\left\{
	\begin{array}{ll}
	&\hat{G}'_{K_1\times K_2}(\hat{\Lambda}_{K_1\times K_2})=\tilde{G}'_{K_1\times K_2}(\hat{\Lambda}_{K_1\times K_2})-r_N=0\ , \\[2mm]
	& \frac{1}{N}\sum_{i=1}^N\rho'\left(u_{K_1}(T_i)^\top\hat{\Lambda}_{K_1\times K_2} v_{K_2}(\boldsymbol{X}_i)\right)m(T_i;\hat{\boldsymbol{\beta}})L'(Y_i-g(T_i;\hat{\boldsymbol{\beta}}))=0\ .
	\end{array}
	\right.
	\end{align}}
where 
\begin{align*}
\tilde{G}_{K_1\times K_2}(\Lambda):=&\frac{1}{N}\sum_{i=1}^N\Bigg\{\rho\left(u_{K_1}(T_i)^\top{\Lambda} v_{K_2}(\boldsymbol{X}_i)\right)-\mathbb{E}[u_{K_1}(T)]^\top\Lambda v_{K_2}(\boldsymbol{X}_i)\\
&\qquad\qquad  -u_{K_1}(T_i)^\top\Lambda  \mathbb{E}[v_{K_2}(\boldsymbol{X})] + \mathbb{E}\left[u_{K_1}(T)\right]^\top\Lambda  \mathbb{E}[v_{K_2}(\boldsymbol{X})]\Bigg\}, \\
r_N:=&\left(\frac{1}{N}\sum_{i=1}^Nu_{K_1}(T_i)-\mathbb{E}[u_{K_1}(T)]\right)\left(\frac{1}{N}\sum_{i=1}^Nv_{K_2}(\boldsymbol{X}_i)-\mathbb{E}[v_{K_2}(\boldsymbol{X})]\right)^\top.
\end{align*}
By computing the second moment and  using Chebyshev's inequality, we can obtain that $r_N=O_p(\sqrt{K}/N)$. We can write the matrix equation \eqref{eq:first_order} in a vector form by introducing the following notation:{\small
	\begin{align*}
	&{\boldsymbol{\theta}}^*=\left(({\Lambda}^*_{K_1\times K_2}e_1)^{\top},...,({\Lambda}^*_{K_1\times K_2}e_{K_2})^{\top},{\boldsymbol{\beta}}^\top_0\right)^\top,\\ &\hat{\boldsymbol{\theta}}=\left((\hat{\Lambda}_{K_1\times K_2}e_1)^{\top},...,(\hat{\Lambda}_{K_1\times K_2}e_{K_2})^{\top},\hat{\boldsymbol{\beta}}^\top\right)^\top,\\
	&\varphi_j(T,\boldsymbol{X};\Lambda)=\Big\{\rho'\left(u_{K_1}(T)^\top\Lambda v_{K_2}(\boldsymbol{X})\right)u_{K_1}(T)v_{K_2}^\top(\boldsymbol{X})-\mathbb{E}[u_{K_1}(T)]\cdot v_{K_2}(\boldsymbol{X})\\
	&\qquad \qquad \qquad \quad    -u_{K_1}(T)\cdot \mathbb{E}[v_{K_2}(\boldsymbol{X})] + \mathbb{E}\left[u_{K_1}(T)\right]\cdot \mathbb{E}[v_{K_2}(\boldsymbol{X})]\Big\}e_j\ ,\  j\in\{1,...,K_2\}, \\
	&h(T,\boldsymbol{X},Y;\boldsymbol{\theta})=\left(\varphi_1(T,\boldsymbol{X};\Lambda)^\top,...,\varphi_{K_2}(T,\boldsymbol{X};\Lambda)^\top, \rho'\left(u_{K_1}(T)^\top{\Lambda}  v_{K_2}(\boldsymbol{X}_i)\right)L'(Y-g(T;{\boldsymbol{\beta}}))m(T;{\boldsymbol{\beta}})^\top\right)^\top,
	\end{align*}}
where $e_{j}$ is the $K_2$-dimensional column vector whose $j^{\text{th}}$ component is $1$ while other components are all of $0$'s. Then \eqref{eq:first_order} becomes  
\begin{align}\label{eq:variance}
\frac{1}{N}\sum_{i=1}^Nh(T_i,\boldsymbol{X}_i,Y_i;\hat{\boldsymbol{\theta}})=O_p(\sqrt{K}/N).
\end{align}
Note that $\{\hat{\boldsymbol{\beta}}-{\boldsymbol{\beta}}_0\}$ is the last $p$-components of $\{\hat{\boldsymbol{\theta}}-\boldsymbol{\theta}^*\}$. Let $\boldsymbol{0}_{p\times K}$ (resp. $I_{p\times p}$) denote a $p\times K$ (resp. $p\times p$) zero (resp. identity) matrix.  By Mean Value Theorem we can have that
\begin{align*}
&\sqrt{N}(\hat{\boldsymbol{\beta}}-{\boldsymbol{\beta}}_0)=\sqrt{N}\cdot \left(\boldsymbol{0}_{p\times K},I_{p\times p}\right)(\hat{\boldsymbol{\theta}}-\boldsymbol{\theta}^*)\\ 
=&\left(\boldsymbol{0}_{p\times K},I_{p\times p}\right) \cdot \mathbb{E}\left[\nabla_{\theta}h(T,\boldsymbol{X},Y;{\boldsymbol{\theta}}^*)\right]^{-1}\cdot \left[-\frac{1}{\sqrt{N}}\sum_{i=1}^Nh(T_i,\boldsymbol{X}_i,Y_i;{\boldsymbol{\theta}}^*)\right]+o_p(1)
\end{align*}
where $\tilde{\boldsymbol{\theta}}$ lies on the line joining $\hat{\boldsymbol{\theta}}$ and ${\boldsymbol{\theta}}^*$, the summands $h(T_i,\boldsymbol{X}_i;{\boldsymbol{\theta}}^*)$ are i.i.d. with mean zero. Then we have
\begin{align*}
&V_{eff}=\lim_{N}Var\left(\sqrt{N}(\hat{\boldsymbol{\beta}}-{\boldsymbol{\beta}}_0)\right)\\
=&\lim_{N}\left(\boldsymbol{0}_{p\times K},I_{p\times p}\right)\cdot \mathbb{E}\left[\nabla_{\theta}h(T,\boldsymbol{X},Y;{\boldsymbol{\theta}}^*)\right]^{-1}\cdot \mathbb{E}\left[h(T,\boldsymbol{X},Y;{\boldsymbol{\theta}}^*)h(T,\boldsymbol{X},Y;{\boldsymbol{\theta}}^*)^\top\right]\\
&\qquad \qquad \qquad \quad \cdot \mathbb{E}\left[\nabla_{\theta}h(T,\boldsymbol{X},Y;{\boldsymbol{\theta}}^*)^\top\right]^{-1}\left(\boldsymbol{0}_{p\times K},I_{p\times p}\right)^\top\ .
\end{align*}
Based on above expression, a simple plug-in estimator for the efficient variance is defined by
\begin{align*}
\hat{V}=&\left(\boldsymbol{0}_{p\times K},I_{p\times p}\right)  \left[\frac{1}{N}\sum_{i=1}^N\nabla_{\theta}h(T_i,\boldsymbol{X}_i,Y_i;\hat{\boldsymbol{\theta}})\right]^{-1}\left[\frac{1}{N}\sum_{i=1}^Nh(T_i,\boldsymbol{X}_i,Y_i;\hat{\boldsymbol{\theta}})h(T_i,\boldsymbol{X}_i,Y_i;\hat{\boldsymbol{\theta}})^\top\right]\\
&\qquad\qquad\qquad \times \left[\frac{1}{N}\sum_{i=1}^N\nabla_{\theta}h(T_i,\boldsymbol{X}_i,Y_i;\hat{\boldsymbol{\theta}})^\top\right]^{-1} \left(\boldsymbol{0}_{p\times K},I_{p\times p}\right)^\top.
\end{align*}
Recall ${\Lambda}^*_{K_1\times K_2}$ defined in \eqref{def:Lambda*},  Lemma 4.2 of supplemental material shows that $\|\hat{\Lambda}_{K_1\times K_2}-{\Lambda}^*_{K_1\times K_2}\|=O_p\left(\sqrt{K/N}\right)$. The proof $\hat{V}\xrightarrow{p} V_{eff}$ is a standard argument through the use of Chebyshev's inequality together with the fact $\|\hat{\boldsymbol{\theta}}-{\boldsymbol{\theta}}^*\|\leq \|\hat{\Lambda}_{K_1\times K_2}-{\Lambda}^*_{K_1\times K_2}\|+\|\hat{\boldsymbol{\beta}}-\boldsymbol{\beta}_0\|=O_p\left(\sqrt{K/N}\right)+O_p(N^{-1/2})=O_p\left(\sqrt{K/N}\right)$.

\begin{table}[th] 
	\begin{center} 
		\caption{Simulation Results on $\beta_{1}$ (DGP-1) \label{table:sim_results_beta1_DGP1}} 
		{\fontsize{9.5pt}{16.5pt} \selectfont 
			\begin{tabular}{c|cccc|cccc|cccc}   \hline  \hline 
				\multicolumn{13}{c}{$\rho = 0.0$} \\ \hline  
				&   \multicolumn{4}{c|}{$N = 100$}   &   \multicolumn{4}{c|}{$N = 500$}    &    \multicolumn{4}{c}{$N = 1000$}    \\ \hline
				&    Bias    &   Stdev    &   RMSE    &   CP   &     Bias    &   Stdev    &   RMSE    &   CP  &   Bias    &   Stdev    &   RMSE  &  CP    \\ \hline
				SW ($K_{2} = 9$)  &  0.019    &   0.866   & 0.866   &  0.872  &  0.019    &   0.527    &   0.523       &  0.951    &  -0.004   &  0.544  &  0.544  &  0.937   \\  
				SW ($K_{2} = 15$)  &  0.093    &  0.979   &  0.983   & 0.862     &  0.017   &  0.666   &  0.666   &  0.961   &  0.008   &   0.630  &  0.630   & 0.980   \\
				CBGPS  &  -0.012    &  0.590    &  0.590  &  0.940  &  -0.009    &   0.276    &   0.276       &  0.943    & 0.006    &  0.198  &  0.198  &  0.944   \\    \hline \hline 
				\multicolumn{13}{c}{$\rho = 0.2$} \\ \hline  
				&   \multicolumn{4}{c|}{$N = 100$}   &   \multicolumn{4}{c|}{$N = 500$}    &    \multicolumn{4}{c}{$N = 1000$}    \\ \hline
				&    Bias    &   Stdev    &   RMSE    &   CP   &     Bias    &   Stdev    &   RMSE    &   CP  &   Bias    &   Stdev    &   RMSE  &  CP    \\ \hline
				SW ($K_{2} = 9$)  &  0.011    &  0.934    &  0.934  &  0.874  & -0.010     &   0.620    &   0.620       &  0.947    &   0.019  &  0.561  & 0.561   &  0.946   \\  
				SW ($K_{2} = 15$)  & -0.018     &  1.200   &  1.200   &  0.872    &   -0.005  &  0.633   &   0.633  &   0.978  & -0.004    &  0.641   &   0.641  & 0.976   \\
				CBGPS  &  0.013    &  0.643    &  0.643  &  0.937  & 0.002     &  0.296     &   0.296       &   0.930   &   0.000  &  0.211  & 0.211   &  0.927   \\    \hline \hline 
				\multicolumn{13}{c}{$\rho = 0.4$} \\ \hline  
				&   \multicolumn{4}{c|}{$N = 100$}   &   \multicolumn{4}{c|}{$N = 500$}    &    \multicolumn{4}{c}{$N = 1000$}    \\ \hline
				&    Bias    &   Stdev    &   RMSE    &   CP   &     Bias    &   Stdev    &   RMSE    &   CP  &   Bias    &   Stdev    &   RMSE  &  CP    \\ \hline
				SW ($K_{2} = 9$) &  0.031    &   1.012   &  1.012  &  0.855  & 0.002     &   0.642    &  0.642        &   0.927   &   -0.021  &  0.637  &  0.638  &  0.942   \\  
				SW ($K_{2} = 15$)  & 0.013     &  0.975   &  0.975   &   0.876   &  0.029   &  0.693   &   0.694  &  0.963   & -0.022    & 0.666    &  0.667   &  0.980  \\
				CBGPS  & 0.001     &  0.634    &  0.634  &  0.944  &  -0.001    &  0.318     &   0.318       &  0.924    &  0.009   &  0.227  &  0.227  &  0.916   \\    \hline \hline 
			\end{tabular} 
		} 
	\end{center} 
	{\fontsize{9.5pt}{12pt} \selectfont 
		The DGP is $T_{i} = X_{1i} + X_{2i} + 0.2 X_{3i} + 0.2 X_{4i} + \xi_{i}$ and $Y_{i} = 1 + X_{1i} + 0.1 X_{2i} + 0.1 X_{3i} + 0.1 X_{4i} + T_{i} + \epsilon_{i}$.
		The covariates follow $\boldsymbol{X}_{i} \stackrel{i.i.d.}{\sim} N(0, \boldsymbol{\Sigma})$, where the diagonal elements of $\boldsymbol{\Sigma}$ are all 1 and the off-diagonal elements are all $\rho \in \{ 0.0, 0.2, 0.4 \}$.
		We report the bias, standard deviation, root mean squared error, and coverage probability based on the 95\% confidence band across $J = 1000$ Monte Carlo samples. 
		'SW' signifies our own stabilized-weight estimator, while 'CBGPS' signifies Fong, Hazlett, and Imai's \citeyearpar{Fong_Hazlett_Imai_2017} parametric covariate balancing generalized propensity score (CBGPS) estimator.
		For SW, the polynomials are $u_{K_{1}} (t_{i}) = [1, t_{i}, t_{i}^{2}]^{\top}$ (i.e. $K_{1} = 3$) and $u_{K_{2}} (\boldsymbol{X}_{i}) = [1, X_{1i}, X_{2i}, X_{3i}, X_{4i}, X_{1i}^{2}, X_{2i}^{2}, X_{3i}^{2}, X_{4i}^{2}]^{\top}$ (i.e. $K_{2} = 9$)
		or $u_{K_{2}} (\boldsymbol{X}_{i}) = [1, X_{1i}, X_{2i}, X_{3i}, X_{4i}, X_{1i}^{2}, X_{2i}^{2}, X_{3i}^{2}, X_{4i}^{2}, X_{1i} X_{2i}, X_{1i} X_{3i}, X_{1i} X_{4i}, X_{2i} X_{3i}, X_{2i} X_{4i}, X_{3i} X_{4i}]^{\top}$ (i.e. $K_{2} = 15$).
		For CBGPS, covariates are chosen to be $\boldsymbol{X}_{i} = [X_{1i}, X_{2i}, X_{3i}, X_{4i}]^{\top}$ for the first step of estimating stabilized weights
		and $\boldsymbol{Z}_{i} = [1, T_{i}, \boldsymbol{X}_{i}^{\top}]^{\top}$ for the second step of estimating average treatment effects.
	}
\end{table}

\clearpage

\begin{table}[th] 
	\begin{center} 
		\caption{Simulation Results on $\beta_{2}$ (DGP-1) \label{table:sim_results_beta2_DGP1}} 
		{\fontsize{9.5pt}{16.5pt} \selectfont 
			\begin{tabular}{c|cccc|cccc|cccc}   \hline  \hline 
				\multicolumn{13}{c}{$\rho = 0.0$} \\ \hline  
				&   \multicolumn{4}{c|}{$N = 100$}   &   \multicolumn{4}{c|}{$N = 500$}    &    \multicolumn{4}{c}{$N = 1000$}    \\ \hline
				&    Bias    &   Stdev    &   RMSE    &   CP   &     Bias    &   Stdev    &   RMSE    &   CP  &   Bias    &   Stdev    &   RMSE  &  CP    \\ \hline
				SW ($K_{2} = 9$) & 0.023     &  0.291    &  0.292  &  0.853  &  0.016    &   0.189    &   0.189       &  0.928    &  0.017   &  0.167  & 0.168   &  0.922   \\  
				SW ($K_{2} = 15$)  &  0.036    &  0.316   &  0.318   &   0.819   &  0.022   & 0.201    &  0.202   & 0.969    &   0.035  &  0.186   & 0.189    &  0.979  \\
				CBGPS  &  0.001    &  0.205    &  0.205  &  0.922  &  -0.007    &   0.099    &  0.099        &  0.914    & -0.001    & 0.072   & 0.072   &  0.906   \\    \hline \hline 
				\multicolumn{13}{c}{$\rho = 0.2$} \\ \hline  
				&   \multicolumn{4}{c|}{$N = 100$}   &   \multicolumn{4}{c|}{$N = 500$}    &    \multicolumn{4}{c}{$N = 1000$}    \\ \hline
				&    Bias    &   Stdev    &   RMSE    &   CP   &     Bias    &   Stdev    &   RMSE    &   CP  &   Bias    &   Stdev    &   RMSE  &  CP    \\ \hline
				SW ($K_{2} = 9$) &   0.031   &   0.290   &  0.292  & 0.846   & 0.031     &  0.186     &    0.188      &   0.926   &   0.027  &  0.174  &  0.176  &  0.917   \\  
				SW ($K_{2} = 15$)  & 0.050     &  0.313   &  0.317   &   0.814   &  0.040   &   0.202  &  0.206   &  0.964   & 0.046    &  0.190   & 0.195   &  0.982  \\
				CBGPS  &  -0.013    &  0.219    & 0.219   & 0.911   &  0.002    &  0.106     &    0.106      & 0.890     &  0.000   &  0.083  &  0.083  &  0.909   \\    \hline \hline 
				\multicolumn{13}{c}{$\rho = 0.4$} \\ \hline  
				&   \multicolumn{4}{c|}{$N = 100$}   &   \multicolumn{4}{c|}{$N = 500$}    &    \multicolumn{4}{c}{$N = 1000$}    \\ \hline
				&    Bias    &   Stdev    &   RMSE    &   CP   &     Bias    &   Stdev    &   RMSE    &   CP  &   Bias    &   Stdev    &   RMSE  &  CP    \\ \hline
				SW ($K_{2} = 9$) &   0.044   &  0.287    &  0.290  &  0.841  &  0.038    &   0.190    &   0.194       &  0.916    &  0.038   &  0.167  & 0.172   &  0.923   \\  
				SW ($K_{2} = 15$)  &   0.050   &  0.300   &  0.304   &   0.811   &  0.042   &  0.201   & 0.205    &  0.955   &  0.037   & 0.188    & 0.192    & 0.975   \\
				CBGPS  &   0.015   &  0.227    & 0.227   &  0.895  &  0.007    &  0.115     &     0.115     &   0.911   &  -0.002   &  0.087  &  0.087  &  0.882   \\    \hline \hline 
			\end{tabular} 
		} 
	\end{center} 
	{\fontsize{9.5pt}{12pt} \selectfont 
		The DGP is $T_{i} = X_{1i} + X_{2i} + 0.2 X_{3i} + 0.2 X_{4i} + \xi_{i}$ and $Y_{i} = 1 + X_{1i} + 0.1 X_{2i} + 0.1 X_{3i} + 0.1 X_{4i} + T_{i} + \epsilon_{i}$.
		The covariates follow $\boldsymbol{X}_{i} \stackrel{i.i.d.}{\sim} N(0, \boldsymbol{\Sigma})$, where the diagonal elements of $\boldsymbol{\Sigma}$ are all 1 and the off-diagonal elements are all $\rho \in \{ 0.0, 0.2, 0.4 \}$.
		We report the bias, standard deviation, root mean squared error, and coverage probability based on the 95\% confidence band across $J = 1000$ Monte Carlo samples. 
		'SW' signifies our own stabilized-weight estimator, while 'CBGPS' signifies Fong, Hazlett, and Imai's \citeyearpar{Fong_Hazlett_Imai_2017} parametric covariate balancing generalized propensity score (CBGPS) estimator.
		For SW, the polynomials are $u_{K_{1}} (t_{i}) = [1, t_{i}, t_{i}^{2}]^{\top}$ (i.e. $K_{1} = 3$) and $u_{K_{2}} (\boldsymbol{X}_{i}) = [1, X_{1i}, X_{2i}, X_{3i}, X_{4i}, X_{1i}^{2}, X_{2i}^{2}, X_{3i}^{2}, X_{4i}^{2}]^{\top}$ (i.e. $K_{2} = 9$)
		or $u_{K_{2}} (\boldsymbol{X}_{i}) = [1, X_{1i}, X_{2i}, X_{3i}, X_{4i}, X_{1i}^{2}, X_{2i}^{2}, X_{3i}^{2}, X_{4i}^{2}, X_{1i} X_{2i}, X_{1i} X_{3i}, X_{1i} X_{4i}, X_{2i} X_{3i}, X_{2i} X_{4i}, X_{3i} X_{4i}]^{\top}$ (i.e. $K_{2} = 15$).
		For CBGPS, covariates are chosen to be $\boldsymbol{X}_{i} = [X_{1i}, X_{2i}, X_{3i}, X_{4i}]^{\top}$ for the first step of estimating stabilized weights
		and $\boldsymbol{Z}_{i} = [1, T_{i}, \boldsymbol{X}_{i}^{\top}]^{\top}$ for the second step of estimating average treatment effects.
	}
\end{table}

\clearpage

\begin{table}[th] 
	\begin{center} 
		\caption{Simulation Results on $\beta_{1}$ (DGP-2) \label{table:sim_results_beta1_DGP4}} 
		{\fontsize{9.5pt}{16.5pt} \selectfont 
			\begin{tabular}{c|cccc|cccc|cccc}   \hline  \hline 
				\multicolumn{13}{c}{$\rho = 0.0$} \\ \hline  
				&   \multicolumn{4}{c|}{$N = 100$}   &   \multicolumn{4}{c|}{$N = 500$}    &    \multicolumn{4}{c}{$N = 1000$}    \\ \hline
				&    Bias    &   Stdev    &   RMSE    &   CP   &     Bias    &   Stdev    &   RMSE    &   CP  &   Bias    &   Stdev    &   RMSE  &  CP    \\ \hline
				SW ($K_{2} = 9$) &   -0.100    &   0.954   &  0.960  &  0.847  &  -0.041    &    0.639   &    0.640      &  0.919    &  0.002   &  0.572  & 0.572   & 0.931    \\ 
				SW ($K_{2} = 15$)  &  -0.105    & 1.050    &  1.055   &  0.852    &   0.010  &  0.690   & 0.690    & 0.956    &  0.072  &  0.716  & 0.720  &  0.975  \\
				CBGPS  &   -0.207   &   0.627   &  0.661  &  0.902  &  -0.173    &   0.268    &    0.319      &  0.869    &   -0.174  &  0.181  &  0.251  &  0.817   \\    \hline \hline 
				\multicolumn{13}{c}{$\rho = 0.2$} \\ \hline  
				&   \multicolumn{4}{c|}{$N = 100$}   &   \multicolumn{4}{c|}{$N = 500$}    &    \multicolumn{4}{c}{$N = 1000$}    \\ \hline
				&    Bias    &   Stdev    &   RMSE    &   CP   &     Bias    &   Stdev    &   RMSE    &   CP  &   Bias    &   Stdev    &   RMSE  &  CP    \\ \hline
				SW ($K_{2} = 9$) &   -0.029   &   1.004   & 1.004   &  0.847  &  -0.008    & 0.651      &   0.651       & 0.908     &   0.018  & 0.593   & 0.593   &  0.910   \\  
				SW ($K_{2} = 15$)  &  -0.109    &  1.126   &  1.131   &  0.839    &  0.004   &  0.727   &  0.727   &   0.965  &  0.033   &  0.675   &  0.676   &  0.956  \\
				CBGPS  &   -0.171   &   0.641   &  0.664  &  0.912  &  -0.181    &   0.260    &   0.317       & 0.882     &  -0.176   &  0.188  &  0.258  &   0.821  \\    \hline \hline 
				\multicolumn{13}{c}{$\rho = 0.4$} \\ \hline  
				&   \multicolumn{4}{c|}{$N = 100$}   &   \multicolumn{4}{c|}{$N = 500$}    &    \multicolumn{4}{c}{$N = 1000$}    \\ \hline
				&    Bias    &   Stdev    &   RMSE    &   CP   &     Bias    &   Stdev    &   RMSE    &   CP  &   Bias    &   Stdev    &   RMSE  &  CP    \\ \hline
				SW ($K_{2} = 9$) &   -0.057   &    0.969  &  0.970  &  0.839  &   -0.029   &  0.641     &  0.641        & 0.914     &   0.016  &  0.605  &  0.605  &  0.923   \\  
				SW ($K_{2} = 15$)  & -0.043     &  1.091   &  1.092   & 0.846     &  0.047   &  0.726   &  0.728   &  0.935   &  0.081   &  0.734   &  0.738   &  0.958  \\
				CBGPS  &  -0.162    &   0.652   &  0.672  & 0.920   &  -0.180    &  0.266     &  0.322        & 0.893     &   -0.182  &  0.194  &  0.266  &   0.800  \\    \hline \hline 
			\end{tabular} 
		} 
	\end{center} 
	{\fontsize{9.5pt}{12pt} \selectfont 
		The DGP is $T_{i} = (X_{1i} + 0.5)^{2} + 0.4 X_{2i} + 0.4 X_{3i} + 0.4 X_{4i} + \xi_{i}$ and $Y_{i} = 0.75 X_{1i}^{2} + 0.2 (X_{2i} + 0.5)^{2} + T_{i} + \epsilon_{i}$.
		The covariates follow $\boldsymbol{X}_{i} \stackrel{i.i.d.}{\sim} N(0, \boldsymbol{\Sigma})$, where the diagonal elements of $\boldsymbol{\Sigma}$ are all 1 and the off-diagonal elements are all $\rho \in \{ 0.0, 0.2, 0.4 \}$.
		We report the bias, standard deviation, root mean squared error, and coverage probability based on the 95\% confidence band across $J = 1000$ Monte Carlo samples. 
		'SW' signifies our own stabilized-weight estimator, while 'CBGPS' signifies Fong, Hazlett, and Imai's \citeyearpar{Fong_Hazlett_Imai_2017} parametric covariate balancing generalized propensity score (CBGPS) estimator.
		For SW, the polynomials are $u_{K_{1}} (t_{i}) = [1, t_{i}, t_{i}^{2}]^{\top}$ (i.e. $K_{1} = 3$) and $u_{K_{2}} (\boldsymbol{X}_{i}) = [1, X_{1i}, X_{2i}, X_{3i}, X_{4i}, X_{1i}^{2}, X_{2i}^{2}, X_{3i}^{2}, X_{4i}^{2}]^{\top}$ (i.e. $K_{2} = 9$)
		or $u_{K_{2}} (\boldsymbol{X}_{i}) = [1, X_{1i}, X_{2i}, X_{3i}, X_{4i}, X_{1i}^{2}, X_{2i}^{2}, X_{3i}^{2}, X_{4i}^{2}, X_{1i} X_{2i}, X_{1i} X_{3i}, X_{1i} X_{4i}, X_{2i} X_{3i}, X_{2i} X_{4i}, X_{3i} X_{4i}]^{\top}$ (i.e. $K_{2} = 15$).
		For CBGPS, covariates are chosen to be $\boldsymbol{X}_{i} = [X_{1i}, X_{2i}, X_{3i}, X_{4i}]^{\top}$ for the first step of estimating stabilized weights
		and $\boldsymbol{Z}_{i} = [1, T_{i}, \boldsymbol{X}_{i}^{\top}]^{\top}$ for the second step of estimating average treatment effects.
	}
\end{table}

\clearpage

\begin{table}[th] 
	\begin{center} 
		\caption{Simulation Results on $\beta_{2}$ (DGP-2) \label{table:sim_results_beta2_DGP4}} 
		{\fontsize{9.5pt}{16.5pt} \selectfont 
			\begin{tabular}{c|cccc|cccc|cccc}  \hline  \hline 
				\multicolumn{13}{c}{$\rho = 0.0$} \\ \hline  
				&   \multicolumn{4}{c|}{$N = 100$}   &   \multicolumn{4}{c|}{$N = 500$}    &    \multicolumn{4}{c}{$N = 1000$}    \\ \hline
				&    Bias    &   Stdev    &   RMSE    &   CP   &     Bias    &   Stdev    &   RMSE    &   CP  &   Bias    &   Stdev    &   RMSE  &  CP    \\ \hline
				SW ($K_{2} = 9$) &   0.033   &  0.288    & 0.290   &  0.810  &   0.038   &   0.187    &   0.191       & 0.913     &   0.040  & 0.165   &  0.170  &  0.930   \\   
				SW ($K_{2} = 15$)  &  0.045    &  0.293   &  0.300   &  0.822    &  0.028   &  0.197   &   0.199  & 0.959    &  0.044   &  0.180   &  0.185   &  0.976  \\
				CBGPS  &  0.121    &   0.189   & 0.224   &  0.814  &  0.146    &  0.090     &    0.172      &  0.470    &   0.151  &  0.069  &  0.166  &  0.269   \\    \hline \hline 
				\multicolumn{13}{c}{$\rho = 0.2$} \\ \hline  
				&   \multicolumn{4}{c|}{$N = 100$}   &   \multicolumn{4}{c|}{$N = 500$}    &    \multicolumn{4}{c}{$N = 1000$}    \\ \hline
				&    Bias    &   Stdev    &   RMSE    &   CP   &     Bias    &   Stdev    &   RMSE    &   CP  &   Bias    &   Stdev    &   RMSE  &  CP    \\ \hline
				SW ($K_{2} = 9$) &  0.030    &   0.280   &  0.282  &  0.801  &   0.025   &    0.173   &  0.175        &  0.905    & 0.025    &  0.162  &  0.164  &  0.918   \\  
				SW ($K_{2} = 15$)  &  0.051    &  0.425   &  0.428   &  0.779    & 0.027    &  0.189   &  0.191   &  0.956   & 0.041    &   0.182  &   0.187  & 0.975   \\
				CBGPS  & 0.123     &  0.203    &  0.237  &  0.805  &   0.146   &    0.096   &     0.174     &  0.488    &  0.152   &  0.071  &  0.168  &   0.279  \\    \hline \hline 
				\multicolumn{13}{c}{$\rho = 0.4$} \\ \hline  
				&   \multicolumn{4}{c|}{$N = 100$}   &   \multicolumn{4}{c|}{$N = 500$}    &    \multicolumn{4}{c}{$N = 1000$}    \\ \hline
				&    Bias    &   Stdev    &   RMSE    &   CP   &     Bias    &   Stdev    &   RMSE    &   CP  &   Bias    &   Stdev    &   RMSE  &  CP    \\ \hline
				SW ($K_{2} = 9$) &  0.048    &   0.273   &  0.277  &  0.821  &  0.041    &   0.178    &   0.182       &  0.910    &   0.038  & 0.170   & 0.174   &  0.916   \\  
				SW ($K_{2} = 15$)  & 0.046     &  0.301   &  0.304   &  0.789    &   0.025  &  0.192   & 0.194    & 0.946    &   0.039  &   0.182  &  0.186   &  0.965  \\
				CBGPS  & 0.121     &  0.209    &  0.242  &  0.811  &  0.145    &   0.098    &    0.176      & 0.514     &  0.156   &  0.080  &  0.176  &  0.304   \\    \hline \hline 
			\end{tabular} 
		} 
	\end{center} 
	{\fontsize{9.5pt}{12pt} \selectfont 
		The DGP is $T_{i} = (X_{1i} + 0.5)^{2} + 0.4 X_{2i} + 0.4 X_{3i} + 0.4 X_{4i} + \xi_{i}$ and $Y_{i} = 0.75 X_{1i}^{2} + 0.2 (X_{2i} + 0.5)^{2} + T_{i} + \epsilon_{i}$.
		The covariates follow $\boldsymbol{X}_{i} \stackrel{i.i.d.}{\sim} N(0, \boldsymbol{\Sigma})$, where the diagonal elements of $\boldsymbol{\Sigma}$ are all 1 and the off-diagonal elements are all $\rho \in \{ 0.0, 0.2, 0.4 \}$.
		We report the bias, standard deviation, root mean squared error, and coverage probability based on the 95\% confidence band across $J = 1000$ Monte Carlo samples. 
		'SW' signifies our own stabilized-weight estimator, while 'CBGPS' signifies Fong, Hazlett, and Imai's \citeyearpar{Fong_Hazlett_Imai_2017} parametric covariate balancing generalized propensity score (CBGPS) estimator.
		For SW, the polynomials are $u_{K_{1}} (t_{i}) = [1, t_{i}, t_{i}^{2}]^{\top}$ (i.e. $K_{1} = 3$) and $u_{K_{2}} (\boldsymbol{X}_{i}) = [1, X_{1i}, X_{2i}, X_{3i}, X_{4i}, X_{1i}^{2}, X_{2i}^{2}, X_{3i}^{2}, X_{4i}^{2}]^{\top}$ (i.e. $K_{2} = 9$)
		or $u_{K_{2}} (\boldsymbol{X}_{i}) = [1, X_{1i}, X_{2i}, X_{3i}, X_{4i}, X_{1i}^{2}, X_{2i}^{2}, X_{3i}^{2}, X_{4i}^{2}, X_{1i} X_{2i}, X_{1i} X_{3i}, X_{1i} X_{4i}, X_{2i} X_{3i}, X_{2i} X_{4i}, X_{3i} X_{4i}]^{\top}$ (i.e. $K_{2} = 15$).
		For CBGPS, covariates are chosen to be $\boldsymbol{X}_{i} = [X_{1i}, X_{2i}, X_{3i}, X_{4i}]^{\top}$ for the first step of estimating stabilized weights
		and $\boldsymbol{Z}_{i} = [1, T_{i}, \boldsymbol{X}_{i}^{\top}]^{\top}$ for the second step of estimating average treatment effects.
	}
\end{table}

\clearpage

\begin{table}[th] 
	\begin{center} 
		\caption{Empirical Results Based on Fong, Hazlett, and Imai's \citeyearpar{Fong_Hazlett_Imai_2018} CBGPS Approach \label{table:empirical_result_CBGPS}} 
		{\fontsize{10.0pt}{16.5pt} \selectfont 
			\begin{tabular}{c|c|c|c}   \hline  \hline 
				&  Covariates   &    Parameter of $T_{i}$    &   Parameter of  $T_{i}^{2}$    \\ \hline  
				Case \#1     &    $\boldsymbol{Z}_{i} = [T_{i}, T_{i}^{2}, 1]^{\top}$      &      $\begin{matrix}   0.088  \\  [-0.804, 0.981]  \end{matrix}$    &   $\begin{matrix}  -8.5 \times 10^{-6}   \\  [-5.4 \times 10^{-5}, 3.7 \times 10^{-5}]   \end{matrix}$    \\ \hline
				Case \#2     &    $\boldsymbol{Z}_{i} = [T_{i}, T_{i}^{2}, \boldsymbol{SD}_{i}^{\top}]^{\top}$    &   $\begin{matrix}  1.333   \\ [0.462, 2.204]  \end{matrix}$    &      $\begin{matrix}   -8.6 \times 10^{-5}   \\ [-1.3 \times 10^{-4}, -4.6 \times 10^{-5}]  \end{matrix}$    \\ \hline
				Case \#3     &    $\boldsymbol{Z}_{i} = [T_{i}, T_{i}^{2}, 1, \boldsymbol{X}_{1i}^{\top}]^{\top}$    &   $\begin{matrix}   -0.545  \\ [-1.373, 0.284]   \end{matrix}$   &     $\begin{matrix}  -2.2 \times 10^{-5}   \\ [-6.6 \times 10^{-5}, 2.1 \times 10^{-5}]   \end{matrix}$  \\ \hline
				Case \#4  & $\boldsymbol{Z}_{i} = [T_{i}, T_{i}^{2}, \boldsymbol{SD}_{i}^{\top}, \boldsymbol{X}_{1i}^{\top}]^{\top}$ & $\begin{matrix} -0.216 \\  [-1.044, 0.611] \end{matrix}$ & $\begin{matrix} 2.7 \times 10^{-5} \\  [-1.4 \times 10^{-5}, 6.8 \times 10^{-5}] \end{matrix}$ \\ \hline \hline
			\end{tabular} 
		} 
	\end{center} 
	{\fontsize{9.5pt}{12pt} \selectfont 
		$\boldsymbol{X}_{1i}$ is a vector of eight covariates used in the generalized propensity score model (cf. Eq. (\ref{eq:pretreat_covariates_first})). 
		$\boldsymbol{SD}_{i} = [SD_{1i}, SD_{2i}, \dots, SD_{24i}]^{\top}$, where $SD_{ji}$ is a binary indicator that equals 1 if zip code $i$ belongs to state $j$ and equals 0 otherwise.
		Any zip code contained in the dataset belongs to one and only one of 24 states.
		In this table we report the CBGPS estimates for the parameters of $T_{i}$ and $T_{i}^{2}$ as well as their asymptotic 95\% confidence bands.
	}
\end{table}

\clearpage

\renewcommand{\thesubfigure}{}
\captionsetup[subfigure]{labelformat=simple, labelsep=period}

\clearpage

%%%%%% Kaiji Motegi added the following three lines in order to handle references %%%%
%\singlespacing
%
%\bibliographystyle{dcu}
%\bibliography{Semiparametric}
%%%%%%%%%%%%%%%%%%%%%%%%%%%%%%%%%%%%%%%%%%%%%%%%%%%%%%%%%%%%%%%%%%%%%%%%

%\bibliographystyle{plain}
%\setcitestyle{aysep={},yysep={;}}
%\bibliography{../bib/privault}

\end{document}

\title{Supplemental Material for \\ ``\textit{A Unified Framework for Efficient Estimation of General
		Treatment Models }''}
\author{ \ \ Chunrong Ai\thanks{
Department of Economics, University of Florida.
E-mail: \texttt{chunrong.ai@warrington.ufl.edu}} -- University of Florida \\
Oliver Linton\thanks{
Faculty of Economics, University of Cambridge.
	E-mail: \texttt{obl20@cam.ac.uk}} -- University of Cambridge \\
Kaiji Motegi\thanks{
Graduate School of Economics, Kobe University. E-mail: \texttt{motegi@econ.kobe-u.ac.jp}} -- Kobe University \\
Zheng Zhang\thanks{
Institute of Statistics and Big Data, Renmin University of China. E-mail: \texttt{zhengzhang@ruc.edu.cn}} -- Renmin University of China}
\date{{\large This draft:} \today
}
\maketitle

\setstretch{1.3}

\newpage
\tableofcontents
\listoftables
%\listoffigures

%%%%%%%%%%%%%%%%%%%%%%%%%%%%%%%%%%%%%%%%%%%%%%%%%%%%%%%%%%%%%%%%%%%%%%%%%%%%%%%%%%%%%%

\newpage

\section{Complete Simulation Results}

In this supplemental material, we present complete simulation results that are partly omitted in the main paper \cite{Ai_Motegi_Zhang_cts_treat}.
See Section \ref{sec:sim_design} for simulation designs and Section \ref{sec:sim_results} for results.

\subsection{Simulation Design \label{sec:sim_design}}

We consider four data generating processes (DGPs):
\begin{description}
\setlength{\leftskip}{1.0cm}
\item[DGP-1] $T_{i} = X_{1i} + X_{2i} + 0.2 X_{3i} + 0.2 X_{4i} + \xi_{i}$ and $Y_{i} = 1 + X_{1i} + 0.1 X_{2i} + 0.1 X_{3i} + 0.1 X_{4i} + T_{i} + \epsilon_{i}$.
\item[DGP-2] $T_{i} = (X_{1i} + 0.5)^{2} + 0.4 X_{2i} + 0.4 X_{3i} + 0.4 X_{4i} + \xi_{i}$ and $Y_{i} = 1 + X_{1i} + 0.1 X_{2i} + 0.1 X_{3i} + 0.1 X_{4i} + T_{i} + \epsilon_{i}$.
\item[DGP-3] $T_{i} = X_{1i} + X_{2i} + 0.2 X_{3i} + 0.2 X_{4i} + \xi_{i}$ and $Y_{i} = 0.75 X_{1i}^{2} + 0.2 (X_{2i} + 0.5)^{2} + T_{i} + \epsilon_{i}$.
\item[DGP-4] $T_{i} = (X_{1i} + 0.5)^{2} + 0.4 X_{2i} + 0.4 X_{3i} + 0.4 X_{4i} + \xi_{i}$ and $Y_{i} = 0.75 X_{1i}^{2} + 0.2 (X_{2i} + 0.5)^{2} + T_{i} + \epsilon_{i}$.
\end{description}
For each DGP, $\xi_{i} \stackrel{i.i.d.}{\sim} N(0, 9)$, $\epsilon_{i} \stackrel{i.i.d.}{\sim} N(0, 25)$, 
and $\boldsymbol{X}_{i} = [X_{1i}, X_{2i}, X_{3i}, X_{4i}]^{T} \stackrel{i.i.d.}{\sim} N(0, \boldsymbol{\Sigma})$, where the diagonal elements of $\boldsymbol{\Sigma}$ are all 1 and the off-diagonal elements are all $\rho \in \{ 0.0, 0.2, 0.4 \}$.
We draw $J = 1000$ Monte Carlo samples with sample size $N \in \{  100, 500, 1000 \}$. 

Our DGPs have similar structures with \cite{Fong_Hazlett_Imai_2018}.
The characteristics of the four DGPs can be summarized as follows.
\begin{description}
\setlength{\leftskip}{1.0cm}
\item[DGP-1] $T_{i}$ is linear in $\boldsymbol{X}_{i}$; $Y_{i}$ is linear in $\boldsymbol{X}_{i}$ and $T_{i}$.
\item[DGP-2] $T_{i}$ is nonlinear in $\boldsymbol{X}_{i}$; $Y_{i}$ is linear in $\boldsymbol{X}_{i}$ and $T_{i}$.
\item[DGP-3] $T_{i}$ is linear in $\boldsymbol{X}_{i}$; $Y_{i}$ is nonlinear in $\boldsymbol{X}_{i}$ and linear in $T_{i}$.
\item[DGP-4] $T_{i}$ is nonlinear in $\boldsymbol{X}_{i}$; $Y_{i}$ is nonlinear in $\boldsymbol{X}_{i}$ and linear in $T_{i}$. 
\end{description}

In the main paper \cite{Ai_Motegi_Zhang_cts_treat}, we discuss only DGP-1 and DGP-4 in order to save space. 
(DGP-4 here is called ``DGP-2'' in the main paper since DGP-2 and DGP-3 here are skipped.)
As shown in the main paper, a parametric version of Fong, Hazlett, and Imai's \citeyearpar{Fong_Hazlett_Imai_2018} covariate balancing generalized propensity score (CBGPS) estimators 
produces no bias under DGP-1 and considerable bias under DGP-4.\footnote{
A non-parametric version of the CBGPS estimators is also biased under a data generating process which is similar to DGP-4,
and the magnitude of the bias is as large as the parametric version \citep[cf. Figure 2,][]{Fong_Hazlett_Imai_2018}.
}
Here we also consider DGP-2 and DGP-3, which serve as intermediate cases where \textit{either} $T_{i}$ is nonlinear in $\boldsymbol{X}_{i}$ \textit{or} $Y_{i}$ is nonlinear in $\boldsymbol{X}_{i}$.

Since $Y_{i}$ is linear in $T_{i}$ under all four DGPs, the link function is always of the form $\mathbb{E} [Y^*(t)] = \beta_{1} + \beta_{2} t$.
It is straightforward to see that the true coefficients are $(\beta_{1}, \beta_{2}) = (1, 1)$ for each DGP.  
$\beta_{2}$ is of greater interest than $\beta_{1}$ since $\beta_{2}$ measures the average treatment effect.

To estimate $(\beta_{1}, \beta_{2})$ via our proposed method, we should decide which variables to include in polynomials with respect to $T_{i}$ and $\boldsymbol{X}_{i}$.
For each DGP, we use $u_{K_{1}} (t_{i}) = [1, t_{i}, t_{i}^{2}]^{T}$ (i.e. $K_{1} = 3$) and $u_{K_{2}} (\boldsymbol{X}_{i}) = [1, X_{1i}, X_{2i}, X_{3i}, X_{4i}, X_{1i}^{2}, X_{2i}^{2}, X_{3i}^{2}, X_{4i}^{2}]^{T}$ (i.e. $K_{2} = 9$)
or $u_{K_{2}} (\boldsymbol{X}_{i}) = [1, X_{1i}, X_{2i}, X_{3i}, X_{4i}, X_{1i}^{2}, X_{2i}^{2}, X_{3i}^{2}, X_{4i}^{2}, X_{1i} X_{2i}, X_{1i} X_{3i}, X_{1i} X_{4i}, X_{2i} X_{3i}, X_{2i} X_{4i}, X_{3i} X_{4i}]^{T}$ (i.e. $K_{2} = 15$).
In the literature of non-parametric estimation of treatment effects, it is common to include the first and second moments of $\boldsymbol{X}_{i}$ with or without cross terms \citep[see e.g.][]{chan2016globally}.
Hence we consider two cases with and without the cross terms of $\boldsymbol{X}_{i}$.

We compare our estimators with the parametric version of Fong, Hazlett, and Imai's \citeyearpar{Fong_Hazlett_Imai_2018} CBGPS estimator.
The non-parametric version of CBGPS is omitted since the parametric and non-parametric versions exhibit similar performance according to the simulation results of \cite{Fong_Hazlett_Imai_2018}.
Computation of the parametric CBGPS estimator involves two steps. 
The first step is the estimation of stabilized weights, and the second step is the estimation of average treatment effects.
Our covariates are chosen to be $\boldsymbol{X}_{i} = [X_{1i}, X_{2i}, X_{3i}, X_{4i}]^{T}$ for the first step 
and $\boldsymbol{Z}_{i} = [1, T_{i}, \boldsymbol{X}_{i}^{T}]^{T}$ for the second step.\footnote{
In extra simulations not reported here, we added some irrelevant covariates to the model, as \cite{Fong_Hazlett_Imai_2018} did, in order to see whether the finite sample performance is sensitive to redundant covariates.
Since the results were not so sensitive to the inclusion of irrelevant covariates, we only report results without the irrelevant covariates here.
}

\subsection{Simulation Results \label{sec:sim_results}}

In this section we present our simulation results.
See Tables \ref{table:sim_results_beta1_DGP1}-\ref{table:sim_results_beta2_DGP1} for DGP-1;
Tables \ref{table:sim_results_beta1_DGP2}-\ref{table:sim_results_beta2_DGP2} for DGP-2;
Tables \ref{table:sim_results_beta1_DGP3}-\ref{table:sim_results_beta2_DGP3} for DGP-3;
Tables \ref{table:sim_results_beta1_DGP4}-\ref{table:sim_results_beta2_DGP4} for DGP-4.
For each table, we report the bias, standard deviation, root mean squared error, and coverage probability based on the 95\% confidence bands.

As discussed in the main paper \cite{Ai_Motegi_Zhang_cts_treat}, both of our stabilized-weight (SW) estimator and Fong, Hazlett, and Imai's \citeyearpar{Fong_Hazlett_Imai_2018} CBGPS estimator are unbiased under DGP-1.
The latter often has smaller bias and standard deviation across all sample sizes $N \in \{  100, 500, 1000 \}$.
That is not a surprising result since DGP-1 satisfies the linearity assumption underpinning the CBGPS estimator.
The coverage probability, however, is sometimes comparable between the two estimators.
See, for example, Table \ref{table:sim_results_beta2_DGP1} ($\beta_{2}$) with $\rho = 0.2$ and $N = 500$,
where the coverage probability is 0.926 for the SW estimator with $K_{2} = 9$ and 0.890 for the CBGPS estimator. 
The simulation results are not sensitive to the correlation coefficient $\rho \in \{ 0.0, 0.2, 0.4 \}$.
The SW estimators with and without the cross terms of $\boldsymbol{X}_{i}$ lead to roughly similar performance.

Similar implications hold for DGP-2.
The CBGPS estimator performs well arguably due to a negligibly small degree of nonlinearity.
There is a nonlinear term $(X_{1i} + 0.5)^{2}$ in the DGP of $T_{i}$, but the DGP of $Y_{i}$ is kept to be linear.
Hence the CBGPS estimator maintains the sharp performance as in DGP-1.

Under DGP-3, the SW estimator begins to be more comparable with the CBGPS estimator.
See, for example, Table \ref{table:sim_results_beta2_DGP3} ($\beta_{2}$) with $\rho = 0.4$ and $N = 1000$,
where the bias is 0.003 for the SW estimator with $K_{2} = 9$ and 0.004 for the CBGPS estimator;
the standard deviation is 0.174 for SW and 0.090 for CBGPS;
the coverage probability is 0.932 for SW and 0.892 for CBGPS.
Under DGP-3, there are two nonlinear terms $0.75 X_{1i}^{2}$ and $0.2 (X_{2i} + 0.5)^{2}$ in the DGP of $Y_{i}$.
It is likely that the CBGPS estimator is negatively affected by those nonlinear features while the SW estimator is more robust against nonlinearity. 

As discussed in the main paper \cite{Ai_Motegi_Zhang_cts_treat}, our SW estimator dominates the CBGPS estimator under DGP-4.
See, for example, Table \ref{table:sim_results_beta1_DGP4} ($\beta_{1}$) with $\rho = 0.4$ and $N = 1000$,
where the bias is 0.016 for the SW estimator with $K_{2} = 9$ and -0.182 for the CBGPS estimator;
the standard deviation is 0.605 for SW and 0.194 for CBGPS;
the coverage probability is 0.923 for SW and 0.800 for CBGPS.
Also see Table \ref{table:sim_results_beta2_DGP4} ($\beta_{2}$) with $\rho = 0.4$ and $N = 1000$,
where the bias is 0.038 for SW and 0.156 for CBGPS;
the standard deviation is 0.170 for SW and 0.080 for CBGPS;
the coverage probability is 0.916 for SW and 0.304 for CBGPS.

Under DGP-4, the CBGPS estimator keeps producing negative bias in $\beta_{1}$ and positive bias in $\beta_{2}$.
The bias is considerably large, and it does not vanish as sample size $N$ increases.
Our estimator, in contrast, produces virtually no bias for any sample size $N \in \{  100, 500, 1000 \}$.
It also achieves accurate coverage probability for larger sample sizes $N \in \{  500, 1000 \}$.
The relative advantage of our estimator stems from the strong degree of nonlinearity contained in DGP-4, where both $T_{i}$ and $Y_{i}$ have nonlinear terms. 

In summary, our estimator is by construction robust against the functional form of underlying DGPs, 
while Fong, Hazlett, and Imai's \citeyearpar{Fong_Hazlett_Imai_2018} estimator is sensitive to model misspecification.
In fact, our approach produces unbiased estimates under all DGPs considered, 
while Fong, Hazlett, and Imai's \citeyearpar{Fong_Hazlett_Imai_2018} approach produces biased estimates under DGP-4.
In reality, the functional form of an underlying DGP is unknown to the researcher.
It is therefore of practical use to employ our estimator in order to accomplish correct inference for dose-response curves.   

\clearpage

\begin{table}[th] 
\begin{center} 
\caption{Simulation Results on $\beta_{1}$ (DGP-1) \label{table:sim_results_beta1_DGP1}} 
{\fontsize{9.5pt}{16.5pt} \selectfont 
\begin{tabular}{c|cccc|cccc|cccc}   \hline  \hline 
\multicolumn{13}{c}{$\rho = 0.0$} \\ \hline  
       &   \multicolumn{4}{c|}{$N = 100$}   &   \multicolumn{4}{c|}{$N = 500$}    &    \multicolumn{4}{c}{$N = 1000$}    \\ \hline
      &    Bias    &   Stdev    &   RMSE    &   CP   &     Bias    &   Stdev    &   RMSE    &   CP  &   Bias    &   Stdev    &   RMSE  &  CP    \\ \hline
SW ($K_{2} = 9$)  &  0.019    &   0.866   & 0.866   &  0.872  &  0.019    &   0.527    &   0.523       &  0.951    &  -0.004   &  0.544  &  0.544  &  0.937   \\  
SW ($K_{2} = 15$)  &  0.093    &  0.979   &  0.983   & 0.862     &  0.017   &  0.666   &  0.666   &  0.961   &  0.008   &   0.630  &  0.630   & 0.980   \\
CBGPS  &  -0.012    &  0.590    &  0.590  &  0.940  &  -0.009    &   0.276    &   0.276       &  0.943    & 0.006    &  0.198  &  0.198  &  0.944   \\    \hline \hline 
\multicolumn{13}{c}{$\rho = 0.2$} \\ \hline  
       &   \multicolumn{4}{c|}{$N = 100$}   &   \multicolumn{4}{c|}{$N = 500$}    &    \multicolumn{4}{c}{$N = 1000$}    \\ \hline
      &    Bias    &   Stdev    &   RMSE    &   CP   &     Bias    &   Stdev    &   RMSE    &   CP  &   Bias    &   Stdev    &   RMSE  &  CP    \\ \hline
SW ($K_{2} = 9$)  &  0.011    &  0.934    &  0.934  &  0.874  & -0.010     &   0.620    &   0.620       &  0.947    &   0.019  &  0.561  & 0.561   &  0.946   \\  
SW ($K_{2} = 15$)  & -0.018     &  1.200   &  1.200   &  0.872    &   -0.005  &  0.633   &   0.633  &   0.978  & -0.004    &  0.641   &   0.641  & 0.976   \\
CBGPS  &  0.013    &  0.643    &  0.643  &  0.937  & 0.002     &  0.296     &   0.296       &   0.930   &   0.000  &  0.211  & 0.211   &  0.927   \\    \hline \hline 
\multicolumn{13}{c}{$\rho = 0.4$} \\ \hline  
      &   \multicolumn{4}{c|}{$N = 100$}   &   \multicolumn{4}{c|}{$N = 500$}    &    \multicolumn{4}{c}{$N = 1000$}    \\ \hline
      &    Bias    &   Stdev    &   RMSE    &   CP   &     Bias    &   Stdev    &   RMSE    &   CP  &   Bias    &   Stdev    &   RMSE  &  CP    \\ \hline
SW ($K_{2} = 9$) &  0.031    &   1.012   &  1.012  &  0.855  & 0.002     &   0.642    &  0.642        &   0.927   &   -0.021  &  0.637  &  0.638  &  0.942   \\  
SW ($K_{2} = 15$)  & 0.013     &  0.975   &  0.975   &   0.876   &  0.029   &  0.693   &   0.694  &  0.963   & -0.022    & 0.666    &  0.667   &  0.980  \\
CBGPS  & 0.001     &  0.634    &  0.634  &  0.944  &  -0.001    &  0.318     &   0.318       &  0.924    &  0.009   &  0.227  &  0.227  &  0.916   \\    \hline \hline 
\end{tabular} 
} 
\end{center} 
{\fontsize{9.5pt}{12pt} \selectfont 
The DGP is $T_{i} = X_{1i} + X_{2i} + 0.2 X_{3i} + 0.2 X_{4i} + \xi_{i}$ and $Y_{i} = 1 + X_{1i} + 0.1 X_{2i} + 0.1 X_{3i} + 0.1 X_{4i} + T_{i} + \epsilon_{i}$.
The covariates follow $\boldsymbol{X}_{i} \stackrel{i.i.d.}{\sim} N(0, \boldsymbol{\Sigma})$, where the diagonal elements of $\boldsymbol{\Sigma}$ are all 1 and the off-diagonal elements are all $\rho \in \{ 0.0, 0.2, 0.4 \}$.
We report the bias, standard deviation, root mean squared error, and coverage probability based on the 95\% confidence band across $J = 1000$ Monte Carlo samples. 
'SW' signifies our own stabilized-weight estimator, while 'CBGPS' signifies Fong, Hazlett, and Imai's \citeyearpar{Fong_Hazlett_Imai_2018} parametric covariate balancing generalized propensity score (CBGPS) estimator.
For SW, the polynomials are $u_{K_{1}} (t_{i}) = [1, t_{i}, t_{i}^{2}]^{T}$ (i.e. $K_{1} = 3$) and $u_{K_{2}} (\boldsymbol{X}_{i}) = [1, X_{1i}, X_{2i}, X_{3i}, X_{4i}, X_{1i}^{2}, X_{2i}^{2}, X_{3i}^{2}, X_{4i}^{2}]^{T}$ (i.e. $K_{2} = 9$)
or $u_{K_{2}} (\boldsymbol{X}_{i}) = [1, X_{1i}, X_{2i}, X_{3i}, X_{4i}, X_{1i}^{2}, X_{2i}^{2}, X_{3i}^{2}, X_{4i}^{2}, X_{1i} X_{2i}, X_{1i} X_{3i}, X_{1i} X_{4i}, X_{2i} X_{3i}, X_{2i} X_{4i}, X_{3i} X_{4i}]^{T}$ (i.e. $K_{2} = 15$).
For CBGPS, covariates are chosen to be $\boldsymbol{X}_{i} = [X_{1i}, X_{2i}, X_{3i}, X_{4i}]^{T}$ for the first step of estimating stabilized weights
and $\boldsymbol{Z}_{i} = [1, T_{i}, \boldsymbol{X}_{i}^{T}]^{T}$ for the second step of estimating average treatment effects.
}
\end{table}

\clearpage

\begin{table}[th] 
\begin{center} 
\caption{Simulation Results on $\beta_{2}$ (DGP-1) \label{table:sim_results_beta2_DGP1}} 
{\fontsize{9.5pt}{16.5pt} \selectfont 
\begin{tabular}{c|cccc|cccc|cccc}   \hline  \hline 
\multicolumn{13}{c}{$\rho = 0.0$} \\ \hline  
       &   \multicolumn{4}{c|}{$N = 100$}   &   \multicolumn{4}{c|}{$N = 500$}    &    \multicolumn{4}{c}{$N = 1000$}    \\ \hline
      &    Bias    &   Stdev    &   RMSE    &   CP   &     Bias    &   Stdev    &   RMSE    &   CP  &   Bias    &   Stdev    &   RMSE  &  CP    \\ \hline
SW ($K_{2} = 9$) & 0.023     &  0.291    &  0.292  &  0.853  &  0.016    &   0.189    &   0.189       &  0.928    &  0.017   &  0.167  & 0.168   &  0.922   \\  
SW ($K_{2} = 15$)  &  0.036    &  0.316   &  0.318   &   0.819   &  0.022   & 0.201    &  0.202   & 0.969    &   0.035  &  0.186   & 0.189    &  0.979  \\
CBGPS  &  0.001    &  0.205    &  0.205  &  0.922  &  -0.007    &   0.099    &  0.099        &  0.914    & -0.001    & 0.072   & 0.072   &  0.906   \\    \hline \hline 
\multicolumn{13}{c}{$\rho = 0.2$} \\ \hline  
       &   \multicolumn{4}{c|}{$N = 100$}   &   \multicolumn{4}{c|}{$N = 500$}    &    \multicolumn{4}{c}{$N = 1000$}    \\ \hline
      &    Bias    &   Stdev    &   RMSE    &   CP   &     Bias    &   Stdev    &   RMSE    &   CP  &   Bias    &   Stdev    &   RMSE  &  CP    \\ \hline
SW ($K_{2} = 9$) &   0.031   &   0.290   &  0.292  & 0.846   & 0.031     &  0.186     &    0.188      &   0.926   &   0.027  &  0.174  &  0.176  &  0.917   \\  
SW ($K_{2} = 15$)  & 0.050     &  0.313   &  0.317   &   0.814   &  0.040   &   0.202  &  0.206   &  0.964   & 0.046    &  0.190   & 0.195   &  0.982  \\
CBGPS  &  -0.013    &  0.219    & 0.219   & 0.911   &  0.002    &  0.106     &    0.106      & 0.890     &  0.000   &  0.083  &  0.083  &  0.909   \\    \hline \hline 
\multicolumn{13}{c}{$\rho = 0.4$} \\ \hline  
      &   \multicolumn{4}{c|}{$N = 100$}   &   \multicolumn{4}{c|}{$N = 500$}    &    \multicolumn{4}{c}{$N = 1000$}    \\ \hline
      &    Bias    &   Stdev    &   RMSE    &   CP   &     Bias    &   Stdev    &   RMSE    &   CP  &   Bias    &   Stdev    &   RMSE  &  CP    \\ \hline
SW ($K_{2} = 9$) &   0.044   &  0.287    &  0.290  &  0.841  &  0.038    &   0.190    &   0.194       &  0.916    &  0.038   &  0.167  & 0.172   &  0.923   \\  
SW ($K_{2} = 15$)  &   0.050   &  0.300   &  0.304   &   0.811   &  0.042   &  0.201   & 0.205    &  0.955   &  0.037   & 0.188    & 0.192    & 0.975   \\
CBGPS  &   0.015   &  0.227    & 0.227   &  0.895  &  0.007    &  0.115     &     0.115     &   0.911   &  -0.002   &  0.087  &  0.087  &  0.882   \\    \hline \hline 
\end{tabular} 
} 
\end{center} 
{\fontsize{9.5pt}{12pt} \selectfont 
The DGP is $T_{i} = X_{1i} + X_{2i} + 0.2 X_{3i} + 0.2 X_{4i} + \xi_{i}$ and $Y_{i} = 1 + X_{1i} + 0.1 X_{2i} + 0.1 X_{3i} + 0.1 X_{4i} + T_{i} + \epsilon_{i}$.
The covariates follow $\boldsymbol{X}_{i} \stackrel{i.i.d.}{\sim} N(0, \boldsymbol{\Sigma})$, where the diagonal elements of $\boldsymbol{\Sigma}$ are all 1 and the off-diagonal elements are all $\rho \in \{ 0.0, 0.2, 0.4 \}$.
We report the bias, standard deviation, root mean squared error, and coverage probability based on the 95\% confidence band across $J = 1000$ Monte Carlo samples. 
'SW' signifies our own stabilized-weight estimator, while 'CBGPS' signifies Fong, Hazlett, and Imai's \citeyearpar{Fong_Hazlett_Imai_2018} parametric covariate balancing generalized propensity score (CBGPS) estimator.
For SW, the polynomials are $u_{K_{1}} (t_{i}) = [1, t_{i}, t_{i}^{2}]^{T}$ (i.e. $K_{1} = 3$) and $u_{K_{2}} (\boldsymbol{X}_{i}) = [1, X_{1i}, X_{2i}, X_{3i}, X_{4i}, X_{1i}^{2}, X_{2i}^{2}, X_{3i}^{2}, X_{4i}^{2}]^{T}$ (i.e. $K_{2} = 9$)
or $u_{K_{2}} (\boldsymbol{X}_{i}) = [1, X_{1i}, X_{2i}, X_{3i}, X_{4i}, X_{1i}^{2}, X_{2i}^{2}, X_{3i}^{2}, X_{4i}^{2}, X_{1i} X_{2i}, X_{1i} X_{3i}, X_{1i} X_{4i}, X_{2i} X_{3i}, X_{2i} X_{4i}, X_{3i} X_{4i}]^{T}$ (i.e. $K_{2} = 15$).
For CBGPS, covariates are chosen to be $\boldsymbol{X}_{i} = [X_{1i}, X_{2i}, X_{3i}, X_{4i}]^{T}$ for the first step of estimating stabilized weights
and $\boldsymbol{Z}_{i} = [1, T_{i}, \boldsymbol{X}_{i}^{T}]^{T}$ for the second step of estimating average treatment effects.
}
\end{table}

\clearpage

\begin{table}[th] 
\begin{center} 
\caption{Simulation Results on $\beta_{1}$ (DGP-2) \label{table:sim_results_beta1_DGP2}} 
{\fontsize{9.5pt}{16.5pt} \selectfont 
\begin{tabular}{c|cccc|cccc|cccc}   \hline  \hline 
\multicolumn{13}{c}{$\rho = 0.0$} \\ \hline  
       &   \multicolumn{4}{c|}{$N = 100$}   &   \multicolumn{4}{c|}{$N = 500$}    &    \multicolumn{4}{c}{$N = 1000$}    \\ \hline
      &    Bias    &   Stdev    &   RMSE    &   CP   &     Bias    &   Stdev    &   RMSE    &   CP  &   Bias    &   Stdev    &   RMSE  &  CP    \\ \hline
SW ($K_{2} = 9$) &   -0.050    &  0.983    &  0.985  &  0.839  &  -0.055    &  0.584     &   0.586       &   0.922   & -0.042    & 0.563   &  0.565  &  0.928   \\ 
SW ($K_{2} = 15$)  & -0.003     &  1.033   &  1.033   &  0.843    &  -0.055   &  0.682   &  0.684   &   0.953  &   -0.089  &   0.735  &   0.740  &  0.965  \\
CBGPS  &   -0.025   &   0.586   &  0.586  &  0.934  &  0.010    &   0.255    &   0.255       &  0.957    &  -0.001   &  0.181  &  0.181  &   0.942  \\    \hline \hline 
\multicolumn{13}{c}{$\rho = 0.2$} \\ \hline  
       &   \multicolumn{4}{c|}{$N = 100$}   &   \multicolumn{4}{c|}{$N = 500$}    &    \multicolumn{4}{c}{$N = 1000$}    \\ \hline
      &    Bias    &   Stdev    &   RMSE    &   CP   &     Bias    &   Stdev    &   RMSE    &   CP  &   Bias    &   Stdev    &   RMSE  &  CP    \\ \hline
SW ($K_{2} = 9$) & -0.051     &   0.937   &  0.938  &  0.868  &  -0.061    &  0.645     &   0.648       &  0.904    &  -0.061   &  0.592  & 0.595   &   0.928  \\  
SW ($K_{2} = 15$)  & -0.100     &  1.086   &  1.091   &  0.839    & -0.062    &  0.737   &  0.739   &  0.962   & -0.076  &  0.722   & 0.726  & 0.960  \\
CBGPS  &   -0.014   &  0.604    &  0.604  &  0.938  &  -0.006    &   0.265    &   0.266       &  0.936    &  -0.008   &  0.185  &  0.185  &   0.950  \\    \hline \hline 
\multicolumn{13}{c}{$\rho = 0.4$} \\ \hline  
      &   \multicolumn{4}{c|}{$N = 100$}   &   \multicolumn{4}{c|}{$N = 500$}    &    \multicolumn{4}{c}{$N = 1000$}    \\ \hline
      &    Bias    &   Stdev    &   RMSE    &   CP   &     Bias    &   Stdev    &   RMSE    &   CP  &   Bias    &   Stdev    &   RMSE  &  CP    \\ \hline
SW ($K_{2} = 9$) &   -0.114   &  0.989    &  0.996  &  0.864  &   -0.052   &   0.664    &    0.666      &  0.921    &  -0.075   & 0.591   &  0.596  &   0.919  \\  
SW ($K_{2} = 15$)  &  -0.111    & 1.064    &   1.070  &  0.851    &  -0.100   &  0.700   &  0.707   & 0.944    &  -0.095   &  0.715   &  0.721   &  0.958  \\
CBGPS  &  -0.006    &  0.625    &  0.625  &  0.923  &  0.011    &   0.271    &   0.271       & 0.934     &  0.004   &  0.190  &  0.190  &  0.937   \\    \hline \hline 
\end{tabular} 
} 
\end{center} 
{\fontsize{9.5pt}{12pt} \selectfont 
The DGP is $T_{i} = (X_{1i} + 0.5)^{2} + 0.4 X_{2i} + 0.4 X_{3i} + 0.4 X_{4i} + \xi_{i}$ and $Y_{i} = 1 + X_{1i} + 0.1 X_{2i} + 0.1 X_{3i} + 0.1 X_{4i} + T_{i} + \epsilon_{i}$.
The covariates follow $\boldsymbol{X}_{i} \stackrel{i.i.d.}{\sim} N(0, \boldsymbol{\Sigma})$, where the diagonal elements of $\boldsymbol{\Sigma}$ are all 1 and the off-diagonal elements are all $\rho \in \{ 0.0, 0.2, 0.4 \}$.
We report the bias, standard deviation, root mean squared error, and coverage probability based on the 95\% confidence band across $J = 1000$ Monte Carlo samples. 
'SW' signifies our own stabilized-weight estimator, while 'CBGPS' signifies Fong, Hazlett, and Imai's \citeyearpar{Fong_Hazlett_Imai_2018} parametric covariate balancing generalized propensity score (CBGPS) estimator.
For SW, the polynomials are $u_{K_{1}} (t_{i}) = [1, t_{i}, t_{i}^{2}]^{T}$ (i.e. $K_{1} = 3$) and $u_{K_{2}} (\boldsymbol{X}_{i}) = [1, X_{1i}, X_{2i}, X_{3i}, X_{4i}, X_{1i}^{2}, X_{2i}^{2}, X_{3i}^{2}, X_{4i}^{2}]^{T}$ (i.e. $K_{2} = 9$)
or $u_{K_{2}} (\boldsymbol{X}_{i}) = [1, X_{1i}, X_{2i}, X_{3i}, X_{4i}, X_{1i}^{2}, X_{2i}^{2}, X_{3i}^{2}, X_{4i}^{2}, X_{1i} X_{2i}, X_{1i} X_{3i}, X_{1i} X_{4i}, X_{2i} X_{3i}, X_{2i} X_{4i}, X_{3i} X_{4i}]^{T}$ (i.e. $K_{2} = 15$).
For CBGPS, covariates are chosen to be $\boldsymbol{X}_{i} = [X_{1i}, X_{2i}, X_{3i}, X_{4i}]^{T}$ for the first step of estimating stabilized weights
and $\boldsymbol{Z}_{i} = [1, T_{i}, \boldsymbol{X}_{i}^{T}]^{T}$ for the second step of estimating average treatment effects.
}
\end{table}

\clearpage

\begin{table}[th] 
\begin{center} 
\caption{Simulation Results on $\beta_{2}$ (DGP-2) \label{table:sim_results_beta2_DGP2}} 
{\fontsize{9.5pt}{16.5pt} \selectfont 
\begin{tabular}{c|cccc|cccc|cccc}   \hline  \hline 
\multicolumn{13}{c}{$\rho = 0.0$} \\ \hline  
       &   \multicolumn{4}{c|}{$N = 100$}   &   \multicolumn{4}{c|}{$N = 500$}    &    \multicolumn{4}{c}{$N = 1000$}    \\ \hline
      &    Bias    &   Stdev    &   RMSE    &   CP   &     Bias    &   Stdev    &   RMSE    &   CP  &   Bias    &   Stdev    &   RMSE  &  CP    \\ \hline
SW ($K_{2} = 9$) &  0.029    &   0.275   &  0.276  &  0.821  &   0.020   &   0.177    &   0.178       &  0.909    &  0.021    &  0.159  &  0.160  & 0.914   \\   
SW ($K_{2} = 15$)  &  0.038    &  0.301   &  0.303   &   0.806   &  0.020   &  0.183   &   0.184  &  0.966   &   0.024  &  0.181   &  0.182   & 0.983   \\
CBGPS  &  0.012    &  0.187    &  0.187  &  0.888  &  -0.001    &  0.086     &  0.086        &   0.918   &  0.000   &  0.060  &  0.060  &  0.905   \\    \hline \hline 
\multicolumn{13}{c}{$\rho = 0.2$} \\ \hline  
       &   \multicolumn{4}{c|}{$N = 100$}   &   \multicolumn{4}{c|}{$N = 500$}    &    \multicolumn{4}{c}{$N = 1000$}    \\ \hline
      &    Bias    &   Stdev    &   RMSE    &   CP   &     Bias    &   Stdev    &   RMSE    &   CP  &   Bias    &   Stdev    &   RMSE  &  CP    \\ \hline
SW ($K_{2} = 9$) &   0.028   &  0.299    &  0.300  &  0.805  &   0.020   &   0.174    &   0.175       &  0.915    &   0.031  & 0.164   & 0.167   &  0.916   \\  
SW ($K_{2} = 15$)  &  0.037    &  0.295   &  0.298   &  0.796    &  0.029   &  0.199   &  0.201   &  0.969   &  0.030   &  0.183   &  0.186   &  0.966  \\
CBGPS  &   -0.009   & 0.191     & 0.191   &  0.874  &   0.000   &   0.087    &   0.087       & 0.897     &   0.001  &  0.066  &  0.066  &  0.902   \\    \hline \hline 
\multicolumn{13}{c}{$\rho = 0.4$} \\ \hline  
      &   \multicolumn{4}{c|}{$N = 100$}   &   \multicolumn{4}{c|}{$N = 500$}    &    \multicolumn{4}{c}{$N = 1000$}    \\ \hline
      &    Bias    &   Stdev    &   RMSE    &   CP   &     Bias    &   Stdev    &   RMSE    &   CP  &   Bias    &   Stdev    &   RMSE  &  CP    \\ \hline
SW ($K_{2} = 9$) &   0.038   &   0.288   & 0.290   &  0.808  &    0.030  &  0.181     &   0.183       & 0.887     &  0.029   &  0.159  &  0.162  &  0.898   \\  
SW ($K_{2} = 15$)  &  0.051    &  0.300   &  0.305   &   0.791   &  0.044   &  0.196   &  0.200   & 0.939    &  0.036   & 0.173    & 0.177    & 0.964   \\
CBGPS  &  -0.009    &  0.192    & 0.192   & 0.884   &   -0.003   &  0.090     &   0.090       & 0.888     &   -0.001  &  0.069  &  0.069  &  0.870   \\    \hline \hline 
\end{tabular} 
} 
\end{center} 
{\fontsize{9.5pt}{12pt} \selectfont 
The DGP is $T_{i} = (X_{1i} + 0.5)^{2} + 0.4 X_{2i} + 0.4 X_{3i} + 0.4 X_{4i} + \xi_{i}$ and $Y_{i} = 1 + X_{1i} + 0.1 X_{2i} + 0.1 X_{3i} + 0.1 X_{4i} + T_{i} + \epsilon_{i}$.
The covariates follow $\boldsymbol{X}_{i} \stackrel{i.i.d.}{\sim} N(0, \boldsymbol{\Sigma})$, where the diagonal elements of $\boldsymbol{\Sigma}$ are all 1 and the off-diagonal elements are all $\rho \in \{ 0.0, 0.2, 0.4 \}$.
We report the bias, standard deviation, root mean squared error, and coverage probability based on the 95\% confidence band across $J = 1000$ Monte Carlo samples. 
'SW' signifies our own stabilized-weight estimator, while 'CBGPS' signifies Fong, Hazlett, and Imai's \citeyearpar{Fong_Hazlett_Imai_2018} parametric covariate balancing generalized propensity score (CBGPS) estimator.
For SW, the polynomials are $u_{K_{1}} (t_{i}) = [1, t_{i}, t_{i}^{2}]^{T}$ (i.e. $K_{1} = 3$) and $u_{K_{2}} (\boldsymbol{X}_{i}) = [1, X_{1i}, X_{2i}, X_{3i}, X_{4i}, X_{1i}^{2}, X_{2i}^{2}, X_{3i}^{2}, X_{4i}^{2}]^{T}$ (i.e. $K_{2} = 9$)
or $u_{K_{2}} (\boldsymbol{X}_{i}) = [1, X_{1i}, X_{2i}, X_{3i}, X_{4i}, X_{1i}^{2}, X_{2i}^{2}, X_{3i}^{2}, X_{4i}^{2}, X_{1i} X_{2i}, X_{1i} X_{3i}, X_{1i} X_{4i}, X_{2i} X_{3i}, X_{2i} X_{4i}, X_{3i} X_{4i}]^{T}$ (i.e. $K_{2} = 15$).
For CBGPS, covariates are chosen to be $\boldsymbol{X}_{i} = [X_{1i}, X_{2i}, X_{3i}, X_{4i}]^{T}$ for the first step of estimating stabilized weights
and $\boldsymbol{Z}_{i} = [1, T_{i}, \boldsymbol{X}_{i}^{T}]^{T}$ for the second step of estimating average treatment effects.
}
\end{table}

\clearpage

\begin{table}[th] 
\begin{center} 
\caption{Simulation Results on $\beta_{1}$ (DGP-3) \label{table:sim_results_beta1_DGP3}} 
{\fontsize{9.5pt}{16.5pt} \selectfont 
\begin{tabular}{c|cccc|cccc|cccc}  \hline  \hline 
\multicolumn{13}{c}{$\rho = 0.0$} \\ \hline  
       &   \multicolumn{4}{c|}{$N = 100$}   &   \multicolumn{4}{c|}{$N = 500$}    &    \multicolumn{4}{c}{$N = 1000$}    \\ \hline
      &    Bias    &   Stdev    &   RMSE    &   CP   &     Bias    &   Stdev    &   RMSE    &   CP  &   Bias    &   Stdev    &   RMSE  &  CP    \\ \hline
SW ($K_{2} = 9$) &   0.005    &  0.832    &  0.832  &  0.888  &  0.065    &   0.572    &   0.576       & 0.937     &  0.096   &  0.570  & 0.579   &  0.950   \\ 
SW ($K_{2} = 15$)  & -0.088     &  0.925   &  0.929   &  0.871    &  0.073   &  0.688   &  0.692   &  0.984   &   0.155  &  0.686   &  0.704   &  0.979  \\
CBGPS  &  -0.030    &  0.606    & 0.606   &  0.949  &   0.007   &   0.302    &     0.302     &  0.915    &  0.016   &  0.234  &  0.235  &  0.932   \\    \hline \hline 
\multicolumn{13}{c}{$\rho = 0.2$} \\ \hline  
       &   \multicolumn{4}{c|}{$N = 100$}   &   \multicolumn{4}{c|}{$N = 500$}    &    \multicolumn{4}{c}{$N = 1000$}    \\ \hline
      &    Bias    &   Stdev    &   RMSE    &   CP   &     Bias    &   Stdev    &   RMSE    &   CP  &   Bias    &   Stdev    &   RMSE  &  CP    \\ \hline
SW ($K_{2} = 9$) &   -0.007   &  0.883    &  0.883  &  0.886  &   0.066   &   0.634    &   0.638       & 0.936     &  0.130   &  0.595  &  0.609  &   0.952  \\  
SW ($K_{2} = 15$)  &  -0.029    &  1.005   &  1.005   &   0.857   &  0.090    &  0.686   &  0.691   &  0.966  &  0.201   &  0.694   &  0.722   & 0.981   \\
CBGPS  &   -0.009   &  0.645    &  0.645  &  0.942  &   0.037   &   0.318    &   0.320   &  0.947  &  0.027    &   0.253  &  0.255  &  0.927       \\    \hline \hline 
\multicolumn{13}{c}{$\rho = 0.4$} \\ \hline  
      &   \multicolumn{4}{c|}{$N = 100$}   &   \multicolumn{4}{c|}{$N = 500$}    &    \multicolumn{4}{c}{$N = 1000$}    \\ \hline
      &    Bias    &   Stdev    &   RMSE    &   CP   &     Bias    &   Stdev    &   RMSE    &   CP  &   Bias    &   Stdev    &   RMSE  &  CP    \\ \hline
SW ($K_{2} = 9$) &   -0.028   &   1.009   &  1.009  & 0.834   &  0.057    &   0.681    &  0.684        &  0.928    &  0.085   &  0.619  &  0.625  &  0.937   \\  
SW ($K_{2} = 15$)  &  -0.017    &  1.006   &  1.006   & 0.872     & 0.065    &  0.684   &  0.687   &   0.965  &   0.111  &  0.745   &  0.753   &   0.978  \\
CBGPS  &  0.030    &  0.662    & 0.663   &  0.955  &   0.078   &  0.341     &    0.350      &  0.936    &   0.051  &  0.281  &  0.286  &  0.914   \\    \hline \hline 
\end{tabular} 
} 
\end{center} 
{\fontsize{9.5pt}{12pt} \selectfont 
The DGP is $T_{i} = X_{1i} + X_{2i} + 0.2 X_{3i} + 0.2 X_{4i} + \xi_{i}$ and $Y_{i} = 0.75 X_{1i}^{2} + 0.2 (X_{2i} + 0.5)^{2} + T_{i} + \epsilon_{i}$.
The covariates follow $\boldsymbol{X}_{i} \stackrel{i.i.d.}{\sim} N(0, \boldsymbol{\Sigma})$, where the diagonal elements of $\boldsymbol{\Sigma}$ are all 1 and the off-diagonal elements are all $\rho \in \{ 0.0, 0.2, 0.4 \}$.
We report the bias, standard deviation, root mean squared error, and coverage probability based on the 95\% confidence band across $J = 1000$ Monte Carlo samples. 
'SW' signifies our own stabilized-weight estimator, while 'CBGPS' signifies Fong, Hazlett, and Imai's \citeyearpar{Fong_Hazlett_Imai_2018} parametric covariate balancing generalized propensity score (CBGPS) estimator.
For SW, the polynomials are $u_{K_{1}} (t_{i}) = [1, t_{i}, t_{i}^{2}]^{T}$ (i.e. $K_{1} = 3$) and $u_{K_{2}} (\boldsymbol{X}_{i}) = [1, X_{1i}, X_{2i}, X_{3i}, X_{4i}, X_{1i}^{2}, X_{2i}^{2}, X_{3i}^{2}, X_{4i}^{2}]^{T}$ (i.e. $K_{2} = 9$)
or $u_{K_{2}} (\boldsymbol{X}_{i}) = [1, X_{1i}, X_{2i}, X_{3i}, X_{4i}, X_{1i}^{2}, X_{2i}^{2}, X_{3i}^{2}, X_{4i}^{2}, X_{1i} X_{2i}, X_{1i} X_{3i}, X_{1i} X_{4i}, X_{2i} X_{3i}, X_{2i} X_{4i}, X_{3i} X_{4i}]^{T}$ (i.e. $K_{2} = 15$).
For CBGPS, covariates are chosen to be $\boldsymbol{X}_{i} = [X_{1i}, X_{2i}, X_{3i}, X_{4i}]^{T}$ for the first step of estimating stabilized weights
and $\boldsymbol{Z}_{i} = [1, T_{i}, \boldsymbol{X}_{i}^{T}]^{T}$ for the second step of estimating average treatment effects.
}
\end{table}

\clearpage

\begin{table}[th] 
\begin{center} 
\caption{Simulation Results on $\beta_{2}$ (DGP-3) \label{table:sim_results_beta2_DGP3}} 
{\fontsize{9.5pt}{16.5pt} \selectfont 
\begin{tabular}{c|cccc|cccc|cccc}   \hline  \hline 
\multicolumn{13}{c}{$\rho = 0.0$} \\ \hline  
       &   \multicolumn{4}{c|}{$N = 100$}   &   \multicolumn{4}{c|}{$N = 500$}    &    \multicolumn{4}{c}{$N = 1000$}    \\ \hline
      &    Bias    &   Stdev    &   RMSE    &   CP   &     Bias    &   Stdev    &   RMSE    &   CP  &   Bias    &   Stdev    &   RMSE  &  CP    \\ \hline
SW ($K_{2} = 9$) &   0.019   &   0.273   &  0.274  &  0.840  &  0.014    &   0.187    &    0.188      &  0.936    &  0.006   &  0.161  &  0.161  &  0.956   \\   
SW ($K_{2} = 15$)  &  0.011    &  0.296   &   0.297  &  0.842    &   0.017  &  0.212   &  0.213   &   0.971  & -0.003    & 0.184    & 0.184    &  0.979  \\
CBGPS  &  -0.000    &  0.224    & 0.224   & 0.905   &  0.005    &   0.100    &   0.100       &   0.918   &  0.001   & 0.073   &  0.073  &  0.915   \\    \hline \hline 
\multicolumn{13}{c}{$\rho = 0.2$} \\ \hline  
       &   \multicolumn{4}{c|}{$N = 100$}   &   \multicolumn{4}{c|}{$N = 500$}    &    \multicolumn{4}{c}{$N = 1000$}    \\ \hline
      &    Bias    &   Stdev    &   RMSE    &   CP   &     Bias    &   Stdev    &   RMSE    &   CP  &   Bias    &   Stdev    &   RMSE  &  CP    \\ \hline
SW ($K_{2} = 9$) & -0.003     &  0.281    &  0.282  &  0.836  &  0.005    &   0.188    &    0.188      &  0.920    &   0.012  &  0.170  & 0.170   & 0.942    \\  
SW ($K_{2} = 15$)  &  0.008    &  0.304   &  0.304   &  0.809    & 0.002    &  0.200   &  0.200   &  0.969   &  0.010   &  0.182   &  0.182   &  0.990  \\
CBGPS  &  0.011    &  0.222    &  0.222  &  0.909  &   0.002   &   0.107    &   0.107       &  0.907    &  -0.005   &  0.081  &  0.082  &   0.912  \\    \hline \hline 
\multicolumn{13}{c}{$\rho = 0.4$} \\ \hline  
      &   \multicolumn{4}{c|}{$N = 100$}   &   \multicolumn{4}{c|}{$N = 500$}    &    \multicolumn{4}{c}{$N = 1000$}    \\ \hline
      &    Bias    &   Stdev    &   RMSE    &   CP   &     Bias    &   Stdev    &   RMSE    &   CP  &   Bias    &   Stdev    &   RMSE  &  CP    \\ \hline
SW ($K_{2} = 9$) &  0.002    &  0.288    & 0.288   &  0.830  &   0.008   &   0.184    &    0.185      &  0.917    &   0.003  &  0.174  &  0.174  &   0.932  \\  
SW ($K_{2} = 15$)  & 0.011     &  0.298   &  0.299   &  0.820    &   0.018  &  0.198   &  0.199   &   0.968  &  0.007   &  0.188   &   0.188  &  0.982  \\
CBGPS  &  0.004    &  0.220    &  0.220  &  0.915  &   0.001   &   0.117    &   0.117       &  0.892    &  0.004   &  0.090  &  0.090  &  0.892   \\    \hline \hline 
\end{tabular} 
} 
\end{center} 
{\fontsize{9.5pt}{12pt} \selectfont 
The DGP is $T_{i} = X_{1i} + X_{2i} + 0.2 X_{3i} + 0.2 X_{4i} + \xi_{i}$ and $Y_{i} = 0.75 X_{1i}^{2} + 0.2 (X_{2i} + 0.5)^{2} + T_{i} + \epsilon_{i}$.
The covariates follow $\boldsymbol{X}_{i} \stackrel{i.i.d.}{\sim} N(0, \boldsymbol{\Sigma})$, where the diagonal elements of $\boldsymbol{\Sigma}$ are all 1 and the off-diagonal elements are all $\rho \in \{ 0.0, 0.2, 0.4 \}$.
We report the bias, standard deviation, root mean squared error, and coverage probability based on the 95\% confidence band across $J = 1000$ Monte Carlo samples. 
'SW' signifies our own stabilized-weight estimator, while 'CBGPS' signifies Fong, Hazlett, and Imai's \citeyearpar{Fong_Hazlett_Imai_2018} parametric covariate balancing generalized propensity score (CBGPS) estimator.
For SW, the polynomials are $u_{K_{1}} (t_{i}) = [1, t_{i}, t_{i}^{2}]^{T}$ (i.e. $K_{1} = 3$) and $u_{K_{2}} (\boldsymbol{X}_{i}) = [1, X_{1i}, X_{2i}, X_{3i}, X_{4i}, X_{1i}^{2}, X_{2i}^{2}, X_{3i}^{2}, X_{4i}^{2}]^{T}$ (i.e. $K_{2} = 9$)
or $u_{K_{2}} (\boldsymbol{X}_{i}) = [1, X_{1i}, X_{2i}, X_{3i}, X_{4i}, X_{1i}^{2}, X_{2i}^{2}, X_{3i}^{2}, X_{4i}^{2}, X_{1i} X_{2i}, X_{1i} X_{3i}, X_{1i} X_{4i}, X_{2i} X_{3i}, X_{2i} X_{4i}, X_{3i} X_{4i}]^{T}$ (i.e. $K_{2} = 15$).
For CBGPS, covariates are chosen to be $\boldsymbol{X}_{i} = [X_{1i}, X_{2i}, X_{3i}, X_{4i}]^{T}$ for the first step of estimating stabilized weights
and $\boldsymbol{Z}_{i} = [1, T_{i}, \boldsymbol{X}_{i}^{T}]^{T}$ for the second step of estimating average treatment effects.
}
\end{table}

\clearpage

\begin{table}[th] 
\begin{center} 
\caption{Simulation Results on $\beta_{1}$ (DGP-4) \label{table:sim_results_beta1_DGP4}} 
{\fontsize{9.5pt}{16.5pt} \selectfont 
\begin{tabular}{c|cccc|cccc|cccc}   \hline  \hline 
\multicolumn{13}{c}{$\rho = 0.0$} \\ \hline  
       &   \multicolumn{4}{c|}{$N = 100$}   &   \multicolumn{4}{c|}{$N = 500$}    &    \multicolumn{4}{c}{$N = 1000$}    \\ \hline
      &    Bias    &   Stdev    &   RMSE    &   CP   &     Bias    &   Stdev    &   RMSE    &   CP  &   Bias    &   Stdev    &   RMSE  &  CP    \\ \hline
SW ($K_{2} = 9$) &   -0.100    &   0.954   &  0.960  &  0.847  &  -0.041    &    0.639   &    0.640      &  0.919    &  0.002   &  0.572  & 0.572   & 0.931    \\ 
SW ($K_{2} = 15$)  &  -0.105    & 1.050    &  1.055   &  0.852    &   0.010  &  0.690   & 0.690    & 0.956    &  0.072  &  0.716  & 0.720  &  0.975  \\
CBGPS  &   -0.207   &   0.627   &  0.661  &  0.902  &  -0.173    &   0.268    &    0.319      &  0.869    &   -0.174  &  0.181  &  0.251  &  0.817   \\    \hline \hline 
\multicolumn{13}{c}{$\rho = 0.2$} \\ \hline  
       &   \multicolumn{4}{c|}{$N = 100$}   &   \multicolumn{4}{c|}{$N = 500$}    &    \multicolumn{4}{c}{$N = 1000$}    \\ \hline
      &    Bias    &   Stdev    &   RMSE    &   CP   &     Bias    &   Stdev    &   RMSE    &   CP  &   Bias    &   Stdev    &   RMSE  &  CP    \\ \hline
SW ($K_{2} = 9$) &   -0.029   &   1.004   & 1.004   &  0.847  &  -0.008    & 0.651      &   0.651       & 0.908     &   0.018  & 0.593   & 0.593   &  0.910   \\  
SW ($K_{2} = 15$)  &  -0.109    &  1.126   &  1.131   &  0.839    &  0.004   &  0.727   &  0.727   &   0.965  &  0.033   &  0.675   &  0.676   &  0.956  \\
CBGPS  &   -0.171   &   0.641   &  0.664  &  0.912  &  -0.181    &   0.260    &   0.317       & 0.882     &  -0.176   &  0.188  &  0.258  &   0.821  \\    \hline \hline 
\multicolumn{13}{c}{$\rho = 0.4$} \\ \hline  
      &   \multicolumn{4}{c|}{$N = 100$}   &   \multicolumn{4}{c|}{$N = 500$}    &    \multicolumn{4}{c}{$N = 1000$}    \\ \hline
      &    Bias    &   Stdev    &   RMSE    &   CP   &     Bias    &   Stdev    &   RMSE    &   CP  &   Bias    &   Stdev    &   RMSE  &  CP    \\ \hline
SW ($K_{2} = 9$) &   -0.057   &    0.969  &  0.970  &  0.839  &   -0.029   &  0.641     &  0.641        & 0.914     &   0.016  &  0.605  &  0.605  &  0.923   \\  
SW ($K_{2} = 15$)  & -0.043     &  1.091   &  1.092   & 0.846     &  0.047   &  0.726   &  0.728   &  0.935   &  0.081   &  0.734   &  0.738   &  0.958  \\
CBGPS  &  -0.162    &   0.652   &  0.672  & 0.920   &  -0.180    &  0.266     &  0.322        & 0.893     &   -0.182  &  0.194  &  0.266  &   0.800  \\    \hline \hline 
\end{tabular} 
} 
\end{center} 
{\fontsize{9.5pt}{12pt} \selectfont 
The DGP is $T_{i} = (X_{1i} + 0.5)^{2} + 0.4 X_{2i} + 0.4 X_{3i} + 0.4 X_{4i} + \xi_{i}$ and $Y_{i} = 0.75 X_{1i}^{2} + 0.2 (X_{2i} + 0.5)^{2} + T_{i} + \epsilon_{i}$.
The covariates follow $\boldsymbol{X}_{i} \stackrel{i.i.d.}{\sim} N(0, \boldsymbol{\Sigma})$, where the diagonal elements of $\boldsymbol{\Sigma}$ are all 1 and the off-diagonal elements are all $\rho \in \{ 0.0, 0.2, 0.4 \}$.
We report the bias, standard deviation, root mean squared error, and coverage probability based on the 95\% confidence band across $J = 1000$ Monte Carlo samples. 
'SW' signifies our own stabilized-weight estimator, while 'CBGPS' signifies Fong, Hazlett, and Imai's \citeyearpar{Fong_Hazlett_Imai_2018} parametric covariate balancing generalized propensity score (CBGPS) estimator.
For SW, the polynomials are $u_{K_{1}} (t_{i}) = [1, t_{i}, t_{i}^{2}]^{T}$ (i.e. $K_{1} = 3$) and $u_{K_{2}} (\boldsymbol{X}_{i}) = [1, X_{1i}, X_{2i}, X_{3i}, X_{4i}, X_{1i}^{2}, X_{2i}^{2}, X_{3i}^{2}, X_{4i}^{2}]^{T}$ (i.e. $K_{2} = 9$)
or $u_{K_{2}} (\boldsymbol{X}_{i}) = [1, X_{1i}, X_{2i}, X_{3i}, X_{4i}, X_{1i}^{2}, X_{2i}^{2}, X_{3i}^{2}, X_{4i}^{2}, X_{1i} X_{2i}, X_{1i} X_{3i}, X_{1i} X_{4i}, X_{2i} X_{3i}, X_{2i} X_{4i}, X_{3i} X_{4i}]^{T}$ (i.e. $K_{2} = 15$).
For CBGPS, covariates are chosen to be $\boldsymbol{X}_{i} = [X_{1i}, X_{2i}, X_{3i}, X_{4i}]^{T}$ for the first step of estimating stabilized weights
and $\boldsymbol{Z}_{i} = [1, T_{i}, \boldsymbol{X}_{i}^{T}]^{T}$ for the second step of estimating average treatment effects.
}
\end{table}

\clearpage

\begin{table}[th] 
\begin{center} 
\caption{Simulation Results on $\beta_{2}$ (DGP-4) \label{table:sim_results_beta2_DGP4}} 
{\fontsize{9.5pt}{16.5pt} \selectfont 
\begin{tabular}{c|cccc|cccc|cccc}  \hline  \hline 
\multicolumn{13}{c}{$\rho = 0.0$} \\ \hline  
       &   \multicolumn{4}{c|}{$N = 100$}   &   \multicolumn{4}{c|}{$N = 500$}    &    \multicolumn{4}{c}{$N = 1000$}    \\ \hline
      &    Bias    &   Stdev    &   RMSE    &   CP   &     Bias    &   Stdev    &   RMSE    &   CP  &   Bias    &   Stdev    &   RMSE  &  CP    \\ \hline
SW ($K_{2} = 9$) &   0.033   &  0.288    & 0.290   &  0.810  &   0.038   &   0.187    &   0.191       & 0.913     &   0.040  & 0.165   &  0.170  &  0.930   \\   
SW ($K_{2} = 15$)  &  0.045    &  0.293   &  0.300   &  0.822    &  0.028   &  0.197   &   0.199  & 0.959    &  0.044   &  0.180   &  0.185   &  0.976  \\
CBGPS  &  0.121    &   0.189   & 0.224   &  0.814  &  0.146    &  0.090     &    0.172      &  0.470    &   0.151  &  0.069  &  0.166  &  0.269   \\    \hline \hline 
\multicolumn{13}{c}{$\rho = 0.2$} \\ \hline  
       &   \multicolumn{4}{c|}{$N = 100$}   &   \multicolumn{4}{c|}{$N = 500$}    &    \multicolumn{4}{c}{$N = 1000$}    \\ \hline
      &    Bias    &   Stdev    &   RMSE    &   CP   &     Bias    &   Stdev    &   RMSE    &   CP  &   Bias    &   Stdev    &   RMSE  &  CP    \\ \hline
SW ($K_{2} = 9$) &  0.030    &   0.280   &  0.282  &  0.801  &   0.025   &    0.173   &  0.175        &  0.905    & 0.025    &  0.162  &  0.164  &  0.918   \\  
SW ($K_{2} = 15$)  &  0.051    &  0.425   &  0.428   &  0.779    & 0.027    &  0.189   &  0.191   &  0.956   & 0.041    &   0.182  &   0.187  & 0.975   \\
CBGPS  & 0.123     &  0.203    &  0.237  &  0.805  &   0.146   &    0.096   &     0.174     &  0.488    &  0.152   &  0.071  &  0.168  &   0.279  \\    \hline \hline 
\multicolumn{13}{c}{$\rho = 0.4$} \\ \hline  
      &   \multicolumn{4}{c|}{$N = 100$}   &   \multicolumn{4}{c|}{$N = 500$}    &    \multicolumn{4}{c}{$N = 1000$}    \\ \hline
      &    Bias    &   Stdev    &   RMSE    &   CP   &     Bias    &   Stdev    &   RMSE    &   CP  &   Bias    &   Stdev    &   RMSE  &  CP    \\ \hline
SW ($K_{2} = 9$) &  0.048    &   0.273   &  0.277  &  0.821  &  0.041    &   0.178    &   0.182       &  0.910    &   0.038  & 0.170   & 0.174   &  0.916   \\  
SW ($K_{2} = 15$)  & 0.046     &  0.301   &  0.304   &  0.789    &   0.025  &  0.192   & 0.194    & 0.946    &   0.039  &   0.182  &  0.186   &  0.965  \\
CBGPS  & 0.121     &  0.209    &  0.242  &  0.811  &  0.145    &   0.098    &    0.176      & 0.514     &  0.156   &  0.080  &  0.176  &  0.304   \\    \hline \hline 
\end{tabular} 
} 
\end{center} 
{\fontsize{9.5pt}{12pt} \selectfont 
The DGP is $T_{i} = (X_{1i} + 0.5)^{2} + 0.4 X_{2i} + 0.4 X_{3i} + 0.4 X_{4i} + \xi_{i}$ and $Y_{i} = 0.75 X_{1i}^{2} + 0.2 (X_{2i} + 0.5)^{2} + T_{i} + \epsilon_{i}$.
The covariates follow $\boldsymbol{X}_{i} \stackrel{i.i.d.}{\sim} N(0, \boldsymbol{\Sigma})$, where the diagonal elements of $\boldsymbol{\Sigma}$ are all 1 and the off-diagonal elements are all $\rho \in \{ 0.0, 0.2, 0.4 \}$.
We report the bias, standard deviation, root mean squared error, and coverage probability based on the 95\% confidence band across $J = 1000$ Monte Carlo samples. 
'SW' signifies our own stabilized-weight estimator, while 'CBGPS' signifies Fong, Hazlett, and Imai's \citeyearpar{Fong_Hazlett_Imai_2018} parametric covariate balancing generalized propensity score (CBGPS) estimator.
For SW, the polynomials are $u_{K_{1}} (t_{i}) = [1, t_{i}, t_{i}^{2}]^{T}$ (i.e. $K_{1} = 3$) and $u_{K_{2}} (\boldsymbol{X}_{i}) = [1, X_{1i}, X_{2i}, X_{3i}, X_{4i}, X_{1i}^{2}, X_{2i}^{2}, X_{3i}^{2}, X_{4i}^{2}]^{T}$ (i.e. $K_{2} = 9$)
or $u_{K_{2}} (\boldsymbol{X}_{i}) = [1, X_{1i}, X_{2i}, X_{3i}, X_{4i}, X_{1i}^{2}, X_{2i}^{2}, X_{3i}^{2}, X_{4i}^{2}, X_{1i} X_{2i}, X_{1i} X_{3i}, X_{1i} X_{4i}, X_{2i} X_{3i}, X_{2i} X_{4i}, X_{3i} X_{4i}]^{T}$ (i.e. $K_{2} = 15$).
For CBGPS, covariates are chosen to be $\boldsymbol{X}_{i} = [X_{1i}, X_{2i}, X_{3i}, X_{4i}]^{T}$ for the first step of estimating stabilized weights
and $\boldsymbol{Z}_{i} = [1, T_{i}, \boldsymbol{X}_{i}^{T}]^{T}$ for the second step of estimating average treatment effects.
}
\end{table}

\clearpage

\section{Assumptions}
\begin{assumption}[\emph{Unconfounded Treatment Assignment}]
	\label{as:TYindep} For all $t\in \mathcal{T}$, given $\boldsymbol{X}$ , $T$
	is independent of $Y^{\ast }(t)$, i.e., $Y^{\ast }(t)\perp T|\boldsymbol{X,}$
	for all $t\in \mathcal{T}$.
\end{assumption}

\begin{assumption}
	\label{as:suppX} The support $\mathcal{X}$ of $\boldsymbol{X}$ is a compact
	subset of $\mathbb{R}^{r}$. The support $\mathcal{T}$ of the treatment
	variable $T$ is a compact subset of $\mathbb{R}$.
\end{assumption}

\begin{assumption}
	\label{as:pi0} There exist two positive constants $\eta_1$ and $\eta_2$ such
	that 
	\begin{equation*}
	0 < \eta_1 \leq \pi_0(t,\boldsymbol{x}) \leq \eta_2 <\infty \ ,\ \forall (t, 
	\boldsymbol{x}) \in \mathcal{T}\times\mathcal{X}\ .
	\end{equation*}
\end{assumption}

\begin{assumption}
	\label{as:smooth_pi} There exist $\Lambda _{K_{1}\times K_{2}}\in \mathbb{R}
	^{K_{1}\times K_{2}}$ and a positive constant $\alpha >0$ such that 
	\begin{equation*}
	\sup_{(t,\boldsymbol{x})\in \mathcal{T}\times \mathcal{X}}\left\vert (\rho
	^{\prime -1}\left( \pi _{0}(t,\boldsymbol{x})\right) -u_{K_{1}}(t)^{\top
	}\Lambda _{K_{1}\times K_{2}}v_{K_{2}}(\boldsymbol{x})\right\vert
	=O(K^{-\alpha }).
	\end{equation*}
\end{assumption}

\begin{assumption}
	\label{as:u&v} For every $K_1$ and $K_2$, the smallest eigenvalues of $%
	\mathbb{E}\left[u_{K_1}(T)u_{K_1}(T)^\top\right]$ and $\mathbb{E}\left[%
	v_{K_2}(\boldsymbol{X})v_{K_2}(\boldsymbol{X})^\top\right]$ are bounded away
	from zero uniformly in $K_1$ and $K_2$.
\end{assumption}

\begin{assumption}
	\label{as:K&N_consistency}There are two sequences of constants $\zeta
	_{1}(K_{1})$ and $\zeta _{2}(K_{2})$ satisfying\newline
	$\sup_{t\in \mathcal{T}}\Vert u_{K_{1}}(t)\Vert \leq \zeta _{1}(K_{1})$ and $%
	\sup_{\boldsymbol{x}\in \mathcal{X}}\Vert v_{K_{2}}(\boldsymbol{x})\Vert
	\leq \zeta _{2}(K_{2})$, $K=K_{1}(N)K_{2}(N)$ and $\zeta (K)=\zeta
	_{1}(K_{1})\zeta _{2}(K_{2})$, such that $\zeta (K)K^{-\alpha }\rightarrow 0$
	and $\zeta (K)\sqrt{K/N}\rightarrow 0$ as $N\rightarrow \infty $.
\end{assumption}

\begin{assumption}
	\label{as:Theta} The parameter space $\Theta\subset\mathbb{R} ^{p}$ is a
	compact set and the true parameter $\boldsymbol{\beta}_0$ is in the interior
	of $\Theta$ , where $p\in\mathbb{N}$.
\end{assumption}

\begin{assumption}
	\label{as:solution} There exists a unique solution $\boldsymbol{\beta }_{0}$
	for the optimization problem 
	\begin{equation*}
	\min_{\boldsymbol{\beta }\in \Theta }\int_{\mathcal{T}}\mathbb{E}\left[
	L(Y^{\ast }(t)-g(t;\boldsymbol{\beta }))\right] dF_{T}(t)\ .
	\end{equation*}
\end{assumption}

\begin{assumption}
	\label{as:EY2}  $\mathbb{E}\left[\sup_{\boldsymbol{\beta} \in
		\Theta }|L\left( Y-g(T;\boldsymbol{\beta} )\right)|^{2}\right]<\infty $.
\end{assumption}

\begin{assumption}
	\label{as:m_smooth} The following conditions hold true:
	
	\begin{enumerate}
		\item $g(t;\boldsymbol{\beta })$ is twice continuously differentiable in $%
		\boldsymbol{\beta }\in \Theta $;
		
		\item $L(Y-g(T;\boldsymbol{\beta }))$ is differentiable in $\boldsymbol{\beta }$ with probability one, i.e., for any directional vector $\boldsymbol{%
			\ \eta }\in \mathbb{R}^{p}$, there exists an integrable random variable $
		L^{\prime }(Y-g(T;\boldsymbol{\beta }))$ such that{\footnotesize \ 
			\begin{equation*}
			\mathbb{P}\left( \lim_{\epsilon \rightarrow 0}\frac{L(Y-g(T;\boldsymbol{\
					\beta }+\epsilon \boldsymbol{\eta }))-L(Y-g(T;\boldsymbol{\beta }))}{
				\epsilon }=L^{\prime }(Y-g(T;\boldsymbol{\beta }))\cdot \left\langle m(T; 
			\boldsymbol{\beta }),\boldsymbol{\eta }\right\rangle _{\mathbb{R}
				^{p}}\right) =1,
			\end{equation*}
		} where $\left\langle \cdot ,\cdot \right\rangle _{\mathbb{R}^{p}}$ is the
		inner product in Euclidean space $\mathbb{R}^{p}$;
		
		\item $\mathbb{E}\left[ L^{\prime }(Y-g(T;\boldsymbol{\beta }_{0}))^{2}
		\right] <\infty $.
	\end{enumerate}
\end{assumption}

\begin{assumption}
	\label{as:first_order} Suppose that 
	\begin{equation*}
	\frac{1}{N}\sum_{i=1}^{N}\hat{\pi}_{K}(T_{i},\boldsymbol{X}_{i})L^{\prime
	}\left( Y_{i}-g(T_{i};\hat{\boldsymbol{\beta }})\right) m(T_{i};\hat{%
		\boldsymbol{\beta }})=0
	\end{equation*}%
	holds with probability approaching one.
\end{assumption}

\begin{assumption}
	\label{as:H} $\mathbb{E}\left[ \pi _{0}(T,\boldsymbol{X})L^{\prime }(Y-g(T;
	\boldsymbol{\beta }))m(T;\boldsymbol{\beta })\right] $ is differentiable
	with respect to $\boldsymbol{		\beta}$ and $H_{0}:=-\nabla _{\beta }\mathbb{E}\left[ \pi
	_{0}(T,\boldsymbol{X})L^{\prime }(Y-g(T;\boldsymbol{\beta }))m(T;\boldsymbol{%
		\beta })\right] \Big|_{\boldsymbol{\beta }=\boldsymbol{\beta }_{0}}$ is
	nonsingular.
\end{assumption}

\begin{assumption}
	\label{as:smooth_varepsilon} $\varepsilon (t,\boldsymbol{x};{\boldsymbol{
			\beta }}_{0}):=\mathbb{E}[L^{\prime }(Y-g(T;\boldsymbol{\beta }_{0}))|T=t, 
	\boldsymbol{X}=\boldsymbol{x}]$ is continuously differentiable in $(t, 
	\boldsymbol{x})$.
\end{assumption}

\begin{assumption}
	\label{as:entropy} \ 
	
	\begin{enumerate}
		\item $\mathbb{E}\left[ \sup_{\boldsymbol{\beta} \in \Theta }|L^{\prime }(Y-g(T;\boldsymbol{\beta}
		))^{2+\delta }\right] <\infty $ for some $\delta >0$;
		
		\item The function class $\{L^{\prime }(y-g(t;\boldsymbol{\beta} )):\boldsymbol{\beta} \in \Theta \}$
		satisfies: 
		\begin{equation*}
		\mathbb{E}\left[ \sup_{\boldsymbol{\beta} _{1}:\Vert \boldsymbol{\beta}_{1}-\boldsymbol{\beta} \Vert <\delta
		}\left\vert L^{\prime }(Y-g(T;\boldsymbol{\beta} _{1}))-L^{\prime }(Y-g(T;\boldsymbol{\beta}
		))\right\vert ^{2}\right] ^{1/2}\leq a\cdot \delta ^{b}
		\end{equation*}%
		for any $\forall \boldsymbol{\beta}\in \Theta $ and any small $\delta >0$ and for some
		finite positive constants $a$ and $b$.
	\end{enumerate}
\end{assumption}

\begin{assumption}
	\label{as:K&N_c} $\zeta (K)\sqrt{K^{4}/N}\rightarrow 0$ and $\sqrt{N}%
	K^{-\alpha}\to 0$ as $N\to \infty$.
\end{assumption}

\section{Efficiency Bound}
\subsection{Proof of Theorem 3.1 \label{sec:proof_efficiency_theorem}}
	Without loss of generality, we only consider the distribution of $(T,\boldsymbol{X},Y)$ to be absolutely continuous with respect to Lebesgue measure, i.e., there exists a density function $f_{T,X,Y}(t,\boldsymbol{x},y)$ such that $dF_{T,X,Y}(t,\boldsymbol{x},y)=f_{T,X,Y}(t,\boldsymbol{x},y)dtd\boldsymbol{x}dy$. For discrete cases, the proof can be established by using a similar argument.\\
	
	We follow the approach of \citet[Section 3.3]{Bickel_Klaassen_Ritov_Wellner_1993} to derive the variance bound of $\boldsymbol{\beta}_0$, see also \cite{tchetgen2012semiparametric}. Let  $\left\{f^{\alpha}_{Y,T,X}(y,t,\boldsymbol{x})\right\}_{\alpha \in\mathbb{R}}$ denote a one dimensional regular parametric submodel with $
	f^{\alpha=0}_{Y,T,X}(y,t,\boldsymbol{x})=f_{Y,T,X}(y,t,\boldsymbol{x})$.
	By definition, $\boldsymbol{\beta}_0$ solves following equation:
	\begin{align}\label{def_beta^*_eff_0}
	\int_{\mathcal{T}}\mathbb{E}\left[ m(t;\boldsymbol{\beta}_0)L'\left(Y^*(t)-g\left(t;\boldsymbol{\beta}_0\right)\right)\right]f_T(t)dt=0 \ .
	\end{align}
	By Assumption \ref{as:TYindep}, \eqref{def_beta^*_eff_0} is equivalent to
	\begin{align} \notag %\label{def_beta^*_eff}
		\int_{\mathcal{T}} \int_{\mathcal{X}} \mathbb{E}\left[m(T;\boldsymbol{\beta}_0)L'\left(Y-g\left(T;\boldsymbol{\beta}_0\right)\right)|T=t,\boldsymbol{X}=\boldsymbol{x}\right]f_{X}(\boldsymbol{x}) f_T(t)d\boldsymbol{x}dt=0  \ .
	\end{align}
	Therefore, the parameter $\boldsymbol{\beta}(\alpha)$ induced by the submodel $f^{\alpha}_{Y,T,X}(y,t,\boldsymbol{x})$ satisfies:
	\begin{align}
	&\int_{\mathcal{T}}\int_{\mathcal{X}}m(t;\boldsymbol{\beta}(\alpha))\cdot \mathbb{E}^{\alpha}\left[L'\left(Y-g\left(t;\boldsymbol{\beta}(\alpha)\right)\right)|T=t,\boldsymbol{X}=\boldsymbol{x}\right]f^{{\alpha}}_{T}(t) f^{{\alpha}}_{X}(\boldsymbol{x})d\boldsymbol{x}dt=0\ , \label{perturbated_model}
	\end{align}
	where $\mathbb{E}^{\alpha}\left[\cdot|T=t,\boldsymbol{X}=\boldsymbol{x}\right]$ denotes taking expectation with respect to the submodel $f^{\alpha}_{Y|T,X}(\cdot|t,\boldsymbol{x})$. \\

	Differentiating both sides of \eqref{perturbated_model} with respect to $\alpha$, evaluating at $\alpha = 0$ and using the condition $Y^*(t)\perp T|\boldsymbol{X}$, we can deduce  that
	\begin{align*}
	0=&\int_{\mathcal{T}}\int_{\mathcal{X}} \frac{\partial}{\partial \alpha} \bigg|_{\alpha=0} \left\{m(t;\boldsymbol{\beta}(\alpha))\mathbb{E}^{\alpha}\left[L'(Y-g(t;\boldsymbol{\beta}(\alpha)))|T=t,\boldsymbol{X}=\boldsymbol{x}\right] f^{{\alpha}}_{T}(t)f^{{\alpha}}_{X}(\boldsymbol{x})\right\}d\boldsymbol{x}dt\\
	=&\int_{\mathcal{T}}\int_{\mathcal{X}}  \mathbb{E}\left[L'(Y-g(t;\boldsymbol{\beta}_0))|T=t,\boldsymbol{X}=\boldsymbol{x}\right] f_{T}(t)f_{X}(\boldsymbol{x})\nabla_{\boldsymbol{\beta}}m(t;\boldsymbol{\beta}_0) d\boldsymbol{x}dt\cdot \frac{\partial}{\partial \alpha} \bigg|_{\alpha=0}\boldsymbol{\beta}(\alpha) \\
	&+\int_{\mathcal{X}\times \mathcal{T}} \mathbb{E}[L'(Y-g(t;\boldsymbol{\beta}_0))|T=t,\boldsymbol{X}=\boldsymbol{x}]m(t;\boldsymbol{\beta}_0)\cdot \frac{\partial }{\partial \alpha}f^{\alpha}_{X}(\boldsymbol{x})\bigg|_{\alpha=0}f_{T}(t)d\boldsymbol{x}dt \notag\\
	&+\int_{\mathcal{Y}\times \mathcal{X}\times  \mathcal{T}} m(t;\boldsymbol{\beta}_0)L'(y-g(t;\boldsymbol{\beta}_0)) \cdot \frac{\partial }{\partial \alpha}f^{\alpha}_{Y|T,X}(y|t,\boldsymbol{x})\bigg|_{\alpha=0}f_{X}(\boldsymbol{x})f_T(t)dyd\boldsymbol{x}dt \notag \\
	& + \int_{\mathcal{X}\times \mathcal{T}}m(t;\boldsymbol{\beta}_0) \cdot  \nabla_{\beta}\mathbb{E}[L'(Y^*(t)-g(t;\boldsymbol{\beta}))|T=t,\boldsymbol{X}=\boldsymbol{x}]\Bigg|_{\beta=\boldsymbol{\beta}_0}  \cdot \frac{\partial}{\partial \alpha} \bigg|_{\alpha=0}\boldsymbol{\beta}(\alpha)\cdot  f_{T}(t)f_{X}(\boldsymbol{x})d\boldsymbol{x}dt \\
	&+ \int_{\mathcal{X}\times \mathcal{T}} \mathbb{E}[L'(Y-g(t;\boldsymbol{\beta}_0))|T=t,\boldsymbol{X}=\boldsymbol{x}]m(t;\boldsymbol{\beta}_0)\cdot \frac{\partial }{\partial \alpha}f^{\alpha}_{T}(t)\bigg|_{\alpha=0}f_{X}(\boldsymbol{x})d\boldsymbol{x}dt \\
	=&\int_{\mathcal{T}}\int_{\mathcal{X}}  \mathbb{E}\left[L'(Y^*(t)-g(t;\boldsymbol{\beta}_0))|\boldsymbol{X}=\boldsymbol{x}\right] f_{T}(t)f_{X}(\boldsymbol{x})\nabla_{\boldsymbol{\beta}}m(t;\boldsymbol{\beta}_0) d\boldsymbol{x}dt\cdot \frac{\partial}{\partial \alpha} \bigg|_{\alpha=0}\boldsymbol{\beta}(\alpha) \\
	&+\int_{\mathcal{X}\times \mathcal{T}} \mathbb{E}[L'(Y-g(t;\boldsymbol{\beta}_0))|T=t,\boldsymbol{X}=\boldsymbol{x}]m(t;\boldsymbol{\beta}_0)\cdot \frac{\partial }{\partial \alpha}f^{\alpha}_{X}(\boldsymbol{x})\bigg|_{\alpha=0}f_{T}(t)d\boldsymbol{x}dt \notag\\
	&  +\int_{\mathcal{Y}\times \mathcal{X}\times  \mathcal{T}} m(t;\boldsymbol{\beta}_0)L'(y-g(t;\boldsymbol{\beta}_0)) \cdot \frac{\partial }{\partial \alpha}f^{\alpha}_{Y|T,X}(y|t,\boldsymbol{x})\bigg|_{\alpha=0}f_{X}(\boldsymbol{x})f_T(t)dyd\boldsymbol{x}dt \notag \\
		&  + \int_{\mathcal{X}\times \mathcal{T}}m(t;\boldsymbol{\beta}_0) \cdot  \nabla_{\beta}\mathbb{E}[L'(Y^*(t)-g(t;\boldsymbol{\beta}))|\boldsymbol{X}=\boldsymbol{x}]\Bigg|_{\boldsymbol{\beta}=\boldsymbol{\beta}_0}  \cdot  f_{T}(t)f_{X}(\boldsymbol{x})d\boldsymbol{x}dt \cdot \frac{\partial}{\partial \alpha} \bigg|_{\alpha=0}\boldsymbol{\beta}(\alpha) \\
	& + \int_{\mathcal{X}\times \mathcal{T}} \mathbb{E}[L'(Y-g(t;\boldsymbol{\beta}_0))|T=t,\boldsymbol{X}=\boldsymbol{x}]m(t;\boldsymbol{\beta}_0)\cdot \frac{\partial }{\partial \alpha}f^{\alpha}_{T}(t)\bigg|_{\alpha=0}f_{X}(\boldsymbol{x})d\boldsymbol{x}dt \\
	=&\int_{\mathcal{T}}  \mathbb{E}[L'(Y^*(t)-g(t;\boldsymbol{\beta}_0))]\cdot  f_{T}(t)\nabla_{\boldsymbol{\beta}}m(t;\boldsymbol{\beta}_0) dt\cdot \frac{\partial}{\partial \alpha} \bigg|_{\alpha=0}\boldsymbol{\beta}(\alpha) \\
	&+\int_{\mathcal{X}\times \mathcal{T}} \mathbb{E}[L'(Y-g(t;\boldsymbol{\beta}_0))|T=t,\boldsymbol{X}=\boldsymbol{x}]m(t;\boldsymbol{\beta}_0)\cdot \frac{\partial }{\partial \alpha}f^{\alpha}_{X}(\boldsymbol{x})\bigg|_{\alpha=0}f_{T}(t)d\boldsymbol{x}dt \notag\\
	&  +\int_{\mathcal{Y}\times \mathcal{X}\times  \mathcal{T}} m(t;\boldsymbol{\beta}_0)\cdot L'(y-g(t;\boldsymbol{\beta}_0)) \cdot \frac{\partial }{\partial \alpha}f^{\alpha}_{Y|T,X}(y|t,\boldsymbol{x})\bigg|_{\alpha=0}f_{X}(\boldsymbol{x})f_T(t)dyd\boldsymbol{x}dt \notag \\
		& +\int_{\mathcal{T}} \nabla_{\beta}\mathbb{E}[L'(Y^*(t)-g(t;\boldsymbol{\beta}))]\bigg|_{\boldsymbol{\beta}=\boldsymbol{\beta}_0}m(t;\boldsymbol{\beta}_0)   \cdot  f_{T}(t)dt \cdot \frac{\partial}{\partial \alpha} \bigg|_{\alpha=0}\boldsymbol{\beta}(\alpha) \\
	&  + \int_{\mathcal{X}\times \mathcal{T}} \mathbb{E}[L'(Y-g(t;\boldsymbol{\beta}_0))|T=t,\boldsymbol{X}=\boldsymbol{x}]m(t;\boldsymbol{\beta}_0)\cdot \frac{\partial }{\partial \alpha}f^{\alpha}_{T}(t)\bigg|_{\alpha=0}f_{X}(\boldsymbol{x})d\boldsymbol{x}dt\\
	=&\nabla_{\beta}\left\{\int_{\mathcal{T}}  \mathbb{E}[L'(Y^*(t)-g(t;\boldsymbol{\beta}))]\cdot  m(t;\boldsymbol{\beta})f_{T}(t) dt\right\}\Bigg|_{\boldsymbol{\beta}=\boldsymbol{\beta}_0}\cdot \frac{\partial}{\partial \alpha} \bigg|_{\alpha=0}\boldsymbol{\beta}(\alpha) \\
	&+\int_{\mathcal{X}\times \mathcal{T}} \mathbb{E}[L'(Y-g(t;\boldsymbol{\beta}_0))|T=t,\boldsymbol{X}=\boldsymbol{x}]m(t;\boldsymbol{\beta}_0)\cdot \frac{\partial }{\partial \alpha}f^{\alpha}_{X}(\boldsymbol{x})\bigg|_{\alpha=0}f_{T}(t)d\boldsymbol{x}dt \notag\\
	&  +\int_{\mathcal{Y}\times \mathcal{X}\times  \mathcal{T}} m(t;\boldsymbol{\beta}_0)\cdot L'(y-g(t;\boldsymbol{\beta}_0)) \cdot \frac{\partial }{\partial \alpha}f^{\alpha}_{Y|T,X}(y|t,\boldsymbol{x})\bigg|_{\alpha=0}f_{X}(\boldsymbol{x})f_T(t)dyd\boldsymbol{x}dt \notag \\
	&  + \int_{\mathcal{X}\times \mathcal{T}} \mathbb{E}[L'(Y-g(t;\boldsymbol{\beta}_0))|T=t,\boldsymbol{X}=\boldsymbol{x}]m(t;\boldsymbol{\beta}_0)\cdot \frac{\partial }{\partial \alpha}f^{\alpha}_{T}(t)\bigg|_{\alpha=0}f_{X}(\boldsymbol{x})d\boldsymbol{x}dt.
	\end{align*}
	Since $H_0=- \nabla_{\beta}\left\{\int_{\mathcal{T}}  \mathbb{E}[L'(Y^*(t)-g(t;\boldsymbol{\beta}))]\cdot  m(t;\boldsymbol{\beta})f_{T}(t) dt\right\}\Bigg|_{\boldsymbol{\beta}=\boldsymbol{\beta}_0}$ is invertible by Assumption \ref{as:EY2}, we get 
	\begin{align}
	&\frac{\partial }{\partial \alpha}\bigg|_{\alpha=0}\boldsymbol{\beta}(\alpha)  =H_0^{-1}\cdot  \bigg\{\int_{\mathcal{X}\times \mathcal{T}} \mathbb{E}[L'(Y-g(t;\boldsymbol{\beta}_0))|T=t,\boldsymbol{X}=\boldsymbol{x}]m(t;\boldsymbol{\beta}_0)\cdot \frac{\partial }{\partial \alpha}f^{\alpha}_{X}(\boldsymbol{x})\bigg|_{\alpha=0}f_{T}(t)d\boldsymbol{x}dt \notag\\
		&\qquad   +\int_{\mathcal{Y}\times \mathcal{X}\times  \mathcal{T}} m(t;\boldsymbol{\beta}_0) \cdot L'(y-g(t;\boldsymbol{\beta}_0)) \cdot \frac{\partial }{\partial \alpha}f^{\alpha}_{Y|T,X}(y|t,\boldsymbol{x})\bigg|_{\alpha=0}f_{X}(\boldsymbol{x})f_T(t)dyd\boldsymbol{x}dt \notag \\
		&\qquad   + \int_{\mathcal{X}\times \mathcal{T}} \mathbb{E}[L'(Y-g(t;\boldsymbol{\beta}_0))|T=t,\boldsymbol{X}=\boldsymbol{x}]m(t;\boldsymbol{\beta}_0)\cdot \frac{\partial }{\partial \alpha}f^{\alpha}_{T}(t)\bigg|_{\alpha=0}f_{X}(\boldsymbol{x})d\boldsymbol{x}dt \bigg\} \notag \ .
	\end{align}
	The efficient influence function of $\boldsymbol{\beta}_0$, denoted by $S_{eff}(Y,T,\boldsymbol{X};\boldsymbol{\beta}_0)$, is a unique function satisfying the following equation:
	\begin{align}\label{def_efficient_influence}
	\frac{\partial }{\partial \alpha}\bigg|_{\alpha=0}\boldsymbol{\beta}(\alpha)=\mathbb{E}\left[S_{eff}(Y,T,\boldsymbol{X};\boldsymbol{\beta}_0)\frac{\partial}{\partial \alpha}\bigg|_{\alpha=0}\log f^{\alpha}_{Y,X,T}(Y,\boldsymbol{X},T)\right] \ .
	\end{align}
	Therefore, to justify our theorem, it suffices to substitute 
	$S_{eff}(Y,T,\boldsymbol{X};\boldsymbol{\beta}_0) = H_0 ^{-1} \psi(Y,T,\boldsymbol{X};\boldsymbol{\beta}_0)$
	into \eqref{def_efficient_influence} and check the validity. 
	Note that
	\begin{align} 
	&\mathbb{E}\left[S_{eff}(Y,T,\boldsymbol{X};\boldsymbol{\beta}_0)\frac{\partial}{\partial \alpha}\bigg|_{\alpha=0}\log f^{\alpha}_{Y,X,T}(Y,\boldsymbol{X},T)\right] \notag\\
	=&H_0 ^{-1}\int_{\mathcal{X}\times\mathcal{T}\times \mathcal{Y}}\psi(y,t,\boldsymbol{x};\boldsymbol{\beta}_0)\frac{\partial }{\partial \alpha}\bigg|_{\alpha=0}  f^{\alpha}_{Y|X,T}(y|\boldsymbol{x},t)f_{T,X}(t,\boldsymbol{x})dyd\boldsymbol{x}dt \label{eff_1}\\
	+&H_0^{-1}\int_{\mathcal{X}\times\mathcal{T}\times \mathcal{Y}}\psi(y,t,\boldsymbol{x};\boldsymbol{\beta}_0) f_{Y|X,T}(y|\boldsymbol{x},t)\frac{\partial }{\partial \alpha}\bigg|_{\alpha=0}f^{\alpha}_{T|X}(t|\boldsymbol{x})f_X(\boldsymbol{x})dyd\boldsymbol{x}dt \label{eff_2}\\
	+&H_0 ^{-1}\int_{\mathcal{X}\times\mathcal{T}\times \mathcal{Y}}\psi(y,t,\boldsymbol{x};\boldsymbol{\beta}_0) f_{Y|X,T}(y|\boldsymbol{x},t)f_{T|X}(t|\boldsymbol{x})\frac{\partial }{\partial \alpha}\bigg|_{\alpha=0}f^{\alpha}_X(\boldsymbol{x})dyd\boldsymbol{x}dt. \label{eff_3}  \end{align}
	For the term \eqref{eff_1}, we have
	\begin{align*}
	\eqref{eff_1}=&H_0 ^{-1}\int_{\mathcal{X}\times\mathcal{T}\times \mathcal{Y}}  \bigg\{ \frac{f_T(t)}{f_{T|X}(t|\boldsymbol{x})}m(t;\boldsymbol{\beta}_0) \cdot  L'(y-g(t;\boldsymbol{\beta}_0))-\frac{f_T(t)}{f_{T|X}(t|\boldsymbol{x})}m(t;\boldsymbol{\beta}_0) \cdot\varepsilon(t,\boldsymbol{x};\boldsymbol{\beta}_0)\\
	&\qquad +\mathbb{E}\left[\varepsilon(T,\boldsymbol{X};\boldsymbol{\beta}_0)\pi_0(T,\boldsymbol{X})m(T;\boldsymbol{\beta}_0)|\boldsymbol{X}=\boldsymbol{x}\right]+\mathbb{E}\left[\varepsilon(T,\boldsymbol{X};\boldsymbol{\beta}_0)\pi_0(T,\boldsymbol{X})m(T;\boldsymbol{\beta}_0)|T=t\right]\bigg\} \\
	& \times \frac{\partial }{\partial \alpha}\bigg|_{\alpha=0}  f^{\alpha}_{Y|X,T}(y|\boldsymbol{x},t)f_{T,X}(t,\boldsymbol{x})dyd\boldsymbol{x}dt\\
	=&H_0 ^{-1}\int_{\mathcal{X}\times\mathcal{T}\times \mathcal{Y}}  \frac{f_T(t)}{f_{T|X}(t|\boldsymbol{x})}m(t;\boldsymbol{\beta}_0) \cdot L'(y-g(t;\boldsymbol{\beta}_0))\cdot  \frac{\partial }{\partial \alpha}\bigg|_{\alpha=0}  f^{\alpha}_{Y|X,T}(y|\boldsymbol{x},t)f_{T,X}(t,\boldsymbol{x})dyd\boldsymbol{x}dt\\
	=&H_0 ^{-1}\int_{\mathcal{X}\times\mathcal{T}\times \mathcal{Y}}m(t;\boldsymbol{\beta}_0)\cdot  L'(y-g(t;\boldsymbol{\beta}_0))\cdot \frac{\partial }{\partial \alpha}\bigg|_{\alpha=0}  f^{\alpha}_{Y|X,T}(y|\boldsymbol{x},t)f_{T}(t)f_{X}(\boldsymbol{x})dyd\boldsymbol{x}dt.
	\end{align*}
	For the term \eqref{eff_2},  we have
		\begin{align*}
		\eqref{eff_2}=&H_0 ^{-1}\int_{\mathcal{X}\times\mathcal{T}\times \mathcal{Y}}  \bigg\{ \frac{f_T(t)}{f_{T|X}(t|\boldsymbol{x})}m(t;\boldsymbol{\beta}_0) \cdot  L'(y-g(t;\boldsymbol{\beta}_0))-\frac{f_T(t)}{f_{T|X}(t|\boldsymbol{x})}m(t;\boldsymbol{\beta}_0) \cdot\varepsilon(t,\boldsymbol{x};\boldsymbol{\beta}_0)\\
		&\qquad +\mathbb{E}\left[\varepsilon(T,\boldsymbol{X};\boldsymbol{\beta}_0)\pi_0(T,\boldsymbol{X})m(T;\boldsymbol{\beta}_0)|\boldsymbol{X}=\boldsymbol{x}\right]+\mathbb{E}\left[\varepsilon(T,\boldsymbol{X};\boldsymbol{\beta}_0)\pi_0(T,\boldsymbol{X})m(T;\boldsymbol{\beta}_0)|T=t\right]\bigg\} \\
		& \times   f_{Y|X,T}(y|\boldsymbol{x},t)\frac{\partial }{\partial \alpha}\bigg|_{\alpha=0}f^{\alpha}_{T|X}(t|\boldsymbol{x})f_{X}(\boldsymbol{x})dyd\boldsymbol{x}dt\\
	=&H_0 ^{-1}\int_{\mathcal{X}\times\mathcal{T}}  \bigg\{\mathbb{E}\left[\varepsilon(T,\boldsymbol{X};\boldsymbol{\beta}_0)\pi_0(T,\boldsymbol{X})m(T;\boldsymbol{\beta}_0)|\boldsymbol{X}=\boldsymbol{x}\right]+\mathbb{E}\left[\varepsilon(T,\boldsymbol{X};\boldsymbol{\beta}_0)\pi_0(T,\boldsymbol{X})m(T;\boldsymbol{\beta}_0)|T=t\right]\bigg\} \\
	& \qquad \cdot  \frac{\partial }{\partial \alpha}\bigg|_{\alpha=0}f^{\alpha}_{T|X}(t|\boldsymbol{x})f_{X}(\boldsymbol{x})d\boldsymbol{x}dt\\
	=&H_0 ^{-1}\int_{\mathcal{X}\times\mathcal{T}}  \mathbb{E}\left[\varepsilon(T,\boldsymbol{X};\boldsymbol{\beta}_0)\pi_0(T,\boldsymbol{X})m(T;\boldsymbol{\beta}_0)|T=t\right]\cdot   \frac{\partial }{\partial \alpha}\bigg|_{\alpha=0}f^{\alpha}_{T|X}(t|\boldsymbol{x})f_{X}(\boldsymbol{x})d\boldsymbol{x}dt\\
	=&H_0 ^{-1}\int_{\mathcal{X}\times\mathcal{T}}  \mathbb{E}\left[\varepsilon(T,\boldsymbol{X};\boldsymbol{\beta}_0)\pi_0(T,\boldsymbol{X})m(T;\boldsymbol{\beta}_0)|T=t\right]\cdot   \frac{\partial }{\partial \alpha}\bigg|_{\alpha=0}f^{\alpha}_{T}(t)dt\\
	=&H_0 ^{-1}\int_{\mathcal{X}\times\mathcal{T}}  \varepsilon(t,\boldsymbol{x};\boldsymbol{\beta}_0)\frac{f_{T}(t)}{f_{T|X}(t|\boldsymbol{x})}m(t;\boldsymbol{\beta}_0)\cdot   \frac{\partial }{\partial \alpha}\bigg|_{\alpha=0}f^{\alpha}_{T}(t)\cdot f_{X|T}(\boldsymbol{x}|t)d\boldsymbol{x}dt\\
	=&H_0 ^{-1}\int_{\mathcal{X}\times\mathcal{T}}  \varepsilon(t,\boldsymbol{x};\boldsymbol{\beta}_0) m(t;\boldsymbol{\beta}_0)\cdot   \frac{\partial }{\partial \alpha}\bigg|_{\alpha=0}f^{\alpha}_{T}(t)\cdot f_{X}(\boldsymbol{x})d\boldsymbol{x}dt,
		\end{align*}
where the first equality holds in accordance with the definition of 	$\int_{\mathcal{Y}}L'(y-g(t;\boldsymbol{\beta}_0))f_{Y|X,T}(y|\boldsymbol{x},t)dy=:\varepsilon(t,\boldsymbol{x};\boldsymbol{\beta}_0)$.	\\
		
For the term \eqref{eff_3}, we  have	
		\begin{align*}
		\eqref{eff_3}=&H_0 ^{-1}\int_{\mathcal{X}\times\mathcal{T}\times \mathcal{Y}}  \bigg\{ \frac{f_T(t)}{f_{T|X}(t|\boldsymbol{x})}m(t;\boldsymbol{\beta}_0) \cdot  L'(y-g(t;\boldsymbol{\beta}_0))-\frac{f_T(t)}{f_{T|X}(t|\boldsymbol{x})}m(t;\boldsymbol{\beta}_0) \cdot\varepsilon(t,\boldsymbol{x};\boldsymbol{\beta}_0)\\
		&\qquad +\mathbb{E}\left[\varepsilon(T,\boldsymbol{X};\boldsymbol{\beta}_0)\pi_0(T,\boldsymbol{X})m(T;\boldsymbol{\beta}_0)|\boldsymbol{X}=\boldsymbol{x}\right]+\mathbb{E}\left[\varepsilon(T,\boldsymbol{X};\boldsymbol{\beta}_0)\pi_0(T,\boldsymbol{X})m(T;\boldsymbol{\beta}_0)|T=t\right]\bigg\} \\
		&\times f_{Y|X,T}(y|\boldsymbol{x},t)f_{T|X}(t|\boldsymbol{x})\frac{\partial }{\partial \alpha}\bigg|_{\alpha=0}f^{\alpha}_{X}(\boldsymbol{x})dyd\boldsymbol{x}dt\\
		=&H_0 ^{-1}\int_{\mathcal{X}\times\mathcal{T}}  \bigg\{\mathbb{E}\left[\varepsilon(T,\boldsymbol{X};\boldsymbol{\beta}_0)\pi_0(T,\boldsymbol{X})m(T;\boldsymbol{\beta}_0)|\boldsymbol{X}=\boldsymbol{x}\right]+\mathbb{E}\left[\varepsilon(T,\boldsymbol{X};\boldsymbol{\beta}_0)\pi_0(T,\boldsymbol{X})m(T;\boldsymbol{\beta}_0)|T=t\right]\bigg\} \\
		&\times f_{T|X}(t|\boldsymbol{x})\cdot  \frac{\partial }{\partial \alpha}\bigg|_{\alpha=0}f^{\alpha}_{X}(\boldsymbol{x})d\boldsymbol{x}dt\\
		=&H_0 ^{-1}\int_{\mathcal{X}\times\mathcal{T}}  \mathbb{E}\left[\varepsilon(T,\boldsymbol{X};\boldsymbol{\beta}_0)\pi_0(T,\boldsymbol{X})m(T;\boldsymbol{\beta}_0)|\boldsymbol{X}=\boldsymbol{x}\right]\cdot   f_{T|X}(t|\boldsymbol{x})\cdot \frac{\partial }{\partial \alpha}\bigg|_{\alpha=0}f^{\alpha}_{X}(\boldsymbol{x})d\boldsymbol{x}dt\\
			=&H_0 ^{-1}\int_{\mathcal{X} }  \mathbb{E}\left[\varepsilon(T,\boldsymbol{X};\boldsymbol{\beta}_0)\pi_0(T,\boldsymbol{X})m(T;\boldsymbol{\beta}_0)|\boldsymbol{X}=\boldsymbol{x}\right]\cdot \frac{\partial }{\partial \alpha}\bigg|_{\alpha=0}f^{\alpha}_{X}(\boldsymbol{x})d\boldsymbol{x}\\
				=&H_0 ^{-1}\int_{\mathcal{X}\times\mathcal{T}}   \varepsilon(t,\boldsymbol{x};\boldsymbol{\beta}_0)m(t;\boldsymbol{\beta}_0) \cdot   f_{T}(t)\cdot \frac{\partial }{\partial \alpha}\bigg|_{\alpha=0}f^{\alpha}_{X}(\boldsymbol{x})d\boldsymbol{x}dt.
		\end{align*}
We have proved \eqref{def_efficient_influence} holds, hence $S_{eff}$ is the efficient influence function of $\boldsymbol{\beta}_0$.

\subsection{Particular Case I: Binary Treatment Effects \label{eff_bound:binary}}
In this section, we show that when $T\in\{0,1\}$, $g(t;\boldsymbol{\beta})=\beta_0+\beta_1\cdot t$ and $L(v)=v^2$,  our general  efficiency bound derived in Theorem 3.1 reduces to the well-known efficiency bound for average treatment effects in \cite{robins1994estimation} and \cite{hahn1998role}. In accordance with our identification condition, $\beta_0$ and $\beta_1$ are identified by minimizing the following loss function 
$$\sum_{t\in\{0,1\}}\mathbb{E}[(Y^*(t)-\beta_0-\beta_1\cdot t)^2]\cdot \mathbb{P}(T=t).$$
The solutions are given by 
$$\beta_0=\mathbb{E}[Y^*(0)], \ \beta_1=\mathbb{E}[Y^*(1)-Y^*(0)].$$
Here $\beta_1$ is the average treatment effects.

\begin{cor}\label{cor_effbound_binary}
	Suppose $T\in\{0,1\}$, $L(v)=v^2$, $g(t;\boldsymbol{\beta})=\beta_0+\beta_1\cdot t$ and the conditions in Theorem 3.1 hold, the efficient influence functions of $\beta_0$ and $\beta_1$ given by Theorem 3.1 reduce to
	\begin{align*}
	&S_{eff}(T,\boldsymbol{X},Y;\beta_0)=\phi_2(T,\bold{X},Y;\beta_0), \\
	&S_{eff}(T,\boldsymbol{X},Y;\beta_1,\beta_0)=\phi_2(T,\bold{X},Y;\beta_0)-\phi_1(T,\bold{X},Y;\beta_1,\beta_0),
	\end{align*}
	where
	\begin{align*}
	&\phi_1(T,\boldsymbol{X},Y;\boldsymbol{\beta})=\frac{T}{\mathbb{P}(T=1|\boldsymbol{X})}\cdot Y^*(1)- \left\{\frac{T}{\mathbb{P}(T=1|\boldsymbol{X})}-1\right\}\cdot \mathbb{E}[Y^*(1)|\boldsymbol{X}] -\beta_0-\beta_1,  \\
	&\phi_2(T,\boldsymbol{X},Y;\boldsymbol{\beta})=\frac{1-T}{\mathbb{P}(T=0|\boldsymbol{X})}\cdot Y^*(0)- \left\{\frac{1-T}{\mathbb{P}(T=0|\boldsymbol{X})}-1\right\}\cdot \mathbb{E}[Y^*(0)|\boldsymbol{X}] -\beta_0,
	\end{align*}
	and they are the same as the efficient influence functions given in \cite{robins1994estimation} and  \cite{hahn1998role}.	
\end{cor}
\begin{proof}
	Using our notation, we have
	\begin{align*}
	&\boldsymbol{\beta}_0=(\beta_0,\beta_1)^{\top}\ ,\ g(t;\boldsymbol{\beta}_0)=\beta_0+\beta_1 \cdot t , \quad m(t;\boldsymbol{\beta}_0)=\begin{bmatrix}
	1 \\ t
	\end{bmatrix}\ ,\ H_0=\mathbb{E}\left[m(T;\boldsymbol{\beta}_0)m(T;\boldsymbol{\beta}_0)^{\top}\right]\ ,\\ &\varepsilon(T,\boldsymbol{X};\boldsymbol{\beta}_0)=T\cdot\left\{ \mathbb{E}[Y^*(1)-Y^*(0)|\boldsymbol{X}]-\beta_1\right\}+ \mathbb{E}[Y^*(0)|\boldsymbol{X}]-\beta_0, \\ 
	&\pi_0(T,\boldsymbol{X})=\frac{T\cdot p+(1-T)\cdot q}{T\cdot \mathbb{P}(T=1|\boldsymbol{X})+T\cdot \mathbb{P}(T=0|\boldsymbol{X})}=\frac{T}{\mathbb{P}(T=1|\boldsymbol{X})}\cdot p +\frac{1-T}{\mathbb{P}(T=0|\boldsymbol{X})}\cdot q,
	\end{align*} 
	where $p=\mathbb{P}(T=1)$ and  $q=\mathbb{P}(T=0)$.
	In accordance with our Theorem 3.1,  the efficient influence function of $(\beta_0,\beta_1)$ is
 \begin{align*}
H^{-1}_0\bigg\{\pi_0 (T,
		\boldsymbol{X})m(T;\boldsymbol{\beta}_0)\left\{ Y-\mathbb{E}[Y|
		\boldsymbol{X},T]\right\} +\mathbb{E}\left[ \varepsilon (T,\boldsymbol{X};
		\boldsymbol{\beta }_{0})\pi_0 (T,\boldsymbol{X})m(T;\boldsymbol{\beta}_0)|
		\boldsymbol{X}\right]\bigg\}\ .
		\end{align*}
	With some computation, we have 
	\begin{align} \label{m_inverse}
		H^{-1}_0 =\begin{bmatrix}
	1& p \\
	p& p
	\end{bmatrix}^{-1}=\frac{1}{pq}\cdot \begin{bmatrix}
	p& -p \\
	-p& 1
	\end{bmatrix}=\begin{bmatrix}
	\frac{1}{q}& -\frac{1}{q} \\
	-\frac{1}{q}& \frac{1}{pq}
	\end{bmatrix}\ .
	\end{align}
and
	\begin{align}
	&\pi_0 (T,
	\boldsymbol{X})m(T;\boldsymbol{\beta}_0)\left\{ Y-\mathbb{E}[Y|%
	\boldsymbol{X},T]\right\} \notag\\
	=&\frac{T}{\mathbb{P}(T=1|\boldsymbol{X})}\cdot p \cdot \begin{bmatrix}
	1 \\ T
	\end{bmatrix}\cdot\bigg\{Y-T\cdot \mathbb{E}[Y^*(1)|\boldsymbol{X}]-(1-T)\cdot \mathbb{E}[Y^*(0)|\boldsymbol{X}]\bigg\}  \notag\\
	&+\frac{1-T}{\mathbb{P}(T=0|\boldsymbol{X})}\cdot q\cdot  \begin{bmatrix}
	1 \\ T
	\end{bmatrix}\cdot\bigg\{Y-T\cdot \mathbb{E}[Y^*(1)|\boldsymbol{X}]-(1-T)\cdot \mathbb{E}[Y^*(0)|\boldsymbol{X}]\bigg\}  \notag\\
	=&\frac{T}{\mathbb{P}(T=1|\boldsymbol{X})}\cdot p \cdot \begin{bmatrix}
	1 \\ 1
	\end{bmatrix}\cdot\bigg\{Y^*(1)- \mathbb{E}[Y^*(1)|\boldsymbol{X}]\bigg\}
	+\frac{1-T}{\mathbb{P}(T=0|\boldsymbol{X})}\cdot q \cdot \begin{bmatrix}
	1 \\ 0
	\end{bmatrix}\cdot\bigg\{Y^*(0)- \mathbb{E}[Y^*(0)|\boldsymbol{X}]\bigg\} \notag\\
	=&\begin{bmatrix}
	\frac{T}{\mathbb{P}(T=1|\boldsymbol{X})}\cdot \left\{Y^*(1)- \mathbb{E}[Y^*(1)|\boldsymbol{X}]\right\}\cdot p+\frac{1-T}{\mathbb{P}(T=0|\boldsymbol{X})}\cdot \left\{Y^*(0)- \mathbb{E}[Y^*(0)|\boldsymbol{X}]\right\}\cdot q \\[4mm]
	\frac{T}{\mathbb{P}(T=1|\boldsymbol{X})}\cdot \left\{Y^*(1)- \mathbb{E}[Y^*(1)|\boldsymbol{X}]\right\}\cdot p
	\end{bmatrix} \label{eff:pi}
	\end{align}
	and
	\begin{align}
	&\mathbb{E}\left[ \varepsilon (T,\boldsymbol{X};
	\boldsymbol{\beta }_{0})\pi_0 (T,\boldsymbol{X})m(T;\boldsymbol{\beta}_0)|%
	\boldsymbol{X}\right]  \notag\\
	=&\mathbb{E}\Bigg[\bigg(T\cdot\left\{ \mathbb{E}[Y^*(1)-Y^*(0)|\boldsymbol{X}]-\beta_1\right\}+ \mathbb{E}[Y^*(0)|\boldsymbol{X}]-\beta_0\bigg)\cdot \frac{T}{\mathbb{P}(T=1|\boldsymbol{X})}\cdot p \cdot \begin{bmatrix}
	1 \\ T
	\end{bmatrix} \bigg|\boldsymbol{X}\Bigg]  \notag\\
	&+\mathbb{E}\Bigg[\bigg(T\cdot\left\{ \mathbb{E}[Y^*(1)-Y^*(0)|\boldsymbol{X}]-\beta_1\right\}+ \mathbb{E}[Y^*(0)|\boldsymbol{X}]-\beta_0\bigg)\cdot \frac{1-T}{\mathbb{P}(T=0|\boldsymbol{X})}\cdot q \cdot \begin{bmatrix}
	1 \\ T
	\end{bmatrix} \bigg|\boldsymbol{X} \Bigg] \notag \\
	=&\mathbb{E}\Bigg[\bigg( \mathbb{E}[Y^*(1)|\boldsymbol{X}]-\beta_1-\beta_0\bigg)\cdot \frac{T}{\mathbb{P}(T=1|\boldsymbol{X})}\cdot p \cdot \begin{bmatrix}
	1 \\ 1
	\end{bmatrix} \bigg|\boldsymbol{X}\Bigg] \notag\\
	&+\mathbb{E}\Bigg[\bigg( \mathbb{E}[Y^*(0)|\boldsymbol{X}]-\beta_0\bigg)\cdot \frac{1-T}{\mathbb{P}(T=0|\boldsymbol{X})}\cdot q \cdot \begin{bmatrix}
	1 \\ 0
	\end{bmatrix} \bigg|\boldsymbol{X} \Bigg]  \notag\\
	=&\begin{bmatrix}
	\bigg(\mathbb{E}[Y^*(1)|\boldsymbol{X}]-\beta_1-\beta_0\bigg)\cdot p +\bigg(\mathbb{E}[Y^*(0)|\boldsymbol{X}]-\beta_1\bigg)\cdot q
	\\[4mm]
	\bigg(\mathbb{E}[Y^*(1)|\boldsymbol{X}]-\beta_1-\beta_0\bigg)\cdot p
	\end{bmatrix}. \label{eff:epsilon}
	\end{align}
	Therefore, with \eqref{m_inverse}, \eqref{eff:pi}, and \eqref{eff:epsilon} we can obtain that
	\begin{align*}
	&\pi_0 (T,
	\boldsymbol{X})m(T;\boldsymbol{\beta}_0)\left\{ Y-\mathbb{E}[Y|%
	\boldsymbol{X},T]\right\}+\mathbb{E}\left[ \varepsilon (T,\boldsymbol{X};%
	\boldsymbol{\beta }_{0})\pi_0 (T,\boldsymbol{X})m(T;\boldsymbol{\beta}_0)|%
	\boldsymbol{X}\right]\\
	=&\begin{pmatrix}
	p\cdot \phi_1(T,\boldsymbol{X},Y;\boldsymbol{\beta}_0)+q\cdot \phi_2(T,\boldsymbol{X},Y;\boldsymbol{\beta}_0) \\[2mm]
	p\cdot \phi_1(T,\boldsymbol{X},Y;\boldsymbol{\beta}_0)
	\end{pmatrix},
	\end{align*}
	and the efficient influence functions of $\beta_1$ and $\beta_2$ are given by
	\begin{align*}
	&\begin{bmatrix}
	\frac{1}{q}& -\frac{1}{q} \\
	-\frac{1}{q}& \frac{1}{pq}
	\end{bmatrix}\cdot \begin{pmatrix}
	p\cdot \phi_1(T,\boldsymbol{X},Y;\boldsymbol{\beta})+q\cdot \phi_2(T,\boldsymbol{X},Y;\boldsymbol{\beta}) \\[2mm]
	p\cdot \phi_1(T,\boldsymbol{X},Y;\boldsymbol{\beta})
	\end{pmatrix}=\begin{pmatrix}
	\phi_2(T,\boldsymbol{X},Y;\boldsymbol{\beta}) \\[2mm]
	\phi_1(T,\boldsymbol{X},Y;\boldsymbol{\beta})- \phi_2(T,\boldsymbol{X},Y;\boldsymbol{\beta})\ .
	\end{pmatrix}.
	\end{align*}
\end{proof}

\subsection{Particular Case II: Multiple Treatment Effects \label{eff_bound:multiple}}
In this section, we show that when $T\in\{0,1,...,J\}$, $J\in\mathbb{N}$, $g(t;\boldsymbol{\beta})=\sum_{j=0}^J \beta_j \cdot I(t=j)$ and $L(v)=v^2$, our general  efficiency bound derived in Theorem 3.1 reduces to the efficiency bound of multi-level treatment effects given in \cite{cattaneo2010efficient}. In accordance with our proposed identification condition, $\{\beta_j\}_{j=0}^J$ are identified by minimizing the following loss function
$$\sum_{j=0}^J\mathbb{E}\left[(Y^*(j)-\beta_j)^2\right]\cdot \mathbb{P}(T=j).$$ 
The solutions are $\beta_j=\mathbb{E}[Y^*(j)]$ for $j\in\{0,...,J\}$.

\begin{cor}
	\label{cor_effbound_multiple}
	Suppose $T\in\{0,1,...,J\}$, $J\in\mathbb{N}$, $g(t;\boldsymbol{\beta})=\sum_{j=0}^J \beta_j \cdot I(t=j)$, $L(v)=v^2$, and the conditions in Theorem 3.1 hold, the efficient influence functions of $\{\beta_j\}_{j=0}^J$  given by Theorem 3.1 reduce to
	\begin{align*}
	S_{eff}(T,\boldsymbol{X},Y;\beta_j)=\frac{I(T=j)}{\mathbb{P}(T=j|\boldsymbol{X})}\cdot \left\{Y^*(j)-\mathbb{E}[Y^*(j)|\boldsymbol{X}]\right\} + \mathbb{E}[Y^*(j)|X]- \beta_j,\ j\in\{0,...,J\},
	\end{align*}
	and they are the same as the efficient influence functions given in \cite{cattaneo2010efficient}.
\end{cor}
\begin{proof}
	Using our notation, we have
	\begin{align*}
	\boldsymbol{\beta}_0=(\beta_0,...,\beta_J)^{\top}, \ g(t;\boldsymbol{\beta}_0)=\sum_{j=0}^J \beta_j \cdot I(t=j), \ m(t;\boldsymbol{\beta}_0)=\begin{bmatrix}
	I(t=0) \\ I(t=1)\\ \vdots \\ I(t=J)
	\end{bmatrix}, \ H_0=\mathbb{E}\left[m(T;\boldsymbol{\beta}_0)m(T;\boldsymbol{\beta}_0)^\top\right].
	\end{align*}
	Then
	\begin{align*}
	\varepsilon(T,\boldsymbol{X};\boldsymbol{\beta}_0)=&\mathbb{E}[Y|T,X]-g(T;\boldsymbol{\beta}_0)\\
	=&\sum_{j=0}^J \mathbb{E}[Y^*(j)|X] \cdot I(t=j)-\sum_{j=0}^J \beta_j \cdot I(T=j)\\
	=&\sum_{j=0}^J\left( \mathbb{E}[Y^*(j)|X]- \beta_j\right) \cdot I(T=j)
	\end{align*}
	and 
	\begin{align*}
	\pi_0(T,X)=\sum_{j=0}^J\frac{I(T=j)}{\mathbb{P}(T=j|\boldsymbol{X})}\cdot p_j, \ \text{where} \ p_j=\mathbb{P}(T=j).
	\end{align*}
	Then we have
	\begin{align*}
	H_0^{-1}=\mathbb{E}\left[m(T;\boldsymbol{\beta}_0)m(T;\boldsymbol{\beta}_0)^\top\right]^{-1}=\begin{bmatrix}
	& p_0^{-1} &  & &  \\
	&  & p_1^{-1} & & \\
	& &  \cdots & & \\
	&  & & & p_J^{-1} 
	\end{bmatrix},
	\end{align*}
	and
	\begin{align}
	&\pi_0 (T,%
	\boldsymbol{X})m(T;\boldsymbol{\beta}_0)\left\{ Y-\mathbb{E}[Y|%
	\boldsymbol{X},T]\right\} \notag\\
	=&\left\{\sum_{j=0}^J\frac{I(T=j)}{\mathbb{P}(T=j|\boldsymbol{X})}\cdot p_j\right\} \cdot \begin{bmatrix}
	I(T=0) \\ I(T=1)\\ \vdots \\I(T=J)
	\end{bmatrix}\cdot\bigg\{Y-\sum_{j=0}^J I(T=j)\cdot \mathbb{E}[Y^*(j)|\boldsymbol{X}]\bigg\}   \notag\\
	=&\begin{bmatrix}
	I(T=0) \\ I(T=1)\\ \vdots \\I(T=J)
	\end{bmatrix} \left\{\sum_{j=0}^J\frac{I(T=j)}{\mathbb{P}(T=j|\boldsymbol{X})}\cdot p_j\cdot Y^*(j)-\sum_{j=0}^J\frac{I(T=j)}{\mathbb{P}(T=j|\boldsymbol{X})}\cdot p_j\cdot \mathbb{E}[Y^*(j)|\boldsymbol{X}] \right\}\notag\\
	=& \begin{bmatrix}
	\frac{I(T=0)}{\mathbb{P}(T=0|\boldsymbol{X})}\cdot p_0\cdot \left\{Y^*(0)-\mathbb{E}[Y^*(0)|\boldsymbol{X}]\right\} \\[2mm]  
	\frac{I(T=1)}{\mathbb{P}(T=1|\boldsymbol{X})}\cdot p_1\cdot \left\{Y^*(1)-\mathbb{E}[Y^*(1)|\boldsymbol{X}]\right\}\\ \vdots \\[2mm] \frac{I(T=J)}{\mathbb{P}(T=J|\boldsymbol{X})}\cdot p_J\cdot \left\{Y^*(j)-\mathbb{E}[Y^*(j)|\boldsymbol{X}]\right\}
	\end{bmatrix}\label{eff:pi_mul}
	\end{align}
	and
	\begin{align*}
	& \varepsilon (T,\boldsymbol{X};%
	\boldsymbol{\beta }_{0})\pi_0 (T,\boldsymbol{X})m(T;\boldsymbol{\beta}_0)\\
	=&\left\{\sum_{j=0}^J\left( \mathbb{E}[Y^*(j)|X]- \beta_j\right) \cdot I(T=j)\right\}\left\{\sum_{j=0}^J\frac{I(T=j)}{\mathbb{P}(T=j|\boldsymbol{X})}\cdot p_j\right\}\begin{bmatrix}
	I(T=0) \\ I(T=1)\\ \vdots \\ I(T=J)
	\end{bmatrix} \\
	=&\begin{bmatrix}
	\frac{I(T=0)}{\mathbb{P}(T=0|\boldsymbol{X})}\cdot p_0\cdot \left\{ \mathbb{E}[Y^*(0)|X]- \beta_0\right\} \\[2mm] \frac{I(T=1)}{\mathbb{P}(T=1|\boldsymbol{X})}\cdot p_1\cdot \left\{ \mathbb{E}[Y^*(1)|X]- \beta_1\right\}\\ \vdots \\ \frac{I(T=J)}{\mathbb{P}(T=J|\boldsymbol{X})}\cdot p_J\cdot \left\{ \mathbb{E}[Y^*(j)|X]- \beta_J\right\}
	\end{bmatrix}
	\end{align*}
	and
	\begin{align}
	\mathbb{E}\left[ \varepsilon (T,\boldsymbol{X};%
	\boldsymbol{\beta }_{0})\pi_0 (T,\boldsymbol{X})m(T;\boldsymbol{\beta}_0)|%
	\boldsymbol{X}\right]= \begin{bmatrix}
	p_0\cdot \left\{ \mathbb{E}[Y^*(0)|X]- \beta_0\right\} \\[2mm]   p_1\cdot \left\{ \mathbb{E}[Y^*(1)|X]- \beta_1\right\}\\ \vdots \\   p_J\cdot \left\{ \mathbb{E}[Y^*(j)|X]- \beta_J\right\}
	\end{bmatrix}. \label{eff:epsilon_mul}
	\end{align}
	From Theorem 3.1, the efficient influence function of $\boldsymbol{\beta }_0=(\beta_0,...,\beta_J)$ is given by 
		\begin{align*}
		&H_0^{-1}\left\{\pi_0 (T,%
		\boldsymbol{X})m(T;\boldsymbol{\beta}_0)\left\{ Y-\mathbb{E}[Y|%
		\boldsymbol{X},T]\right\}+	\mathbb{E}\left[ \varepsilon (T,\boldsymbol{X};%
		\boldsymbol{\beta }_{0})\pi_0 (T,\boldsymbol{X})m(T;\boldsymbol{\beta}_0)|%
		\boldsymbol{X}\right]\right\}\\
		=&\begin{bmatrix}
		\frac{I(T=0)}{\mathbb{P}(T=0|\boldsymbol{X})}\cdot \left\{Y^*(0)-\mathbb{E}[Y^*(0)|\boldsymbol{X}]\right\} + \mathbb{E}[Y^*(0)|X]- \beta_0\\[2mm]  
		\frac{I(T=1)}{\mathbb{P}(T=1|\boldsymbol{X})} \cdot \left\{Y^*(1)-\mathbb{E}[Y^*(1)|\boldsymbol{X}]\right\}  +\mathbb{E}[Y^*(1)|X]- \beta_1\\ \vdots \\[2mm] \frac{I(T=J)}{\mathbb{P}(T=J|\boldsymbol{X})} \cdot \left\{Y^*(j)-\mathbb{E}[Y^*(j)|\boldsymbol{X}]\right\}+ \mathbb{E}[Y^*(j)|X]- \beta_J
		\end{bmatrix},
		\end{align*}
	which is the same as the efficient influence function developed in  Corollary 1 of \cite{cattaneo2010efficient}.
\end{proof}

\subsection{Particular Case III: Quantile Treatment Effects}
In this section, we show that when $T\in\{0,1\}$ is a binary treatment variable, $L(v)=v(\tau-I(v\leq 0))$ is the check function with  $\tau\in(0,1)$, and $g(t;\boldsymbol{\beta}_0)=\beta_0\cdot (1-t)+\beta_1\cdot t$, where $\boldsymbol{\beta}_0=(\beta_0,\beta_1)$,  our general efficiency bound derived in Theorem 3.1 reduces to the efficiency bound of quantile treatment effects given in \cite{Firpo2007Efficient}. In accordance with our identification condition, $\beta_0$ and $\beta_1$ are identified by minimizing the following loss function   
\begin{align*}
 \sum_{j\in\{0,1\}}\mathbb{P}(T=j)\cdot \mathbb{E}\left[(Y^*(j)-\beta_j)\left\{\tau-I(Y^*(j)\leq \beta_j) \right\}\right].
\end{align*}
The solutions are  $\beta_0=\inf\{q: \mathbb{P}(Y^*(0)\leq q)\geq \tau\}$ and $\beta_1=\inf\{q: \mathbb{P}(Y^*(1)\leq q)\geq \tau\}$, which are the $\tau^{th}$ quantiles of potential outcomes.

\begin{cor}\label{cor_effbound_QTE}
Let $T\in\{0,1\}$, $f_{Y^*(1)}$ and $f_{Y^*(0)}$ be the probability densities of the potential outcomes $Y^*(1)$ and $Y^*(0)$ respectively,  $g(t;\boldsymbol{\beta}_0)=\beta_0\cdot (1-t)+\beta_1\cdot t$, $L(v)=v(\tau-I(v\leq  0))$, and the conditions in Theorem 3.1 hold, then  the efficient influence function of $\boldsymbol{\beta}_0$ given by Theorem 3.1 reduces to
\begin{align*}
S_{eff}(Y,T,\boldsymbol{X};\boldsymbol{\beta}_0)
=\begin{bmatrix}
\frac{1-T}{\mathbb{P}(T=0|\boldsymbol{X})}\cdot \left\{\frac{\tau-I(Y^*(0)\leq \beta_0) }{f_{Y^*(0)}(\beta_0)}\right\}-  \left(\frac{1-T}{\mathbb{P}(T=0|\boldsymbol{X})}-1\right)\cdot \mathbb{E}\left[\frac{\tau-I(Y^*(0)\leq \beta_0)}{f_{Y^*(0)}(\beta_0)} \big|\boldsymbol{X}\right]\\[2mm]
\frac{T}{\mathbb{P}(T=1|\boldsymbol{X})} \cdot \left\{\frac{\tau-I(Y^*(1)\leq \beta_1)}{f_{Y^*(1)}(\beta_1)} \right\}-  \left(\frac{T}{\mathbb{P}(T=1|\boldsymbol{X})}-1\right)\cdot \mathbb{E}\left[\frac{\tau-I(Y^*(1)\leq \beta_1)}{f_{Y^*(1)}(\beta_1)} \big|\boldsymbol{X}\right]
\end{bmatrix},
\end{align*}
 which is the same as the efficient influence function given in \cite{Firpo2007Efficient}.	
\end{cor}
\begin{proof}
Using our notation, we have
\begin{align*}
&\boldsymbol{\beta}_0=(\beta_0,\beta_1)^{\top},\ g(t;\boldsymbol{\beta}_0)=\beta_0\cdot (1-t)+\beta_1 \cdot t , \ m(t;\boldsymbol{\beta}_0)=\begin{bmatrix}
1-t \\ t
\end{bmatrix},\\
&L(v)=v(\tau-I(v\leq 0)), \ L'(v)=\tau-I(v\leq 0)\ \text{a.s.}, \\
 &\varepsilon(T,\boldsymbol{X};\boldsymbol{\beta}_0)=T\cdot \mathbb{E}[\tau-I(Y^*(1)\leq \beta_1)|\boldsymbol{X}]+(1-T)\cdot \mathbb{E}[\tau-I(Y^*(0)\leq \beta_0)|\boldsymbol{X}], \\ 
&\pi_0(T,\boldsymbol{X})=\frac{T}{\mathbb{P}(T=1|\boldsymbol{X})}\cdot p +\frac{1-T}{\mathbb{P}(T=0|\boldsymbol{X})}\cdot q\ , \ p=\mathbb{P}(T=1), \ q=\mathbb{P}(T=0).
\end{align*}
 Direct computation yields 
\begin{align*}
&\pi_0(T,\boldsymbol{X})m(T;\boldsymbol{\beta}_0)L'(Y-g(T;\boldsymbol{\beta}_0))\\
=&\left\{\frac{T}{\mathbb{P}(T=1|\boldsymbol{X})}\cdot p +\frac{1-T}{\mathbb{P}(T=0|\boldsymbol{X})}\cdot q\right\}\cdot \begin{bmatrix}
1-T \\ T
\end{bmatrix}\cdot \bigg\{\tau-I(Y\leq \beta_0\cdot (1-T)+\beta_1\cdot T )\bigg\}\\
=& \begin{bmatrix}
\frac{1-T}{\mathbb{P}(T=0|\boldsymbol{X})}\cdot q\cdot \left\{\tau-I(Y^*(0)\leq \beta_0) \right\} \\[2mm] 
\frac{T}{\mathbb{P}(T=1|\boldsymbol{X})}\cdot p\cdot \left\{\tau-I(Y^*(1)\leq \beta_1) \right\}
\end{bmatrix}
\end{align*}
and
\begin{align*}
\pi_0(T,\boldsymbol{X})m(T;\boldsymbol{\beta}_0)\varepsilon(T,\boldsymbol{X};\boldsymbol{\beta}_0)=\begin{bmatrix}
\frac{1-T}{\mathbb{P}(T=0|\boldsymbol{X})}\cdot q\cdot \mathbb{E}\left[\tau-I(Y^*(0)\leq \beta_0)|\boldsymbol{X} \right] \\[2mm] 
\frac{T}{\mathbb{P}(T=1|\boldsymbol{X})}\cdot p\cdot \mathbb{E}\left[\tau-I(Y^*(1)\leq \beta_1) |\boldsymbol{X}\right]
\end{bmatrix}
\end{align*}
and 
\begin{align*}
\mathbb{E}\left[\pi_0(T,\boldsymbol{X})m(T;\boldsymbol{\beta}_0)\varepsilon(T,\boldsymbol{X};\boldsymbol{\beta}_0)|\boldsymbol{X}\right]=\begin{bmatrix}
 q\cdot \mathbb{E}\left[\tau-I(Y^*(0)\leq \beta_0)|\boldsymbol{X} \right] \\[2mm] 
p\cdot  \mathbb{E}\left[\tau-I(Y^*(1)\leq \beta_1) |\boldsymbol{X}\right]
\end{bmatrix}
\end{align*}
and 
\begin{align*}
H_0=&\nabla_{\boldsymbol{\beta}}\mathbb{E}\left[\pi_0(T,\boldsymbol{X})m(T;\boldsymbol{\beta})L'(Y-g(T;\boldsymbol{\beta}))\right]
=\begin{bmatrix}
-q\cdot f_{Y^*(0)}(\beta_0) & 0 \\
0 &-p\cdot f_{Y^*(1)}(\beta_1) 
\end{bmatrix}.
\end{align*}
Therefore, by Theorem 3.1,  the efficient influence function of $\boldsymbol{\beta}_0$ is 
\begin{align*}
&S_{eff}(Y,T,\boldsymbol{X};\boldsymbol{\beta}_0)\\
=&H_0^{-1}\cdot \bigg\{ \pi_0 (T,
\boldsymbol{X})m(T;\boldsymbol{\beta }_{0})L'(Y-g(T;\boldsymbol{\beta}_0))-\pi_0 (T,\boldsymbol{X})m(T;\boldsymbol{\beta }_{0})\varepsilon (T,\boldsymbol{X};
\boldsymbol{\beta }_{0})\\
& \qquad \qquad+\mathbb{E}\left[ \varepsilon (T,\boldsymbol{X};
\boldsymbol{\beta }_{0})\pi_0 (T,\boldsymbol{X})m(T;\boldsymbol{\beta }_{0})|
\boldsymbol{X}\right] \bigg\}\\
=&\begin{bmatrix}
q^{-1}\cdot \frac{1}{f_{Y^*(0)}(\beta_0)} & 0 \\
0 &p^{-1}\cdot \frac{1}{f_{Y^*(1)}(\beta_1)}
\end{bmatrix}\\
&\times  \begin{bmatrix}
\frac{1-T}{\mathbb{P}(T=0|\boldsymbol{X})}\cdot q\cdot \left\{\tau-I(Y^*(0)\leq \beta_0) \right\}-q\cdot \left(\frac{1-T}{\mathbb{P}(T=0|\boldsymbol{X})}-1\right)\cdot \mathbb{E}\left[\tau-I(Y^*(0)\leq \beta_0) |\boldsymbol{X}\right]\\[2mm]
\frac{T}{\mathbb{P}(T=1|\boldsymbol{X})}\cdot p\cdot \left\{\tau-I(Y^*(1)\leq \beta_1) \right\}-p\cdot \left(\frac{T}{\mathbb{P}(T=1|\boldsymbol{X})}-1\right)\cdot \mathbb{E}\left[\tau-I(Y^*(1)\leq \beta_1) |\boldsymbol{X}\right]
\end{bmatrix}\\
=&\begin{bmatrix}
\frac{1-T}{\mathbb{P}(T=0|\boldsymbol{X})}\cdot \left\{\frac{\tau-I(Y^*(0)\leq \beta_0) }{f_{Y^*(0)}(\beta_0)}\right\}-  \left(\frac{1-T}{\mathbb{P}(T=0|\boldsymbol{X})}-1\right)\cdot \mathbb{E}\left[\frac{\tau-I(Y^*(0)\leq \beta_0)}{f_{Y^*(0)}(\beta_0)} \bigg|\boldsymbol{X}\right]\\[2mm]
\frac{T}{\mathbb{P}(T=1|\boldsymbol{X})} \cdot \left\{\frac{\tau-I(Y^*(1)\leq \beta_1)}{f_{Y^*(1)}(\beta_1)} \right\}-  \left(\frac{T}{\mathbb{P}(T=1|\boldsymbol{X})}-1\right)\cdot \mathbb{E}\left[\frac{\tau-I(Y^*(1)\leq \beta_1)}{f_{Y^*(1)}(\beta_1)} \bigg|\boldsymbol{X}\right]
\end{bmatrix},
\end{align*}
which coincides with efficiency bound derived in \cite{Firpo2007Efficient}.
\end{proof}

\section{Convergence Rate of Estimated Stabilized Weights \label{sec:key_lemmas}}
In this section, we establish the convergence rate of estimated stabilized weights $\hat{\pi}_K(T,\boldsymbol{X})$. Let $G_{K_1\times K_2}^*$, $\Lambda_{K_1\times K_2}^*$ and $\pi_K^*(t,\boldsymbol{x})$  be the theoretical
counterparts of $\hat{G}_{K_1\times K_2}$, $\hat{\Lambda}_{K_1\times K_2}$ and $\hat{\pi}_K(t,\boldsymbol{x})$ respectively: 
\begin{align*}
G_{K_1\times K_2}^*(\Lambda):= & \mathbb{E}[\hat{G}_{K_1\times
	K_2}(\Lambda)]=\mathbb{E}\left[\rho\left(u_{K_1}(T)^{\top}\Lambda v_{K_2}(%
\boldsymbol{X})\right)\right]-\mathbb{E}[u_{K_1}(T)^{\top}]\cdot\Lambda\cdot 
\mathbb{E}[v_{K_2}(\boldsymbol{X})], \\[2mm]
\Lambda_{K_1\times K_2}^* := & \arg\max G_{K_1\times K_2}^*(\Lambda), \\[2mm]
\pi_K^*(t,\boldsymbol{x}):= &\rho'\left(u_{K_1}(t)^{\top}\Lambda_{K_1\times K_2}^*v_{K_2}(\boldsymbol{x})\right).
\end{align*}
As discussed in Appendix A.3,  we assume the sieve basises $u_{K_1}(T)$ and $v_{K_2}(\boldsymbol{X})$ are orthonormalized, i.e.,
\begin{align}\label{eq:orthonormal_basis}
\mathbb{E}\left[u_{K_1}(T)u_{K_1}^{\top}(T)\right]=I_{K_1\times K_1}, \quad \mathbb{E}\left[v_{K_2}(\boldsymbol{X})v_{K_2}^{\top}(\boldsymbol{X})\right]=I_{K_2\times K_2} . 
\end{align}
Let
\begin{align*}
& \zeta_1(K_1) :=\sup_{t\in\mathcal{T}}\|u_{K_1}(t)\|\ , \
\zeta_2(K_2):=\sup_{\boldsymbol{x}\in\mathcal{X}}\|v_{K_2}(\boldsymbol{x})\| \ , \ K=K_1\cdot K_2 \ , \ \zeta(K)=\zeta_1(K_1)\zeta_2(K_2).
\end{align*}
We also recall the following property satisfied by $\pi_0(T,\boldsymbol{X})$: for any integrable functions $u(t)$ and $v(\boldsymbol{X})$,
	\begin{equation}
	\mathbb{E}\left[ \pi _{0}(T,\boldsymbol{X})u(T)v(\boldsymbol{X})\right] =%
	\mathbb{E}[u(T)]\cdot \mathbb{E}[v(\boldsymbol{X})]. \label{moment1}
	\end{equation}%
\subsection{Lemma $\ref{lemma_pi^*}$}
The first lemma states that $\pi_K^*(t,\boldsymbol{x})$ is arbitrarily close to the true stabilized weights $\pi_0(t,\boldsymbol{x})$.
\begin{lemma}\label{lemma_pi^*}
 Under Assumption \ref{as:suppX}-\ref{as:K&N_consistency}, we have
	$$ \sup_{(t,\boldsymbol{x}) \in \mathcal{T} \times \mathcal{X}}|\pi_0(t,\boldsymbol{x}) - \pi^*_K(t,\boldsymbol{x}) | = O\left(K^{-\alpha}\zeta(K)\right),$$
	and
	$$ \mathbb{E}\left[|\pi_0(T,\boldsymbol{X}) - \pi^*_K(T,\boldsymbol{X}) |^2\right] = O\left(K^{-2\alpha}\right),$$
	and
	$$\frac{1}{N}\sum_{i=1}^N|\pi_0(T_i,\boldsymbol{X}_i) - \pi^*_K(T_i,\boldsymbol{X}_i) |^2=O_p\left(K^{-2\alpha}\right).$$
\end{lemma}
\begin{proof} By Assumption \ref{as:pi0}, $\pi_0(t,\boldsymbol{x})\in [\eta_1, \eta_2], \ \forall (t,\boldsymbol{x}) \in \mathcal{T}\times \mathcal{X}$ and $(\rho')^{-1}$ is strictly decreasing.
	Define $$ \overline{\gamma} := \sup_{(t,\boldsymbol{x}) \in \mathcal{T}\times\mathcal{X}} (\rho')^{-1}\left(\pi_0(t,\boldsymbol{x})\right)
	\leq (\rho')^{-1}(\eta_1) ~~ \text{and} ~~
	\underline{\gamma} := \inf_{(t,\boldsymbol{x}) \in \mathcal{T}\times\mathcal{X}} (\rho')^{-1}\left(\pi_0(t,\boldsymbol{x})\right)
	\geq(\rho')^{-1}(\eta_2),$$
	which are two finite constants. 
	By Assumptions \ref{as:smooth_pi},
	there exist a constant $C> 0$ and a $K_1\times K_2$ matrix $\Lambda_{K_1\times K_2} \in \mathbb{R}^{K_1\times K_2}$ such that 
	\begin{align*}
	\sup_{(t,\boldsymbol{x})\in \mathcal{T}\times \mathcal{X}}\left|(\rho')^{-1}\left(\pi_0(t,\boldsymbol{x})\right) - u_{K_1}(t)^{\top}\Lambda_{K_1\times K_2}v_{K_2}(\boldsymbol{x})\right|  < CK^{-\alpha},
	\end{align*}
	which implies
	\begin{align}\label{eq:resubmit_lamdau_K}
	u_{K_1}(t)^{\top}\Lambda_{K_1\times K_2}v_{K_2}(\boldsymbol{x}) \in & \left((\rho')^{-1}\left(\pi_0(t,\boldsymbol{x})\right) -  CK^{-\alpha}, (\rho')^{-1}\left(\pi_0(t,\boldsymbol{x})\right) + CK^{-\alpha} \right)  \\
	\subset & \left[\underline{\gamma} -  CK^{-\alpha}, \overline{\gamma} +  CK^{-\alpha}\right], \  \forall (t,\boldsymbol{x}) \in \mathcal{T}\times \mathcal{X}, \notag
	\end{align}
	and 
	\begin{align*}
	&\rho'\left(u_{K_1}(t)^{\top}\Lambda_{K_1\times K_2}v_{K_2}(\boldsymbol{x}) +CK^{-\alpha}\right) - \rho'(u_{K_1}(t)^{\top}\Lambda_{K_1\times K_2}v_{K_2}(\boldsymbol{X}))\\
	< &\pi_0(t,\boldsymbol{x}) - \rho'\left(u_{K_1}(t)^{\top}\Lambda_{K_1\times K_2}v_{K_2}(\boldsymbol{x})\right) \\
	< & \rho'\left(u_{K_1}(t)^{\top}\Lambda_{K_1\times K_2}v_{K_2}(\boldsymbol{x})-CK^{-\alpha}\right) - \rho'(u_{K_1}(t)^{\top}\Lambda_{K_1\times K_2}v_{K_2}(\boldsymbol{x})) \ , \forall (t,\boldsymbol{x})\in \mathcal{T}\times \mathcal{X}.
	\end{align*}
	Let $\Gamma_1:= [\underline{\gamma} -1, \overline{\gamma} + 1]$, 
	by Mean Value Theorem, for large enough $K$, there exist
	\begin{align*} 
	\xi_1(t,\boldsymbol{x}) \in & \left(u_{K_1}(t)^{\top}\Lambda_{K_1\times K_2}v_{K_2}(\boldsymbol{x}), u_{K_1}(t)^{\top}\Lambda_{K_1\times K_2}v_{K_2}(\boldsymbol{x}) + CK^{-\alpha}\right)\\
	\subset &\left[\underline{\gamma} -CK^{-\alpha}, \overline{\gamma} + 2CK^{-\alpha}\right]\subset\Gamma_1 \ , \\
	\xi_2(t,\boldsymbol{x}) \in& \left(u_{K_1}(t)^{\top}\Lambda_{K_1\times K_2}v_{K_2}(\boldsymbol{x}) -CK^{-\alpha}, u_{K_1}(t)^{\top}\Lambda_{K_1\times K_2}v_{K_2}(\boldsymbol{x})\right)\\
	\subset & \left[\underline{\gamma} - 2CK^{-\alpha}, \overline{\gamma} + CK^{-\alpha}\right]\subset \Gamma_1,
	\end{align*}
	such that
	\begin{align*}
	\rho'\left(u_{K_1}(t)^{\top}\Lambda_{K_1\times K_2}v_{K_2}(\boldsymbol{x}) + CK^{-\alpha}\right) - \rho'\left(u_{K_1}(t)^{\top}\Lambda_{K_1\times K_2}v_{K_2}(\boldsymbol{x}) \right)= \rho''(\xi_1(t,x))CK^{-\alpha}\geq -a_1CK^{-\alpha}
	\end{align*}
	and
	\begin{align*}
 \rho'\left(u_{K_1}(t)^{\top}\Lambda_{K_1\times K_2}v_{K_2}(\boldsymbol{x}) -  CK^{-\alpha}\right) - \rho'\left(u_{K_1}(t)^{\top}\Lambda_{K_1\times K_2}v_{K_2}(\boldsymbol{x}) \right)= -\rho''(\xi_2(t,\boldsymbol{x}))CK^{-\alpha} \leq a_2CK^{-\alpha},
	\end{align*}
	where $ -a_1 :=\inf_{\gamma \in \Gamma_1} \rho''(\gamma)$ and $a_2 := \sup_{\gamma \in \Gamma_1}\left( -\rho''(\gamma)\right)$.
	Let $a := \max\{a_1, a_2\}$, we have
	\begin{align} \label{eq:SW1}
	\sup_{(t,\boldsymbol{x})\in \mathcal{T}\times \mathcal{X}}\left| {\pi_0(t,\boldsymbol{x})} - \rho'\left(u_{K_1}(t)^{\top}\Lambda_{K_1\times K_2}v_{K_2}(\boldsymbol{x}) \right)\right| < aCK^{-\alpha}.
	\end{align}

	For some fixed $C_2 > 0$ (to be chosen later), define
	$$ \Upsilon_{K_1\times K_2} := \left\{\Lambda \in \mathbb{R}^{K_1\times K_2}: \|\Lambda - \Lambda_{K_1\times K_2}\| \leq  C_2K^{-\alpha}\right\}.$$
	For sufficiently large $K_1$ and $K_2$,  
	we have that $\forall \Lambda \in \Upsilon_{K_1\times K_2}$, $\forall (t,\boldsymbol{x}) \in \mathcal{T}\times\mathcal{X}$,
	\begin{align*} 
	&\left|u_{K_1}(t)^{\top}\Lambda v_{K_2}(\boldsymbol{x}) - u_{K_1}(t)^{\top}\Lambda_{K_1\times K_2}v_{K_2}(\boldsymbol{x})\right|\\ 
	\leq &\|\Lambda - \Lambda_{K_1\times K_2}\| \cdot \sup_{\boldsymbol{x}\in\mathcal{X}}\|v_{K_2}(\boldsymbol{x}) \| \cdot \sup_{t\in\mathcal{T}}\|u_{K_1}(t)\|
	\leq C_2K^{-\alpha}\zeta_1(K_1)\zeta_2(K_2) .
	\end{align*}
	Then in light of \eqref{eq:resubmit_lamdau_K} and Assumption \ref{as:K&N_c},   for large enough $K_1$ and $K_2$, $\forall\Lambda \in \Upsilon_{K_1\times K_2}$ and $\forall (t,\boldsymbol{x}) \in \mathcal{T}\times\mathcal{X}$, we can deduce that
	\begin{align}\label{eq:gamma3}
	& u_{K_1}(t)^{\top}\Lambda v_{K_2}(\boldsymbol{x})  \in \left(u_{K_1}(t)^{\top}\Lambda_{K_1\times K_2}v_{K_2}(\boldsymbol{x}) - C_2K^{-\alpha}\zeta_1(K_1)\zeta_2(K_2), \right. \\
	&\left. \quad \quad \quad \quad \quad \quad \quad \quad \quad u_{K_1}(t)^{\top}\Lambda_{K_1\times K_2}v_{K_2}(\boldsymbol{x})+ C_2K^{-\alpha}\zeta_1(K_1)\zeta_2(K_2)\right) \notag \\ 
	& ~~~~~~~~~~~~~~~~~~~~~\subset  \left[\underline{\gamma} - CK^{-\alpha} -  C_2K^{-\alpha}\zeta_1(K_1)\zeta_2(K_2), \right. \notag\\
	&~~~~~~~~~~~~~~~~~~~~~~~~~~~\left. \overline{\gamma} + CK^{-\alpha} +  C_2K^{-\alpha}\zeta_1(K_1)\zeta_2(K_2)\right]\subset \Gamma_1.\notag
	\end{align} 
By definition
	\begin{align*}
G_{K_1\times K_2}^*\left(\Lambda\right)= \mathbb{E}\left[\rho\left(u_{K_1}(T)^{\top}\Lambda v_{K_2}(\boldsymbol{X})\right)\right] -\mathbb{E}[u_{K_1}(T)]^{\top}\Lambda \mathbb{E}[v_{K_2}(\boldsymbol{X})],
	\end{align*}
 is a strictly concave function of $\Lambda$. By  \eqref{moment1}, the formula $\tr(AB)=\tr(BA)$ for matrices $A$ and $B$,  the facts $\mathbb{E}\left[v_{K_2}(\boldsymbol{X})v_{K_2}(\boldsymbol{X})^\top\right]=I_{K_2\times K_2}$ and $\mathbb{E}\left[u_{K_1}(T)u_{K_1}(T)^\top\right]=I_{K_1\times K_1}$, we can deduce that{\footnotesize
\begin{align}
&\|\nabla {G}^*_{K_1\times K_2}(\Lambda_{K_1\times K_2})\|^2\notag\\
=&\left\|\mathbb{E}\left[\rho'\left(u_{K_1}(T)^{\top}\Lambda_{K_1\times K_2} v_{K_2}(\boldsymbol{X})\right)u_{K_1}(T)v_{K_2}(\boldsymbol{X})^\top\right] -\mathbb{E}[u_{K_1}(T)]\mathbb{E}[v_{K_2}(\boldsymbol{X})]^{\top}\right\|^2 \notag\\
=&\left\|\mathbb{E}\left[\rho'\left(u_{K_1}(T)^{\top}\Lambda_{K_1\times K_2} v_{K_2}(\boldsymbol{X})\right)u_{K_1}(T)v_{K_2}(\boldsymbol{X})^\top\right] -\mathbb{E}[\pi_0(T,\boldsymbol{X})u_{K_1}(T)v_{K_2}(\boldsymbol{X})]^{\top}\right\|^2\quad \text{(by  \eqref{moment1})} \notag\\
=&\left\|\mathbb{E}\left[\sqrt{\pi_0(T,\boldsymbol{X})}\frac{\left\{\rho'\left(u_{K_1}(T)^{\top}\Lambda_{K_1\times K_2} v_{K_2}(\boldsymbol{X})\right)-\pi_0(T,\boldsymbol{X})\right\}}{\sqrt{\pi_0(T,\boldsymbol{X})}}u_{K_1}(T)v_{K_2}(\boldsymbol{X})^\top\right] \right\|^2  \notag\\
=&\tr\Bigg\{\mathbb{E}\left[\sqrt{\pi_0(T,\boldsymbol{X})}\frac{\left\{\rho'\left(u_{K_1}(T)^{\top}\Lambda_{K_1\times K_2} v_{K_2}(\boldsymbol{X})\right)-\pi_0(T,\boldsymbol{X})\right\}}{\sqrt{\pi_0(T,\boldsymbol{X})}}u_{K_1}(T)v_{K_2}(\boldsymbol{X})^\top\right] \notag \\
&\quad \times \mathbb{E}\left[\sqrt{\pi_0(T,\boldsymbol{X})}\frac{\left\{\rho'\left(u_{K_1}(T)^{\top}\Lambda_{K_1\times K_2} v_{K_2}(\boldsymbol{X})\right)-\pi_0(T,\boldsymbol{X})\right\}}{\sqrt{\pi_0(T,\boldsymbol{X})}}v_{K_2}(\boldsymbol{X})u_{K_1}(T)^\top\right]\Bigg\}  \notag\\
=&\tr\Bigg\{\mathbb{E}\left[\sqrt{\pi_0(T,\boldsymbol{X})}\frac{\left\{\rho'\left(u_{K_1}(T)^{\top}\Lambda_{K_1\times K_2} v_{K_2}(\boldsymbol{X})\right)-\pi_0(T,\boldsymbol{X})\right\}}{\sqrt{\pi_0(T,\boldsymbol{X})}}u_{K_1}(T)v_{K_2}(\boldsymbol{X})^\top\right] \cdot \mathbb{E}\left[u_{K_2}(\boldsymbol{X})u_{K_2}(\boldsymbol{X})^\top\right]\notag \\
&\quad \times \mathbb{E}\left[\sqrt{\pi_0(T,\boldsymbol{X})}\frac{\left\{\rho'\left(u_{K_1}(T)^{\top}\Lambda_{K_1\times K_2} v_{K_2}(\boldsymbol{X})\right)-\pi_0(T,\boldsymbol{X})\right\}}{\sqrt{\pi_0(T,\boldsymbol{X})}}v_{K_2}(\boldsymbol{X})u_{K_1}(T)^\top\right]\cdot \mathbb{E}\left[u_{K_1}(T)u_{K_1}(T)^\top\right]\Bigg\}  \notag\\
=&\mathbb{E}\Bigg[\tr\Bigg\{u_{K_1}(T)^\top\cdot \mathbb{E}\left[\sqrt{\pi_0(T,\boldsymbol{X})}\frac{\left\{\rho'\left(u_{K_1}(T)^{\top}\Lambda_{K_1\times K_2} v_{K_2}(\boldsymbol{X})\right)-\pi_0(T,\boldsymbol{X})\right\}}{\sqrt{\pi_0(T,\boldsymbol{X})}}u_{K_1}(T)v_{K_2}(\boldsymbol{X})^\top\right] \cdot \mathbb{E}\left[u_{K_2}(\boldsymbol{X})u_{K_2}(\boldsymbol{X})^\top\right]\notag \\
&\qquad \times \mathbb{E}\left[\sqrt{\pi_0(T,\boldsymbol{X})}\frac{\left\{\rho'\left(u_{K_1}(T)^{\top}\Lambda_{K_1\times K_2} v_{K_2}(\boldsymbol{X})\right)-\pi_0(T,\boldsymbol{X})\right\}}{\sqrt{\pi_0(T,\boldsymbol{X})}}v_{K_2}(\boldsymbol{X})u_{K_1}(T)^\top\right]\cdot u_{K_1}(T)\Bigg\} \Bigg]  \notag\\
=&\mathbb{E}\Bigg[\pi_0(T,\boldsymbol{X}) \cdot u_{K_1}(T)^\top\cdot \mathbb{E}\left[\sqrt{\pi_0(T,\boldsymbol{X})}\frac{\left\{\rho'\left(u_{K_1}(T)^{\top}\Lambda_{K_1\times K_2} v_{K_2}(\boldsymbol{X})\right)-\pi_0(T,\boldsymbol{X})\right\}}{\sqrt{\pi_0(T,\boldsymbol{X})}}u_{K_1}(T)v_{K_2}(\boldsymbol{X})^\top\right] \cdot u_{K_2}(\boldsymbol{X})\notag \\
&\qquad\times   \cdot u_{K_2}(\boldsymbol{X})^\top \mathbb{E}\left[\sqrt{\pi_0(T,\boldsymbol{X})}\frac{\left\{\rho'\left(u_{K_1}(T)^{\top}\Lambda_{K_1\times K_2} v_{K_2}(\boldsymbol{X})\right)-\pi_0(T,\boldsymbol{X})\right\}}{\sqrt{\pi_0(T,\boldsymbol{X})}}v_{K_2}(\boldsymbol{X})u_{K_1}(T)^\top\right]\cdot u_{K_1}(T) \Bigg] \quad \text{(by  \eqref{moment1})}  \notag\\
=&\mathbb{E}\Bigg[\bigg|{\pi_0(T,\boldsymbol{X})}^{\frac{1}{4}}u_{K_1}(T) \mathbb{E}\left[\sqrt{\pi_0(T,\boldsymbol{X})}\frac{\left\{\rho'\left(u_{K_1}(T)^{\top}\Lambda_{K_1\times K_2} v_{K_2}(\boldsymbol{X})\right)-\pi_0(T,\boldsymbol{X})\right\}}{\sqrt{\pi_0(T,\boldsymbol{X})}}u_{K_1}(T)v_{K_2}(\boldsymbol{X})^\top\right] {\pi_0(T,\boldsymbol{X})}^{\frac{1}{4}}v_{K_2}(\boldsymbol{X}) \bigg|^2\Bigg]\label{eq:nablaG^*_LambdaK}.
\end{align}}
Note that the term in the last expression {\footnotesize
$${\pi_0(T,\boldsymbol{X})}^{\frac{1}{4}}u_{K_1}(T)\cdot \mathbb{E}\left[\sqrt{\pi_0(T,\boldsymbol{X})}\frac{\left\{\rho'\left(u_{K_1}(T)^{\top}\Lambda_{K_1\times K_2} v_{K_2}(\boldsymbol{X})\right)-\pi_0(T,\boldsymbol{X})\right\}}{\sqrt{\pi_0(T,\boldsymbol{X})}}u_{K_1}(T)v_{K_2}(\boldsymbol{X})^\top\right] {\pi_0(T,\boldsymbol{X})}^{\frac{1}{4}}v_{K_2}(\boldsymbol{X})$$}
is the $L^2(dF_{T,X})$-projection of $\frac{\left\{\rho'\left(u_{K_1}(T)^{\top}\Lambda_{K_1\times K_2} v_{K_2}(\boldsymbol{X})\right)-\pi_0(T,\boldsymbol{X})\right\}}{\sqrt{\pi_0(T,\boldsymbol{X})}}$ on the space spanned by {\footnotesize$\{{\pi_0(T,\boldsymbol{X})}^{\frac{1}{4}}u_{K_1}(T)$, ${\pi_0(T,\boldsymbol{X})}^{\frac{1}{4}}v_{K_2}(\boldsymbol{X})\}$}, which implies that{\footnotesize
\begin{align}
	&\mathbb{E}\Bigg[\bigg|{\pi_0(T,\boldsymbol{X})}^{\frac{1}{4}}u_{K_1}(T) \mathbb{E}\left[\sqrt{\pi_0(T,\boldsymbol{X})}\frac{\left\{\rho'\left(u_{K_1}(T)^{\top}\Lambda_{K_1\times K_2} v_{K_2}(\boldsymbol{X})\right)-\pi_0(T,\boldsymbol{X})\right\}}{\sqrt{\pi_0(T,\boldsymbol{X})}}u_{K_1}(T)v_{K_2}(\boldsymbol{X})^\top\right]  {\pi_0(T,\boldsymbol{X})}^{\frac{1}{4}}v_{K_2}(\boldsymbol{X}) \bigg|^2\Bigg]\notag\\
	\leq &\mathbb{E}\left[\bigg|\frac{\left\{\rho'\left(u_{K_1}(T)^{\top}\Lambda_{K_1\times K_2} v_{K_2}(\boldsymbol{X})\right)-\pi_0(T,\boldsymbol{X})\right\}}{\sqrt{\pi_0(T,\boldsymbol{X})}}\bigg|^2\right]. \label{eq:L^2}
\end{align}}
Now, with \eqref{eq:nablaG^*_LambdaK}, \eqref{eq:L^2}, we can obtain that
\begin{align}
&\|\nabla {G}^*_{K_1\times K_2}(\Lambda_{K_1\times K_2})\| \notag\\
\leq& \mathbb{E}\left[\bigg|\frac{\left\{\rho'\left(u_{K_1}(T)^{\top}\Lambda_{K_1\times K_2} v_{K_2}(\boldsymbol{X})\right)-\pi_0(T,\boldsymbol{X})\right\}}{\sqrt{\pi_0(T,\boldsymbol{X})}}\bigg|^2\right]^{\frac{1}{2}}\notag\\
\leq & \frac{1}{\sqrt{\eta_1}} \sup_{(t,\boldsymbol{x})\in\mathcal{T}\times\mathcal{X}}\left|\rho'\left(u_{K_1}(t)^{\top}\Lambda_{K_1\times K_2} v_{K_2}(\boldsymbol{x})\right)-\pi_0(t,\boldsymbol{x})\right| \qquad (\text{by Assumption \ref{as:pi0}}) \notag \\
\leq & \frac{aC}{\sqrt{\eta_1}}\cdot K^{-\alpha} \qquad (\text{by \eqref{eq:SW1}}. \label{order:nabla_G_K*}
\end{align}
Note that for any $\Lambda \in \partial \Upsilon_{K_1\times K_2}$, i.e. $\|\Lambda - \Lambda_{K_1\times K_2}\|=  C_2K^{-\alpha}$, by Mean Value Theorem and the fact $\rho''(y)=-\rho'(y)$, we can deduce that{\small
	\begin{align*}
&	\quad {G}^*_{K_1\times K_2}(\Lambda) - {G}^*_{K_1\times K_2}(\Lambda_{K_1\times K_2})\\
	&= \sum_{j=1}^{K_2}(\lambda_j - \lambda_j^K)^{\top} \frac{\partial}{\partial \lambda_i}{G}^*_{K_1\times K_2}(\lambda_1^K,\ldots,\lambda_{K_2}^K)\\
	&\qquad  + \sum_{l=1}^{K_2}\sum_{j=1}^{K_2}\frac{1}{2}(\lambda_j - \lambda_j^K)^{\top} \frac{\partial^2}{\partial \lambda_i \partial \lambda_l}{G}^*_{K_1\times K_2}(\bar{\lambda}^K_1,\ldots,\bar{\lambda}^K_{K_2})(\lambda_l - \lambda_l^K) \notag\\
		& \leq  \left\|\Lambda - \Lambda_{K_1\times K_2}\right\|\|\nabla{G}^*_{K_1\times K_2}(\Lambda_{K_1\times K_2})\|  \\
		&
		+\frac{1}{2} \sum_{l=1}^{K_2}\sum_{j=1}^{K_2}(\lambda_j - \lambda_j^K)^{\top} \mathbb{E}\left[\rho''\left(u_{K_1}^\top(T)\bar{\Lambda}_{K_1\times K_2}v_{K_2}(\boldsymbol{X})\right)u_{K_1}(T)u_{K_1}(T)^\top v_{K_2,j}(\boldsymbol{X})v_{K_2,l}(\boldsymbol{X})\right](\lambda_l - \lambda_l^K) \notag\\
			& =  \left\|\Lambda - \Lambda_{K_1\times K_2}\right\|\|\nabla{G}^*_{K_1\times K_2}(\Lambda_{K_1\times K_2})\|  \\
			&
		\quad	-\frac{1}{2} \sum_{l=1}^{K_2}\sum_{j=1}^{K_2}(\lambda_j - \lambda_j^K)^{\top} \mathbb{E}\left[\frac{\rho'\left(u_{K_1}^\top(T)\bar{\Lambda}_{K_1\times K_2}v_{K_2}(\boldsymbol{X})\right)}{\pi_0(T,\boldsymbol{X})}\pi_0(T,\boldsymbol{X})u_{K_1}(T)u_{K_1}(T)^\top v_{K_2,j}(\boldsymbol{X})v_{K_2,l}(\boldsymbol{X})\right](\lambda_l - \lambda_l^K) \notag\\
	& \leq \left\|\Lambda - \Lambda_{K_1\times K_2}\right\|\|\nabla{G}^*_{K_1\times K_2}(\Lambda_{K_1\times K_2})\|  \\
	&\quad-\frac{a_3}{2\eta_2} \sum_{l=1}^{K_2}\sum_{j=1}^{K_2}(\lambda_j - \lambda_j^K)^{\top} \mathbb{E}\left[\pi_0(T,\boldsymbol{X})u_{K_1}(T)u_{K_1}(T)^\top v_{K_2,j}(\boldsymbol{X})v_{K_2,l}(\boldsymbol{X})\right](\lambda_l - \lambda_l^K) \notag \quad \text{(by $a_3=\inf_{y\in\Gamma_1}\{\rho'(y)\}$)}\\
		& = \left\|\Lambda - \Lambda_{K_1\times K_2}\right\|\|\nabla{G}^*_{K_1\times K_2}(\Lambda_{K_1\times K_2})\|   \\
		&\quad-\frac{a_3}{2\eta_2} \sum_{l=1}^{K_2}\sum_{j=1}^{K_2}(\lambda_j - \lambda_j^K)^{\top} \mathbb{E}\left[u_{K_1}(T)u_{K_1}(T)^\top\right]\mathbb{E}\left[ v_{K_2,j}(\boldsymbol{X})v_{K_2,l}(\boldsymbol{X})\right](\lambda_l - \lambda_l^K)   \quad \text{(by  \eqref{moment1})}\notag\\
			& = \left\|\Lambda - \Lambda_{K_1\times K_2}\right\|\|\nabla{G}^*_{K_1\times K_2}(\Lambda_{K_1\times K_2})\|  -\frac{a_3}{2\eta_2} \sum_{l=1}^{K_2}\sum_{j=1}^{K_2}(\lambda_j - \lambda_j^K)^{\top} \mathbb{E}\left[ v_{K_2,j}(\boldsymbol{X})v_{K_2,l}(\boldsymbol{X})\right](\lambda_l - \lambda_l^K)  \notag\\
				& = \left\|\Lambda - \Lambda_{K_1\times K_2}\right\|\|\nabla{G}^*_{K_1\times K_2}(\Lambda_{K_1\times K_2})\|  -\frac{a_3}{2\eta_2} \sum_{j=1}^{K_2}(\lambda_j - \lambda_j^K)^{\top} (\lambda_j - \lambda_j^K) \notag\\
	&=  \left\|\Lambda - \Lambda_{K_1\times K_2}\right\|\|\nabla{G}^*_{K_1\times K_2}(\Lambda_{K_1\times K_2})\|
	- \frac {a_3}{2\eta_2} \left\|\Lambda - \Lambda_{K_1\times K_2}\right\|^2   \notag\\
	& = \left\|\Lambda - \Lambda_{K_1\times K_2}\right\|\left(\|\nabla {G}^*_{K_1\times K_2}(\Lambda_{K_1\times K_2})\| -
	\frac {a_3}{2\eta_2} \left\|\Lambda - \Lambda_{K_1\times K_2}\right\|\right)\\
	&\leq  \left\|\Lambda - \Lambda_{K_1\times K_2}\right\|\left(\frac{aC}{\sqrt{\eta_1}}K^{-\alpha}-\frac {a_3}{2\eta_2}\cdot C_2K^{-\alpha}\right), \quad \text{(by \eqref{order:nabla_G_K*})}
	\end{align*}}
where $\bar{\Lambda}_{K_1\times K_2}=(\bar{\lambda}_1^K,...,\bar{\lambda}_{K_2}^K)$ lies on the line joining  ${\Lambda}=(\lambda_1,...,\lambda_{K_2})$ and ${\Lambda}_{K_1\times K_2}=(\lambda_1^K,...,\lambda_{K_2}^K)$, which implies $u_{K_1}^\top(t)\bar{\Lambda}_{K_1\times K_2}v_{K_2}(\boldsymbol{x})\in \Gamma_1$ by \eqref{eq:gamma3};  $a_3=\inf_{y\in\Gamma_1}\{\rho'(y)\}>0$ is a finite positive constant; 	
the fourth and fifth equalities follow from $\mathbb{E}\left[u_{K_1}(T)u_{K_1}(T)^\top\right]=I_{K_1\times K_1}$ and $\mathbb{E}\left[v_{K_2}(\boldsymbol{X})v_{K_2}(\boldsymbol{X})^\top\right]=I_{K_2\times K_2}$ respectively. Therefore, by choosing
$$C_2>\frac{2\eta_2}{a_3}\cdot \frac{aC}{{\sqrt{\eta_1}}},$$
we can obtain the following conclusion:
	\begin{align}\label{eq:concave}
	G^*_{K_1\times K_2}(\Lambda_{K_1\times K_2}) > G^*_{K_1\times K_2}(\Lambda) \ ,~~ \forall \Lambda \in \partial\Upsilon_{K_1\times K_2}\ .
	\end{align}
	Since $G^*_{K_1\times K_2}$ is continuous, \eqref{eq:concave} implies that there exists a local maximum of $G^*_{K_1\times K_2}$ in the interior of $\Upsilon_{K_1\times K_2}$. Note that $G^*_{K_1\times K_2}$ is strictly concave with a unique global maximum point $\Lambda^*_{K_1\times K_2}$, therefore we can claim that 
	\begin{align}\label{Lambda_interior}
	\Lambda^*_{K_1\times K_2} \in \Upsilon_{K_1\times K_2}^\circ, \ \text{i.e.}\ \|\Lambda^*_{K_1\times K_2}-\Lambda_{K_1\times K_2}\|=O(K^{-\alpha}) \ .
	\end{align}
	
	By Mean Value Theorem, \eqref{eq:gamma3} and \eqref{Lambda_interior},  we can deduce that
	\begin{align} \notag %\label{eq:rho1g3}
	&|\rho'\left(u_{K_1}(t)\Lambda_{K_1\times K_2}v_{K_2}(\boldsymbol{x})\right) - \rho'\left(u_{K_1}(t)\Lambda^*_{K_1\times K_2}v_{K_2}(\boldsymbol{x})\right)| \\
	= &|\rho''(\xi^*(t,\boldsymbol{x}))|\left|u_{K_1}(t)\Lambda_{K_1\times K_2}v_{K_2}(\boldsymbol{x})-u_{K_1}(t)\Lambda^*_{K_1\times K_2}v_{K_2}(\boldsymbol{x})\right| \notag\\
	\leq & -\rho''(\xi^*(t,\boldsymbol{x})) \times \|\Lambda_{K_1\times K_2}-\Lambda_{K_1\times K_2}^*  \| \times \sup_{t\in\mathcal{T}}\|u_{K_1}(t)\| \times \sup_{\boldsymbol{x}\in \mathcal{X}}\|v_{K_2}(\boldsymbol{x})\| \notag \\
	\leq & a_2C_2K^{-\alpha}\zeta_1(K_1)\zeta_2(K_2) \ ,\notag
	\end{align}
	where  $ a_2 = \sup_{\gamma \in \Gamma_1} \{ -\rho''(\gamma) \} <\infty$ is a finite positive constant, and $\xi^*(t,\boldsymbol{x})$
	lies between the point $u_{K_1}(t)^{\top} \Lambda_{K_1\times K_2}^*v_{K_2}(\boldsymbol{x})$ and $u_{K_1}(t)^{\top} \Lambda_{K_1\times K_2}v_{K_2}(\boldsymbol{x})$ 
	(note \eqref{eq:gamma3}  implies   $\xi^*(t,\boldsymbol{x}) \in \Gamma_1 $ for all $(t,\boldsymbol{x}) \in \mathcal{T}\times \mathcal{X}$ and large enough $K$). 
	Therefore, using the triangle inequality, and Assumption \ref{as:K&N_c}, we can have 
	\begin{align}
	&\sup_{(t,\boldsymbol{x}) \in \mathcal{T}\times \mathcal{X}}\left|{\pi_0(t,\boldsymbol{x})} - \pi^*_K(t,\boldsymbol{x})\right| \notag \\
	\leq &~ \sup_{(t,\boldsymbol{x}) \in \mathcal{T}\times \mathcal{X}}\left|{\pi_0(t,\boldsymbol{x})} -  \rho'\left(u_{K_1}(t)\Lambda_{K_1\times K_2}v_{K_2}(\boldsymbol{x})\right)\right|\notag\\
	&+ \sup_{(t,\boldsymbol{x}) \in \mathcal{T}\times \mathcal{X}}\left|\rho'\left(u_{K_1}(t)\Lambda_{K_1\times K_2}v_{K_2}(\boldsymbol{x})\right) -  \rho'\left(u_{K_1}(t)\Lambda^*_{K_1\times K_2}v_{K_2}(\boldsymbol{x})\right)\right| \notag \\
	\leq &~ aCK^{-\alpha} + a_2C_2K^{-\alpha}\zeta_1(K_1)\zeta_2(K_2)=O\left(K^{-\alpha}\zeta(K)\right), \notag 
	\end{align} 
where $\zeta(K)=\zeta_1(K_1)\zeta_2(K_2)$. \\	
	
We next prove $\mathbb{E}\left[\left|{\pi_0(T,\boldsymbol{X})} - \pi^*_K(T,\boldsymbol{X})\right|^2\right]=O\left(K^{-2\alpha}\right)$. By Assumption \ref{as:K&N_c}, we can deduce that 
{\footnotesize
\begin{align*}
&\mathbb{E}\left[\left|{\pi_0(T,\boldsymbol{X})} - \pi^*_K(T,\boldsymbol{X})\right|^2\right]\\
\leq &2\cdot \mathbb{E}\left[\left|{\pi_0(T,\boldsymbol{X})} -  \rho'\left(u_{K_1}(T)\Lambda_{K_1\times K_2}v_{K_2}(\boldsymbol{X})\right)\right|^2\right]+2\cdot \mathbb{E}\left[\left|\rho'\left(u_{K_1}(T)\Lambda^*_{K_1\times K_2}v_{K_2}(\boldsymbol{X})\right) -  \rho'\left(u_{K_1}(T)\Lambda_{K_1\times K_2}v_{K_2}(\boldsymbol{X})\right)\right|^2\right]\\
\leq & 2\cdot  \sup_{(t,\boldsymbol{x}) \in \mathcal{T}\times \mathcal{X}}\left|{\pi_0(t,\boldsymbol{x})} -  \rho'\left(u_{K_1}(t)\Lambda_{K_1\times K_2}v_{K_2}(\boldsymbol{x})\right)\right|^2+2\sup_{\gamma\in \Gamma_1}|\rho''(\gamma)|^2\cdot  \mathbb{E}\left[\left|u_{K_1}^\top(T)\left\{\Lambda^*_{K_1\times K_2}-\Lambda_{K_1\times K_2}\right\}v_{K_2}(\boldsymbol{X}) \right|^2\right]\\
\leq & O(K^{-2\alpha})+ O(1)\cdot \mathbb{E}\left[\left|u_{K_1}^\top(T)\left\{\Lambda^*_{K_1\times K_2}-\Lambda_{K_1\times K_2}\right\}v_{K_2}(\boldsymbol{X}) \right|^2\right]. 
\end{align*}}
We next compute the order of $\mathbb{E}\left[\left|u_{K_1}^\top(T)\left\{\Lambda^*_{K_1\times K_2}-\Lambda_{K_1\times K_2}\right\}v_{K_2}(\boldsymbol{X}) \right|^2\right]$.  Note that $\mathbb{E}[u_{K_1}(T)u_{K_1}(T)^\top]=I_{K_1\times K_1}$, $\mathbb{E}[v_{K_2}(\boldsymbol{X})v_{K_2}(\boldsymbol{X})^\top]=I_{K_2\times K_2}$, \eqref{moment1}, \eqref{Lambda_interior} and Assumption \ref{as:pi0}, we can deduce that {\footnotesize
		\begin{align}
		&\mathbb{E}\left[\left|u_{K_1}^\top(T)\left\{\Lambda^*_{K_1\times K_2}-\Lambda_{K_1\times K_2}\right\}v_{K_2}(\boldsymbol{X}) \right|^2\right]\notag\\
		=& \mathbb{E}\left[ u_{K_1}^\top(T)\left\{\Lambda^*_{K_1\times K_2}-\Lambda_{K_1\times K_2}\right\}v_{K_2}(\boldsymbol{X})v_{K_2}(\boldsymbol{X})^{\top}\left\{\Lambda^*_{K_1\times K_2}-\Lambda_{K_1\times K_2}\right\}^\top u_{K_1}(T)\right]  \notag\\
		=& \mathbb{E}\left[\frac{1}{\pi_0(T,\boldsymbol{X})}\pi_0(T,\boldsymbol{X}) u_{K_1}^\top(T)\left\{\Lambda^*_{K_1\times K_2}-\Lambda_{K_1\times K_2}\right\}v_{K_2}(\boldsymbol{X})v_{K_2}(\boldsymbol{X})^{\top}\left\{\Lambda^*_{K_1\times K_2}-\Lambda_{K_1\times K_2}\right\}^\top u_{K_1}(T)\right] \notag\\
		\leq& \frac{1}{\eta_1}\cdot \mathbb{E}\left[\pi_0(T,\boldsymbol{X}) u_{K_1}^\top(T)\left\{\Lambda^*_{K_1\times K_2}-\Lambda_{K_1\times K_2}\right\}v_{K_2}(\boldsymbol{X})v_{K_2}(\boldsymbol{X})^{\top}\left\{\Lambda^*_{K_1\times K_2}-\Lambda_{K_1\times K_2}\right\}^\top u_{K_1}(T)\right] \notag\\
		=&\frac{1}{\eta_1}\cdot \int_{\mathcal{T}} u_{K_1}^\top(t)\left\{\Lambda^*_{K_1\times K_2}-\Lambda_{K_1\times K_2}\right\}\mathbb{E}\left[v_{K_2}(\boldsymbol{X})v_{K_2}(\boldsymbol{X})^{\top}\right]\left\{\Lambda^*_{K_1\times K_2}-\Lambda_{K_1\times K_2}\right\}^\top u_{K_1}(t)dF_T(t)\quad (\text{by \eqref{moment1}}) \notag\\
		=& \frac{1}{\eta_1}\cdot \int_{\mathcal{T}} u_{K_1}^\top(t)\left\{\Lambda^*_{K_1\times K_2}-\Lambda_{K_1\times K_2}\right\}\cdot\left\{\Lambda^*_{K_1\times K_2}-\Lambda_{K_1\times K_2}\right\}^\top u_{K_1}(t)dF_T(t) \notag\\
		=&\frac{1}{\eta_1}\cdot \int_{\mathcal{T}} \tr\Bigg(\left\{\Lambda^*_{K_1\times K_2}-\Lambda_{K_1\times K_2}\right\}\cdot\left\{\Lambda^*_{K_1\times K_2}-\Lambda_{K_1\times K_2}\right\}^\top u_{K_1}(t)u_{K_1}^\top(t)\Bigg)dF_T(t) \notag\\
		=&\frac{1}{\eta_1}\cdot \tr\Bigg(\left\{\Lambda^*_{K_1\times K_2}-\Lambda_{K_1\times K_2}\right\}\cdot\left\{\Lambda^*_{K_1\times K_2}-\Lambda_{K_1\times K_2}\right\}^\top\Bigg) \notag\\
		\leq &  \frac{1}{\eta_1}\cdot\|\Lambda^*_{K_1\times K_2}-\Lambda_{K_1\times K_2}\|^2= O(K^{-2\alpha}). \quad (\text{by \eqref{Lambda_interior}}) \label{eq:u_lam_v}
		\end{align}}	
Therefore, we can obtain
\begin{align*}
\mathbb{E}\left[\left|{\pi_0(T,\boldsymbol{X})} - \pi^*_K(T,\boldsymbol{X})\right|^2\right]=O\left(K^{-2\alpha}\right).
\end{align*}

We finally prove $N^{-1}\sum_{i=1}^N\left|{\pi_0(T_i,\boldsymbol{X}_i)} - \pi^*_K(T_i,\boldsymbol{X}_i)\right|^2=O_p\left(K^{-2\alpha}\right)$. Note that by \eqref{eq:u_lam_v}, we can have{\footnotesize
\begin{align*}
&\mathbb{E}\left[\left\{\frac{1}{N}\sum_{i=1}^N  \left|u_{K_1}^\top(T_i)\left\{\Lambda^*_{K_1\times K_2}-\Lambda_{K_1\times K_2}\right\}v_{K_2}(\boldsymbol{X}_i) \right|^2-\mathbb{E}\left[\left|u_{K_1}^\top(T)\left\{\Lambda^*_{K_1\times K_2}-\Lambda_{K_1\times K_2}\right\}v_{K_2}(\boldsymbol{X}) \right|^2\right]\right\}^2\right]\\
\leq &\frac{1}{N}\cdot \mathbb{E}\left[\left|u_{K_1}^\top(T)\left\{\Lambda^*_{K_1\times K_2}-\Lambda_{K_1\times K_2}\right\}v_{K_2}(\boldsymbol{X}) \right|^4\right]\\
\leq& \frac{1}{N}\cdot \mathbb{E}\left[\left|u_{K_1}^\top(T)\left\{\Lambda^*_{K_1\times K_2}-\Lambda_{K_1\times K_2}\right\}v_{K_2}(\boldsymbol{X}) \right|^2\right]\cdot \sup_{(t,\boldsymbol{x})\in\mathcal{T}\times\mathcal{X}}\left|u_{K_1}^\top(t)\left\{\Lambda^*_{K_1\times K_2}-\Lambda_{K_1\times K_2}\right\}v_{K_2}(\boldsymbol{x}) \right|^2\\
\leq &	\frac{1}{N}\cdot O(K^{-2\alpha})\cdot  \zeta_1(K_1)^2\zeta_2(K_2)^2\cdot \left\|\Lambda^*_{K_1\times K_2}-\Lambda_{K_1\times K_2}\right\|^2\leq \frac{1}{N}\cdot \zeta(K)^2\cdot O( K^{-4\alpha}),
\end{align*}}
then in light of  Chebyshev's inequality and Assumption \ref{as:K&N_consistency}, we have 
\begin{align}\label{eq:uv-E[uv]}
&\frac{1}{N}\sum_{i=1}^N  \left|u_{K_1}^\top(T_i)\left\{\Lambda^*_{K_1\times K_2}-\Lambda_{K_1\times K_2}\right\}v_{K_2}(\boldsymbol{X}_i) \right|^2-\mathbb{E}\left[\left|u_{K_1}^\top(T)\left\{\Lambda^*_{K_1\times K_2}-\Lambda_{K_1\times K_2}\right\}v_{K_2}(\boldsymbol{X}) \right|^2\right]\notag \\
=&O_p\left(\frac{\zeta(K)}{\sqrt{N}}K^{-2\alpha}\right)=o_p\left(K^{-2\alpha}\right).
\end{align}
With \eqref{Lambda_interior}, \eqref{eq:u_lam_v}, \eqref{eq:uv-E[uv]}, and Assumption \ref{as:pi0}, we can deduce that 
	\begin{align*}
	&\frac{1}{N}\sum_{i=1}^N\left|{\pi_0(T_i,\boldsymbol{X}_i)} - \pi^*_K(T_i,\boldsymbol{X}_i)\right|^2\\
	\leq &\frac{2}{N}\sum_{i=1}^N  \left|{\pi_0(T_i,\boldsymbol{X}_i)} -  \rho'\left(u_{K_1}(T_i)\Lambda_{K_1\times K_2}v_{K_2}(\boldsymbol{X}_i)\right)\right|^2\\ &+\frac{2}{N}\sum_{i=1}^N  \left|\rho'\left(u_{K_1}(T_i)\Lambda^*_{K_1\times K_2}v_{K_2}(\boldsymbol{X}_i)\right) -  \rho'\left(u_{K_1}(T_i)\Lambda_{K_1\times K_2}v_{K_2}(\boldsymbol{X}_i)\right)\right|^2 \\
	\leq &2\sup_{(t,\boldsymbol{x}) \in \mathcal{T}\times \mathcal{X}}\left|{\pi_0(t,\boldsymbol{x})} -  \rho'\left(u_{K_1}(t)\Lambda_{K_1\times K_2}v_{K_2}(\boldsymbol{x})\right)\right|^2\\
	&+\sup_{\gamma\in \Gamma_1}|\rho''(\gamma)|^2\cdot \frac{2}{N}\sum_{i=1}^N  \left|u_{K_1}^\top(T_i)\left\{\Lambda^*_{K_1\times K_2}-\Lambda_{K_1\times K_2}\right\}v_{K_2}(\boldsymbol{X}_i) \right|^2  \\
	\leq &2\sup_{(t,\boldsymbol{x}) \in \mathcal{T}\times \mathcal{X}}\left|{\pi_0(t,\boldsymbol{x})} -  \rho'\left(u_{K_1}(t)\Lambda_{K_1\times K_2}v_{K_2}(\boldsymbol{x})\right)\right|^2\\
	&+2\cdot \sup_{\gamma\in \Gamma_1}|\rho''(\gamma)|^2\cdot \mathbb{E}\left[ \left|u_{K_1}^\top(T)\left\{\Lambda^*_{K_1\times K_2}-\Lambda_{K_1\times K_2}\right\}v_{K_2}(\boldsymbol{X}) \right|^2\right]+o_p\left(K^{-2\alpha}\right) \\
	 =& O(K^{-2\alpha})+ O(K^{-2\alpha})+o_p\left(K^{-2\alpha}\right) = O_p\left(K^{-2\alpha}\right). \quad (\text{by \eqref{Lambda_interior}})
	\end{align*}

\end{proof}

\subsection{Lemma $\ref{lemma_pi^hat}$}
\begin{lemma}\label{lemma_pi^hat}  Under Assumption \ref{as:suppX}-\ref{as:K&N_consistency}, we have 
	$$ \left\|\hat{\Lambda}_{K_1\times K_2}- \Lambda_{K_1\times K_2}^*\right\| = O_p\left(\sqrt{\frac {K} {N}} \right)\ .$$
\end{lemma}
\begin{proof}
	Define $$ \hat{S}_N := \frac {1} {N} \sum_{i=1}^N \sum_{l=1}^{K_2}\sum_{j=1}^{K_2}(\lambda_j-\lambda_j^*)^{\top} \pi_0(T_i,\boldsymbol{X}_i)u_{K_1}(T_i)u_{K_1}(T_i)^{\top}(\lambda_l-\lambda_l^*)v_{K_2,j}(\boldsymbol{X}_i)v_{K_2,l}(\boldsymbol{X}_i),$$
	where $\lambda_j$ and $\lambda^*_j$ are the $j$-th column of $\Lambda$ and $\Lambda_{K_1\times K_2}^*$ respectively.
	Since $\hat{S}_N$ is symmetric,  using  \eqref{moment1} and the facts that $\mathbb{E}\left[u_{K_1}(T)u_{K_1}(T)^{\top }\right]=I_{K_1\times K_1}$ and $\mathbb{E}\left[v_{K_2}(\boldsymbol{X})v_{K_2}(\boldsymbol{X})^{\top}\right]=I_{K_2\times K_2}$, we can have
	\begin{align*}
	\mathbb{E}\left[\hat{S}_N\right]=&\sum_{l=1}^{K_2}\sum_{j=1}^{K_2}(\lambda_j-\lambda_j^*)^{\top} \mathbb{E}\left[\pi_0(T,\boldsymbol{X})u_{K_1}(T)u_{K_1}(T)^{\top}v_{K_2,j}(\boldsymbol{X})v_{K_2,l}(\boldsymbol{X})\right](\lambda_l-\lambda_l^*)\\
	=&\sum_{l=1}^{K_2}\sum_{j=1}^{K_2}(\lambda_j-\lambda_j^*)^{\top} \mathbb{E}\left[u_{K_1}(T)u_{K_1}(T)^{\top}\right]\mathbb{E}[v_{K_2,j}(\boldsymbol{X})v_{K_2,l}(\boldsymbol{X})](\lambda_l-\lambda_l^*)\\
	=&\sum_{j=1}^{K_2}(\lambda_j-\lambda_j^*)^{\top}(\lambda_j-\lambda_j^*)=\left\|{\Lambda}-{\Lambda}^*_{K_1\times K_2}\right\|.
	\end{align*}  
	Then we can further deduce that
	\begin{align*}
	&\mathbb{E}\left[\left|\hat{S}_N - \left\|{\Lambda}-{\Lambda}^*_{K_1\times K_2}\right\| \right|^2\right]\notag\\
	= & \mathbb{E}[\hat{S}_N^2] - 2\mathbb{E}[\hat{S}_N]\left\|{\Lambda}-{\Lambda}^*_{K_1\times K_2}\right\| + \left\|{\Lambda}-{\Lambda}^*_{K_1\times K_2}\right\|^2\notag \\
	= &\frac{N}{N^2}\cdot \mathbb{E}\left[\left(\sum_{l=1}^{K_2}\sum_{j=1}^{K_2}(\lambda_j-\lambda_j^*)^{\top} \pi_0(T,\boldsymbol{X})u_{K_1}(T)u_{K_1}(T)^{\top}(\lambda_l-\lambda_l^*)v_{K_2,j}(\boldsymbol{X})v_{K_2,l}(\boldsymbol{X})\right)^2\right]\\
	&+2\cdot \frac{C_N^2}{N^2}\cdot \mathbb{E}\left[\sum_{l=1}^{K_2}\sum_{j=1}^{K_2}(\lambda_j-\lambda_j^*)^{\top} \pi_0(T,\boldsymbol{X})u_{K_1}(T)u_{K_1}(T)^{\top}(\lambda_l-\lambda_l^*)v_{K_2,j}(\boldsymbol{X})v_{K_2,l}(\boldsymbol{X})\right]^2 \notag \\
	&-\left\|{\Lambda}-{\Lambda}^*_{K_1\times K_2}\right\|^2 \notag \\
		= &\frac{1}{N}\mathbb{E}\left[\left(\sum_{l=1}^{K_2}\sum_{j=1}^{K_2}(\lambda_j-\lambda_j^*)^{\top} \pi_0(T,\boldsymbol{X})u_{K_1}(T)u_{K_1}(T)^{\top}(\lambda_l-\lambda_l^*)v_{K_2,j}(\boldsymbol{X})v_{K_2,l}(\boldsymbol{X})\right)^2\right]\\
		&+\frac{N(N-1)}{N^2}\cdot \mathbb{E}\left[\hat{S}_N\right]^2 -\left\|{\Lambda}-{\Lambda}^*_{K_1\times K_2}\right\|^2 \notag \\
	=&\frac{1}{N}\mathbb{E}\left[\left(\sum_{l=1}^{K_2}\sum_{j=1}^{K_2}(\lambda_j-\lambda_j^*)^{\top} \pi_0(T,\boldsymbol{X})u_{K_1}(T)u_{K_1}(T)^{\top}(\lambda_l-\lambda_l^*)v_{K_2,j}(\boldsymbol{X})v_{K_2,l}(\boldsymbol{X})\right)^2\right]\\
	& \quad -\frac{1}{N}\left\|{\Lambda}-{\Lambda}^*_{K_1\times K_2}\right\|^2\\
	< & \frac{1}{N}\mathbb{E}\left[\left(\sum_{l=1}^{K_2}\sum_{j=1}^{K_2}(\lambda_j-\lambda_j^*)^{\top} \pi_0(T,\boldsymbol{X})u_{K_1}(T)u_{K_1}(T)^{\top}(\lambda_l-\lambda_l^*)v_{K_2,j}(\boldsymbol{X})v_{K_2,l}(\boldsymbol{X})\right)^2\right].
	\end{align*}
	In light of the fact that 
	$$ 0 \leq  y^{\top} \left\{\pi_0(t,\boldsymbol{x})u_{K_1}(t)u_{K_1}(t)^{\top} \right\}y  \leq \eta_2 \zeta_1(K_1)^2 y^{\top}y \ , \ \forall y\in\mathbb{R}^{K_1},\ \forall (t,\boldsymbol{x})\in \mathcal{T}\times \mathcal{X} \ ,$$
	we can deduce that
	\begin{align*}
	&\sum_{l=1}^{K_2}\sum_{j=1}^{K_2}(\lambda_j-\lambda_j^*)^{\top} \left\{ \pi_0(T,\boldsymbol{X})u_{K_1}(T)u_{K_1}(T)^{\top}\right\}(\lambda_l-\lambda_l^*)v_{K_2,j}(\boldsymbol{X})v_{K_2,l}(\boldsymbol{X})\\
	=&\left[\sum_{j=1}^{K_2}v_{K_2,j}(\boldsymbol{X})(\lambda_j-\lambda_j^*)^{\top}\right] \cdot \left\{ \pi_0(T,\boldsymbol{X})u_{K_1}(T)u_{K_1}(T)^{\top}\right\}\cdot \left[\sum_{l=1}^{K_2}(\lambda_l-\lambda_l^*)v_{K_2,l}(\boldsymbol{X})\right] \\
	\leq & \eta_2 \cdot \|u_{K_1}(T)\|^2\cdot \left\|\sum_{i=1}^{K_2}(\lambda_i-\lambda_i^*)^{\top}v_{K_2,i}(\boldsymbol{X}) \right\|^2\\
	\leq &  \eta_2 \cdot \|u_{K_1}(T)\|^2\cdot \left(\sum_{i=1}^{K_2}\|\lambda_i-\lambda_i^*\|^2\right)\left(\sum_{i=1}^{K_2}v_{K_2,i}(\boldsymbol{X})^2\right)\\
	=& \eta_2 \cdot \|u_{K_1}(T)\|^2\cdot \left\|\Lambda-\Lambda^*_{K_1\times K_2}\right\|^2\|v_{K_2}(\boldsymbol{X})\|^2 .
	\end{align*}
	Therefore, we can obtain that{\small
	\begin{align}\label{eq:resubmit}
	&\mathbb{E}\left[\left|\hat{S}_N - \left\|{\Lambda}-{\Lambda}^*_{K_1\times K_2}\right\| \right|^2\right] \notag\\
	\leq  
	& \frac{1}{N} \eta^2_2 \cdot \mathbb{E}\left[ \|u_{K_1}(T)\|^4\cdot \|v_{K_2}(\boldsymbol{X})\|^4  \right]\cdot \left\|\Lambda-\Lambda^*_{K_1\times K_2}\right\|^4  \notag\\
	\leq&  \frac{1}{N} \eta^2_2 \cdot \zeta_1(K_1)^2\cdot \zeta_2(K_2)^2\cdot   \mathbb{E}\left[ \|u_{K_1}(T)\|^2\cdot \|v_{K_2}(\boldsymbol{X})\|^2  \right]\cdot \left\|\Lambda-\Lambda^*_{K_1\times K_2}\right\|^4 \notag\\
	=&  \frac{1}{N} \eta^2_2 \cdot \zeta_1(K_1)^2\cdot \zeta_2(K_2)^2\cdot   \mathbb{E}\left[\frac{1}{\pi_0(T,\boldsymbol{X})}\cdot\pi_0(T,\boldsymbol{X}) \|u_{K_1}(T)\|^2\cdot \|v_{K_2}(\boldsymbol{X})\|^2  \right]\cdot \left\|\Lambda-\Lambda^*_{K_1\times K_2}\right\|^4 \notag\\
	\leq& \frac{1}{N} \frac{\eta^2_2}{\eta_1} \cdot \zeta_1(K_1)^2\cdot \zeta_2(K_2)^2\cdot   \mathbb{E}\left[\pi_0(T,\boldsymbol{X}) \|u_{K_1}(T)\|^2\cdot \|v_{K_2}(\boldsymbol{X})\|^2  \right]\cdot \left\|\Lambda-\Lambda^*_{K_1\times K_2}\right\|^4\notag \quad  (\text{by Assumption}\ \ref{as:pi0})\\
	=&\frac{1}{N} \frac{\eta^2_2}{\eta_1} \cdot \zeta_1(K_1)^2\cdot \zeta_2(K_2)^2\cdot   \mathbb{E}\left[ \|u_{K_1}(T)\|^2 \right]\cdot\mathbb{E}\left[ \|v_{K_2}(\boldsymbol{X})\|^2  \right]\cdot \left\|\Lambda-\Lambda^*_{K_1\times K_2}\right\|^4\notag \quad (\text{by} \eqref{moment1})\\
	=&\frac{1}{N} \frac{\eta^2_2}{\eta_1} \cdot \zeta_1(K_1)^2\cdot \zeta_2(K_2)^2\cdot  K_1\cdot K_2\cdot \left\|\Lambda-\Lambda^*_{K_1\times K_2}\right\|^4  \notag (\text{since}\ \mathbb{E}[ \|u_{K_1}(T)\|^2]=K_1\ \text{and}\ \mathbb{E}[ \|v_{K_2}(\boldsymbol{X})\|^2]=K_2)\\
	=&\frac{1}{N} \frac{\eta^2_2}{\eta_1} \cdot \zeta(K)^2\cdot K\cdot \left\|\Lambda-\Lambda^*_{K_1\times K_2}\right\|^4\ . \quad (\text{since}\ \zeta(K)=\zeta_1(K_1)\zeta_2(K_2)\ \text{and} \ K=K_1\cdot K_2)
	\end{align}}
	Considering the event set 
	$$E_{N}:= \left\{\hat{S}_N > \frac{1}{2}\left\|{\Lambda}-{\Lambda}^*_{K_1\times K_2}\right\|^2,\ \Lambda\neq \Lambda_{K_1\times K_2}^*\right\}\ ,$$
	by Chebyshev's inequality,  \eqref{eq:resubmit}, and Assumption \ref{as:K&N_c} we can get
	\begin{align*}
	&\mathbb{P}\left(\left|\hat{S}_N- \left\|{\Lambda}-{\Lambda}^*_{K_1\times K_2}\right\|^2\right|\geq \frac{1}{2} \left\|{\Lambda}-{\Lambda}^*_{K_1\times K_2}\right\|^2,\ \Lambda\neq \Lambda_{K_1\times K_2}^*\right)\\
	&\leq \frac{4\mathbb{E}\left[\left|\hat{S}_N - \left\|{\Lambda}-{\Lambda}^*_{K_1\times K_2}\right\| \right|^2\right]}{\left\|\Lambda-\Lambda^*_{K_1\times K_2}\right\|^4} \\
	& \leq\frac{4}{N} \frac{\eta^2_2}{\eta_1}\cdot \zeta(K)^2\cdot K \leq O\left( \frac{\zeta(K)^2K}{N}\right)=o(1),
	\end{align*}
which implies that for any $\epsilon>0$, there exists $N_0(\epsilon)\in\mathbb{N}$ such that $N>N_0(\epsilon)$ large enough 
	\begin{align}\label{eq:resubmit_E}
	\mathbb{P}\left((E_{N})^c\right)<\mathbb{P}\left(\left|\hat{S}_N- \left\|{\Lambda}-{\Lambda}^*_{K_1\times K_2}\right\|^2\right|\geq \frac{1}{2} \left\|{\Lambda}-{\Lambda}^*_{K_1\times K_2}\right\|^2,\ \Lambda\neq \Lambda_{K_1\times K_2}^*\right) <\frac{\epsilon}{2} \ .
	\end{align}
	
	Note that
	\begin{align*}
	&\frac{\partial}{\partial {\lambda_j}}\hat{G}_{K_1\times K_2}(\lambda_1,\ldots,\lambda_{K_2})\\
	=&\frac{1}{N}\sum_{i=1}^N\rho'\left(u_{K_1}^{\top}(T_i)\Lambda v_{K_2}(\boldsymbol{X}_i)\right)u_{K_1}(T_i)v_{K_2,j}(\boldsymbol{X}_i)-\frac{1}{N^2}\sum_{i=1}^N\sum_{l=1}^Nu_{K_1}(T_i)v_{K_2,j}(\boldsymbol{X}_l)\\ 
	=& \frac{1}{N}\sum_{i=1}^N \left\{\rho'\left(u_{K_1}^{\top}(T_i)\Lambda v_{K_2}(\boldsymbol{X}_i)\right)u_{K_1}(T_i)v_{K_2,j}(\boldsymbol{X}_i)-\mathbb{E}[v_{K_2,j}(\boldsymbol{X})]u_{K_1}(T_i)\right\} \\
	&-\frac{1}{N}\sum_{i=1}^Nu_{K_1}(T_i)\left\{\frac{1}{N}\sum_{l=1}^Nv_{K_2,j}(\boldsymbol{X}_l)-\mathbb{E}[v_{K_2,j}(\boldsymbol{X})] \right\}
	\end{align*}
	and
	\begin{align*}
	&\frac{\partial}{\partial {\lambda_j}}{G}^*_{K_1\times K_2}(\lambda_1,\ldots,\lambda_{K_2})=\mathbb{E}\left[\rho'\left(u_{K_1}^{\top}(T_i)\Lambda v_{K_2}(\boldsymbol{X}_i)\right)u_{K_1}(T_i)v_{K_2,j}(\boldsymbol{X}_i)\right]-\mathbb{E}[u_{K_1}(T)]\cdot \mathbb{E}[v_{K_2,j}(\boldsymbol{X})] .
	\end{align*} 
	Since $\Lambda_{K_1\times K_2}^*$ is the unique maximizer  of $G^*_{K_1\times K_2}(\cdot)$, then for each $j\in \{1,\ldots,K_2\}$, 
	\begin{align*} 
	&\frac{\partial}{\partial {\lambda_j}}{G}^*_{K_1\times K_2}(\lambda_1^*,\ldots,\lambda^*_{K_2})\\
	=&\mathbb{E}\left[\rho'\left(u_{K_1}^{\top}(T)\Lambda_{K_1\times K_2}^* v_{K_2}(\boldsymbol{X})\right)u_{K_1}(T)v_{K_2,j}(\boldsymbol{X})\right]-\mathbb{E}[u_{K_1}(T)]\mathbb{E}[v_{K_2,j}(\boldsymbol{X})]=0.
	\end{align*}
	Therefore, for large enough $K$, we can deduce that{\small
	\begin{align} \label{eq:EG2<}
	&\mathbb{E}\left[\|\nabla\hat{G}_{K_1\times K_2}(\Lambda^*_{K_1\times K_2})\|^2\right] = \sum_{j=1}^{K_2}\mathbb{E}\left[\left\|\frac{\partial}{\partial {\lambda_j}}\hat{G}_{K_1\times K_2}(\lambda_1^*,\ldots,\lambda^*_{K_2})\right\|^2\right]  \\
	\leq & 2 \sum_{j=1}^{K_2}\mathbb{E}\left[\left\|\frac{1}{N}\sum_{i=1}^N \left\{\rho'\left(u_{K_1}^{\top}(T_i)\Lambda^*_{K_1\times K_2} v_{K_2}(\boldsymbol{X}_i)\right)u_{K_1}(T_i)v_{K_2,j}(\boldsymbol{X}_i)-\mathbb{E}[v_{K_2,j}(\boldsymbol{X})]u_{K_1}(T_i)\right\}  \right\|^2\right]\notag \\
	&+ 2\sum_{j=1}^{K_2}\mathbb{E}\left[\left\|\frac{1}{N}\sum_{i=1}^Nu_{K_1}(T_i)\left\{\frac{1}{N}\sum_{l=1}^Nv_{K_2,j}(\boldsymbol{X}_l)-\mathbb{E}[v_{K_2,j}(\boldsymbol{X})] \right\}\right\|^2\right] \notag \\
	=&   \frac{2}{N^2} \sum_{j=1}^{K_2}\sum_{i=1}^N \mathbb{E}\left[\left\|\rho'\left(u_{K_1}^{\top}(T_i)\Lambda^*_{K_1\times K_2} v_{K_2}(\boldsymbol{X}_i)\right)u_{K_1}(T_i)v_{K_2,j}(\boldsymbol{X}_i)-\mathbb{E}[v_{K_2,j}(\boldsymbol{X})]u_{K_1}(T_i)\right\|^2\right] \notag \\
	&+ 2\sum_{j=1}^{K_2}\mathbb{E}\left[\left\|\frac{1}{N}\sum_{i=1}^Nu_{K_1}(T_i)\left\{\frac{1}{N}\sum_{l=1}^Nv_{K_2,j}(\boldsymbol{X}_l)-\mathbb{E}[v_{K_2,j}(\boldsymbol{X})] \right\}\right\|^2\right] \notag \\
	\leq &  \frac{2}{N^2} \sum_{j=1}^{K_2}\sum_{i=1}^N \mathbb{E}\left[\left\|\rho'\left(u_{K_1}^{\top}(T_i)\Lambda^*_{K_1\times K_2} v_{K_2}(\boldsymbol{X}_i)\right)u_{K_1}(T_i)v_{K_2,j}(\boldsymbol{X}_i)-\mathbb{E}[v_{K_2,j}(\boldsymbol{X})]u_{K_1}(T_i)\right\|^2\right] \notag \\
	&+ 2\sum_{j=1}^{K_2}\mathbb{E}\left[\left|\frac{1}{N}\sum_{l=1}^Nv_{K_2,j}(\boldsymbol{X}_l)-\mathbb{E}[v_{K_2,j}(\boldsymbol{X})]\right|^2\right]\cdot \mathbb{E}\left[\left\|\frac{1}{N}\sum_{i=1}^Nu_{K_1}(T_i)\right\|^2\right] \notag \\
	\leq & \frac{4}{N} \sum_{j=1}^{K_2} \left\{ \mathbb{E}\left[\left\|\rho'\left(u_{K_1}^{\top}(T)\Lambda^*_{K_1\times K_2} v_{K_2}(\boldsymbol{X})\right)u_{K_1}(T)v_{K_2,j}(\boldsymbol{X})\right\|^2\right]\right.  +\mathbb{E}[v_{K_2,j}(\boldsymbol{X})^2]\mathbb{E}\left[\|u_{K_1}(T)\|^2\right] \bigg\}\notag \\
	&+ \frac{2}{N}\sum_{j=1}^{K_2}\mathbb{E}\left[v_{K_2,j}(\boldsymbol{X})^2\right]\cdot \mathbb{E}\left[\left\|u_{K_1}(T)\right\|^2\right] \notag \\
	= & \frac{4}{N} \sum_{j=1}^{K_2} \left\{ \mathbb{E}\left[\frac{\left|\rho'\left(u_{K_1}^{\top}(T)\Lambda^*_{K_1\times K_2} v_{K_2}(\boldsymbol{X})\right)\right|^2}{\pi_0(T,\bold{X})}\cdot \pi_0(T,\bold{X})\cdot \left\|u_{K_1}(T)v_{K_2,j}(\boldsymbol{X})\right\|^2\right]\right.  +\mathbb{E}[v_{K_2,j}(\boldsymbol{X})^2]\mathbb{E}\left[\|u_{K_1}(T)\|^2\right] \bigg\}\notag \\
	&+ \frac{2}{N}\sum_{j=1}^{K_2}\mathbb{E}\left[v_{K_2,j}(\boldsymbol{X})^2\right]\cdot \mathbb{E}\left[\left\|u_{K_1}(T)\right\|^2\right] \notag \\
	\leq &  \frac{4}{N} \sum_{j=1}^{K_2} \left\{ \frac{\left(\sup_{\gamma\in\Gamma_1}\rho'(\gamma)\right)^2}{\eta_1}\cdot \mathbb{E}\left[ \pi_0(T,\bold{X})\cdot \left\|u_{K_1}(T)v_{K_2,j}(\boldsymbol{X})\right\|^2\right]\right.  +\mathbb{E}[v_{K_2,j}(\boldsymbol{X})^2]\mathbb{E}\left[\|u_{K_1}(T)\|^2\right] \bigg\}\notag \\
	&+ \frac{2}{N}\sum_{j=1}^{K_2}\mathbb{E}\left[v_{K_2,j}(\boldsymbol{X})^2\right]\cdot \mathbb{E}\left[\left\|u_{K_1}(T)\right\|^2\right] \notag\\
	= &  \frac{4}{N} \sum_{j=1}^{K_2} \left\{ \frac{\left(\sup_{\gamma\in\Gamma_1}\rho'(\gamma)\right)^2}{\eta_1}\cdot \mathbb{E}\left[  v_{K_2,j}(\boldsymbol{X})^2\right] \mathbb{E}\left[  \left\|u_{K_1}(T)\right\|^2\right]\right.  +\mathbb{E}[v_{K_2,j}(\boldsymbol{X})^2]\mathbb{E}\left[\|u_{K_1}(T)\|^2\right] \bigg\}\notag \\
	&+ \frac{2}{N}\sum_{j=1}^{K_2}\mathbb{E}\left[v_{K_2,j}(\boldsymbol{X})^2\right]\cdot \mathbb{E}\left[\left\|u_{K_1}(T)\right\|^2\right] \notag\\
	\leq & \frac{1}{N}\left\{ \frac{4}{\eta_1}\left(\sup_{\gamma\in\Gamma_1}\rho'(\gamma)\right)^2+4+2\right\}\cdot \mathbb{E}\left[\left\|u_{K_1}(T)\right\|^2\right]\sum_{j=1}^{K_2}\mathbb{E}\left[v_{K_2,j}(\boldsymbol{X})^2\right] \notag \\
	=&\frac{1}{N}\left\{ \frac{4}{\eta_1}\left(\sup_{\gamma\in\Gamma_1}\rho'(\gamma)\right)^2+6\right\} K_1K_2=  C_4^2 \frac{K}{N}, \notag
	\end{align}}
	where the last inequality follows by Assumption \ref{as:K&N_c} and $C_4:=\sqrt{ \frac{4}{\eta_1}\left(\sup_{\gamma\in\Gamma_1}\rho'(\gamma)\right)^2+6}$ is a finite universal constant.\\
	
	Let $\epsilon>0$, fix $\dps C_5(\epsilon) > 0$ (to be chosen later) and define
	\begin{align} \notag %\label{eq:EG10<} 
	\hat{\Upsilon}_{K_1\times K_2}(\epsilon):= \left\{\Lambda \in \mathbb{R}^{K_1\times K_2}:  \|\Lambda - \Lambda_{K_1\times K_2}^*\|
	\leq C_5(\epsilon) C_4 \sqrt{\frac {K} {N}} \right\}.
	\end{align}
	For $\forall \Lambda \in \hat{\Upsilon}_{K_1\times K_2}(\epsilon), \forall
	(t,\boldsymbol{x}) \in\mathcal{T}\times \mathcal{X}$, we can have 
	\begin{align*}
	& \left|u_{K_1}(t)^{\top}\Lambda v_{K_2}(\boldsymbol{x}) - u_{K_1}(t)^{\top}\Lambda^*_{K_1\times K_2}v_{K_2}(\boldsymbol{x})\right|\\
	\leq& \|\Lambda - \Lambda_{K_1\times K_2}^* \|\sup_{t\in\mathcal{T}}\|u_{K_1}(t)\| \sup_{\boldsymbol{x}\in\mathcal{X}}\|v_{K_2}(\boldsymbol{x})\| \leq C_5(\epsilon)C_4\sqrt{\frac {K} {N}} \zeta_1(K_1)\zeta_2(K_2) ,
	\end{align*}
	thus for large enough $N$, in accordance with Assumption \ref{as:K&N_c} and \eqref{eq:resubmit_lamdau_K}, we have 
	\begin{align}
	&u_{K_1}(t)^{\top}\Lambda v_{K_2}(\boldsymbol{x}) 
	\in \Bigg[  u_{K_1}(t)^{\top}\Lambda^*_{K_1\times K_2}v_{K_2}(\boldsymbol{x})
	- C_5(\epsilon)C_4 \zeta_1(K_1)\zeta_2(K_2) \sqrt{\frac {K} {N}},  \notag\\
	& \quad \quad \quad \quad \quad \quad \quad \quad \quad  u_{K_1}(t)^{\top}\Lambda^*_{K_1\times K_2}v_{K_2}(\boldsymbol{x})
	+ C_5(\epsilon)C_4 \zeta_1(K_1)\zeta_2(K_2) \sqrt{\frac {K} {N}} \Bigg] \notag \\
	&\quad \quad \quad \quad \quad  \quad \quad \subset  \Bigg[  \underline{\gamma} - CK^{-\alpha} -  C_5(\epsilon)C_4 \zeta_1(K_1)\zeta_2(K_2) \sqrt{\frac {K} {N}}, \notag \\
	&~~~~~~~~~~~~~~~~~~~~~~~~~~~~\quad \quad \overline{\gamma} + CK^{-\alpha} +  C_5(\epsilon)C_4 \zeta_1(K_1)\zeta_2(K_2) \sqrt{\frac {K} {N}}\Bigg] \subset \Gamma_2(\epsilon)\ , \label{Gamma_2_epsilon}
	\end{align}
	where $\Gamma_2(\epsilon):= \left[\underline{\gamma}-1-C_5(\epsilon), \overline{\gamma}+1+C_5(\epsilon)\right]$ 
	is a compact set and independent of $(t,\boldsymbol{x})$. \\
	
	For any $\Lambda \in \partial \hat{\Upsilon}_{K_1\times K_2}(\epsilon)$, there exists
	$\bar{\Lambda}$ on the line joining  $\Lambda$ and $\Lambda_{K_1\times K_2}^*$ such that
	\begin{align*}
	\hat{G}_{K_1\times K_2}(\Lambda) =& \hat{G}_{K_1\times K_2}(\Lambda_{K_1\times K_2}^*) + \sum_{j=1}^{K_2}(\lambda_j - \lambda_j^*)^{\top} \frac{\partial}{\partial \lambda_i}\hat{G}_{K_1\times K_2}(\lambda_1^*,\ldots,\lambda_{K_2}^*)\\
	&+  \frac{1}{2}\sum_{l=1}^{K_2}\sum_{j=1}^{K_2}(\lambda_j - \lambda_j^*)^{\top} \frac{\partial^2}{\partial \lambda_i \partial \lambda_l}\hat{G}_{K_1\times K_2}(\bar{\lambda}_1,\ldots,\bar{\lambda}_{K_2})(\lambda_l - \lambda_l^*)\ ,
	\end{align*}
	where $\bar{\lambda}_j$ denotes the $j$-th column of $\bar{\Lambda}$.
	For the second order term in above equality, note that $u_{K_1}^{\top}(t)\bar{\Lambda}v_{K_2}(\boldsymbol{x})\in\Gamma_2(\epsilon)$ for all $(t,\boldsymbol{x})\in\mathcal{T}\times\mathcal{X}$, we can further deduce that
	\begin{align}\label{eq:resubmit_secondorder}
	&\sum_{l=1}^{K_2}\sum_{j=1}^{K_2}(\lambda_j - \lambda_j^*)^{\top} \frac{\partial^2}{\partial \lambda_i \partial \lambda_l}\hat{G}_{K_1\times K_2}(\bar{\lambda}_1,\ldots,\bar{\lambda}_{K_2})(\lambda_l - \lambda_l^*) \\
	= &\frac {1} {N} \sum_{i=1}^N\sum_{j=1}^{K_2}\sum_{l=1}^{K_2}(\lambda_j-\lambda^*_{j})^{\top}u_{K_1}(T_i)\rho''\left(u_{K_1}^{\top}(T_i)\bar{\Lambda}v_{K_2}(\boldsymbol{X}_i)\right)(\lambda_l - \lambda_l^*)^{\top}u_{K_1}(T_i)v_{K_2,j}(\boldsymbol{X}_i)v_{K_2,l}(\boldsymbol{X}_i) \notag\\
	\leq & - \frac {\bar{b}(\epsilon)} {N} \sum_{i=1}^N\sum_{j=1}^{K_2}\sum_{l=1}^{K_2} (\lambda_j-\lambda^*_{j})^{\top}u_{K_1}(T_i) u_{K_1}(T_i)^{\top}(\lambda_l - \lambda_l^*)v_{K_2,j}(\boldsymbol{X}_i)v_{K_2,l}(\boldsymbol{X}_i)\notag\\
	= & - \frac {\bar{b}(\epsilon)} {N} \sum_{i=1}^N\sum_{j=1}^{K_2}\sum_{l=1}^{K_2} \frac{1}{\pi_0(T_i,\boldsymbol{X}_i)} (\lambda_j-\lambda^*_{j})^{\top} \pi_0(T_i,\boldsymbol{X}_i)u_{K_1}(T_i) u_{K_1}(T_i)^{\top}(\lambda_l - \lambda_l^*)v_{K_2,j}(\boldsymbol{X}_i)v_{K_2,l}(\boldsymbol{X}_i)\notag\\
	\leq & - \frac {\bar{b}(\epsilon)} {N \eta_2} \sum_{i=1}^N\sum_{j=1}^{K_2}\sum_{l=1}^{K_2} (\lambda_j-\lambda^*_{j})^{\top}\pi_0(T_i,\boldsymbol{X}_i) u_{K_1}(T_i) u_{K_1}(T_i)^{\top}(\lambda_l - \lambda_l^*)v_{K_2,j}(\boldsymbol{X}_i)v_{K_2,l}(\boldsymbol{X}_i)\notag\\
	=& -\frac{\bar{b}(\epsilon)}{\eta_2}\hat{S}_N\ , \notag
	\end{align}
	where $-\bar{b}(\epsilon):=
	\sup_{\gamma \in \Gamma_2(\epsilon)} \rho''(\gamma)<\infty$ for each fixed $\epsilon$. 
	Therefore, on the event $E_{N}$ and for large enough $N$, we can deduce that for any $\Lambda\in \partial\hat{\Upsilon}_{K_1\times K_2}(\epsilon)$,
	\begin{align}\label{eq:resubmit_convex}
	\text{on the event}\  E_N:	&\quad \ \hat{G}_{K_1\times K_2}(\Lambda) - \hat{G}_{K_1\times K_2}(\Lambda_{K_1\times K_2}^*)\\
	&= \sum_{j=1}^{K_2}(\lambda_j - \lambda_j^*)^{\top} \frac{\partial}{\partial \lambda_i}\hat{G}_{K_1\times K_2}(\lambda_1^*,\ldots,\lambda_{K_2}^*) \notag\\
	&\quad + \sum_{l=1}^{K_2}\sum_{j=1}^{K_2}\frac{1}{2}(\lambda_j - \lambda_j^*)^{\top} \frac{\partial^2}{\partial \lambda_i \partial \lambda_l}\hat{G}_{K_1\times K_2}(\bar{\lambda}_1,\ldots,\bar{\lambda}_{K_2})(\lambda_l - \lambda_l^*) \notag\\
	& \leq \left\|\Lambda - \Lambda_{K_1\times K_2}^*\right\|\|\nabla\hat{G}_{K_1\times K_2}(\Lambda_{K_1\times K_2}^*)\|  
	-\frac{\bar{b}(\epsilon)}{2\eta_2} \hat{S}_N \ \text{(by \eqref{eq:resubmit_secondorder})} \notag\\
	&\leq  \left\|\Lambda - \Lambda_{K_1\times K_2}^*\right\|\|\nabla\hat{G}_{K_1\times K_2}(\Lambda_{K_1\times K_2}^*)\|
	- \frac {\bar{b}(\epsilon)}{4\eta_2} \left\|\Lambda - \Lambda_{K_1\times K_2}^*\right\|^2   \notag\\
	& = \left\|\Lambda - \Lambda_{K_1\times K_2}^*\right\|\left(\|\nabla\hat{G}_{K_1\times K_2}(\Lambda_{K_1\times K_2}^*)\| -
	\frac {\bar{b}(\epsilon)}{4\eta_2} \left\|\Lambda - \Lambda_{K_1\times K_2}^*\right\|\right) \ , \notag
	\end{align}
	where  the second inequality follows from definition of the event $E_{N}$. \\
	
	Note that for sufficiently large $N$, by Chebyshev's inequality and \eqref{eq:EG2<} we have
	\begin{align} \label{eq:G^1<}
	&\mathbb{P}\left\{\|\nabla\hat{G}_{K_1\times K_2}(\Lambda_{K_1\times K_2}^*)\| \geq
	\frac {\bar{b}(\epsilon)}{4\eta_2} \left\|\Lambda - \Lambda_{K_1\times K_2}^*\right\|\right\}\\ \leq&\frac{16\eta_2^2}{\bar{b}(\epsilon)^2}\cdot \frac{\mathbb{E}\left[\left\|\nabla\hat{G}_{K_1\times K_2}(\Lambda_{K_1\times K_2}^*)\right\|^2\right]}{ \left\|\Lambda - \Lambda_{K_1\times K_2}^*\right\|^2}
	\leq\frac{16\eta_2^2}{\bar{b}(\epsilon)^2C_5^2(\epsilon)} \leq \frac{\epsilon}{2}\ ,\notag
	\end{align}
	where the last inequality holds by choosing $$C_5(\epsilon) \geq \sqrt{\frac{32\eta_2^2}{\bar{b}(\epsilon)^2\epsilon}}\ .$$
	
	Therefore, for sufficiently large $N$, by \eqref{eq:resubmit_E} and \eqref{eq:G^1<} we can derive
	\begin{align} \label{eq:resubmit_events}
	&\mathbb{P}\left( (E_{N})^c \ \text{or} \  \|\nabla\hat{G}_{K_1\times K_2}(\Lambda^*_{K_1\times K_2})\| \geq
	\frac {\bar{b}(\epsilon)}{2\eta_2} \left\|\Lambda - \Lambda_{K_1\times K_2}^*\right\|\right)
	\leq \frac {\epsilon} {2} + \frac {\epsilon} {2}  = \epsilon \notag \\
	\Rightarrow &\mathbb{P}\left( E_{N} \ \text{and} \  \|\nabla\hat{G}_{K_1\times K_2}(\Lambda^*_{K_1\times K_2})\| <
	\frac {\bar{b}(\epsilon)}{2\eta_2} \left\|\Lambda - \Lambda_{K_1\times K_2}^*\right\|\right) 
	> 1-\epsilon.
	\end{align}
	With  \eqref{eq:resubmit_convex} and \eqref{eq:resubmit_events}, we can obtain that
	$$\mathbb{P}\left\{\hat{G}_{K_1\times K_2}(\Lambda) - \hat{G}_{K_1\times K_2}(\Lambda_{K_1\times K_2}^*) < 0 ,~~ \forall \Lambda \in \partial\hat{\Upsilon}_{K_1\times K_2}(\epsilon) \right\} \geq 1 - \epsilon\ .$$
	Note that the event $\left\{ \hat{G}_{K_1\times K_2}(\Lambda_{K_1\times K_2}^*) > \hat{G}_{K_1\times K_2}(\Lambda) ,~ \forall \Lambda \in \partial\hat{\Upsilon}_{K_1\times K_2}(\epsilon) \right\}$ implies that there exists a local maximizer in the interior of   $\hat{\Upsilon}_{K_1\times K_2}(\epsilon)$. Since $\hat{G}_{K_1\times K_2}(\cdot)$ is strictly concave and  $\hat{\Lambda}_{K_1\times K_2} $ is the unique  global maximizer of $\hat{G}_{K_1\times K_2}$, then 
	\begin{align}\label{eq:Lambdahat_in_Upsilonhat}
\mathbb{P}\left(\hat{\Lambda}_{K_1\times K_2} \in \hat{\Upsilon}_{K_1\times K_2}(\epsilon)\right)>1-\epsilon ,
	\end{align}
	 i.e.
	$ \left\|\hat{\Lambda}_{K_1\times K_2}- \Lambda_{K_1\times K_2}^*\right\| = O_p\left(\sqrt{\frac {K} {N}} \right)$.
\end{proof}

\subsection{Corollary $\ref{cor:pi^-pi*}$}
The next corollary states that $\hat{\pi}_K(t,\boldsymbol{x})$ is arbitrarily close to  ${\pi}^*_K(t,\boldsymbol{x})$.
\begin{cor}\label{cor:pi^-pi*}
 Under Assumption \ref{as:suppX}-\ref{as:K&N_consistency}, we have 
	$$\sup_{(t,\boldsymbol{x})\in\mathcal{T}\times\mathcal{X}}|\hat{\pi}_K(t,\boldsymbol{x})-{\pi}^*_K(t,\boldsymbol{x})|=O_p\left(\zeta(K)\sqrt{\frac{K}{N}}\right),$$
	and
	$$\int_{\mathcal{T}\times \mathcal{X}}|\hat{\pi}_K(t,\boldsymbol{x})-{\pi}^*_K(t,\boldsymbol{x})|^2dF_{T,X}(t,\boldsymbol{x})=O_p\left(\frac{K}{N}\right),$$
	and
$$\frac{1}{N}\sum_{i=1}^N|\hat{\pi}_K(T_i,\boldsymbol{X}_i)-{\pi}^*_K(T_i,\boldsymbol{X}_i)|^2=O_p\left(\frac{K}{N}\right).$$
\end{cor}
\begin{proof}
	From the proof of Lemma \ref{lemma_pi^hat}, we know the facts $\mathbb{P}\left(\hat{\Lambda}_{K_1\times K_2}\in \hat{\Upsilon}_{K_1\times K_2}(\epsilon)\right)>1-\epsilon$ and  \eqref{Gamma_2_epsilon}. Then  for any element $\tilde{\Lambda}_{K_1\times K_2}$  lying on the line joining $\hat{\Lambda}_{K_1\times K_2}$ and $\Lambda_{K_1\times K_2}^*$, we can have that $\mathbb{P}(u_{K_1}(t)^{\top}\tilde{\Lambda}_{K_1\times K_2} v_{K_2}(\boldsymbol{x})\in \Gamma_{2}(\epsilon)$  for all $(t,\boldsymbol{x})\in \mathcal{T}\times\mathcal{X})\geq 1-\epsilon$, which implies
	\begin{align}\label{eq:rho''_tilde_Op(1)}
	\sup_{(t,\boldsymbol{x})\in\mathcal{T}\times\mathcal{X}}|\rho''(u_{K_1}(t)\tilde{\Lambda}_{K_1\times K_2}v_{K_2}(\boldsymbol{x}))|=O_p(1).
	\end{align}
	 Using  Mean Value Theorem, Lemma \ref{lemma_pi^*}, and \eqref{eq:rho''_tilde_Op(1)}, we can obtain that
	\begin{align} \notag 
	&\sup_{(t,\boldsymbol{x})\in\mathcal{T}\times\mathcal{X}}|\hat{\pi}_K(t,\boldsymbol{x})-{\pi}^*_K(t,\boldsymbol{x})|\\
	=&\sup_{(t,\boldsymbol{x})\in\mathcal{T}\times\mathcal{X}}|\rho'\left(u_{K_1}(t)\hat{\Lambda}_{K_1\times K_2}v_{K_2}(\boldsymbol{x})\right) - \rho'\left(u_{K_1}(t)\Lambda^*_{K_1\times K_2}v_{K_2}(\boldsymbol{x})\right)| \notag \\
	\leq  &\sup_{(t,\boldsymbol{x})\in\mathcal{T}\times\mathcal{X}}|\rho''(u_{K_1}(t)\tilde{\Lambda}_{K_1\times K_2}v_{K_2}(\boldsymbol{x}))|\sup_{(t,\boldsymbol{x})\in\mathcal{T}\times\mathcal{X}}\left|u_{K_1}(t)\hat{\Lambda}_{K_1\times K_2}v_{K_2}(\boldsymbol{x})-u_{K_1}(t)\Lambda^*_{K_1\times K_2}v_{K_2}(\boldsymbol{x})\right| \notag\\
	\leq & O_p(1) \cdot \|\hat{\Lambda}_{K_1\times K_2}-\Lambda_{K_1\times K_2}^*  \| \cdot \sup_{t\in\mathcal{T}}\|u_{K_1}(t)\| \cdot\sup_{\boldsymbol{x}\in \mathcal{X}}\|v_{K_2}(\boldsymbol{x})\| \notag \\
	\leq & O_p(1)\cdot O_p\left(\sqrt{\frac {K} {N}} \right) \zeta_1(K_1)\cdot \zeta_2(K_2)=O_p\left(\zeta(K)\sqrt{\frac{K}{N}}\right).\notag
	\end{align} 
	
Note that by Mean Value Theorem and \eqref{eq:rho''_tilde_Op(1)}, we can deduce that  
		\begin{align*}
		&\int_{\mathcal{T}\times \mathcal{X}}|\hat{\pi}_K(t,\boldsymbol{x})-{\pi}^*_K(t,\boldsymbol{x})|^2dF_{T,X}(t,\boldsymbol{x})\\
		\leq&\sup_{(t,\boldsymbol{x})\in\mathcal{T}\times\mathcal{X}}|\rho''(u_{K_1}(t)\tilde{\Lambda}_{K_1\times K_2}v_{K_2}(\boldsymbol{x}))|^2\int_{\mathcal{T}\times \mathcal{X}}\left|u_{K_1}(t)\left\{\hat{\Lambda}_{K_1\times K_2}-\Lambda^*_{K_1\times K_2}\right\}v_{K_2}(\boldsymbol{x})\right|^2dF_{T,X}(t,\boldsymbol{x})\\
		\leq &O_p(1)\cdot \int_{\mathcal{T}\times \mathcal{X}}\left|u_{K_1}(t)\left\{\hat{\Lambda}_{K_1\times K_2}-\Lambda^*_{K_1\times K_2}\right\}v_{K_2}(\boldsymbol{x})\right|^2dF_{T,X}(t,\boldsymbol{x}) .
		\end{align*}
We estimate $\int_{\mathcal{T}\times \mathcal{X}}\left|u_{K_1}(t)\left\{\hat{\Lambda}_{K_1\times K_2}-\Lambda^*_{K_1\times K_2}\right\}v_{K_2}(\boldsymbol{x})\right|^2dF_{T,X}(t,\boldsymbol{x})$.	Note that $\mathbb{E}[u_{K_1}(T)[u_{K_1}(T)^\top]=I_{K_1\times K_1}$, $\mathbb{E}[v_{K_2}(\boldsymbol{X})v_{K_2}(\boldsymbol{X})^\top]=I_{K_2\times K_2}$, \eqref{moment1} and Assumption \ref{as:pi0}, we can deduce that{\small
		\begin{align}
		&\int_{\mathcal{T}\times \mathcal{X}}\left|u_{K_1}(t)\left\{\hat{\Lambda}_{K_1\times K_2}-\Lambda^*_{K_1\times K_2}\right\}v_{K_2}(\boldsymbol{x})\right|^2dF_{T,X}(t,\boldsymbol{x}) \notag\\
		\leq &   \int_{\mathcal{T}\times \mathcal{X}}u_{K_1}^\top(t)\left\{\hat{\Lambda}_{K_1\times K_2}-\Lambda^*_{K_1\times K_2}\right\}v_{K_2}(\boldsymbol{x})v_{K_2}(\boldsymbol{x})^\top\left\{\hat{\Lambda}_{K_1\times K_2}-\Lambda^*_{K_1\times K_2}\right\}^\top u_{K_1}(t) dF_{T,X}(t,\boldsymbol{x}) \notag\\
		=&  \int_{\mathcal{T}\times \mathcal{X}}\frac{1}{\pi_0(t,\boldsymbol{x})}\pi_0(t,\boldsymbol{x}) u_{K_1}^\top(t)\left\{\hat{\Lambda}_{K_1\times K_2}-\Lambda^*_{K_1\times K_2}\right\}v_{K_2}(\boldsymbol{x})v_{K_2}(\boldsymbol{x})^\top\left\{\hat{\Lambda}_{K_1\times K_2}-\Lambda^*_{K_1\times K_2}\right\}^\top u_{K_1}(t) dF_{T,X}(t,\boldsymbol{x}) \notag\\
	\leq &    \frac{1}{\eta_1} \int_{\mathcal{T}\times \mathcal{X}}\pi_0(t,\boldsymbol{x})\cdot u_{K_1}^\top(t)\left\{\hat{\Lambda}_{K_1\times K_2}-\Lambda^*_{K_1\times K_2}\right\}v_{K_2}(\boldsymbol{x})v_{K_2}(\boldsymbol{x})^\top\left\{\hat{\Lambda}_{K_1\times K_2}-\Lambda^*_{K_1\times K_2}\right\}^\top u_{K_1}(t) dF_{T,X}(t,\boldsymbol{x})\notag \\
	= & \frac{1}{\eta_1} \int_{\mathcal{T}} u_{K_1}^\top(t)\left\{\hat{\Lambda}_{K_1\times K_2}-\Lambda^*_{K_1\times K_2}\right\}\left(\int_{\mathcal{X}}v_{K_2}(\boldsymbol{x})v_{K_2}(\boldsymbol{x})^\top dF_X(x)\right)\left\{\hat{\Lambda}_{K_1\times K_2}-\Lambda^*_{K_1\times K_2}\right\}^\top u_{K_1}(t) dF_{T}(t) \notag\\
		= & \frac{1}{\eta_1} \int_{\mathcal{T}} u_{K_1}^\top(t)\left\{\hat{\Lambda}_{K_1\times K_2}-\Lambda^*_{K_1\times K_2}\right\}\left\{\hat{\Lambda}_{K_1\times K_2}-\Lambda^*_{K_1\times K_2}\right\}^\top u_{K_1}(t) dF_{T}(t) \notag\\
			= & \frac{1}{\eta_1} \tr\Bigg( \left\{\hat{\Lambda}_{K_1\times K_2}-\Lambda^*_{K_1\times K_2}\right\}\left\{\hat{\Lambda}_{K_1\times K_2}-\Lambda^*_{K_1\times K_2}\right\}^\top \int_{\mathcal{T}}u_{K_1}(t)u_{K_1}^\top(t) dF_{T}(t)\Bigg) \notag\\
				= &  \frac{1}{\eta_1} \tr\Bigg( \left\{\hat{\Lambda}_{K_1\times K_2}-\Lambda^*_{K_1\times K_2}\right\}\left\{\hat{\Lambda}_{K_1\times K_2}-\Lambda^*_{K_1\times K_2}\right\}^\top \Bigg)\notag \\
=&  \frac{1}{\eta_1}\cdot  \left\|\hat{\Lambda}_{K_2\times K_2}-\Lambda^*_{K_1\times K_2}\right\|^2 =O_p\left(\frac{K}{N}\right). \label{eq:u_Lambda_v}
		\end{align}}
Then we obtain 
\begin{align*}
\int_{\mathcal{T}\times \mathcal{X}}|\hat{\pi}_K(t,\boldsymbol{x})-{\pi}^*_K(t,\boldsymbol{x})|^2dF_{T,X}(t,\boldsymbol{x})=O_p\left(\frac{K}{N}\right).
\end{align*}

Similar to \eqref{eq:uv-E[uv]}, we have 
{\small
	\begin{align}\label{eq:uv-E[uv]_hat}
	&\frac{1}{N}\sum_{i=1}^N  \left|u_{K_1}^\top(T_i)\left\{\hat{\Lambda}_{K_1\times K_2}-\Lambda^*_{K_1\times K_2}\right\}v_{K_2}(\boldsymbol{X}_i) \right|^2-\int_{\mathcal{T}\times \mathcal{X}}\left|u_{K_1}(t)\left\{\hat{\Lambda}_{K_1\times K_2}-\Lambda^*_{K_1\times K_2}\right\}v_{K_2}(\boldsymbol{x})\right|^2dF_{T,X}(t,\boldsymbol{x})\notag \\
	=&O_p\left(\frac{\zeta(K)}{\sqrt{N}}\cdot \|\hat{\Lambda}_{K_1\times K_2}-\Lambda^*_{K_1\times K_2}\|^2\right)=O_p\left(\frac{\zeta(K)}{\sqrt{N}}\cdot \frac{K}{N}\right)=o_p\left(\frac{K}{N}\right).
	\end{align}}
where the last equality holds in light of Assumption \ref{as:K&N_consistency}. Hence, with \eqref{eq:u_Lambda_v} and \eqref{eq:uv-E[uv]_hat}, we have
	\begin{align*}
	&\frac{1}{N}\sum_{i=1}^N|\hat{\pi}_K(T_i,\boldsymbol{X}_i)-{\pi}^*_K(T_i,\boldsymbol{X}_i)|^2 \\
	\leq&\sup_{(t,\boldsymbol{x})\in\mathcal{T}\times\mathcal{X}}|\rho''(u_{K_1}(t)\tilde{\Lambda}_{K_1\times K_2}v_{K_2}(\boldsymbol{x}))|^2\cdot \frac{1}{N}\sum_{i=1}^N\left|u_{K_1}(T_i)\left\{\hat{\Lambda}_{K_1\times K_2}-\Lambda^*_{K_1\times K_2}\right\}v_{K_2}(\boldsymbol{X}_i)\right|^2\\
	\leq &O_p(1)\cdot \int_{\mathcal{T}\times \mathcal{X}}\left|u_{K_1}(t)\left\{\hat{\Lambda}_{K_1\times K_2}-\Lambda^*_{K_1\times K_2}\right\}v_{K_2}(\boldsymbol{x})\right|^2dF_{T,X}(t,\boldsymbol{x})+o_p\left(\frac{K}{N}\right) \\
	\leq &O_p\left(\frac{K}{N}\right)+o_p\left(\frac{K}{N}\right)=O_p\left(\frac{K}{N}\right).
	\end{align*}
\end{proof}

\section{Efficient Estimation}
\subsection{Proof of Theorem 5.17 \label{sec:proof_main_theorem}}
Since we have proved $\|\hat{\boldsymbol{\beta}}-\boldsymbol{\beta}_0\|\xrightarrow{p}0$ in Theorem 10, 
we now begin to prove the asymptotic efficiency of $\hat{\boldsymbol{\beta}}$.  By Assumption \ref{as:first_order}, $\hat{\boldsymbol{\beta}}$ is a unique solution of the following equation:
\begin{align}\label{eq:beta^hat}
\frac{1}{N}\sum_{i=1}^{N}\hat{\pi}_{K}(T_{i},\boldsymbol{X}_{i})m(T_{i};%
\hat{\boldsymbol{\beta }})L'\left\{ Y_{i}-g\left( T_{i};\hat{\boldsymbol{\beta 
	}}\right) \right\} =0 ,
\end{align}
with probability approaching to one. Note that $L'(\cdot)$ may be a non-differentiable function, e.g. $L'(v)=\tau-I(v\leq 0)$ in quantile regression, we cannot simply apply Mean Value Theorem on \eqref{eq:beta^hat} to obtain the expression for $\sqrt{N}(\hat{\boldsymbol{\beta}}-\boldsymbol{\beta}_0)$. To solve this problem, we define 
\begin{align*}
f(\boldsymbol{\beta}):=\mathbb{E}\left[\pi_0 (T,\boldsymbol{X})L'(Y
-g(T;\boldsymbol{\beta} ))m(T;\boldsymbol{\beta})\right],
\end{align*} 	
which is a differentiable function in $\boldsymbol{\beta}$ and by definition $f(\boldsymbol{\beta}_0)=0$. Using Mean Value Theorem, we can obtain that
\begin{align*}
0=\sqrt{N}f(\boldsymbol{\beta}_0)=\sqrt{N}f(\hat{\boldsymbol{\beta}})-\nabla_{\beta}f(\tilde{\boldsymbol{\beta}})\cdot \sqrt{N}(\hat{\boldsymbol{\beta}}-\boldsymbol{\beta}_0)\ ,
\end{align*}
where $\tilde{\boldsymbol{\beta}}$ lies on the line joining $\hat{\boldsymbol{\beta}}$ and $\boldsymbol{\beta}_0$. Because $\nabla_{\beta}f(\boldsymbol{\beta})$ is continuous in $\boldsymbol{\beta}$ at $\boldsymbol{\beta}_0$, and $\|\hat{\boldsymbol{\beta}}-{\boldsymbol{\beta}}_0\|\xrightarrow{p}0$, then we have
\begin{align*}
\sqrt{N}(\hat{\boldsymbol{\beta}}-\boldsymbol{\beta}_0)=&\nabla_{\beta}f(\boldsymbol{\beta}_0)^{-1}\cdot \sqrt{N}f(\hat{\boldsymbol{\beta}}).
\end{align*}
Define the empirical process:
\begin{align*}
\mu_N(\boldsymbol{\beta})=\frac{1}{\sqrt{N}}\sum_{i=1}^N \bigg\{\hat{\pi}_K (T_{i},\boldsymbol{X}_{i})L'(Y
_{i} -g(T_{i};\boldsymbol{\beta} ))m(T_i;\boldsymbol{\beta})-\mathbb{E}\left[\pi_0 (T,\boldsymbol{X})L'(Y
-g(T;\boldsymbol{\beta} ))m(T;\boldsymbol{\beta})\right]\bigg\}\ ,
\end{align*}
and by Assumption \ref{as:first_order} we can have
	\begin{align*}
	&\sqrt{N}(\hat{\boldsymbol{\beta}}-\boldsymbol{\beta}_0)\\
	=&\nabla_{\beta}f(\boldsymbol{\beta}_0)^{-1}\cdot \Bigg\{ \sqrt{N}f(\hat{\boldsymbol{\beta}})-\frac{1}{\sqrt{N}}\sum_{i=1}^N\hat{\pi}_K (T_{i},\boldsymbol{X}_{i})L'(Y
	_{i} -g(T_{i};\hat{\boldsymbol{\beta}} ))m(T_i;\hat{\boldsymbol{\beta}})\\ &\qquad \qquad \qquad +\frac{1}{\sqrt{N}}\sum_{i=1}^N\hat{\pi}_K (T_{i},\boldsymbol{X}_{i})L'(Y
	_{i} -g(T_{i};\hat{\boldsymbol{\beta}} ))m(T_i;\hat{\boldsymbol{\beta}}) \Bigg\}\\
	=&-\nabla_{\beta}f(\boldsymbol{\beta}_0)^{-1}\cdot \mu_N(\hat{\boldsymbol{\beta}})+o_p(1)\\
	=&H_0^{-1}\cdot \bigg\{\left(\mu_N(\hat{\boldsymbol{\beta}}) -\mu_N(\boldsymbol{\beta}_0)\right) +\mu_N(\boldsymbol{\beta}_0)\bigg\}+o_p(1).
	\end{align*}
By Assumption \ref{as:m_smooth} and \ref{as:entropy}, Theorems 4 and 5 of \cite{andrews1994empirical}, we can conclude that $\mu_N(\cdot)$ is stochastically equicontinuous, which implies $\mu_N(\hat{\boldsymbol{\beta}}) -\mu_N(\boldsymbol{\beta}_0)\xrightarrow{p}0$. We also note that $\mathbb{E}[\pi_0 (T,\boldsymbol{X})L'(Y
-g(T;\boldsymbol{\beta}_0 ))m(T;\boldsymbol{\beta}_0)]=0$,  then 
\begin{align}\label{eq:beta^hat-beta0}
\sqrt{N}(\hat{\boldsymbol{\beta}}-\boldsymbol{\beta}_0)=H_0^{-1}\frac{1}{\sqrt{N}}\sum_{i=1}^N \hat{\pi}_K (T_{i},\boldsymbol{X}_{i})L'(Y
_{i} -g(T_{i},\beta_0 ))m(T_i;\boldsymbol{\beta}_0)+o_p(1).
\end{align}
	We next claim the following important Lemma \ref{asy:equivalent}, and leave its proof to Section \ref{sec:proof_equivalent}. 
	\begin{lemma}
		\label{asy:equivalent} Under Assumption \ref{as:TYindep}-\ref{as:K&N_c}, we have 
		\begin{align}  \label{asy:equivalent_1}
		\frac{1}{\sqrt{N}}\sum_{i=1}^N\hat{\pi}_K(T_i,\boldsymbol{X}_i)m(T_i;{%
			\boldsymbol{\beta}}_0)L'\left\{Y_i-g\left(T_i;{\boldsymbol{\beta}}%
		_0\right)\right\}=\frac{1}{\sqrt{N}}\sum_{i=1}^N\psi(Y_i,T_i,\boldsymbol{X}%
		_i;\boldsymbol{\beta}_0) +o_p(1),
		\end{align}
		where \begin{align*}
			&\psi (Y,T,\boldsymbol{X};\boldsymbol{\beta }_{0}):= \pi_0 (T,
			\boldsymbol{X})m(T;\boldsymbol{\beta }_{0})L'(Y-g(T;\boldsymbol{\beta}_0))-\pi_0 (T,\boldsymbol{X})m(T;\boldsymbol{\beta }_{0})\varepsilon (T,\boldsymbol{X};
			\boldsymbol{\beta }_{0}) \\
			&\qquad \qquad\qquad+\mathbb{E}\left[ \varepsilon (T,\boldsymbol{X};
			\boldsymbol{\beta }_{0})\pi_0 (T,\boldsymbol{X})m(T;\boldsymbol{\beta }_{0})|
			\boldsymbol{X}\right] +\mathbb{E}\left[ \varepsilon (T,\boldsymbol{X};
			\boldsymbol{\beta }_{0})\pi_0 (T,\boldsymbol{X})m(T;\boldsymbol{\beta }_{0})|
			T\right] \ ,
		\end{align*}  
		and 	$\varepsilon (T,\boldsymbol{X};\boldsymbol{\beta }_{0}):=  \mathbb{E}
		[L'(Y-g(T;\boldsymbol{\beta}_0))|T,\boldsymbol{X}]$.
	\end{lemma}
  Lemma \ref{asy:equivalent} is the most important
	step for establishing the efficiency of our proposed estimator. A key technique in proving Lemma \ref
	{asy:equivalent} is a use of a weighted least square projection of 
	$L'(Y-g(T;\boldsymbol{\beta}_0))$ onto the space linearly spanned by the approximation
	basis $\{u_{K_{1}}(T),v_{K_{2}}(\boldsymbol{X})\}$.  
	
	Combining \eqref{eq:beta^hat-beta0} and Lemma \ref%
	{asy:equivalent}, we can obtain the asymptotic  expression for  $\sqrt{N}(\hat{%
		\boldsymbol{\beta }}-\boldsymbol{\beta }_{0})$: 
	\begin{align*}
	\sqrt{N}(\hat{\boldsymbol{\beta }}-\boldsymbol{\beta }_{0})= H_0^{-1}\frac{1}{\sqrt{N}}\sum_{i=1}^{N}\psi (T_{i},\boldsymbol{X}%
	_{i},Y_{i};{\boldsymbol{\beta }}_{0})+o_{p}(1) = \frac{1}{\sqrt{N}}\sum_{i=1}^{N}S_{eff}(T_{i},\boldsymbol{X}_{i},Y_{i};{%
		\boldsymbol{\beta }}_{0})+o_{p}(1),
	\end{align*}
	which leads to our Theorem 5.17.

	\subsection{Proof of Lemma \ref{asy:equivalent} \label{sec:proof_equivalent}}
	Before proving Lemma \ref{asy:equivalent}, we prepare some preliminary notation and results that will be used later. Since $\hat{\Lambda}_{K_1\times K_2}$ is a unique maximizer of the concave function $\hat{G}_{K_1\times K_2}$, then
	\begin{align*}
	\frac{1}{N}\sum_{i=1}^N\rho'\left(u_{K_1}(T_i)^{\top}\hat{\Lambda}_{K_1\times K_2}v_{K_2}(\boldsymbol{X}_i)\right)u_{K_1}(T_i)v_{K_2}(\boldsymbol{X}_i)^{\top}-\frac{1}{N^2}\sum_{i=1}^N\sum_{l=1}^Nu_{K_1}(T_l)v_{K_2}(\boldsymbol{X}_i)^{\top}=0.
	\end{align*}
	Using Mean Value Theorem, we can have
	\begin{align}\label{aa}
	&\frac{1}{N}\sum_{i=1}^N\rho'\left(u_{K_1}(T_i)^{\top}{\Lambda}^*_{K_1\times K_2}v_{K_2}(\boldsymbol{X}_i)\right)u_{K_1}(T_i)v_{K_2}(\boldsymbol{X}_i)^{\top} \notag\\
	+& \frac{1}{N}\sum_{i=1}^N\rho''\left(u_{K_1}(T_i)^{\top}\tilde{\Lambda}_{K_1\times K_2}v_{K_2}(\boldsymbol{X}_i)\right)u_{K_1}(T_i)u_{K_1}(T_i)^{\top} \left\{\hat{\Lambda}_{K_1\times K_2}-{\Lambda}^*_{K_1\times K_2}\right\}v_{K_2}(\boldsymbol{X}_i)v_{K_2}(\boldsymbol{X}_i)^{\top} \notag\\
	=&\frac{1}{N^2}\sum_{i=1}^N\sum_{l=1}^N u_{K_1}(T_l)v_{K_2}(\boldsymbol{X}_i)^{\top} \ ,
	\end{align}
	where  $\tilde{\Lambda}_{K_1\times K_2}$ lies on the line joining from $\hat{\Lambda}_{K_1\times K_2}$ to ${\Lambda}^*_{K_1\times K_2}$. We define the following notation:  
	\begin{align}
	&\hat{A}_{K_1\times K_2} := \hat{\Lambda}_{K_1\times K_2}-\Lambda_{K_1\times K_2}^* , \label{def:A} \\
	&\tilde{A}_{K_1\times K_2} := \tilde{\Lambda}_{K_1\times K_2}-\Lambda_{K_1\times K_2}^*  , \label{def:A_tilde}
	\end{align}
	and
	\begin{align}\label{def:A*}
	&A_{K_1\times K_2}^*:= \nabla\hat{G}_{K_1\times K_2}\left(\Lambda_{K_1\times K_2}^*\right) \notag \\
	=&\frac{1}{N}\sum_{i=1}^N\rho'\left(u_{K_1}(T_i)^{\top}\Lambda_{K_1\times K_2}^* v_{K_2}(\boldsymbol{X}_i)\right)u_{K_1}(T_i)v_{K_2}(\boldsymbol{X}_i)^{\top}-\left(\frac{1}{N}\sum_{l=1}^N u_{K_1}(T_l)\right)\left(\frac{1}{N}\sum_{i=1}^N v_{K_2}(\boldsymbol{X}_i)^{\top}\right).
	\end{align}
In light of \eqref{eq:EG2<} we have 
	$$\left\|A^*_{K_1\times K_2}\right\|= O_p\left(\sqrt{\frac{K}{N}}\right). $$
	From \eqref{aa}, $A^*_{K_1\times K_2}$ can also be written as  
		\begin{align} \label{eq:A^*_{K_1timesK_2}}
		A^*_{K_1\times K_2}=-\frac{1}{N}\sum_{i=1}^N\rho''\left(u_{K_1}(T_i)^{\top}\tilde{\Lambda}_{K_1\times K_2}v_{K_2}(\boldsymbol{X}_i)\right)u_{K_1}(T_i)u_{K_1}(T_i)^{\top} \left\{\hat{\Lambda}_{K_1\times K_2}-\Lambda_{K_1\times K_2}^*\right\}v_{K_2}(\boldsymbol{X}_i)v_{K_2}(\boldsymbol{X}_i)^{\top}.
		\end{align}
	
	We now start to prove Lemma \ref{asy:equivalent}. We decompose  $\frac{1}{\sqrt{N}}\sum_{i=1}^N\hat{\pi}_K(T_i,\boldsymbol{X}_i)\left\{Y_i-g\left(T_i;\boldsymbol{\beta}_0\right)\right\}m(T_i;\boldsymbol{\beta}_0)$ as follows:
		\begin{align}
		&\frac{1}{\sqrt{N}}\sum_{i=1}^N\hat{\pi}_K(T_i,\boldsymbol{X}_i)L'\left\{Y_i-g\left(T_i;\boldsymbol{\beta}_0\right)\right\}m(T_i;\boldsymbol{\beta}_0)\notag\\
		=&\label{eq:WK} \frac {1} {\sqrt{N}} \sum_{i=1}^N \Bigg\{ \left(\hat{\pi}_K(T_i,\boldsymbol{X}_i) - \pi_K^*(T_i,\boldsymbol{X}_i)\right)L'\left\{Y_i-g\left(T_i;\boldsymbol{\beta}_0\right)\right\}m(T_i;\boldsymbol{\beta}_0)   \\ 
		&\quad \quad \quad  \quad \quad -\int_{\mathcal{T}}\int_{\mathcal{X}} \left(\hat{\pi}_K(t,\boldsymbol{x}) - \pi_K^*(t,\boldsymbol{x})\right)\varepsilon(\boldsymbol{x},t;\boldsymbol{\beta}_0)m(t;\boldsymbol{\beta}_0) dF_{X,T}(\boldsymbol{x},t)\Bigg\} \notag \\[2mm]
		\label{eq:VK} &~ + \frac {1} {\sqrt{N}} \sum_{i=1}^N \Bigg\{\left(\pi_K^*(T_i,\boldsymbol{X}_i) - {\pi_0(T_i,\boldsymbol{X}_i)}\right)L'\left\{Y_i-g\left(T_i;\boldsymbol{\beta}_0\right)\right\}m(T_i;\boldsymbol{\beta}_0) \\
		&\quad \quad \quad  \quad \quad  -  \int_{\mathcal{T}}\int_{\mathcal{X}} m(t;\boldsymbol{\beta}_0)\varepsilon(t,\boldsymbol{x};\boldsymbol{\beta}_0)\left(\pi_K^*(t,\boldsymbol{x}) - {\pi_0(t,\boldsymbol{x})}\right) dF_{X,T}(\boldsymbol{x},t) \Bigg\} \notag \\
		\label{eq:Lemma1}&~ + \sqrt{N}\int_{\mathcal{T}}\int_{\mathcal{X}} m(t;\boldsymbol{\beta}_0)\varepsilon(t,\boldsymbol{x};\boldsymbol{\beta}_0)\left(\pi_K^*(t,\boldsymbol{x}) - {\pi_0(t,\boldsymbol{x})}\right) dF_{X,T}(\boldsymbol{x},t)  \\[2mm]
		&~ \label{eq:tauto}+  {\sqrt{N}}\int_{\mathcal{T}}\int_{\mathcal{X}} \left(\hat{\pi}_K(t,\boldsymbol{x}) - \pi_K^*(t,\boldsymbol{x})\right)\varepsilon(\boldsymbol{x},t;\boldsymbol{\beta}_0)m(t;\boldsymbol{\beta}_0) dF_{X,T}(\boldsymbol{x},t) \\
		& \qquad  - \sqrt{N}\int_{\mathcal{T}}\int_{\mathcal{X}} \varepsilon(t,\boldsymbol{x};\boldsymbol{\beta}_0)\rho''\left(u_{K_1}^{\top}(t)\tilde{\Lambda}_{K_1\times K_2}v_{K_2}(\boldsymbol{x})\right)u_{K_1}^{\top}(t)\hat{A}_{K_1\times K_2}v_{K_2}(\boldsymbol{x}) m(t;\boldsymbol{\beta}_0) dF_{X,T}(\boldsymbol{x},t) \notag \\[4mm]
		& \label{eq:Q} +\sqrt{N} \int_{\mathcal{T}}\int_{\mathcal{X}} \varepsilon(t,\boldsymbol{x};\boldsymbol{\beta}_0)\rho''\left(u_{K_1}^{\top}(t)\tilde{\Lambda}_{K_1\times K_2}v_{K_2}(\boldsymbol{x})\right)u_{K_1}^{\top}(t)\hat{A}_{K_1\times K_2}v_{K_2}(\boldsymbol{x}) m(t;\boldsymbol{\beta}_0) dF_{X,T}(\boldsymbol{x},t)\\ 
		&\qquad - \sqrt{N} \int_{\mathcal{T}}\int_{\mathcal{X}} \varepsilon(t,\boldsymbol{x};\boldsymbol{\beta}_0)\rho''\left(u_{K_1}^{\top}(t){\Lambda}^*_{K_1\times K_2}v_{K_2}(\boldsymbol{x})\right)u_{K_1}^{\top}(t)A^*_{K_1\times K_2}v_{K_2}(\boldsymbol{x}) m(t;\boldsymbol{\beta}_0) dF_{X,T}(\boldsymbol{x},t)\notag \\[2mm]
		& \label{eq:proj}+{\sqrt{N}}\int_{\mathcal{X}} \varepsilon(t,\boldsymbol{x};\boldsymbol{\beta}_0)\rho''\left(u_{K_1}^{\top}(t){\Lambda}^*_{K_1\times K_2}v_{K_2}(\boldsymbol{x})\right)u_{K_1}^{\top}(t)A^*_{K_1\times K_2}v_{K_2}(\boldsymbol{x}) m(t;\boldsymbol{\beta}_0) dF_{X,T}(\boldsymbol{x},t)\\
		&\qquad  +\frac{1}{\sqrt{N}}\sum_{i=1}^N \bigg\{ \pi_0(T_i,\boldsymbol{X}_i)m(T_i;\boldsymbol{\beta}_0)\varepsilon(T_i,\boldsymbol{X}_i;\boldsymbol{\beta}_0)-\mathbb{E}\left[\pi_0(T,\boldsymbol{X})m(T;\boldsymbol{\beta}_0)\varepsilon(T,\boldsymbol{x};\boldsymbol{\beta}_0)|\boldsymbol{X}=\boldsymbol{X}_i\right]\notag\\
		&\qquad\qquad \qquad \qquad  -\mathbb{E}\left[\pi_0(T,\boldsymbol{X})m(T;\boldsymbol{\beta}_0)\varepsilon(T,\boldsymbol{x};\boldsymbol{\beta}_0)|T=T_i\right]\bigg\} \notag \\[2mm]
		&~ + \frac {1} {\sqrt{N}} \sum_{i=1}^N \bigg\{ \pi_0(T_i,\boldsymbol{X}_i)L'\left\{Y_i-g\left(T_i;\boldsymbol{\beta}_0\right)\right\}m(T_i;\boldsymbol{\beta}_0)-\pi_0(T_i,\boldsymbol{X}_i)m(T_i;\boldsymbol{\beta}_0)\varepsilon(T_i,\boldsymbol{X}_i;\boldsymbol{\beta}_0)\label{eq:Normal}\\
		&  \quad \quad \quad   +\mathbb{E}\left[\pi_0(T,\boldsymbol{X})m(T;\boldsymbol{\beta}_0)\varepsilon(T,\boldsymbol{X};\boldsymbol{\beta}_0)|\boldsymbol{X}=\boldsymbol{X}_i\right]+\mathbb{E}\left[\pi_0(T,\boldsymbol{X})m(T;\boldsymbol{\beta}_0)\varepsilon(T,\boldsymbol{x};\boldsymbol{\beta}_0)|T=T_i\right] \bigg\}, \notag
		\end{align}
		where $\hat{A}_{K_1\times K_2}$ and $A_{K_1\times K_2}^*$ are defined in \eqref{def:A} and \eqref{eq:A^*_{K_1timesK_2}}. We show that the terms \eqref{eq:WK}-\eqref{eq:proj} are all of $o_p(1)$, while the term \eqref{eq:Normal} is  asymptotically normal.\\
		
		\noindent \textbf{\emph{For term \eqref{eq:WK}}:}
		
		\noindent Denoting \eqref{eq:WK} by $W_K$ and applying Mean Value Theorem twice, we can obtain 
		\begin{align*}
		W_K = &\frac {1} {\sqrt{N}} \sum_{i=1}^N \Bigg[L'\left\{Y_i-g\left(T_i;\boldsymbol{\beta}_0\right)\right\}m(T_i;\boldsymbol{\beta}_0)\rho''\left(u_{K_1}^{\top}(T_i)\tilde{\Lambda}_{K_1\times K_2}v_{K_2}(\boldsymbol{X}_i)\right)u_{K_1}(T_i)^{\top}\hat{A}_{K_1\times K_2}v_{K_2}(\boldsymbol{X}_i) \\
		&-\int_{\mathcal{T}} \int_{\mathcal{X}} m(t;\boldsymbol{\beta}_0)\varepsilon(t,\boldsymbol{x};\boldsymbol{\beta}_0) \rho''\left(u_{K_1}^{\top}(t)\tilde{\Lambda}_{K_1\times K_2}v_{K_2}(\boldsymbol{x})\right)u_{K_1}^{\top}(t)\hat{A}_{K_1\times K_2}v_{K_2}(\boldsymbol{x})dF_{X,T}(\boldsymbol{x},t)\Bigg]\\
		=&\frac {1} {\sqrt{N}} \sum_{i=1}^N \Bigg[L'\left\{Y_i-g\left(T_i;\boldsymbol{\beta}_0\right)\right\}m(T_i;\boldsymbol{\beta}_0)\rho''\left(u_{K_1}^{\top}(T_i){\Lambda}^*_{K_1\times K_2}v_{K_2}(\boldsymbol{X}_i)\right) u_{K_1}(T_i)^{\top}\hat{A}_{K_1\times K_2}v_{K_2}(\boldsymbol{X}_i) \\
		&- \int_{\mathcal{T}}\int_{\mathcal{X}} m(t;\boldsymbol{\beta}_0) \varepsilon(t,\boldsymbol{x};\boldsymbol{\beta}_0)\rho''\left(u_{K_1}^{\top}(t){\Lambda}^*_{K_1\times K_2}v_{K_2}(\boldsymbol{x})\right)u_{K_1}^{\top}(t)\hat{A}_{K_1\times K_2}v_{K_2}(\boldsymbol{x})dF_{X,T}(\boldsymbol{x},t)\Bigg]\\
		&+\frac {1} {\sqrt{N}} \sum_{i=1}^N \Bigg[L'\left\{Y_i -g\left(T_i;\boldsymbol{\beta}_0\right)\right\}m(T_i;\boldsymbol{\beta}_0)\rho'''\left(\xi_3(T_i,\boldsymbol{X}_i)\right)\left\{u_{K_1}(T_i)^{\top}\tilde{A}_{K_1\times K_2}v_{K_2}(\boldsymbol{X}_i)\right\}\\
		& \quad \quad \quad  \quad \quad  \times u_{K_1}(T_i)^{\top}\hat{A}_{K_1\times K_2}v_{K_2}(\boldsymbol{X}_i)\bigg] \\
		&-\sqrt{N}\int_{\mathcal{T}} \int_{\mathcal{X}} m(t;\boldsymbol{\beta}_0)\varepsilon(t,\boldsymbol{x};\boldsymbol{\beta}_0)\rho'''\left(\xi_3(t,\boldsymbol{x})\right)\left\{u_{K_1}(t)^{\top}\tilde{A}_{K_1\times K_2}v_{K_2}(\boldsymbol{x})\right\} \\
		&\qquad \qquad \qquad\times  u_{K_1}^{\top}(t)\hat{A}_{K_1\times K_2}v_{K_2}(\boldsymbol{x})dF_{X,T}(\boldsymbol{x},t)\\
		= &  W_{1K} + W_{2K} + W_{3K},
		\end{align*}
		where $\tilde{A}_{K_1\times K_2}$ is defined in \eqref{def:A_tilde}, and{\footnotesize
			\begin{align*}
			W_{1K} := &~\frac {1} {\sqrt{N}} \sum_{i=1}^N \Bigg[L'\left\{Y_i-g\left( T_i;\boldsymbol{\beta}_0\right)\right\}m(T_i;\boldsymbol{\beta}_0)\rho''\left(u_{K_1}^{\top}(T_i){\Lambda}^*_{K_1\times K_2}v_{K_2}(\boldsymbol{X}_i)\right) u_{K_1}(T_i)^{\top}\hat{A}_{K_1\times K_2}v_{K_2}(\boldsymbol{X}_i) \\
			&\qquad \qquad - \int_{\mathcal{T}}\int_{\mathcal{X}} m(t;\boldsymbol{\beta}_0) \varepsilon(t,\boldsymbol{x};\boldsymbol{\beta}_0)\rho''\left(u_{K_1}^{\top}(t){\Lambda}^*_{K_1\times K_2}v_{K_2}(\boldsymbol{x})\right)u_{K_1}^{\top}(t)\hat{A}_{K_1\times K_2}v_{K_2}(\boldsymbol{x})dF_{X,T}(\boldsymbol{x},t)\Bigg]\ ,\\
			W_{2K} := &~\frac {1} {\sqrt{N}} \sum_{i=1}^N \Bigg[L'\left\{Y_i-g\left(T_i;\boldsymbol{\beta}_0\right)\right\}m(T_i;\boldsymbol{\beta}_0)\rho'''\left(\xi_3(T_i,\boldsymbol{X}_i)\right)\left\{u_{K_1}(T_i)^{\top}\tilde{A}_{K_1\times K_2}v_{K_2}(\boldsymbol{X}_i)\right\} u_{K_1}(T_i)^{\top}\hat{A}_{K_1\times K_2}v_{K_2}(\boldsymbol{X}_i)\bigg]\ , \\
			W_{3K} := &~ -{\sqrt{N}}  \int_{\mathcal{T}}\int_{\mathcal{X}} m(t;\boldsymbol{\beta}_0)\varepsilon(t,\boldsymbol{x};\boldsymbol{\beta}_0)\rho'''\left(\xi_3(t,\boldsymbol{x})\right)\left\{u_{K_1}(t)^{\top}\tilde{A}_{K_1\times K_2}v_{K_2}(\boldsymbol{x})\right\}  u_{K_1}^{\top}(t)\hat{A}_{K_1\times K_2}v_{K_2}(\boldsymbol{x})dF_{X,T}(\boldsymbol{x},t)\ ,
			\end{align*}}
		and $\xi_3(t,\boldsymbol{x})$ lies between $u_{K_1}(t)\tilde{\Lambda}_{K_1\times K_2}^{\top}v_{K_2}(\boldsymbol{x})$ and $u_{K_1}(t){\Lambda}^*_{K_1\times K_2}v_{K_2}(\boldsymbol{x})$.  \\
		
		For the term $W_{1K}$, we denote its $k^{th}$ component by $W_{1K,k}$, $k=1,\ldots,p$, and we also let  $m_k(T_i;\boldsymbol{\beta}_0)$ be the $k^{th}$ component of $m(T_i;\boldsymbol{\beta}_0)$, i.e.,
		\begin{align*}
		W_{1K,k} &:=  \frac {1} {\sqrt{N}} \sum_{i=1}^N \Bigg[L'\left\{Y_i-g\left(T_i;\boldsymbol{\beta}_0\right)\right\}m_{k}(T_i;\boldsymbol{\beta}_0)\rho''\left(u_{K_1}^{\top}(T_i){\Lambda}^*_{K_1\times K_2}v_{K_2}(\boldsymbol{X}_i)\right) u_{K_1}(T_i)^{\top}\hat{A}_{K_1\times K_2}v_{K_2}(\boldsymbol{X}_i) \\
		&- \int_{\mathcal{T}}\int_{\mathcal{X}} m_k(t;\boldsymbol{\beta}_0) \varepsilon(t,\boldsymbol{x};\boldsymbol{\beta}_0)\rho''\left(u_{K_1}^{\top}(t){\Lambda}^*_{K_1\times K_2}v_{K_2}(\boldsymbol{x})\right)u_{K_1}^{\top}(t)\hat{A}_{K_1\times K_2}v_{K_2}(\boldsymbol{x})dF_{X,T}(\boldsymbol{x},t)\Bigg]\\
		=&\tr \bigg\{ \frac {1} {\sqrt{N}} \sum_{i=1}^N \Bigg[L'\left\{Y_i-g\left(T_i;\boldsymbol{\beta}_0\right)\right\}m_{k}(T_i;\boldsymbol{\beta}_0)\rho''\left(u_{K_1}^{\top}(T_i){\Lambda}^*_{K_1\times K_2}v_{K_2}(\boldsymbol{X}_i)\right) v_{K_2}(\boldsymbol{X}_i)u_{K_1}(T_i)^{\top} \\
		&- \int_{\mathcal{T}}\int_{\mathcal{X}} m_k(t;\boldsymbol{\beta}_0) \varepsilon(t,\boldsymbol{x};\boldsymbol{\beta}_0)\rho''\left(u_{K_1}^{\top}(t){\Lambda}^*_{K_1\times K_2}v_{K_2}(\boldsymbol{x})\right)v_{K_2}(\boldsymbol{x})u_{K_1}^{\top}(t)dF_{X,T}(\boldsymbol{x},t)\Bigg]\hat{A}_{K_1\times K_2} \bigg\}\\
		=& \tr \left\{U_{K_2\times K_1}(k) \hat{A}_{K_1\times K_2}\right\} ,
		\end{align*}
		where
		\begin{align*}
		U_{K_2\times K_1}(k):= &  \frac {1} {\sqrt{N}} \sum_{i=1}^N \Bigg[L'\left\{Y_i-g\left(T_i;\boldsymbol{\beta}_0\right)\right\}m_{k}(T_i;\boldsymbol{\beta}_0)\rho''\left(u_{K_1}^{\top}(T_i){\Lambda}^*_{K_1\times K_2}v_{K_2}(\boldsymbol{X}_i)\right)v_{K_2}(\boldsymbol{X}_i)u_{K_1}(T_i)^{\top} \\
		&- \int_{\mathcal{T}} \int_{\mathcal{X}} m_k(t;\boldsymbol{\beta}_0) \varepsilon (t,\boldsymbol{x};\boldsymbol{\beta}_0)\rho''\left(u_{K_1}^{\top}(t){\Lambda}^*_{K_1\times K_2}v_{K_2}(\boldsymbol{x})\right)v_{K_2}(\boldsymbol{x})u_{K_1}^{\top}(t)dF_{X,T}(\boldsymbol{x},t)\Bigg].
		\end{align*}
	We compute the second moment of $U_{K_2\times K_1}(k)$ to get that 
			\begin{align*}
			&\mathbb{E}\left[\|U_{K_2\times K_1}(k)\|^2\right] =  \mathbb{E}[\tr\{(U_{K_2\times K_1}(k))^{\top}U_{K_2\times K_1}(k)\}]\\
			=& \mathbb{E}\left[L'\left\{Y -g\left(T;\boldsymbol{\beta}_0\right)\right\}^2m_{k}(T;\boldsymbol{\beta}_0)^2\rho''\left(u_{K_1}^{\top}(T){\Lambda}^*_{K_1\times K_2}v_{K_2}(\boldsymbol{X})\right)^2\|v_{K_2}(\boldsymbol{X})\|^2\|
			u_{K_1}(T)\|^2\right] \\
			&~ -\tr \bigg\{ \mathbb{E}[m_k(T;\boldsymbol{\beta}_0)\varepsilon(T,\boldsymbol{X};\boldsymbol{\beta}^*)\rho''\left(u_{K_1}^{\top}(T){\Lambda}^*_{K_1\times K_2}v_{K_2}(\boldsymbol{X})\right)
			u_{K_1}(T)v^{\top}_{K_2}(\boldsymbol{X})]\\ 
			& \quad \quad \quad \quad \times \mathbb{E}[m_k(T;\boldsymbol{\beta}_0)\varepsilon(T,\boldsymbol{X};\boldsymbol{\beta}_0)\rho''\left(u_{K_1}^{\top}(T){\Lambda}^*_{K_1\times K_2}v_{K_2}(\boldsymbol{X})\right)v_{K_2}(\boldsymbol{X})u_{K_1}(T)^{\top}] \bigg\} \\
			\leq &\mathbb{E}\left[L'\left\{Y-g\left(T;\boldsymbol{\beta}_0\right)\right\}^2m_{k}(T;\boldsymbol{\beta}_0)^2\rho''\left(u_{K_1}^{\top}(T){\Lambda}^*_{K_1\times K_2}v_{K_2}(\boldsymbol{X})\right)^2\|v_{K_2}(\boldsymbol{X})\|^2\|u_{K_1}(T)\|^2\right] \\
			\leq &  \mathbb{E}\left[L'\left\{Y -g\left(T;\boldsymbol{\beta}_0\right)\right\}^2\right]\left( \sup_{t\in\mathcal{T}}m_k(t;\boldsymbol{\beta}_0)^2\right) \cdot a_3\cdot \mathbb{E}\left[\|v_{K_2}(\boldsymbol{X})\|^2\|u_{K_1}(T)\|^2\right]\\
			= &\mathbb{E}\left[L'\left\{Y -g\left(T;\boldsymbol{\beta}_0\right)\right\}^2\right]\left( \sup_{t\in\mathcal{T}}m_k(t;\boldsymbol{\beta}_0)^2\right) \cdot a_3\cdot \mathbb{E}\left[\frac{1}{\pi_0(T,\boldsymbol{X})}\cdot \pi_0(T,\boldsymbol{X})\cdot\|v_{K_2}(\boldsymbol{X})\|^2\|u_{K_1}(T)\|^2\right]\\
			\leq  &\mathbb{E}\left[L'\left\{Y -g\left(T;\boldsymbol{\beta}_0\right)\right\}^2\right]\left( \sup_{t\in\mathcal{T}}m_k(t;\boldsymbol{\beta}_0)^2\right) \cdot a_3\cdot \frac{1}{\eta_1}\cdot \mathbb{E}\left[ \pi_0(T,\boldsymbol{X})\cdot\|v_{K_2}(\boldsymbol{X})\|^2\|u_{K_1}(T)\|^2\right]\\
			=  &\mathbb{E}\left[L'\left\{Y -g\left(T;\boldsymbol{\beta}_0\right)\right\}^2\right]\left( \sup_{t\in\mathcal{T}}m_k(t;\boldsymbol{\beta}_0)^2\right) \cdot a_3\cdot \frac{1}{\eta_1}\cdot \mathbb{E}\left[ \|v_{K_2}(\boldsymbol{X})\|^2\right] \mathbb{E}\left[\|u_{K_1}(T)\|^2\right] \quad (\text{by using}\  \eqref{moment1})\\
			\leq& O(1) \cdot O(K_2)\cdot O(K_1)=O(K), 
			\end{align*}
		where $\dps a_3 := \sup_{\gamma \in \Gamma_1} |\rho''(\gamma)|^2 < +\infty$, the second inequality follows from this definition and the fact that
		$u_{K_1}^{\top}(t){\Lambda}^*_{K_1\times K_2}v_{K_2}(\boldsymbol{x}) \in \Gamma_1,\ \forall (t,\boldsymbol{x}) \in \mathcal{T}\times\mathcal{X}$ when $K$ is large enough; the third inequality follows from Assumption \ref{as:pi0}; the forth inequality follows from Assumption \ref{as:EY2} and the facts
		\begin{align}\label{eq:E[u^2]}
		&\mathbb{E}[\|u_{K_1}(T)\|^2]=\mathbb{E}[\tr(u_{K_1}(T)u_{K_1}^\top(T))]=\tr(I_{K_1\times K_1})=K_1, \\
		&\mathbb{E}[\|v_{K_2}(\boldsymbol{X})\|^2]=\mathbb{E}[\tr(v_{K_2}(\boldsymbol{X})v_{K_2}^\top(\boldsymbol{X}))]=\tr(I_{K_2\times K_2})=K_2. \label{eq:E[v^2]}
		\end{align} 
		Then in light of  Chebyshev's inequality, Lemma \ref{lemma_pi^hat} and  Assumption \ref{as:K&N_c}, we have $$|W_{1K,k}|\leq \|U_{K_2\times K_1}\|\|\hat{A}_{K_1\times K_2}\|  =O_p(\sqrt{K})O_p\left(\sqrt{\frac{K}{N}}\right)=O_p\left(\sqrt{\frac{K^2}{N}}\right) ,$$ which implies 
		\begin{align} \notag %\label{eq:resubmit_W1K}
		\|W_{1K}\|^2= \sum_{k=1}^{p}|W_{1K,k}|^2= O_p\left({\frac{K^2}{N}}\right) .
		\end{align}
		
		For the term  $W_{3K}$, since  $\xi_3(t,\boldsymbol{x})$ lies between $u_{K_1}(t)^{\top}\Lambda_{K_1\times K_2}^*v_{K_2}(\boldsymbol{x})$ and $u_{K_1}(t)^{\top}\tilde{\Lambda}_{K_1\times K_2}^*v_{K_2}(\boldsymbol{x})$, which implies $\xi_3(t,\boldsymbol{x})$ lies between $u_{K_1}(t)^{\top}\Lambda_{K_1\times K_2}^*v_{K_2}(\boldsymbol{x})$ and $u_{K_1}(t)^{\top}\hat{\Lambda}_{K_1\times K_2}^*v_{K_2}(\boldsymbol{x})$. Then in light of \eqref{Gamma_2_epsilon} and \eqref{eq:Lambdahat_in_Upsilonhat}, we have $\mathbb{P}\left(\xi_3(t,\boldsymbol{x})\in \Gamma_2(\epsilon),\  \forall (t,\boldsymbol{x})\in \mathcal{T}\times \mathcal{X}\right)>1-\epsilon$, therefore,
		\begin{align}\label{xi_3}
		\sup_{(t,\boldsymbol{x})\in\mathcal{T}\times \mathcal{X}}|\rho'''\left(\xi_3(t,\boldsymbol{x})\right)|=O_p(1) \ .
		\end{align} 
		With \eqref{eq:u_Lambda_v}, \eqref{xi_3}, the fact $\|\tilde{A}_{K_1\times K_2}\|\leq \|\hat{A}_{K_1\times K_2}\|$, Lemma \ref{lemma_pi^hat}, and Assumption \ref{as:K&N_c}, we can derive that{\footnotesize
		\begin{align}
		&\left\|W_{3K}  \right\| =\bigg\|{\sqrt{N}}\int_{\mathcal{T}}\int_{\mathcal{X}} m(t;\boldsymbol{\beta}_0)\varepsilon(t,\boldsymbol{x};\boldsymbol{\beta}_0)\rho'''\left(\xi_3(t,\boldsymbol{x})\right)\left\{u_{K_1}(t)^{\top}\tilde{A}_{K_1\times K_2}v_{K_2}(\boldsymbol{x})\right\}  u_{K_1}^{\top}(t)\hat{A}_{K_1\times K_2}v_{K_2}(\boldsymbol{x})dF_{X,T}(\boldsymbol{x},t) \bigg\| \notag\\
 		\leq&\sqrt{N}\sup_{(t,\boldsymbol{x})\in\mathcal{T}\times \mathcal{X}}\left|\rho'''\left(\xi_3(t,\boldsymbol{x})\right)\right| \sup_{t\in\mathcal{T}}\|m(t;\boldsymbol{\beta}_0)\|\cdot\sup_{(t,\boldsymbol{x})\in\mathcal{T}\times\mathcal{X}}|\varepsilon(t,\boldsymbol{x};\boldsymbol{\beta}_0)| \notag\\
 		&\qquad \qquad \cdot \int_{\mathcal{T}}\int_{\mathcal{X}} \left|u_{K_1}(t)^{\top}\tilde{A}_{K_1\times K_2}v_{K_2}(\boldsymbol{x})\right|\cdot \left|  u_{K_1}^{\top}(t)\hat{A}_{K_1\times K_2}v_{K_2}(\boldsymbol{x})\right|dF_{X,T}(\boldsymbol{x},t) \notag\\
 		\leq & \sqrt{N}\cdot O_p(1)\cdot O(1)\cdot O(1)\cdot \left\{\int_{\mathcal{T}}\int_{\mathcal{X}} \left|u_{K_1}(t)^{\top}\tilde{A}_{K_1\times K_2}v_{K_2}(\boldsymbol{x})\right|^2dF_{X,T}(\boldsymbol{x},t)\right\}^{\frac{1}{2}} \notag\\
 		&\qquad \qquad  \cdot \left\{\int_{\mathcal{T}}\int_{\mathcal{X}}\left|  u_{K_1}^{\top}(t)\hat{A}_{K_1\times K_2}v_{K_2}(\boldsymbol{x})\right|^2dF_{X,T}(\boldsymbol{x},t)\right\}^{\frac{1}{2}}\notag\\
 		=&  \sqrt{N}\cdot O_p(1)\cdot O(1)\cdot O(1)\cdot O_p\left(\sqrt{\frac{K}{N}}\right)\cdot O_p\left(\sqrt{\frac{K}{N}}\right)=O_p\left(\sqrt{\frac{K^2}{N}}\right) \qquad \text{(by (\eqref{eq:u_Lambda_v}))} . \label{order:W_3K}
		\end{align}}

	For the term $W_{2K}$, we can deduce that
	 {\footnotesize
			\begin{align*}
			&\bigg\|\frac {1} {\sqrt{N}} \sum_{i=1}^N \Bigg[L'\left\{Y_i -g\left(T_i;\boldsymbol{\beta}_0\right)\right\}m(T_i;\boldsymbol{\beta}_0)\rho'''\left(\xi_3(T_i,\boldsymbol{X}_i)\right)  \left\{u_{K_1}(T_i)^{\top}\tilde{A}_{K_1\times K_2}v_{K_2}(\boldsymbol{X}_i)\right\} u_{K_1}(T_i)^{\top}\hat{A}_{K_1\times K_2}v_{K_2}(\boldsymbol{X}_i)\bigg]\bigg\|\\
			\leq & \left\{ \frac {1} {\sqrt{N}} \sum_{i=1}^N |L'(Y_i-g\left(T_i;\boldsymbol{\beta}_0\right))|\cdot \|u_{K_1}(T_i)\|^2 \|v_{K_2}(\boldsymbol{X}_i)\|^2\right\}\cdot \sup_{t\in\mathcal{T}}\|m(t;\boldsymbol{\beta}_0)\|\cdot  \sup_{(t,\boldsymbol{x})\in\mathcal{T}\times \mathcal{X}}|\rho'''\left(\xi_3(t,\boldsymbol{x})\right)|\cdot \|\hat{A}_{K_1\times K_2}\|^2\\
			\leq& \sqrt{N} \left\{ \frac {1} {N} \sum_{i=1}^N |L'(Y_i-g\left(T_i;\boldsymbol{\beta}_0\right))|^2\right\}^{\frac{1}{2}} \left\{\frac{1}{N}\sum_{i=1}^N \|u_{K_1}(T_i)\|^4 \|v_{K_2}(\boldsymbol{X}_i)\|^4\right\}^{\frac{1}{2}} \sup_{t\in\mathcal{T}}\|m(t;\boldsymbol{\beta}_0)\| \sup_{(t,\boldsymbol{x})\in\mathcal{T}\times \mathcal{X}}|\rho'''\left(\xi_3(t,\boldsymbol{x})\right)|\cdot \|\hat{A}_{K_1\times K_2}\|^2\\
			\leq& \sqrt{N}\cdot O_p(1)\cdot \{\zeta_1(K_1)\zeta_2(K_2)\}\cdot \left\{\frac{1}{N}\sum_{i=1}^N \|u_{K_1}(T_i)\|^2 \|v_{K_2}(\boldsymbol{X}_i)\|^2\right\}^{\frac{1}{2}}\cdot O(1)\cdot O_p(1)\cdot O_p\left(\frac{K}{N}\right)\\
			\leq &\sqrt{N}\cdot O_p(1)\cdot\zeta(K)\cdot \left\{\mathbb{E}\left[ \|u_{K_1}(T)\|^2\|v_{K_2}(\boldsymbol{X})\|^2\right]+O_p\left(\zeta(K)\sqrt{\frac{K}{N}}\right) \right\}^{\frac{1}{2}}\cdot O(1)\cdot O_p(1)\cdot O_p\left(\frac{K}{N}\right)\\
			\leq &\sqrt{N}\cdot O_p(1)\cdot\zeta(K)\cdot O_p(\sqrt{K})\cdot O(1)\cdot O_p(1)\cdot O_p\left(\frac{K}{N}\right)=O_p\left(\zeta(K)\sqrt{\frac{K^3}{N}}\right)
			\end{align*}}
		where the fourth inequality follows from  the fact that{\footnotesize
			\begin{align*}
			&\mathbb{E}\left[\left(\frac {1} {N} \sum_{i=1}^N  \|u_{K_1}(T_i)\|^2 \|v_{K_2}(\boldsymbol{X}_i)\|^2-\mathbb{E}\left[  \|u_{K_1}(T)\|^2 \|v_{K_2}(\boldsymbol{X})\|^2\right]\right)^2\right]\\
			= & \frac{1}{N}\cdot \mathbb{E}\left[\left(\|u_{K_1}(T)\|^2 \|v_{K_2}(\boldsymbol{X})\|^2-\mathbb{E}\left[\|u_{K_1}(T)\|^2 \|v_{K_2}(\boldsymbol{X})\|^2\right]\right)^2\right]\\
			\leq &\frac{1}{N}\cdot  \mathbb{E}\left[ \|u_{K_1}(T)\|^4 \|v_{K_2}(\boldsymbol{X})\|^4\right]\leq\frac{1}{N}\cdot \zeta_1(K_1)^2\zeta_2(K_2)^2\cdot \mathbb{E}\left[ \|u_{K_1}(T)\|^2 \|v_{K_2}(\boldsymbol{X})\|^2\right]\\
			= &\frac{1}{N}\cdot \zeta_1(K_1)^2\zeta_2(K_2)^2\cdot \mathbb{E}\left[\frac{1}{\pi_0(T,\boldsymbol{X})}\cdot \pi_0(T,\boldsymbol{X}) \|u_{K_1}(T)\|^2 \|v_{K_2}(\boldsymbol{X})\|^2\right]\\
			\leq &\frac{1}{N}\cdot \frac{1}{\eta_1}\cdot \zeta_1(K_1)^2\zeta_2(K_2)^2\cdot \mathbb{E}\left[ \pi_0(T,\boldsymbol{X}) \|u_{K_1}(T)\|^2 \|v_{K_2}(\boldsymbol{X})\|^2\right]\\
			=&\frac{1}{N}\cdot \frac{1}{\eta_1}\cdot \zeta_1(K_1)^2\zeta_2(K_2)^2\cdot \mathbb{E}\left[   \|u_{K_1}(T)\|^2 \right]\cdot  \mathbb{E}\left[    \|v_{K_2}(\boldsymbol{X})\|^2\right] =O\left(\frac{K}{N}\zeta(K)^2\right) .
			\end{align*}}
Therefore, we can obtain that
\begin{align*}
\eqref{eq:WK}=W_{1K}+W_{2K}+W_{3K}=O_p\left(\sqrt{\frac{K^2}{N}}\right)+O_p\left(\zeta(K)\sqrt{\frac{K^3}{N}}\right)+O_p\left(\sqrt{\frac{K^2}{N}}\right)=O_p\left(\zeta(K)\sqrt{\frac{K^3}{N}}\right)\ .
\end{align*}
		Finally, it follows that the term \eqref{eq:WK} is of $o_p(1)$ in light of Assumption \ref{as:K&N_c}.\\

		\noindent \textbf{\emph{For term \eqref{eq:VK}}:}
		Note that
		\begin{align*}
		&\mathbb{E}\left[\Bigg\|\frac {1} {\sqrt{N}} \sum_{i=1}^N \Bigg\{\left(\pi_K^*(T_i,\boldsymbol{X}_i) - {\pi_0(T_i,\boldsymbol{X}_i)}\right)m(T_i;\boldsymbol{\beta}_0)L'\left\{Y_i-g\left(T_i;\boldsymbol{\beta}_0\right)\right\}\right.\\
		& \left. \quad \quad \quad \quad \quad \quad \quad \quad - \mathbb{E}\left[m(T;\boldsymbol{\beta}_0)\varepsilon(T,\boldsymbol{X};\boldsymbol{\beta}_0)\left(\pi_K^*(T,\boldsymbol{X}) - {\pi_0(T,\boldsymbol{X})}\right)\right] \Bigg\}\Bigg\|^2\right]\\
		\leq & \mathbb{E}\left[\left(\pi_K^*(T,\boldsymbol{X}) - {\pi_0(T,\boldsymbol{X})}\right)^2\|m(T;\boldsymbol{\beta}_0)\|^2\cdot L'\left\{Y-g\left(T;\boldsymbol{\beta}_0\right)\right\}^2\right]\\
		\leq &\sup_{(t,\boldsymbol{x})\in\mathcal{T}\times\mathcal{X}}\left|\pi_K^*(t,\boldsymbol{x}) - {\pi_0(t,\boldsymbol{x})}\right|^2\cdot \sup_{t\in\mathcal{T}}\|m(t;\boldsymbol{\beta}_0)\|^2\cdot \mathbb{E}\left[L'\left\{Y-g\left(T;\boldsymbol{\beta}_0\right)\right\}^2\right]\\
		\leq & O(\zeta(K)^2K^{-2\alpha}) \ ,
		\end{align*}
		where the last equality follows from Lemma \ref{lemma_pi^*}.
		Then by Chebyshev's inequality, we can claim that the term \eqref{eq:VK} is of $O_p(\zeta(K)K^{-\alpha})$.\\
		
		\noindent\textbf{\emph{For term \eqref{eq:Lemma1}}:}	By Lemma \ref{lemma_pi^*} and Assumption \ref{as:K&N_c}, we can deduce that 
		\begin{align*}
		&\left\|{\sqrt{N}}\cdot \mathbb{E}\left[m(T;\boldsymbol{\beta}_0)\varepsilon(T,\boldsymbol{X};\boldsymbol{\beta}_0)\left(\pi_K^*(T,\boldsymbol{X}) - {\pi_0(T,\boldsymbol{X})}\right)\right]\right\|\\
		\leq& \sqrt{N} \sup_{t\in\mathcal{T}}\|m(t;\boldsymbol{\beta}_0)\|\cdot \mathbb{E}[|\varepsilon(T,\boldsymbol{X};\boldsymbol{\beta}_0)|^2]^{\frac{1}{2}}\cdot \mathbb{E}\left[|\pi_K^*(T,\boldsymbol{X}) - {\pi_0(T,\boldsymbol{X})}|^2\right]^{\frac{1}{2}}=O\left(\sqrt{N}K^{-\alpha}\right) \ .
		\end{align*} 
	
		\noindent \textbf{\emph{For term \eqref{eq:tauto}}:}	By Mean Value Theorem and the  definition of $\hat{A}_{K_1\times K_2}$ in \eqref{def:A}, the term \eqref{eq:tauto} is exactly equal to zero.
		\\
		
		\noindent \textbf{\emph{For term \eqref{eq:Q}}:}	We can telescope \eqref{eq:Q} as follows:
\begin{align}
			&{\sqrt{N}}\int_{\mathcal{T}}\int_{\mathcal{X}}m(t;\boldsymbol{\beta}_0)\varepsilon(t,\boldsymbol{x};\boldsymbol{\beta}_0)\rho''\left(u_{K_1}^{\top}(t)\tilde{\Lambda}_{K_1\times K_2}v_{K_2}(\boldsymbol{x})\right) u_{K_1}^{\top}(t)\hat{A}_{K_1\times K_2}v_{K_2}(\boldsymbol{x})dF_{X,T}(\boldsymbol{x},t) \notag\\ 
			&-{\sqrt{N}}\int_{\mathcal{T}}\int_{\mathcal{X}}m(t;\boldsymbol{\beta}_0)\varepsilon(t,\boldsymbol{x};\boldsymbol{\beta}_0)\rho''\left(u_{K_1}^{\top}(t){\Lambda}^*_{K_1\times K_2}v_{K_2}(\boldsymbol{x})\right) u_{K_1}^{\top}(t)A^*_{K_1\times K_2}v_{K_2}(\boldsymbol{x})dF_{X,T}(\boldsymbol{x},t) \notag\\
			=&{\sqrt{N}}\int_{\mathcal{T}}\int_{\mathcal{X}}m(t;\boldsymbol{\beta}_0)\varepsilon(t,\boldsymbol{x};\boldsymbol{\beta}_0)\left\{\rho''\left(u_{K_1}^{\top}(t)\tilde{\Lambda}_{K_1\times K_2}v_{K_2}(\boldsymbol{x})\right)-\rho''\left(u_{K_1}^{\top}(t){\Lambda}^*_{K_1\times K_2}v_{K_2}(\boldsymbol{x})\right) \right\} \notag\\
			&\quad \quad \quad \quad \quad  \quad \quad \quad \quad \quad \qquad \qquad  \times u_{K_1}^{\top}(t)\hat{A}_{K_1\times K_2}v_{K_2}(\boldsymbol{x})dF_{X,T}(\boldsymbol{x},t)\label{A_star1}\\ 
			&+{\sqrt{N}}\int_{\mathcal{T}}\int_{\mathcal{X}}m(t;\boldsymbol{\beta}_0)\varepsilon(t,\boldsymbol{x};\boldsymbol{\beta}_0)\rho''\left(u_{K_1}^{\top}(t){\Lambda}^*_{K_1\times K_2}v_{K_2}(\boldsymbol{x})\right)  u_{K_1}^{\top}(t)\notag\\
			&\quad \quad \quad \quad \quad  \quad \quad \quad \quad \quad \qquad \qquad  \times\left\{\hat{A}_{K_1\times K_2}-A^*_{K_1\times K_2}\right\}v_{K_2}(\boldsymbol{x}) dF_{X,T}(\boldsymbol{x},t)  .\label{A_star2}
			\end{align}
		For the term \eqref{A_star1}, by Mean Value Theorem, {\footnotesize
		\begin{align*}
		\eqref{A_star1}=&{\sqrt{N}}\int_{\mathcal{T}}\int_{\mathcal{X}}m(t;\boldsymbol{\beta}_0)\varepsilon(t,\boldsymbol{x};\boldsymbol{\beta}_0)\rho'''\left(\xi_3(t,\boldsymbol{x})\right)\left\{u_{K_1}^{\top}(t)\tilde{A}_{K_1\times K_2}v_{K_2}(\boldsymbol{x})\right\}  \left\{u_{K_1}^{\top}(t)\hat{A}_{K_1\times K_2}v_{K_2}(\boldsymbol{x})\right\}dF_{X,T}(\boldsymbol{x},t)\\
		=&-W_{3K} \ , 
		\end{align*}}
		which is $O_p\left(\sqrt{\frac{K^2}{N}}\right)$ from \eqref{order:W_3K}.\\
		
		For the term \eqref{A_star2}, we first compute the probability order of $\|A^*_{K_1\times K_2}-\hat{A}_{K_1\times K_2}\|$. Using \eqref{eq:A^*_{K_1timesK_2}}, the fact $\rho''(v)=-\rho'(v)$ and Mean Value Theorem, we have{\footnotesize
			\begin{align}
			&A_{K_1\times K_2}^*-\hat{A}_{K_1\times K_2} \notag\\
			=&-\frac{1}{N}\sum_{i=1}^N\rho''\left(u_{K_1}^{\top}(T_i){\Lambda}^*_{K_1\times K_2}v_{K_2}(\boldsymbol{X}_i)\right)u_{K_1}(T_i)u_{K_1}(T_i)^{\top}\hat{A}_{K_1\times K_2}v_{K_2}(\boldsymbol{X}_i)v^{\top}_{K_2}(\boldsymbol{X}_i)\notag \\
			&-\frac{1}{N}\sum_{i=1}^N\rho'''\left(\xi_3(T_i,\boldsymbol{X}_i)\right)\left\{u_{K_1}(T)^{\top}\tilde{A}_{K_1\times K_2}v_{K_2}(\boldsymbol{X}_i)\right\}u_{K_1}(T_i)u_{K_1}(T_i)^{\top}\hat{A}_{K_1\times K_2}v_{K_2}(\boldsymbol{X}_i)v^{\top}_{K_2}(\boldsymbol{X}_i)\notag\\
			&-\hat{A}_{K_1\times K_2} \notag \\
			=&\frac{1}{N}\sum_{i=1}^N\left\{\rho'\left(u_{K_1}^{\top}(T_i){\Lambda}^*_{K_1\times K_2}v_{K_2}(\boldsymbol{X}_i)\right)u_{K_1}(T_i)u_{K_1}(T_i)^{\top} \hat{A}_{K_1\times K_2}v_{K_2}(\boldsymbol{X}_i)v^{\top}_{K_2}(\boldsymbol{X}_i) -\hat{A}_{K_1\times K_2} \right\} \label{A_difference_1} \\
			-&\frac{1}{N}\sum_{i=1}^N\rho'''\left(\xi_3(T_i,\boldsymbol{X}_i)\right)\left\{u_{K_1}(T_i)^{\top}\tilde{A}_{K_1\times K_2}v_{K_2}(\boldsymbol{X}_i)\right\}u_{K_1}(T_i)u_{K_1}(T_i)^{\top}\hat{A}_{K_1\times K_2}v_{K_2}(\boldsymbol{X}_i)v^{\top}_{K_2}(\boldsymbol{X}_i) .\label{A_difference_2}
			\end{align}}
		For the term \eqref{A_difference_1}, by \eqref{moment1} we can write $\hat{A}_{K_1\times K_2}$ as
		$$ \hat{A}_{K_1\times K_2}=\mathbb{E}_{T,\boldsymbol{X}}\left[\pi_0(T,\boldsymbol{X})u_{K_1}(T)u_{K_1}(T)^{\top}\hat{A}_{K_1\times K_2}v_{K_2}(\boldsymbol{X})v^{\top}_{K_2}(\boldsymbol{X})\right], $$
		where $\mathbb{E}_{T,\boldsymbol{X}}[\cdot]$ denotes taking expectation with respect to $(T,\boldsymbol{X})$.
		We telescope \eqref{A_difference_1} as follows:
		\begin{align}
		&\frac{1}{N}\sum_{i=1}^N\bigg\{\rho'\left(u_{K_1}^{\top}(T_i){\Lambda}^*_{K_1\times K_2}v_{K_2}(\boldsymbol{X}_i)\right)u_{K_1}(T_i)u_{K_1}(T_i)^{\top}\hat{A}_{K_1\times K_2}v_{K_2}(\boldsymbol{X}_i)v^{\top}_{K_2}(\boldsymbol{X}_i)- \hat{A}_{K_1\times K_2}\bigg\} \notag\\
		=&\frac{1}{N}\sum_{i=1}^N\bigg\{\left\{\rho'\left(u_{K_1}^{\top}(T_i){\Lambda}^*_{K_1\times K_2}v_{K_2}(\boldsymbol{X}_i)\right)-\pi_0(T_i,\boldsymbol{X}_i)\right\}u_{K_1}(T_i)u_{K_1}(T_i)^{\top}\hat{A}_{K_1\times K_2}v_{K_2}(\boldsymbol{X}_i)v^{\top}_{K_2}(\boldsymbol{X}_i)\bigg\}  \label{A_diff_1_1}\\
		&-\frac{1}{N}\sum_{i=1}^N\bigg\{\pi_0(T_i,\boldsymbol{X}_i)u_{K_1}(T_i)u_{K_1}(T_i)^{\top}\hat{A}_{K_1\times K_2}v_{K_2}(\boldsymbol{X}_i)v^{\top}_{K_2}(\boldsymbol{X}_i)  \notag \\ 
		&\quad \quad \quad \quad \quad \quad-\mathbb{E}_{T,\boldsymbol{X}}\left[\pi_0(T,\boldsymbol{X})u_{K_1}(T)u_{K_1}(T)^{\top}\hat{A}_{K_1\times K_2}v_{K_2}(\boldsymbol{X})v^{\top}_{K_2}(\boldsymbol{X})\right] \bigg\}. \label{A_diff_1_2} 
		\end{align}
		For the term \eqref{A_diff_1_1}, by Lemmas \ref{lemma_pi^*} and \ref{lemma_pi^hat}, we have that{\small
		\begin{align*}
		&\bigg\|\frac{1}{N}\sum_{i=1}^N \left\{\rho'\left(u_{K_1}^{\top}(T_i){\Lambda}^*_{K_1\times K_2}v_{K_2}(\boldsymbol{X}_i)\right)-\pi_0(T_i,\boldsymbol{X}_i)\right\}u_{K_1}(T_i)u_{K_1}(T_i)^{\top} \hat{A}_{K_1\times K_2}v_{K_2}(\boldsymbol{X}_i)v^{\top}_{K_2}(\boldsymbol{X}_i)\bigg\|\\
		\leq & \frac{1}{N}\sum_{i=1}^N \left|\rho'\left(u_{K_1}^{\top}(T_i){\Lambda}^*_{K_1\times K_2}v_{K_2}(\boldsymbol{X}_i)\right)-\pi_0(T_i,\boldsymbol{X}_i)\right|\cdot \|u_{K_1}(T_i)\|^2 \cdot \|v_{K_2}(\boldsymbol{X}_i)\|^2\cdot  \|\hat{A}_{K_1\times K_2}\|\\
		=&\bigg\{\mathbb{E}\left[\left|\rho'\left(u_{K_1}^{\top}(T){\Lambda}^*_{K_1\times K_2}v_{K_2}(\boldsymbol{X})\right)-\pi_0(T,\boldsymbol{X})\right|\cdot \|u_{K_1}(T)\|^2 \cdot \|v_{K_2}(\boldsymbol{X})\|^2\right]+O_p\left(\zeta(K)\sqrt{{\frac{K}{N}}}\right)\bigg\}\cdot  \|\hat{A}_{K_1\times K_2}\|\\
		\leq & \left\{\sup_{(t,\boldsymbol{x})\in\mathcal{T}\times \mathcal{X}}\left|\pi_K^*(t,\boldsymbol{x})-\pi_0(t,\boldsymbol{x})\right|\cdot \frac{1}{\eta_1}\cdot \mathbb{E}\left[\pi_0(T,\boldsymbol{X})\|u_{K_1}(T)\|^2\|v_{K_2}(\boldsymbol{X})\|^2\right] +O_p\left(\zeta(K)\sqrt{{\frac{K}{N}}}\right)  \right\}\cdot  \|\hat{A}_{K_1\times K_2}\| \\
		= & \left\{\sup_{(t,\boldsymbol{x})\in\mathcal{T}\times \mathcal{X}}\left|\pi_K^*(t,\boldsymbol{x})-\pi_0(t,\boldsymbol{x})\right|\cdot \frac{1}{\eta_1}\cdot \mathbb{E}\left[\|u_{K_1}(T)\|^2\right]\cdot \mathbb{E}\left[\|v_{K_2}(\boldsymbol{X})\|^2\right] +O_p\left(\zeta(K)\sqrt{{\frac{K}{N}}}\right)  \right\}\cdot  \|\hat{A}_{K_1\times K_2}\| \\
		\leq & \left\{O\left(K^{-\alpha}\zeta(K)\right) \cdot O(K_1)\cdot O(K_2)+O_p\left(\zeta(K)\sqrt{{\frac{K}{N}}}\right)\right\}\cdot O_p\left(\sqrt{\frac{K}{N}}\right)\\
		\leq &  O_p\left(N^{-\frac{1}{2}}\zeta(K)\cdot K^{\frac{3}{2}-\alpha}\right) .
		\end{align*}}
		For the term \eqref{A_diff_1_2}, define the linear map $\mathcal{J}(\cdot):\mathbb{R}^{{K_1\times K_2}}\rightarrow \mathbb{R}$ by {\footnotesize
			\begin{align*}
			\mathcal{J}(M):= \frac{1}{N}\sum_{i=1}^N\bigg\{\pi_0(T_i,\boldsymbol{X}_i)u_{K_1}(T_i)u_{K_1}(T_i)^{\top}Mv_{K_2}(\boldsymbol{X}_i)v^{\top}_{K_2}(\boldsymbol{X}_i)  -\mathbb{E}_{T,\boldsymbol{X}}\left[\pi_0(T,\boldsymbol{X})u_{K_1}(T)u_{K_1}(T)^{\top}Mv_{K_2}(\boldsymbol{X})v^{\top}_{K_2}(\boldsymbol{X})\right] \bigg\},
			\end{align*}}
		then $ \eqref{A_diff_1_2}=\mathcal{J}(\hat{A}_{K_1\times K_2})$.
		For any fixed $M \in \mathbb{R}^{{K_1\times K_2}}$, by \eqref{moment1} and
		$M=\mathbb{E}[\pi_0(T,\boldsymbol{X})u_{K_1}(T)u_{K_1}(T)^{\top}M$ $\cdot v_{K_2}(\boldsymbol{X})v^{\top}_{K_2}(\boldsymbol{X})]$,
		then we have
			\begin{align*}
			&\mathbb{E}\left[\mathcal{J}(M)^2\right]\\=&\frac{1}{N}\cdot\mathbb{E}\bigg[\bigg\|\pi_0(T,\boldsymbol{X})u_{K_1}(T)u_{K_1}(T)^{\top}Mv_{K_2}(\boldsymbol{X})v^{\top}_{K_2}(\boldsymbol{X})-\mathbb{E}\left[\pi_0(T,\boldsymbol{X})u_{K_1}(T)u_{K_1}(T)^{\top}Mv_{K_2}(\boldsymbol{X})v^{\top}_{K_2}(\boldsymbol{X})\right]\bigg\|^2\bigg]\\
			\leq &\frac{1}{N}\cdot\mathbb{E}\bigg[\bigg\|\pi_0(T,\boldsymbol{X})u_{K_1}(T)u_{K_1}(T)^{\top}Mv_{K_2}(\boldsymbol{X})v^{\top}_{K_2}(\boldsymbol{X})\bigg\|^2\bigg]\\
			\leq &\frac{1}{N} \cdot \eta_2\cdot \mathbb{E}\bigg[ \pi_0(T,\boldsymbol{X})\cdot \|u_{K_1}(T)\|^4\|v_{K_2}(\boldsymbol{X})\|^4\bigg]\cdot \left\|M\right\|^2\\
			= &\frac{1}{N} \cdot \eta_2\cdot \mathbb{E}[  \|u_{K_1}(T)\|^4]\cdot \mathbb{E}[\|v_{K_2}(\boldsymbol{X})\|^4]\cdot \left\|M\right\|^2 \\
			\leq & \frac{1}{N} \cdot \eta_2\cdot \zeta_1(K)^2\cdot \zeta_2(K)^2\cdot \mathbb{E}[  \|u_{K_1}(T)\|^2]\cdot \mathbb{E}[\|v_{K_2}(\boldsymbol{X})\|^2]\cdot \left\|M\right\|^2\\
			=& \left\|M\right\|^2\cdot O\left(\zeta(K)^2\frac{K}{N}\right).
			\end{align*}
		Using Chebyshev's inequality we have  
		$$ |\mathcal{J}(M)| =\|M\| O_p\left(\zeta(K)\sqrt{\frac{K}{N}}\right)\ ,$$
		then in light of Lemma \ref{lemma_pi^hat},
		$$\eqref{A_diff_1_2}=\mathcal{J}(\hat{A}_{K_1\times K_2})=\|\hat{A}_{K_1\times K_2}\| O_p\left(\zeta(K)\sqrt{\frac{K}{N}}\right)= O_p\left(\zeta(K)\frac{K}{N}\right)\ .$$
		Therefore, 
		$$\eqref{A_difference_1}=\eqref{A_diff_1_1}+\eqref{A_diff_1_2}= O_p\left(N^{-\frac{1}{2}}\zeta(K)\cdot K^{\frac{3}{2}-\alpha}\right)+O_p\left(\zeta(K)\frac{K}{N}\right) \ .$$

		For the term \eqref{A_difference_2}, we can deduce that
		\begin{align*}
		&\bigg\|\frac{1}{N}\sum_{i=1}^N\rho'''\left(\xi_3(T_i,\boldsymbol{X}_i)\right)\left\{u_{K_1}(T_i)^{\top}\tilde{A}_{K_1\times K_2}v_{K_2}(\boldsymbol{X}_i)\right\}   \left\{u_{K_1}(T_i)u_{K_1}(T_i)^{\top}\hat{A}_{K_1\times K_2}v_{K_2}(\boldsymbol{X}_i)v^{\top}_{K_2}(\boldsymbol{X}_i)\right\}\bigg\| \\
		\leq & \sup_{(t,\boldsymbol{x})\in \mathcal{T}\times \mathcal{X}}\left|\rho'''\left(\xi_3(t,\boldsymbol{x})\right)\right|\cdot  \|\hat{A}_{K_1\times K_2}\|^2 \cdot  \frac{1}{N}\sum_{i=1}^N\|u_{K_1}(T_i)\|^3\cdot \|v_{K_2}(\boldsymbol{X}_i)\|^3  \\
	\leq & \sup_{(t,\boldsymbol{x})\in \mathcal{T}\times \mathcal{X}}\left|\rho'''\left(\xi_3(t,\boldsymbol{x})\right)\right|\cdot  \|\hat{A}_{K_1\times K_2}\|^2 \cdot \zeta_1(K_1)\cdot \zeta_2(K_2)\cdot \frac{1}{N}\sum_{i=1}^N\|u_{K_1}(T_i)\|^2\cdot \|v_{K_2}(\boldsymbol{X}_i)\|^2  \\
		\leq & \sup_{(t,\boldsymbol{x})\in \mathcal{T}\times \mathcal{X}}\left|\rho'''\left(\xi_3(t,\boldsymbol{x})\right)\right|\cdot \|\hat{A}_{K_1\times K_2}\|^2\cdot \zeta(K) \left\{\mathbb{E}\left[\|u_{K_1}(T)\|^2\|v_{K_2}(\boldsymbol{X})\|^2\right]+O_p\left(\zeta(K)\sqrt{\frac{K}{N}}\right)\right\}\\
		\leq & O_p(1)\cdot O_p\left(\frac{K}{N}\right)\cdot \zeta(K) \cdot O(K)=O_p\left(\zeta(K)\frac{K^2}{N}\right) \ ,
		\end{align*}
		where the fourth inequality follows from \eqref{xi_3} and Lemma \ref{lemma_pi^hat}. Now, we can obtain
		\begin{align}\label{A_K-A_K^*}
		\|\hat{A}_{K_1\times K_2}-A_{K_1\times K_2}^*\|&=\eqref{A_difference_1}+\eqref{A_difference_2}=O_p\left(N^{-\frac{1}{2}}\zeta(K)K^{\frac{3}{2}-\alpha}\right)+O_p\left(\zeta(K)\frac{K}{N}\right)+O_p\left(\zeta(K)\frac{K^2}{N}\right) \notag\\
		&=O_p\left(N^{-\frac{1}{2}}\zeta(K)\cdot K^{\frac{3}{2}-\alpha}\right)+O_p\left(\zeta(K)\frac{K^{2}}{N}\right)  \ .
		\end{align}
		Using \eqref{A_K-A_K^*}, Assumptions \ref{as:EY2} and \ref{as:K&N_c},  for large enough $N$, we  have
		{\footnotesize\begin{align*}
			&\eqref{A_star2}=\bigg\|{\sqrt{N}}\int_{\mathcal{T}}\int_{\mathcal{X}}m(t;\boldsymbol{\beta}_0)\varepsilon(t,\boldsymbol{x};\boldsymbol{\beta}_0)\rho''\left(u_{K_1}^{\top}(t){\Lambda}^*_{K_1\times K_2}v_{K_2}(\boldsymbol{x})\right)u_{K_1}^{\top}(t)  \left\{\hat{A}_{K_1\times K_2}-A^*_{K_1\times K_2}\right\}v_{K_2}(\boldsymbol{x}) dF_{X,T}(\boldsymbol{x},t) \bigg\|\\
			\leq & \sqrt{N}   \sup_{t\in\mathcal{T}}\| m(t;\boldsymbol{\beta}_0)\|\sup_{\gamma\in\Gamma_1}|\rho''\left(\gamma\right)|\cdot \mathbb{E}\left[\left|\varepsilon(T,\boldsymbol{X};\boldsymbol{\beta}_0)\right|^2\right]^{\frac{1}{2}}\cdot \left[\int_{\mathcal{T}\times\mathcal{X}}\left(u_{K_1}(t)\left\{\hat{A}_{K_1\times K_2}-A^*_{K_1\times K_2}\right\}v_{K_2}(\boldsymbol{x})\right)^2dF_{T,X}(t,\boldsymbol{x})\right]^{\frac{1}{2}} \\
			\leq& \sqrt{N}\cdot O(1)\cdot O(1)\cdot O(1)\cdot O(1)\cdot O(	\|\hat{A}_{K_1\times K_2}-A_{K_1\times K_2}^*\|)\\
			\leq &O_p\left(\zeta(K)\cdot K^{\frac{3}{2}-\alpha}\right)+O_p\left(\zeta(K)\frac{K^{2}}{\sqrt{N}}\right),
		\end{align*}}	
	where the second inequality holds since by using the same argument of establishing \eqref{eq:u_Lambda_v}, we have
	$$\int_{\mathcal{T}\times\mathcal{X}}\left(u_{K_1}(t)\left\{\hat{A}_{K_1\times K_2}-A^*_{K_1\times K_2}\right\}v_{K_2}(\boldsymbol{x})\right)^2dF_{T,X}(t,\boldsymbol{x})= O(	\|\hat{A}_{K_1\times K_2}-A_{K_1\times K_2}^*\|).$$
	Therefore, we can obtain that
	\begin{align*}
	\eqref{eq:Q}=\eqref{A_star1}+\eqref{A_star2}=&O_p\left(\sqrt{\frac{K^2}{N}}\right)+O_p\left(\zeta(K)\cdot K^{\frac{3}{2}-\alpha}\right)+O_p\left(\zeta(K)\frac{K^{2}}{\sqrt{N}}\right)\\
	=&O_p\left(\zeta(K)\cdot K^{\frac{3}{2}-\alpha}\right)+O_p\left(\zeta(K)\frac{K^{2}}{\sqrt{N}}\right).
	\end{align*}

		\noindent \textbf{\emph{For term  \eqref{eq:proj}}:} By the definition of $A^*_{K_1\times K_2}$ in \eqref{def:A*},  we have{\footnotesize
			\begin{align}
		\eqref{eq:proj}	&=\frac{1}{\sqrt{N}}\sum_{i=1}^N\bigg\{\int_{\mathcal{T}}\int_{\mathcal{X}}m(t;\boldsymbol{\beta}_0)\varepsilon(t,\boldsymbol{x};\boldsymbol{\beta}_0)\rho''\left(u_{K_1}^{\top}(t)\Lambda_{K_1\times K_2}^*v_{K_2}(\boldsymbol{x})\right)u_{K_1}^{\top}(t)\label{projection_1}\\
			& \qquad  \times \left\{u_{K_1}(T_i)\rho'\left(u_{K_1}(T_i)\Lambda_{K_1\times K_2}^*v_{K_2}(\boldsymbol{X}_i)\right)v^{\top}_{K_2}(\boldsymbol{X}_i)\right\}v_{K_2}(\boldsymbol{x})dF_{X,T}(\boldsymbol{x},t)+ m(T_i;\boldsymbol{\beta}_0)\varepsilon(T_i,\boldsymbol{X}_i;\boldsymbol{\beta}_0)\pi_0(T_i,\boldsymbol{X}_i)\bigg\} \notag \\
			& -\frac{1}{\sqrt{N}}\sum_{i=1}^N\left\{\int_{\mathcal{T}}\int_{\mathcal{X}} m(t;\boldsymbol{\beta}_0)\varepsilon(t,\boldsymbol{x};\boldsymbol{\beta}^*)\rho''\left(u_{K_1}^{\top}(t)\Lambda_{K_1\times K_2}^*v^{\top}_{K_2}(\boldsymbol{x})\right)u_{K_1}^{\top}(t)\right.\label{projection_2}\left(\frac{1}{N}\sum_{l=1}^Nu_{K_1}(T_l)\right)\\
			& \quad \quad  \times \left(\frac{1}{N}\sum_{j=1}^Nv^{\top}_{K_2}(\boldsymbol{X}_j)\right)v_{K_2}(\boldsymbol{x})dF_{X,T}(\boldsymbol{x},t) + \mathbb{E}\left[\pi_0(T,\boldsymbol{X})m(T;\boldsymbol{\beta}_0)\varepsilon(T,\boldsymbol{X};\boldsymbol{\beta}_0)|\boldsymbol{X}=\boldsymbol{X}_i\right]\notag\\
			&\qquad \qquad + \mathbb{E}\left[\pi_0(T,\boldsymbol{X})m(T;\boldsymbol{\beta}_0)\varepsilon(T,\boldsymbol{X};\boldsymbol{\beta}_0)|T=T_i\right]\bigg\}. \notag
			\end{align}}
		We shall show that both \eqref{projection_1} and \eqref{projection_2} are of $o_p(1)$.	Noting  $\rho''=-\rho'$, we can telescope \eqref{projection_1} as follows:
		{\footnotesize
	\begin{align}
		 \eqref{projection_1}=&\frac{1}{\sqrt{N}}\sum_{i=1}^N\bigg\{\int_{\mathcal{T}}\int_{\mathcal{X}}m(t;\boldsymbol{\beta}_0)\varepsilon(t,\boldsymbol{x};\boldsymbol{\beta}_0)\rho'\left(u_{K_1}^{\top}(t)\Lambda_{K_1\times K_2}^*v_{K_2}(\boldsymbol{x})\right)u_{K_1}^{\top}(t)\label{Projection_1_1}\\
		 & \qquad \qquad \qquad  \times \left\{u_{K_1}(T_i)\bigg[-\rho'\left(u_{K_1}(T_i)\Lambda_{K_1\times K_2}^*v_{K_2}(\boldsymbol{X}_i)\right)+\pi_0(T_i,\boldsymbol{X}_i)\bigg]v^{\top}_{K_2}(\boldsymbol{X}_i)\right\}v_{K_2}(\boldsymbol{x})dF_{X,T}(\boldsymbol{x},t)\bigg\}\notag\\
		 &-\frac{1}{\sqrt{N}}\sum_{i=1}^N\bigg\{\int_{\mathcal{T}}\int_{\mathcal{X}}m(t;\boldsymbol{\beta}_0)\varepsilon(t,\boldsymbol{x};\boldsymbol{\beta}_0)\left\{\rho'\left(u_{K_1}^{\top}(t)\Lambda_{K_1\times K_2}^*v_{K_2}(\boldsymbol{x})\right)-\pi_0(t,\boldsymbol{x})\right\}u_{K_1}^{\top}(t) \label{Projection_1_2}\\
		 & \qquad \qquad \qquad  \times \left\{u_{K_1}(T_i)\pi_0(T_i,\boldsymbol{X}_i)v^{\top}_{K_2}(\boldsymbol{X}_i)\right\}v_{K_2}(\boldsymbol{x})dF_{X,T}(\boldsymbol{x},t)\bigg\}\notag\\
		&- \frac{1}{\sqrt{N}}\sum_{i=1}^N\bigg\{\int_{\mathcal{T}}\int_{\mathcal{X}}m(t;\boldsymbol{\beta}_0)\varepsilon(t,\boldsymbol{x};\boldsymbol{\beta}_0)\pi_0(t,\boldsymbol{x})u_{K_1}^{\top}(t) \left\{u_{K_1}(T_i)\pi_0(T_i,\boldsymbol{X}_i)v^{\top}_{K_2}(\boldsymbol{X}_i)\right\}v_{K_2}(\boldsymbol{x})dF_{X,T}(\boldsymbol{x},t)\notag\\
		&\qquad \qquad \qquad +m(T_i;\boldsymbol{\beta}_0)\varepsilon(T_i,\boldsymbol{X}_i;\boldsymbol{\beta}_0)\pi_0(T_i,\boldsymbol{X}_i)\bigg\}\label{Projection_1_3}\ .
	\end{align}	
}
We shall show that	\eqref{Projection_1_1}, \eqref{Projection_1_2} and \eqref{Projection_1_3} are all of $o_p(1)$. Note that second moment of \eqref{Projection_1_1} is	
	{\footnotesize
\begin{align*}
&\mathbb{E}[|\eqref{Projection_1_1}|^2]=\mathbb{E}\Bigg[\Bigg|\int_{\mathcal{T}}\int_{\mathcal{X}}m(t;\boldsymbol{\beta}_0)\varepsilon(t,\boldsymbol{x};\boldsymbol{\beta}_0)\rho'\left(u_{K_1}^{\top}(t)\Lambda_{K_1\times K_2}^*v_{K_2}(\boldsymbol{x})\right)u_{K_1}^{\top}(t)\\
& \qquad \qquad  \times \left\{u_{K_1}(T_i)\bigg[-\rho'\left(u_{K_1}(T_i)\Lambda_{K_1\times K_2}^*v_{K_2}(\boldsymbol{X}_i)\right)+\pi_0(T_i,\boldsymbol{X}_i)\bigg]v^{\top}_{K_2}(\boldsymbol{X}_i)\right\}v_{K_2}(\boldsymbol{x})dF_{X,T}(\boldsymbol{x},t)\Bigg|^2\Bigg]\\
= &\mathbb{E}\Bigg[\Bigg|\int_{\mathcal{T}}\int_{\mathcal{X}}\pi_0(t,\boldsymbol{x})\cdot m(t;\boldsymbol{\beta}_0)\varepsilon(t,\boldsymbol{x};\boldsymbol{\beta}_0)\bigg[\frac{\rho'\left(u_{K_1}^{\top}(t)\Lambda_{K_1\times K_2}^*v_{K_2}(\boldsymbol{x})\right)}{\pi_0(t,\boldsymbol{x})}\bigg]u_{K_1}^{\top}(t)\\
& \qquad \qquad  \times \left\{u_{K_1}(T_i)\bigg[-\rho'\left(u_{K_1}(T_i)\Lambda_{K_1\times K_2}^*v_{K_2}(\boldsymbol{X}_i)\right)+\pi_0(T_i,\boldsymbol{X}_i)\bigg]v^{\top}_{K_2}(\boldsymbol{X}_i)\right\}v_{K_2}(\boldsymbol{x})dF_{X,T}(\boldsymbol{x},t)\Bigg|^2\Bigg]\\
\leq&\mathbb{E}\Bigg[\Bigg|\int_{\mathcal{T}}\int_{\mathcal{X}}\pi_0(t,\boldsymbol{x})\cdot m(t;\boldsymbol{\beta}_0)\varepsilon(t,\boldsymbol{x};\boldsymbol{\beta}_0)\bigg[\frac{\rho'\left(u_{K_1}^{\top}(t)\Lambda_{K_1\times K_2}^*v_{K_2}(\boldsymbol{x})\right)}{\pi_0(t,\boldsymbol{x})}\bigg]u_{K_1}^{\top}(t) \left\{u_{K_1}(T_i)v^{\top}_{K_2}(\boldsymbol{X}_i)\right\}v_{K_2}(\boldsymbol{x})dF_{X,T}(\boldsymbol{x},t)\Bigg|^2\Bigg]\\
&\qquad \times\sup_{(t,\boldsymbol{x})\in\mathcal{T}\times\mathcal{X}} \left\{-\rho'\left(u_{K_1}^{\top}(t)\Lambda_{K_1\times K_2}^*v_{K_2}(\boldsymbol{x})\right)+\pi_0(t,\boldsymbol{x})\right\}^2 \\
=& \left\{ \mathbb{E}\Bigg[\Bigg| m(T_i;\boldsymbol{\beta}_0)\varepsilon(T_i,\boldsymbol{X}_i;\boldsymbol{\beta}_0)\bigg[\frac{\rho'\left(u_{K_1}^{\top}(T_i)\Lambda_{K_1\times K_2}^*v_{K_2}(\boldsymbol{X}_i)\right)}{\pi_0(T_i,\boldsymbol{X}_i)}\bigg]\Bigg|^2\Bigg]+o(1)\right\}\times\sup_{(t,\boldsymbol{x})\in\mathcal{T}\times\mathcal{X}} \left\{-\pi_K^*(t,\boldsymbol{x})+\pi_0(t,\boldsymbol{x})\right\}^2 \\
=&O(1)\cdot O(K^{-\frac{2s}{r}}\zeta(K)^2)=O(K^{-\frac{2s}{r}}\zeta(K)^2)\ ,
\end{align*}}	
where the third equality holds because
$$\int_{\mathcal{T}}\int_{\mathcal{X}}\pi_0(t,\boldsymbol{x})\cdot m(t;\boldsymbol{\beta}_0)\varepsilon(t,\boldsymbol{x};\boldsymbol{\beta}_0)\bigg[\frac{\rho'\left(u_{K_1}^{\top}(t)\Lambda_{K_1\times K_2}^*v_{K_2}(\boldsymbol{x})\right)}{\pi_0(t,\boldsymbol{x})}\bigg]u_{K_1}^{\top}(t) \left\{u_{K_1}(T)v^{\top}_{K_2}(\boldsymbol{X})\right\}v_{K_2}(\boldsymbol{x})dF_{X,T}(\boldsymbol{x},t) $$		
is the weighted $L^2$-projection of 	$m(t;\boldsymbol{\beta}_0)\varepsilon(t,\boldsymbol{x};\boldsymbol{\beta}_0)\bigg[\frac{\rho'\left(u_{K_1}^{\top}(t)\Lambda_{K_1\times K_2}^*v_{K_2}(\boldsymbol{x})\right)}{\pi_0(t,\boldsymbol{x})}\bigg]$ on the space linearly spanned by $\{u_{K_1}(t),v_{K_2}(\boldsymbol{x})\}$ with the weighted measure $\pi_0(t,\boldsymbol{x})dF_{T,X}(t,\boldsymbol{x})$. Similarly, we can also show \eqref{Projection_1_2} and \eqref{Projection_1_3} are  of $o_p(1)$. Therefore, \eqref{projection_1} is of $o_p(1)$.\\
		
For the term \eqref{projection_2}, since $\rho''(v)=-\rho'(v)$ and the fact $\mathbb{E}\left[\pi_0(T,\boldsymbol{X})m(T;\boldsymbol{\beta}_0)\varepsilon(T,\boldsymbol{X};\boldsymbol{\beta}_0)\right]=0$,  we telescope it as follows:{\footnotesize
			\begin{align}
		\eqref{projection_2}	=& {\sqrt{N}}\int_{\mathcal{T}}\int_{\mathcal{X}}m(t;\boldsymbol{\beta}_0)\varepsilon(t,\boldsymbol{x};\boldsymbol{\beta}_0)\rho'\left(u_{K_1}^{\top}(t)\Lambda_{K_1\times K_2}^*v_{K_2}(\boldsymbol{x})\right)u_{K_1}^{\top}(t) \left( \frac{1}{N}\sum_{l=1}^N u_{K_1}(T_l)-\mathbb{E}\left[u_{K_1}(T)\right]\right)  \notag\\
			& \quad \quad \quad \times  \left(\frac{1}{N}\sum_{j=1}^Nv^{\top}_{K_2}(\boldsymbol{X}_j)-\mathbb{E}[v^{\top}_{K_2}(\boldsymbol{X})]\right)v_{K_2}(\boldsymbol{x})dF_{X,T}(\boldsymbol{x},t) \label{projection_2_1} \\
			+&\frac{1}{\sqrt{N}}\sum_{i=1}^N\left\{\int_{\mathcal{T}}\int_{\mathcal{X}}m(t;\boldsymbol{\beta}_0)\varepsilon(t,\boldsymbol{x};\boldsymbol{\beta}_0)\rho'\left(u_{K_1}^{\top}(t)\Lambda_{K_1\times K_2}^*v_{K_2}(\boldsymbol{x})\right)u_{K_1}^{\top}(t)\mathbb{E}\left[ u_{K_1}(T)\right]v^{\top}_{K_2}(\boldsymbol{X}_i)v_{K_2}(\boldsymbol{x})dF_{X,T}(\boldsymbol{x},t)\right.  \notag\\
			&\qquad \qquad \qquad  -\mathbb{E}\left[\pi_0(T,\boldsymbol{X})m(T;\boldsymbol{\beta}_0)\varepsilon(T,\boldsymbol{X};\boldsymbol{\beta}_0)|\boldsymbol{X}=\boldsymbol{X}_i\right]\bigg\} \label{projection_2_4} \\
				+&\frac{1}{\sqrt{N}}\sum_{l=1}^N \bigg\{ \int_{\mathcal{T}}\int_{\mathcal{X}}m(t;\boldsymbol{\beta}_0)\varepsilon(t,\boldsymbol{x};\boldsymbol{\beta}_0)\rho'\left(u_{K_1}^{\top}(t)\Lambda_{K_1\times K_2}^*v_{K_2}(\boldsymbol{x})\right) u_{K_1}^{\top}(t)u_{K_1}(T_l) \mathbb{E}[v^{\top}_{K_2}(\boldsymbol{X})]v_{K_2}(\boldsymbol{x})dF_{X,T}(\boldsymbol{x},t)  \notag\\
				&\qquad \qquad\qquad  -\mathbb{E}\left[\pi_0(T,\boldsymbol{X})m(T;\boldsymbol{\beta}_0)\varepsilon(T,\boldsymbol{X};\boldsymbol{\beta}_0)|T=T_i\right] \bigg\} \label{projection_2_3} \\
			&-\frac{1}{\sqrt{N}}\sum_{i=1}^N\bigg\{\int_{\mathcal{T}}\int_{\mathcal{X}}m(t;\boldsymbol{\beta}_0)\varepsilon(t,\boldsymbol{x};\boldsymbol{\beta}_0)\rho'\left(u_{K_1}^{\top}(t)\Lambda_{K_1\times K_2}^*v_{K_2}(\boldsymbol{x})\right)u_{K_1}^{\top}(t)\mathbb{E}\left[u_{K_1}(T)\right] \mathbb{E}[v^{\top}_{K_2}(\boldsymbol{X})]v_{K_2}(\boldsymbol{x})dF_{X,T}(\boldsymbol{x},t)\notag\\
			&\qquad \qquad \qquad -\mathbb{E}\left[\pi_0(T,\boldsymbol{X})m(T;\boldsymbol{\beta}_0)\varepsilon(T,\boldsymbol{X};\boldsymbol{\beta}_0)\right] \bigg\}  . \label{projection_2_5}
			\end{align}}
		For the term \eqref{projection_2_1},   since 
		\begin{align*}
		&\left\|\frac{1}{N}\sum_{l=1}^N u_{K_1}(T_l)-\mathbb{E}\left[u_{K_1}(T)\right]\right\|=O_p\left(\sqrt{\frac{K_1}{N}}\right) \ ,
		\\ 
		&\left\|\frac{1}{N}\sum_{j=1}^N v_{K_2}(\boldsymbol{X}_j)-\mathbb{E}\left[v_{K_2}(\boldsymbol{X})\right]\right\|=O_p\left(\sqrt{\frac{K_2}{N}}\right) \ ,\\
		& \sup_{(t,\boldsymbol{x})\in\mathcal{T}\times \mathcal{X}}\left|\rho'\left(u_{K_1}^{\top}(t)\Lambda_{K_1\times K_2}^*v_{K_2}(\boldsymbol{x})\right)\right|=O(1) \ ,
		\end{align*}
		and by Assumptions  \ref{as:m_smooth}, \ref{as:EY2}, and \ref{as:K&N_c}, we can deduce that 
		\begin{align*}
		\eqref{projection_2_1}=\sqrt{N}\cdot O(\zeta(K)) O_p\left(\sqrt{\frac{K_1}{N}}\right)O_p\left(\sqrt{\frac{K_2}{N}}\right)=O_p\left(\zeta(K)\sqrt{\frac{K}{N}}\right) =o_p(1) \ .
		\end{align*}

	For the term  \eqref{projection_2_4}, noting the fact that $\mathbb{E}\left[\pi_0(T,\boldsymbol{X})m(T;\boldsymbol{\beta}_0)\varepsilon(T,\boldsymbol{X};\boldsymbol{\beta}_0)|\boldsymbol{X}\right]=\int_{\mathcal{T}}m(t;\boldsymbol{\beta}_0)\varepsilon(t,\boldsymbol{X};\boldsymbol{\beta}_0)$ $dF_T(t)$,  we can rewrite  \eqref{projection_2_4} as follows: 
			\begin{align}
			\eqref{projection_2_4}=&\frac{1}{\sqrt{N}}\sum_{j=1}^N\left\{\int_{\mathcal{T}}\int_{\mathcal{X}}m(t;\boldsymbol{\beta}_0)\varepsilon(t,\boldsymbol{x};\boldsymbol{\beta}_0)\frac{\pi_K^*(t,\boldsymbol{x})}{\pi_0(t,\boldsymbol{x})}u_{K_1}^{\top}(t)\mathbb{E}\left[ u_{K_1}(T)\right]\right.  v^{\top}_{K_2}(\boldsymbol{X}_j)v_{K_2}(\boldsymbol{x})dF_{X}(\boldsymbol{x})dF_T(t)\notag \\
			&\qquad \qquad\qquad   - \int_{\mathcal{T}}m(t;\boldsymbol{\beta}_0)\varepsilon(t,\boldsymbol{X}_j;\boldsymbol{\beta}_0) dF_T(t)\Bigg\}. \notag
			\end{align}
	 By computing the second moment of \eqref{projection_2_4}, we can obtain that{\footnotesize
			\begin{align*}
		&\mathbb{E}\bigg[\bigg\|\int_{\mathcal{T}}\int_{\mathcal{X}}m(t;\boldsymbol{\beta}_0)\varepsilon(t,\boldsymbol{x};\boldsymbol{\beta}_0)\frac{\pi_K^*(t,\boldsymbol{x})}{\pi_0(t,\boldsymbol{x})}u_{K_1}^{\top}(t)\mathbb{E}\left[ u_{K_1}(T)\right] v^{\top}_{K_2}(\boldsymbol{X})v_{K_2}(\boldsymbol{x})dF_{X}(\boldsymbol{x})dF_T(t)   - \int_{\mathcal{T}}m(t;\boldsymbol{\beta}_0)\varepsilon(t,\boldsymbol{X};\boldsymbol{\beta}_0) dF_T(t)\bigg\|^2  \bigg]\\
			\leq &\mathbb{E}\bigg[\bigg\|\int_{\mathcal{T}}\int_{\mathcal{X}}m(t;\boldsymbol{\beta}_0)\varepsilon(t,\boldsymbol{x};\boldsymbol{\beta}_0)\frac{\pi_K^*(t,\boldsymbol{x})}{\pi_0(t,\boldsymbol{x})}u_{K_1}^{\top}(t) u_{K_1}(T^*) v^{\top}_{K_2}(\boldsymbol{X}^*)v_{K_2}(\boldsymbol{x})dF_{X}(\boldsymbol{x})dF_T(t)   - m(T^*;\boldsymbol{\beta}_0)\varepsilon(T^*,\boldsymbol{X}^*;\boldsymbol{\beta}_0) \bigg\|^2  \bigg]\\
		\leq & 2\cdot \mathbb{E}\bigg[\bigg\| \int_{\mathcal{T}}\int_{\mathcal{X}}m(t;\boldsymbol{\beta}_0)\varepsilon(t,\boldsymbol{x};\boldsymbol{\beta}_0)u_{K_1}^{\top}(t)u_{K_1}(T^*) v^{\top}_{K_2}(\boldsymbol{X}^*)v_{K_2}(\boldsymbol{x})dF_{X}(\boldsymbol{x})dF_T(t)   -m(T^*;\boldsymbol{\beta}_0)\varepsilon(T^*,\boldsymbol{X}^*;\boldsymbol{\beta}_0)  \bigg\|^2\bigg] \\
			&+2\cdot \mathbb{E}\bigg[\bigg\| \int_{\mathcal{T}}\int_{\mathcal{X}}m(t;\boldsymbol{\beta}_0)\varepsilon(t,\boldsymbol{x};\boldsymbol{\beta}_0)\frac{\pi^*_K(t,\boldsymbol{x})-\pi_0(t,\boldsymbol{x})}{\pi_0(t,\boldsymbol{x})} u_{K_1}^{\top}(t)u_{K_1}(T^*) v^{\top}_{K_2}(\boldsymbol{X}^*)v_{K_2}(\boldsymbol{x})dF_{X}(\boldsymbol{x})dF_T(t)\bigg\|^2\bigg]\\
			=& o(1),
			\end{align*}}
		where $T^*\sim F_T$,  $\boldsymbol{X}^*\sim F_X$, and $T^*$ is independent of $\boldsymbol{X}^*$; the first inequality holds by Jensen's inequality; the last equality follows from Lemma \ref{lemma_pi^*} and the fact that
		\begin{align*}
		\int_{\mathcal{T}}\int_{\mathcal{X}}m(t;\boldsymbol{\beta}_0)\varepsilon(t,\boldsymbol{x};\boldsymbol{\beta}_0)u_{K_1}^{\top}(t)u_{K_1}(T^*) v^{\top}_{K_2}(\boldsymbol{X}^*)v_{K_2}(\boldsymbol{x})dF_{X}(\boldsymbol{x})dF_T(t)  \end{align*}
		is the $L^2$-projection of $m(T^*;\boldsymbol{\beta}_0)\varepsilon(T^*,\boldsymbol{X}^*;\boldsymbol{\beta}_0)$	on the space spanned by $\{u_{K_1}(T^*),v_{K_2}(\boldsymbol{X}^*)\}$, which implies {\footnotesize
			\begin{align*}
			&\mathbb{E}\bigg[\bigg\| \int_{\mathcal{T}}\int_{\mathcal{X}}m(t;\boldsymbol{\beta}_0)\varepsilon(t,\boldsymbol{x};\boldsymbol{\beta}_0)u_{K_1}^{\top}(t)u_{K_1}(T^*) v^{\top}_{K_2}(\boldsymbol{X}^*)v_{K_2}(\boldsymbol{x})dF_{X}(\boldsymbol{x})dF_T(t) -m(T^*;\boldsymbol{\beta}_0)\varepsilon(T^*,\boldsymbol{X}^*;\boldsymbol{\beta}_0)  \bigg\|^2\bigg]\rightarrow 0.
			\end{align*}}
		Thus \eqref{projection_2_4} is of $o_p(1)$ by Chebyshev's inequality. Similar argument can be applied to show that both \eqref{projection_2_3} and \eqref{projection_2_5} are of $o_p(1)$. Therefore, we can have that
		\begin{align*}
		|\eqref{projection_2}|\leq |\eqref{projection_2_1}|+|\eqref{projection_2_4}|+|\eqref{projection_2_3}|=o_p(1) \ .
		\end{align*}
		Then, we can obtain that
		\begin{align*}
		|\eqref{eq:proj}|\leq |\eqref{projection_1}|+|\eqref{projection_2}|=o_p(1)\ .
		\end{align*}
Summing up all orders \eqref{eq:WK}-\eqref{eq:proj} and using Assumption \ref{as:K&N_c}, we have{\small
\begin{align*}
&\eqref{eq:WK}+\eqref{eq:VK}+\eqref{eq:Lemma1}+\eqref{eq:tauto}+\eqref{eq:Q}+\eqref{eq:proj}\\
=&O_p\left(\zeta(K)\sqrt{\frac{K^3}{N}}\right)+ O_p\left(\zeta(K)K^{-\alpha}\right)+ O(\sqrt{N}K^{-\alpha})+0+ \left\{O_p\left(\zeta(K)\cdot K^{\frac{3}{2}-\alpha}\right)+O_p\left(\zeta(K)\frac{K^{2}}{\sqrt{N}}\right)\right\}+o_p(1)\\
=&o_p(1).
\end{align*}}

\section{Variance Estimation}
\cite{masry1996multivariate}  studies the strong consistency of  kernel regression estimation.  The following conditions are imposed so that Theorem 6 of \cite{masry1996multivariate} applies. Let 
$\boldsymbol{u}=(u_1,...,u_{r+1})$ and $K(\boldsymbol{u})=\prod_{j=1}^{r+1}k(u_j)$.\\

\noindent\textbf{Condition 1}. The kernel $K(\boldsymbol{u})\in L_1$  satisfies $\|\boldsymbol{u}\|K(\boldsymbol{u})\in L_{1}$ and  $\|\boldsymbol{u}\|^2K(\boldsymbol{u})\in L_{1}$.

\noindent\textbf{Condition 2}. The density function $f_{Y,X,T}(y,\boldsymbol{x},t)$ is uniformly bounded away from zero and above, and also it is uniformly continuous on $\mathbb{R}^{r+1}$.

\noindent\textbf{Condition 3}. (a) The kernel $K(\cdot)$ is bounded with compact support; (b) let $H_j(\boldsymbol{u}):=\boldsymbol{u}^jK(\boldsymbol{u})$,  and $|H_j(\boldsymbol{u})-H_j(\boldsymbol{v})|\leq C\|\boldsymbol{u}-\boldsymbol{v}\|$ for all $j$ with $0\leq j\leq 3$.

\noindent\textbf{Condition 4}. The functions $f_{Y,X,T}(y,\boldsymbol{x},t)$, $\mathbb{E}\left[\pi_0(T,\boldsymbol{X})L'(Y-g(T;\boldsymbol{\beta}))|T=t,\boldsymbol{X}=\boldsymbol{x}\right]$, $\mathbb{E}[\pi_0(T,\boldsymbol{X})L'(Y-g(T;\boldsymbol{\beta}))|T=t]$ and $\mathbb{E}\left[\pi_0(T,\boldsymbol{X})L'(Y-g(T;\boldsymbol{\beta}))|\boldsymbol{X}=\boldsymbol{x}\right]$ are twice differentiable, and the derivatives are Lipschitz continuous and uniformly bounded.

\noindent\textbf{Condition 5}. $\mathbb{E}\left[\sup_{\boldsymbol{		\beta}\in \Theta}|L'(Y-g(T;\boldsymbol{\beta}))|^{\sigma}\right]<\infty$ for some $\sigma>2$.

\noindent\textbf{Condition 6}.The bandwidths $h_1\asymp \cdots \asymp h_r \asymp h_Y \asymp h_T\asymp h_N$ go to zero slowly enough such that
\begin{align*}
\frac{N^{1-2/\sigma}h_N^{r+1}}{(\log N)\{(\log N)(\log \log N)^{1+\sigma}\}^{2/\sigma}}\to \infty \ \text{as} \ N\to \infty. 
\end{align*}

\section{Some Extensions}
\subsection{Proof of Theorem 7.1}
	\textbf{(Proof of Consistency)}.		
	Let
	\begin{align*}
	\hat{\gamma}=\left[\sum_{i=1}^Nu_{K_1}(T_i)u_{K_1}(T_i)^{\top}\right]^{-1}\left[\sum_{i=1}^Nu_{K_1}(T_i)\hat{\pi}_K(T_i,\boldsymbol{X}_i)Y_i\right]
	\end{align*}	
	then $\hat{\theta}_K(t)=\hat{\gamma}^{\top}u_{K_1}(t)$. By assumption, there exists $\gamma^*\in\mathbb{R}^{K_1}$ such that
	\begin{align}\label{eq:approximate_error}
	\sup_{t\in\mathcal{T}}\left|\mathbb{E}[\pi_0(T,X)Y|T=t]-(\gamma^*)^{\top}u_{K_1}(t)\right|=O(K_1^{-\tilde{\alpha}}).
	\end{align}
	 We first claim that
	\begin{align}\label{claim:gamma_hat-star}
	\|\hat{\gamma}-\gamma^*\|=O_p\left(\zeta(K)\sqrt{\frac{K}{N}}+\zeta(K)K^{-\alpha}+K_1^{-\tilde{\alpha}}\right)\ ,
	\end{align}
	and the proof will be established later. With the claim \eqref{claim:gamma_hat-star}, we first show that $\int_{\mathcal{T}}|\hat{\theta}_K(t)-\theta(t)|^2dF_T(t)=O_p\left(\frac{\zeta(K)^2K}{N}+\zeta(K)^2K^{-2\alpha}+K_1^{-2\tilde{\alpha}}\right)$. Note that
	\begin{align*}
	&\int_{\mathcal{T}}[\hat{\theta}_K(t)-\theta(t)]^2dF_T(t)\\
	=&\int_{\mathcal{T}}[\hat{\gamma}^{\top}u_{K_1}(t)-(\gamma^*)^{\top}u_{K_1}(t)+(\gamma^*)^{\top}u_{K_1}(t)-\theta(t)]^2dF_T(t)\\
	\leq & 2(\hat{\gamma}-\gamma^*)^{\top}\left[\int_{\mathcal{T}}u_{K_1}(t)u_{K_1}(t)^{\top}dF_T(t)\right](\hat{\gamma}-\gamma^*)+2\int_{\mathcal{T}}[(\gamma^*)^{\top}u_{K_1}(t)-\theta(t)]^2dF_T(t)\\
	\leq &2 \|\hat{\gamma}-\gamma^*\|^2\cdot \lambda_{\max}\left(\mathbb{E}[u_{K_1}(T)u_{K_1}(T)^{\top}]\right)+2\sup_{t\in\mathcal{T}}|(\gamma^*)^{\top}u_{K_1}(t)-\theta(t)|^2\\
	=&O_p\left(\zeta(K)^2\frac{K}{N}+\zeta(K)^2K^{-2\alpha}+K_1^{-2\tilde{\alpha}}\right).
	\end{align*}

	With the claim \eqref{claim:gamma_hat-star}, we next show that $\sup_{t\in\mathcal{T}}|\hat{\theta}_K(t)-\theta(t)|=O_p[\zeta_1(K_1)(\zeta(K)\sqrt{K/N}+\zeta(K)K^{-\alpha}+K_1^{-\alpha})]$. Note that
	\begin{align*}
	&\sup_{t\in\mathcal{T}}|\hat{\theta}_K(t)-\theta(t)|\\
	=&\sup_{t\in\mathcal{T}}\left|\hat{\gamma}^{\top}u_{K_1}(t)-(\gamma^*)^{\top}u_{K_1}(t)+(\gamma^*)^{\top}u_{K_1}(t)-\theta(t)\right|\\
	\leq & \sup_{t\in\mathcal{T}}\|u_{K_1}(t)\|\cdot \left\|\hat{\gamma}-\gamma^*\right\|+\sup_{t\in\mathcal{T}}|(\gamma^*)^{\top}u_{K_1}(t)-\theta(t)|\\
	\leq &\zeta_1(K_1)\cdot \left\{O_p\left(\sup_{(t,x)\in\mathcal{T}\times\mathcal{X}}|\hat{\pi}_K(t,x)-\pi_0(t,x)|\right)+O(K_1^{-\tilde{\alpha}})\right\}+O(K_1^{-\tilde{\alpha}}) \\
	\leq &O_p\left[\zeta_1(K_1)\left(\zeta(K)\sqrt{\frac{K}{N}}+\zeta(K)K^{-\alpha}+K_1^{-\alpha}\right)\right]+O(K_1^{-\tilde{\alpha}})\\
	=&O_p\left[\zeta_1(K_1)\left(\zeta(K)\sqrt{\frac{K}{N}}+\zeta(K)K^{-\alpha}+K_1^{-\tilde{\alpha}}\right)\right].
	\end{align*}
	
	Finally, we turn back to prove the claim \eqref{claim:gamma_hat-star}. Note that
	\begin{align*}
	\hat{\gamma}-\gamma^* =&\left[\sum_{i=1}^Nu_{K_1}(T_i)u_{K_1}(T_i)^{\top}\right]^{-1}\left[\sum_{i=1}^N\hat{\pi}_K(T_i,\boldsymbol{X}_i)u_{K_1}(T_i)Y_i\right]-\gamma^*\\
	=&\left[\sum_{i=1}^Nu_{K_1}(T_i)u_{K_1}(T_i)^{\top}\right]^{-1}\left[\sum_{i=1}^Nu_{K_1}(T_i)\left\{\hat{\pi}_K(T_i,\boldsymbol{X}_i)-{\pi}_0(T_i,\boldsymbol{X}_i)\right\}Y_i\right]\\
	&+\left[\sum_{i=1}^Nu_{K_1}(T_i)u_{K_1}(T_i)^{\top}\right]^{-1}\left[\sum_{i=1}^Nu_{K_1}(T_i)\{\pi_0(T_i,\boldsymbol{X}_i)Y_i-\mathbb{E}[\pi_0(T_i,\boldsymbol{X}_i)Y_i|T_i]\}\right]\\
	&+\left[\sum_{i=1}^Nu_{K_1}(T_i)u_{K_1}(T_i)^{\top}\right]^{-1}\left[\sum_{i=1}^Nu_{K_1}(T_i)\left\{\mathbb{E}[{\pi}_0(T_i,\boldsymbol{X}_i)Y_i|T_i]-(\gamma^*)^{\top}u_{K_1}(T_i)\right\}\right]\\
	=&A_{1N}+A_{2N}+A_{3N}
	\end{align*}
	where
	\begin{align*}
	&A_{1N}=\left[\sum_{i=1}^Nu_{K_1}(T_i)u_{K_1}(T_i)^{\top}\right]^{-1}\left[\sum_{i=1}^Nu_{K_1}(T_i)\left\{\hat{\pi}_K(T_i,\boldsymbol{X}_i)-{\pi}_0(T_i,\boldsymbol{X}_i)\right\}Y_i\right] , \\
	&A_{2N}=\left[\sum_{i=1}^Nu_{K_1}(T_i)u_{K_1}(T_i)^{\top}\right]^{-1}\left[\sum_{i=1}^Nu_{K_1}(T_i)\{\pi_0(T_i,\boldsymbol{X}_i)Y_i-\mathbb{E}[\pi_0(T_i,\boldsymbol{X}_i)Y_i|T_i]\}\right] ,\\
	&A_{3N}=\left[\sum_{i=1}^Nu_{K_1}(T_i)u_{K_1}(T_i)^{\top}\right]^{-1}\left[\sum_{i=1}^Nu_{K_1}(T_i)\left\{\mathbb{E}[{\pi}_0(T_i,\boldsymbol{X}_i)Y_i|T_i]-(\gamma^*)^{\top}u_{K_1}(T_i)\right\}\right] .
	\end{align*}
	We first compute the probability order of $A_{1N}$. We use the following notation:
	\begin{align*}
	&\hat{H}_N:=\left(\left\{\hat{\pi}_K(T_1,X_1)-{\pi}_0(T_1,X_1)\right\}Y_1,...,\left\{\hat{\pi}_K(T_N,X_N)-{\pi}_0(T_N,X_N)\right\}Y_N\right)^\top \ ,\\
	&U_{N\times K_1}:=(u_{K_1}(T_1),...,u_{K_1}(T_N))^\top\ , \\
	&\hat{\Phi}_{K_1\times K_1}:=\frac{1}{N}\sum_{i=1}^Nu_{K_1}(T)u_{K_1}^{\top}(T).
	\end{align*} 
	Then we can obtain that
	\begin{align*}
	\|A_{1N}\|^2=&\left\|\left[\sum_{i=1}^Nu_{K_1}(T_i)u_{K_1}(T_i)^{\top}\right]^{-1}\left[\sum_{i=1}^Nu_{K_1}(T_i)\left\{\hat{\pi}_K(T_i,\boldsymbol{X}_i)-{\pi}_0(T_i,\boldsymbol{X}_i)\right\}Y_i\right]\right\|^2\\
	=& N^{-2}\text{tr}\left(\hat{\Phi}_{K_1\times K_1}^{-1}U_{N\times K_1}^{\top}\hat{H}_N\hat{H}_N^{\top}U_{N\times K_1}\hat{\Phi}_{K_1\times K_1}^{-1}\right)\\
	=& N^{-2}\text{tr}\left(U_{N\times K_1}^{\top}\hat{H}_N\hat{H}_N^{\top}U_{N\times K_1}\hat{\Phi}_{K_1\times K_1}^{-1}\hat{\Phi}_{K_1\times K_1}^{-1}\right)\\
	=& N^{-2}\text{tr}\left(\hat{\Phi}_{K_1\times K_1}^{-1/2}U_{N\times K_1}^{\top}\hat{H}_N\hat{H}_N^{\top}U_{N\times K_1}\hat{\Phi}_{K_1\times K_1}^{-1/2}\hat{\Phi}_{K_1\times K_1}^{-1}\right)\\
	\leq &\lambda_{\max}(\hat{\Phi}_{K_1\times K_1}^{-1})N^{-2}\text{tr}\left(\hat{\Phi}_{K_1\times K_1}^{-1/2}U_{N\times K_1}^{\top}\hat{H}_N\hat{H}_N^{\top}U_{N\times K_1}\hat{\Phi}_{K_1\times K_1}^{-1/2}\right)\\
	=&\lambda_{\max}(\hat{\Phi}_{K_1\times K_1}^{-1})N^{-1}\text{tr}\left(\hat{H}_N\hat{H}_N^{\top}U_{N\times K_1}(U^{\top}_{N\times K_1}U_{N\times K_1})^{-1}U_{N\times K_1}^{\top}\right)\\
	\leq &[\lambda_{\min}(\hat{\Phi}_{K_1\times K_1})]^{-1}N^{-1}\|\hat{H}_N\|^2\\
	=&[\lambda_{\min}(\hat{\Phi}_{K_1\times K_1})]^{-1}\cdot \frac{1}{N}\sum_{i=1}^N\left\{\hat{\pi}_K(T_i,\boldsymbol{X}_i)-{\pi}_0(T_i,\boldsymbol{X}_i)\right\}^2Y_i^2\\
	\leq &[\lambda_{\min}(\hat{\Phi}_{K_1\times K_1})]^{-1}\sup_{(t,x)\in\mathcal{T}\times \mathcal{X}}|\hat{\pi}_K(t,x)-{\pi}_0(t,x)|^2\cdot \frac{1}{N}\sum_{i=1}^NY_i^2\\
	\leq & O_p(1)\cdot O_p\left(\zeta(K)^2K^{-2\alpha}+\frac{\zeta(K)^2K}{N}\right) \cdot O_p(1)\\
	=&O_p\left(\zeta(K)^2K^{-2\alpha}+\frac{\zeta(K)^2K}{N}\right),
	\end{align*}
	where the first inequality follows from the fact that $tr(AB)\leq \lambda_{\max}(B)tr(A)$ for any symmetric matrix $B$ and positive semidefinite matrix $A$, 
	the second inequality follows from the same fact and the fact that $U_{N\times K_1}(U^{\top}_{N\times K_1}U_{N\times K_1})^{-1}U_{N\times K_1}^{\top}$ is a projection matrix with maximum eigenvalue 1, 
	and the fourth inequality follows from the facts that $|\lambda_{\min}(\hat{\Phi}_{K_1\times K_1})|^{-1}=O_p(1)$, 
	$\sup_{(t,x)\in\mathcal{T}\times \mathcal{X}}|\hat{\pi}_K(t,x)-{\pi}_0(t,x)|=O_p\left(\zeta(K)K^{-\alpha}+\zeta(K)\sqrt{K/N}\right)$ and $N^{-1}\sum_{i=1}^NY_i^2=O_p(1)$. \\
	
	Next, we compute the probability order of $A_{2N}$.  We can deduce that
	\begin{align*}
	\|A_{2N}\|^2=&\left\|\left[\sum_{i=1}^Nu_{K_1}(T_i)u_{K_1}(T_i)^{\top}\right]^{-1}\left[\sum_{i=1}^Nu_{K_1}(T_i)\varepsilon_i\right]\right\|^2\\
	=&N^{-2}\text{tr}\left(\hat{\Phi}_{K_1\times K_1}^{-1}U_{N\times K_1}^{\top}\mathcal{E}_N\mathcal{E}_N^{\top}U_{N\times K_1}\hat{\Phi}_{K_1\times K_1}^{-1}\right)\\
	=&N^{-2}\text{tr}\left(U_{N\times K_1}^{\top}\mathcal{E}_N\mathcal{E}_N^{\top}U_{N\times K_1}\hat{\Phi}_{K_1\times K_1}^{-1}\hat{\Phi}_{K_1\times K_1}^{-1}\right)\\
	\leq &[\lambda_{\min}(\hat{\Phi}_{K_1\times K_1})]^{-2}N^{-2}\|U_{N\times K_1}^{\top}\mathcal{E}_N\|^2=O_p\left(\frac{K_1}{N}\right)\ ,
	\end{align*}
	where the last equality follows that $|\lambda_{\min}(\hat{\Phi}_{K_1\times K_1})|^{-1}=O_p(1)$ and $N^{-2}\|U_{N\times K_1}^{\top}\mathcal{E}_N\|^2=O_p(K_1/N)$ by Markov's inequality. \\
	
	We finally compute the probability order of $A_{3N}$. We define the notation
	\begin{align*}
	R_N(\gamma^*)=\left(\left\{\mathbb{E}[{\pi}_0(T_1,X_1)Y_1|T_1]-(\gamma^*)^{\top}u_{K_1}(T_1)\right\},...,\left\{\mathbb{E}[{\pi}_0(T_N,X_N)Y_N|T_N]-(\gamma^*)^{\top}u_{K_1}(T_N)\right\}\right)^{\top}\ ,
	\end{align*}
	then it follows that with probability approaching to 1,
	\begin{align*}
	\|A_{3N}\|^2=&\left\|\left[\sum_{i=1}^Nu_{K_1}(T_i)u_{K_1}(T_i)^{\top}\right]^{-1}\left[\sum_{i=1}^Nu_{K_1}(T_i)\left\{\mathbb{E}[{\pi}_0(T_i,\boldsymbol{X}_i)Y_i|T_i]-(\gamma^*)^{\top}u_{K_1}(T_i)\right\}\right]\right\|^2\\
	=&N^{-2}\left\|\hat{\Phi}_{K_1\times K_1}^{-1}U_{N\times K_1}^{\top}R_N(\gamma^*)\right\|^2\\
	=& N^{-2}\text{tr}\left(\hat{\Phi}_{K_1\times K_1}^{-1}U_{N\times K_1}^{\top}R_N(\gamma^*)R_N(\gamma^*)^{\top}U_{N\times K_1}\hat{\Phi}_{K_1\times K_1}^{-1}\right)\\
	=& N^{-2}\text{tr}\left(U_{N\times K_1}^{\top}R_N(\gamma^*)R_N(\gamma^*)^{\top}U_{N\times K_1}\hat{\Phi}_{K_1\times K_1}^{-1}\hat{\Phi}_{K_1\times K_1}^{-1}\right)\\
	=& N^{-2}\text{tr}\left(\hat{\Phi}_{K_1\times K_1}^{-1/2}U_{N\times K_1}^{\top}R_N(\gamma^*)R_N(\gamma^*)^{\top}U_{N\times K_1}\hat{\Phi}_{K_1\times K_1}^{-1/2}\hat{\Phi}_{K_1\times K_1}^{-1}\right)\\
	\leq &\lambda_{\max}(\hat{\Phi}_{K_1\times K_1}^{-1})N^{-2}\text{tr}\left(\hat{\Phi}_{K_1\times K_1}^{-1/2}U_{N\times K_1}^{\top}R_N(\gamma^*)R_N(\gamma^*)^{\top}U_{N\times K_1}\hat{\Phi}_{K_1\times K_1}^{-1/2}\right)\\
	=&\lambda_{\max}(\hat{\Phi}_{K_1\times K_1}^{-1})N^{-1}\text{tr}\left(R_N(\gamma^*)R_N(\gamma^*)^{\top}U_{N\times K_1}(U^{\top}_{N\times K_1}U_{N\times K_1})^{-1}U_{N\times K_1}^{\top}\right)\\
	\leq &[\lambda_{\min}(\hat{\Phi}_{K_1\times K_1})]^{-1}N^{-1}\|R_N(\gamma^*)\|^2\\
	=&[\lambda_{\min}(\hat{\Phi}_{K_1\times K_1})]^{-1}\cdot \frac{1}{N}\sum_{i=1}^N\left\{\mathbb{E}[{\pi}_0(T_i,\boldsymbol{X}_i)Y_i|T_i]-(\gamma^*)^{\top}u_{K_1}(T_i)\right\}^2=O_p(K_1^{-2\tilde{\alpha}}),
	\end{align*}
	where the first inequality follows from the fact that $tr(AB)\leq \lambda_{\max}(B)tr(A)$ for any symmetric matrix $B$ and positive semidefinite matrix $A$, the second inequality follows from the same fact and the fact that $U_{N\times K_1}(U^{\top}_{N\times K_1}U_{N\times K_1})^{-1}U_{N\times K_1}^{\top}$ is a projection matrix with maximum eigenvalue 1, and the last equality follows from the fact that $|\lambda_{\min}(\hat{\Phi}_{K_1\times K_1})|^{-1}=O_p(1)$ and the fact that $\frac{1}{N}\sum_{i=1}^N\left\{\mathbb{E}[{\pi}_0(T_i,\boldsymbol{X}_i)Y_i|T_i]-(\gamma^*)^{\top}u_{K_1}(T_i)\right\}^2\leq \sup_{t\in\mathcal{T}}|\mathbb{E}[{\pi}_0(T,X)Y|T]-(\gamma^*)^{\top}u_{K_1}(t)|^2=O(K_1^{-2\tilde{\alpha}})$. Thus we complete the proof of \eqref{claim:gamma_hat-star}.\\
	
	\textbf{(Proof of Asymptotic Normality)}. We have the following decomposition for $\hat{\theta}(t)-\theta(t)$:	
	\begin{align*}
	&\hat{\theta}_K(t)-\theta(t)\\
	=&u_{K_1}(t)^{\top}(\hat{\gamma}-\gamma^*)+[(\gamma^*)^{\top}u_{K_1}(t)-\theta(t)]\\
	=&u_{K_1}(t)^{\top}\left[\frac{1}{N}\sum_{i=1}^Nu_{K_1}(T_i)u_{K_1}(T_i)^{\top}\right]^{-1}\left[\frac{1}{N}\sum_{i=1}^Nu_{K_1}(T_i)\bigg\{\hat{\pi}_K(T_i,\boldsymbol{X}_i)Y_i-\mathbb{E}\left[{\pi}_0(T_i,\boldsymbol{X}_i)Y_i|T_i\right]\bigg\}\right]\\
	+&u_{K_1}(t)^{\top}\left[\frac{1}{N}\sum_{i=1}^Nu_{K_1}(T_i)u_{K_1}(T_i)^{\top}\right]^{-1}\left[\frac{1}{N}\sum_{i=1}^Nu_{K_1}(T_i)\cdot \bigg\{\mathbb{E}\left[{\pi}_0(T_i,\boldsymbol{X}_i)Y_i|T_i\right]-(\gamma^*)^{\top}u_{K_1}(T_i)\bigg\}\right]\\
	&+\bigg[(\gamma^*)^{\top}u_{K_1}(t)-\theta(t)\bigg]\\
	=&b_{1N}(t)+b_{2N}(t)+b_{3N}(t)\ ,
	\end{align*}
	where 
	\begin{align*}
	&b_{1N}(t)=u_{K_1}(t)^{\top}\left[\frac{1}{N}\sum_{i=1}^Nu_{K_1}(T_i)u_{K_1}(T_i)^{\top}\right]^{-1}\left[\frac{1}{N}\sum_{i=1}^Nu_{K_1}(T_i)\bigg\{\hat{\pi}_K(T_i,\boldsymbol{X}_i)Y_i-\mathbb{E}\left[{\pi}_0(T_i,\boldsymbol{X}_i)Y_i|T_i\right]\bigg\}\right]\\
	& b_{2N}(t)=u_{K_1}(t)^{\top}\left[\frac{1}{N}\sum_{i=1}^Nu_{K_1}(T_i)u_{K_1}(T_i)^{\top}\right]^{-1}\left[\frac{1}{N}\sum_{i=1}^Nu_{K_1}(T_i)\cdot \bigg\{\mathbb{E}\left[{\pi}_0(T_i,\boldsymbol{X}_i)Y_i|T_i\right]-(\gamma^*)^{\top}u_{K_1}(T_i)\bigg\}\right]\\
	&b_{3N}(t)=(\gamma^*)^{\top}u_{K_1}(t)-\theta(t).
	\end{align*}
	Then we have that
	\begin{align*}
	&\sqrt{N}{V}_K(t)^{-1/2}\left[\hat{\theta}_K(t)-\theta(t)\right]\\
	=&\sqrt{N}{V}_K(t)^{-1/2}b_{1N}(t)+\sqrt{N}{V}_K(t)^{-1/2}b_{2N}(t)+\sqrt{N}{V}_K(t)^{-1/2}b_{3N}(t).
	\end{align*} 
	We shall show that $b_{1N}(t)$ contributes to the asymptotic variance; and  $b_{2N}(t)+b_{3N}(t)$ contributes to the asymptotic bias which is asymptotically negligible. Thus to complete the proof of asymptotic normality, it is sufficient to prove the following results:
	\begin{itemize}
		\item[(i)] $V_K\geq c \|u_{K_1}(t)\|^2$ for some $c>0$;
		\item[(ii)] $\sqrt{N}V_K^{-1/2}b_{1N}(t)\xrightarrow{d}N(0,1)$;
		\item[(iii)] $\sqrt{N}V_K^{-1/2}b_{2N}(t)=o_p(1)$;
		\item[(iv)] $\sqrt{N}V_K^{-1/2}b_{3N}(t)=o_p(1)$.
	\end{itemize}
	
	We first prove  Result (i). Note that $\lambda_{\min}(\Sigma_{K_1\times K_1})\geq \underline{c}_{\sigma^2}\lambda_{\min}(\Phi_{K_1\times K_1})=\underline{c}_{\sigma^2}\lambda_{\min}(I_{K_1\times K_1})\geq \underline{c}_{\sigma^2}$, we can have
	\begin{align*}
	V_K=&u_{K_1}^{\top}(t) \Phi_{K_1\times K_1}^{-1}\Sigma_{K_1\times K_1}\Phi_{K_1\times K_1}^{-1}u_{K_1}(t)\\
	\geq & \lambda_{\min}(\Sigma_{K_1\times K_1}) u_{K_1}^{\top}(t) \Phi_{K_1\times K_1}^{-1}\Phi_{K_1\times K_1}^{-1}u_{K_1}(t)\\
	\geq  & \underline{c}_{\sigma^2} \|u_{K_1}(t)\|^2.
	\end{align*}

	For the claim (ii). Let
	\begin{align*}
	&\tilde{b}_{1N}(t)=u_{K_1}(t)^{\top}\left[\frac{1}{N}\sum_{i=1}^Nu_{K_1}(T_i)u_{K_1}(T_i)^{\top}\right]^{-1}\\
	&\quad\times\Bigg[\frac{1}{N}\sum_{i=1}^Nu_{K_1}(T_i)\bigg\{{\pi}_0(T_i,\boldsymbol{X}_i)Y_i-\mathbb{E}\left[{\pi}_0(T_i,\boldsymbol{X}_i)Y_i|T_i,\boldsymbol{X}_i\right]+\mathbb{E}\left[{\pi}_0(T_i,\boldsymbol{X}_i)Y_i|\boldsymbol{X}_i\right]-\mathbb{E}\left[{\pi}_0(T_i,\boldsymbol{X}_i)Y_i\right]\bigg\}\Bigg]\ .
	\end{align*}
	Similar to the proof of Lemma \ref{asy:equivalent}, we can have that 
	\begin{align*}
 \sqrt{N}{V}_K(t)^{-1/2}\cdot ({b}_{1N}(t)-\tilde{b}_{1N}(t))=o_p(1)\ .
	\end{align*}
 Then
	\begin{align*}
	&\sqrt{N}{V}_K(t)^{-1/2}b_{1N}(t)\\
	=&\sqrt{N}{V}_K(t)^{-1/2}\tilde{b}_{1N}(t)+o_p(1)\\
	=&\sqrt{N}{V}_K(t)^{-1/2}u_{K_1}(t)^{\top}\hat{\Phi}_{K_1\times K_1}^{-1}N^{-1}U_{N\times K_1}^{\top}\mathcal{E}_N\\
	=&\sqrt{N}{V}_K(t)^{-1/2}u_{K_1}(t)^{\top}{\Phi}_{K_1\times K_1}^{-1}N^{-1}U_{N\times K_1}^{\top}\mathcal{E}_N+\sqrt{N}{V}_K(t)^{-1/2}u_{K_1}(t)^{\top}[\hat{\Phi}_{K_1\times K_1}^{-1}-{\Phi}_{K_1\times K_1}^{-1}]N^{-1}U_{N\times K_1}^{\top}\mathcal{E}_N\\
	=&B_{1N,1}(t)+B_{1N,2}(t)\ .
	\end{align*}
	For $B_{1N,1}(t)$, we can simply apply the Liapounov CLT and show that $B_{1N,1}(t)\xrightarrow{d}N(0,1)$. For $B_{1N,2}(t)$, let $\boldsymbol{T}=(T_1,...,T_N)$, we can obtain that
	\begin{align*}
	&\mathbb{E}\left[B_{1N,2}(t)^2|\boldsymbol{T}\right]\\
	=&N^{-1}{V}_K(t)u_{K_1}(t)^{\top}[\hat{\Phi}_{K_1\times K_1}^{-1}-{\Phi}_{K_1\times K_1}^{-1}]\cdot \mathbb{E}\left[U_{N\times K_1}^{\top}\mathcal{E}_N\mathcal{E}_N^{\top}U_{N\times K_1}|\boldsymbol{T}\right]\cdot [\hat{\Phi}_{K_1\times K_1}^{-1}-{\Phi}_{K_1\times K_1}^{-1}]u_{K_1}(t)\\
	=&\lambda_{\max}\left(N^{-1} \mathbb{E}\left[U_{N\times K_1}^{\top}\mathcal{E}_N\mathcal{E}_N^{\top}U_{N\times K_1}|\boldsymbol{T}\right]\right){V}_K(t)u_{K_1}(t)^{\top}[\hat{\Phi}_{K_1\times K_1}^{-1}-{\Phi}_{K_1\times K_1}^{-1}]\cdot [\hat{\Phi}_{K_1\times K_1}^{-1}-{\Phi}_{K_1\times K_1}^{-1}]u_{K_1}(t)\\
	\leq &\lambda_{\max}\left(N^{-1} \mathbb{E}\left[U_{N\times K_1}^{\top}\mathcal{E}_N\mathcal{E}_N^{\top}U_{N\times K_1}|\boldsymbol{T}\right]\right)\left\{V_{K}^{-1}\|u_{K_1}(t)\|^2\right\}\cdot \|\hat{\Phi}_{K_1\times K_1}^{-1}-{\Phi}_{K_1\times K_1}^{-1}\|^2\\
	=&O_p(1)O_p(1)o_p(1)=o_p(1)
	\end{align*}
	where we use the fact that $N^{-1} \mathbb{E}\left[U_{N\times K_1}^{\top}\mathcal{E}_N\mathcal{E}_N^{\top}U_{N\times K_1}|\boldsymbol{T}\right]=N^{-1}\sum_{i=1}^Nu_{K_1}(T_i)u_{K_1}(T_i)^{\top}\sigma^2(T_i)$ has bounded maximum eigenvalue. Therefore, $B_{1N,1}(t)=o_p(1)$ by the conditional Chebyshev's inequality. Thus (ii)  holds. \\
	
	For (iii), by Cauchy-Schwarz's inequality, we can obtain that
	\begin{align*}
	&\sqrt{N}V_{K}^{-1/2}|b_{2N}(t)|\\
	=&N^{-1/2}V_{K}^{-1/2}\left|u_{K_1}(t)^{\top}\hat{\Phi}_{K_1\times K_1}^{-1}U^{\top}_{ N\times K_1}R_N(\gamma^*)\right|\\
	\leq &V_{K}^{-1/2}\left\{u_{K_1}(t)^{\top}\hat{\Phi}_{K_1\times K_1}^{-1}\left(N^{-1}U^{\top}_{N\times K_1}U_{N\times K_1}\right)\hat{\Phi}_{K_1\times K_1}^{-1}u_{K_1}(t)\right\}^{\frac{1}{2}} \left\{R_N(\gamma^*)^{\top}R_N(\gamma^*)\right\}^{\frac{1}{2}}\\
	\leq &V_{K}^{-1/2}\left\{u_{K_1}(t)^{\top}\hat{\Phi}_{K_1\times K_1}^{-1}u_{K_1}(t)\right\}^{\frac{1}{2}} \left\{R_N(\gamma^*)^{\top}R_N(\gamma^*)\right\}^{\frac{1}{2}}\\
	\leq & \{V_{K}^{-1/2}\|u_{K_1}(t)\|\}\cdot |\lambda_{\max}(\Phi_{K_1\times K_1}^{-1})|^{\frac{1}{2}}\cdot O(N^{\frac{1}{2}}\cdot K_1^{-\tilde{\alpha}})\\
	=& O(1)\cdot O_p(1)\cdot o_p(1)=o_p(1)\ .
	\end{align*}
	Similarly, we can show show that $\sqrt{N}V_{K}^{-1/2}|b_{3N}(t)|=o_p(1)$. This completes the proof of the Theorem.

	\subsection{Proof Theorem 7.2}
	Similar to the proof of Lemma \ref{asy:equivalent}, let $\mu(t,x)=\mathbb{E}[Y|T=t,X=x]$,	we decompose  $\sqrt{N}(\hat{\psi}_K-\psi)$ as follows:
	\begin{align}
	&\sqrt{N}(\hat{\psi}_K-\psi)\notag\\
	=&\frac{1}{\sqrt{N}}\sum_{i=1}^N\left\{\hat{\pi}_K(T_i,\boldsymbol{X}_i)Y_i-\psi\right\}\notag\\
	=&\label{eq:WK_psi} \frac {1} {\sqrt{N}} \sum_{i=1}^N \Bigg\{ \left(\hat{\pi}_K(T_i,\boldsymbol{X}_i) - \pi_K^*(T_i,\boldsymbol{X}_i)\right)Y_i  -\int_{\mathcal{T}}\int_{\mathcal{X}} \left(\hat{\pi}_K(t,\boldsymbol{x}) - \pi_K^*(t,\boldsymbol{x})\right)\mu(\boldsymbol{x},t) dF_{X,T}(\boldsymbol{x},t)\Bigg\}  \\[2mm]
	\label{eq:VK_psi} &~ + \frac {1} {\sqrt{N}} \sum_{i=1}^N \Bigg\{\left(\pi_K^*(T_i,\boldsymbol{X}_i) - {\pi_0(T_i,\boldsymbol{X}_i)}\right)Y_i -   \int_{\mathcal{T}}\int_{\mathcal{X}} \mu(t,\boldsymbol{x})\left(\pi_K^*(t,\boldsymbol{x}) - {\pi(t,\boldsymbol{x})}\right) dF_{X,T}(\boldsymbol{x},t) \Bigg\} \\
	\label{eq:Lemma1_psi}&~ + \sqrt{N}\int_{\mathcal{T}}\int_{\mathcal{X}} \mu(t,\boldsymbol{x})\left(\pi_K^*(t,\boldsymbol{x}) - {\pi_0(t,\boldsymbol{x})}\right) dF_{X,T}(\boldsymbol{x},t)  \\[2mm]
	&~ \label{eq:tauto_psi}+  {\sqrt{N}}\int_{\mathcal{T}}\int_{\mathcal{X}} \left(\hat{\pi}_K(t,\boldsymbol{x}) - \pi_K^*(t,\boldsymbol{x})\right)\mu(\boldsymbol{x},t)dF_{X,T}(\boldsymbol{x},t) \\
	& \qquad  - \sqrt{N}\int_{\mathcal{T}}\int_{\mathcal{X}} \mu(t,\boldsymbol{x})\rho''\left(u_{K_1}^{\top}(t)\tilde{\Lambda}_{K_1\times K_2}v_{K_2}(\boldsymbol{x})\right)u_{K_1}^{\top}(t)\hat{A}_{K_1\times K_2}v_{K_2}(\boldsymbol{x}) m(t;\boldsymbol{\beta}_0) dF_{X,T}(\boldsymbol{x},t) \notag \\[4mm]
	& \label{eq:Q_psi} +\sqrt{N} \int_{\mathcal{T}}\int_{\mathcal{X}} \varepsilon(t,\boldsymbol{x};\boldsymbol{\beta}_0)\rho''\left(u_{K_1}^{\top}(t)\tilde{\Lambda}_{K_1\times K_2}v_{K_2}(\boldsymbol{x})\right)u_{K_1}^{\top}(t)\hat{A}_{K_1\times K_2}v_{K_2}(\boldsymbol{x}) dF_{X,T}(\boldsymbol{x},t)\\ 
	&\qquad - \sqrt{N} \int_{\mathcal{T}}\int_{\mathcal{X}} \mu(t,\boldsymbol{x})\rho''\left(u_{K_1}^{\top}(t){\Lambda}^*_{K_1\times K_2}v_{K_2}(\boldsymbol{x})\right)u_{K_1}^{\top}(t)A^*_{K_1\times K_2}v_{K_2}(\boldsymbol{x})  dF_{X,T}(\boldsymbol{x},t)\notag \\[4mm]
	& \label{eq:proj_psi}+{\sqrt{N}}\int_{\mathcal{X}} \mu(t,\boldsymbol{x})\rho''\left(u_{K_1}^{\top}(t){\Lambda}^*_{K_1\times K_2}v_{K_2}(\boldsymbol{x})\right)u_{K_1}^{\top}(t)A^*_{K_1\times K_2}v_{K_2}(\boldsymbol{x}) dF_{X,T}(\boldsymbol{x},t)\\
	&\qquad  +\frac{1}{\sqrt{N}}\sum_{i=1}^N \bigg\{ \pi_0(T_i,\boldsymbol{X}_i)\mu(T_i,\boldsymbol{X}_i)-\mathbb{E}\left[\pi_0(T,\boldsymbol{X})\mu(T,\boldsymbol{X})|\boldsymbol{X}=\boldsymbol{X}_i\right]\notag\\
	&\qquad\qquad\qquad\qquad  -\mathbb{E}\left[\pi_0(T,\boldsymbol{X})\mu(T,\boldsymbol{X})|T=T_i\right]+\mathbb{E}\left[\pi_0(T,\boldsymbol{X})\mu(T,\boldsymbol{X})\right]\bigg\} \notag \\[4mm]
	&~ + \frac {1} {\sqrt{N}} \sum_{i=1}^N \bigg\{ \pi_0(T_i,\boldsymbol{X}_i)Y_i-\pi_0(T_i,\boldsymbol{X}_i)\mu(T_i,\boldsymbol{X}_i)+\mathbb{E}\left[\pi_0(T,\boldsymbol{X})\mu(T,\boldsymbol{X})|T=T_i\right]\label{eq:Normal_psi}\\
	&  \quad \quad \quad  \quad \quad \quad \quad+\mathbb{E}\left[\pi_0(T,\boldsymbol{X})\mu(T,\boldsymbol{X})|\boldsymbol{X}=\boldsymbol{X}_i\right]-\mathbb{E}\left[\pi_0(T,\boldsymbol{X})\mu(T,\boldsymbol{X})\right]-\psi \bigg\}\ .\notag
	\end{align}
	Using the similar argument for showing that \eqref{eq:WK}-\eqref{eq:proj} are all $o_p(1)$ in the proof of Lemma \ref{asy:equivalent}, we can obtain that the terms \eqref{eq:WK_psi}-\eqref{eq:proj_psi} are all $o_p(1)$, while the term \eqref{eq:Normal_psi} is  asymptotically normal and attains the efficiency bound.

\singlespacing
\bibliographystyle{econometrica}
\bibliography{Semiparametric}